\documentclass[12pt,a4paper]{book}
\pdfoutput=1
\usepackage[utf8]{inputenc}
\usepackage{graphicx}
\usepackage{geometry}
\usepackage{amssymb}
\usepackage{amsfonts}
\usepackage{amsmath}
\usepackage{perpage}
\usepackage{epigraph}
\usepackage[titletoc,title]{appendix}
\usepackage{amsthm}
\usepackage{setspace}

\usepackage{hyperref}

\theoremstyle{definition}

\newtheorem*{remark*}{Remark}
\newtheorem{claim}{Claim}

\newtheorem*{definition}{Definition}

\newcommand*{\maketoc}{
  \ifdefined\pdfbookmark
     \phantomsection
     \pdfbookmark[1]{\contentsname}{Contents}
  \else
  \fi

  \tableofcontents
  \clearpage
}

\newcommand{\maketitlepage}{{
  \thispagestyle{empty}
  \vspace*{0in}
  \begin{center}
    \LARGE The Entropic Dynamics of Relativistic Quantum Fields in Curved Space-time
  \end{center}
  \vspace{.6in}
  \vspace{.6in}
  \begin{center}
    Selman Ipek
  \end{center}
  \vspace{.6in}
  \vspace{.6in}
  \begin{center}
 
    A Dissertation \\
Submitted to the University at Albany, State University of New York \\
    in Partial Fulfillment of \\
    the Requirements for the Degree of \\
    Doctor of Philosophy
  \end{center}
  \vspace{.3in}
  \vspace{.3in}
  \begin{center}
 
    College of Arts and Sciences\\
    Department of Physics\\
    2021\\
  \end{center}
  \vspace{.3in}
  \vspace{.3in}
  \clearpage
  }}

\newcommand*{\frontnumber}{
  \pagenumbering{roman}
  }
\newcommand*{\mainnumber}{\pagenumbering{arabic}}

\newcommand*{\makeabstract}{
  \newpage
  \addcontentsline{toc}{section}{Abstract}
  \begin{center}
  \Large \textbf{Abstract}
  \end{center}
It has often been the case in history that the laws of physics have been used as the framework for understanding and implementing information processing. The tacit assumption is that the laws of physics are fundamental and that the notion of information is derived from these laws. Here we take the opposite view: the laws of physics are an application of the rules for processing information. The inferential framework of entropic dynamics (ED) has previously been developed by A. Caticha for the purposes of understanding and deriving quantum theory, in much the same way that E.T. Jaynes used the MaxEnt approach to derive the formalism of statistical mechanics. In this thesis we apply the ED framework to construct a quantum dynamics for scalar fields in space-time.

Our work progresses in stages. We begin by considering a toy model consisting of many interacting particles, resulting in the familiar Schr\"{o}dinger equation for non-relativistic particles. Using a similar methodology, we construct a theory of quantum scalar fields in flat space-time that is relativistic, but not manifestly so. Here we also discuss a novel way in which the ED of quantum scalar fields appears to evade the so-called Wallstrom objection. To go further towards constructing a manifestly covariant quantum ED of fields on a curved space-time, both fixed and dynamical, we borrow from the ``many-time" approaches of P. Weiss, P. Dirac, K. Kucha\v{r}, and C. Teitelboim. For a fixed background the result is a manifestly covariant ED of scalar fields that is in the spirit of the covariant quantum theories proposed by S. Tomonaga and J. Schwinger. However, the formalism is sufficiently flexible so as to allow for the possibility of modeling the back reaction of the quantum matter fields on a fully dynamical classical background. The simplest realization of this classical-quantum interaction shares some formal similarity to semi-classical gravity models, and the semi-classical Einstein equations, in particular. We consider such a theory and discuss its plausibility as a candidate for a quantum gravity theory.

  \clearpage
  }

\def\makeacknowledgements{
  \ifx\@acknowledgements\undefined
  \else
    \ifdefined\phantomsection
     \phantomsection
    \else
    \fi

    \addcontentsline{toc}{section}{Acknowledgments}
    \begin{center}
      \Large \textbf{Acknowledgments}
    \end{center}
  There are many people without whom I would not be in a position to complete this doctorate. The first, and foremost, are my family and friends, whose \textit{pestering} love and friendship may have caused this thesis to take longer than necessary, but without whom it would have never been started, nor completed. I would especially like to thank those who took the daily brunt of the emotional roller-coaster that we call a doctoral candidacy --- my parents, Havzi and Nadide, my grandmother Munevver, and my grandmother Emine, may she rest in peace, but also my sister Sevim, brother-in-law Mark, my vibrant, but rambunctious, nephew Armin, as well as the newest member of our family, my niece Aleyna Nur. I would also like to further thank my friends and family for their patience and understanding during these past several years; I apologize for the number of times I had to turn down your thoughtful and friendly advances so I could pursue my own studies. Perhaps most of all, however, I would like to thank my parents specifically. It is only because of their support, encouragement, and personal sacrifice that I have had the privilege to pursue the things I want. I could never repay their kindness and generosity.

My deepest thanks also go to my advisor, Ariel Caticha. It was his contagious enthusiasm that excited me to pursue physics as this level in the first place, and it was his steadfast support and encouragement that helped me to finish. His basic approach to physics and beyond has made an indelible mark on me as a physicist and person --- it's been a great ride.

Finally, if it takes a village to raise a child, then certainly it takes an entire department to graduate a doctoral candidate. I would like to kindly thank the entire physics department for their support --- personally, academically, monetarily, logistically, and in most ways imaginable. This includes everyone from the staff throughout the years --- Paul, Leslie, Sandra, Ben, David, and others --- to the professors --- especially professors Earle, Knuth, Ernst, and Lunin from whom I have learned so much --- and to my fellow students and colleagues, as well. I would also like to thank my defense committee for their kind interest in my work; you have volunteered your time for my personal advancement, and for this I am very appreciative. A special thanks is also extended to my research group, including Ahmad, Daniel, Felipe, Kevin, Mohammad, Nick, Pedro, Shahid, and many more, with whom I have had many enlightening conversations on physics, inference, and much more. It's been such a pleasure, thank you to all.

    \clearpage
  \fi
  }
  
\def\makededication{
  \ifx\@dedication\undefined
  \else
    \vspace*{1.5in}
    \begin{flushright}
      To my family, thank you.
    \end{flushright}
    \clearpage

  }

\newcommand*{\makefrontmatter}{
\frontnumber\maketitlepage\makeabstract
\makeacknowledgements\makededication\maketoc\mainnumber
}

\begin{document}

\makefrontmatter

\chapter{An Introduction}\label{Ch Intro}
Quantum mechanics is weird. This much we can all agree on. Nonetheless, despite this, we can also agree that the past century has been nothing short of a marvelous success story for quantum theory:
\begin{center}
\epigraph{\textit{``And what a triumph it was, in the old sense of the word: a glorious victory parade, full of wonderful things brought back from far places to make the spectator gasp with awe and laugh with joy."}}{--- \textup{Sidney Coleman}, \cite{Coleman 1988}}
\end{center}

That is, on one hand, quantum theory has allowed us to make remarkable progress in physics --- from particle physics and the standard model, to optics and materials physics. But on the other hand, there are still unresolved issues needing to be addressed --- what does it \textit{mean} to `quantize'? How is `the measurement problem' resolved? What \textit{is} the significance of the quantum state $\Psi$?
\begin{center}
\epigraph{\textit{``Ask anyone today working on foundational questions in quantum theory and you are likely to hear that there is still no consensus on many of these questions - all the while, of course, everybody seems to be in perfect agreement on how to apply the quantum formalism when it comes to making experimental predictions."}}{--- \textup{Schlosshauer et al}, \cite{Schlosshauer et al 2013}}
\end{center}

The situation is itself a contradiction: although there is no uniquely compelling `story' or `interpretation' that we can tell to make sense of quantum theory,\footnote{The dominant interpretations are, of course: the so-called Copenhagen interpretation \cite{Bohr 1928}, the De Broglie-Bohm theory \cite{Bohm 1952} (related to this, see \cite{Madelung}\cite{Takabayasi}), and the `many-worlds' or relative state formulation \cite{Everett 1957}. Informal polling (see e.g., \cite{Schlosshauer et al 2013}\cite{Tegmark 1998}) seems to imply that none of these interpretations are widely favored above the others.} physicists seem to be making do with it all the same. Indeed, the simple fact is that, if pushed, many would likely admit that \cite{Fuchs Peres 2000} ``quantum theory needs no `interpretation'." It is justified on empirical grounds. Or, to put it even more bluntly, we should all just: ``shut up and calculate!"

This state of affairs indicates a rather nihilistic view of physicists towards `interpretations' of quantum theory, and `quantum foundations' more generally. If the goal has been to convince physicists of a particular interpretation, then this enterprise has largely been a failure \cite{Grinbaum 2007}. But this does not mean that understanding quantum theory is a pursuit that should be abandoned. Not in the least. What it does suggest, however, is that new methods are necessary for clarifying these issues. Interpretations are, after all, largely \textit{post hoc} efforts --- from the answer, retroactively attach a meaning. The approach ventures on philosophical means rather than constructive and scientific ones.
\subsubsection*{Entropic dynamics and quantum reconstruction--- }
\epigraph{\textit{``...quantum mechanics will cease to look puzzling only when we will be able to derive the formalism of the theory from a set of simple physical assertions (``postulates'', ``principles'') about the world. Therefore, we should not try to append a reasonable interpretation to the quantum mechanics formalism, but rather to derive the formalism from a set of experimentally motivated postulates."}}{--- \textup{Carlo Rovelli}, \cite{Rovelli 1996}}

The view that we take here is that the proper means for understanding quantum theory is to first construct it from some basic \textit{a priori} defined principles and concepts. This is the basis of the quantum \textit{reconstruction} program.

In the last half century or so, there has been a surge of interest in this type of endeavor, coming from a variety of directions. Some proposals, for example, entertain the notion that quantum theory might emerge from some (presumably more palatable) sub-quantum theory (see e.g., \cite{Nelson 1964}-\cite{tHooft 2016}). More recently, a wave of efforts have featured information theoretic concepts as unifying principles for quantum reconstruction \cite{Wootters 1981}-\cite{Reginatto Hall 2012}. In any case, a crucial choice must be made as to which principles, postulates, ideas, etc., must be regarded as \textit{sacred} and immutable --- this is the definitive step on the basis of which a reconstruction will live or die.

Enter Entropic Dynamics (ED), which is a reconstruction of quantum theory in which the principles of inference and information processing are held as sacrosanct. In ED, probabilities are Bayesian \cite{Laplace}\cite{Jaynes 2003}, they are manipulated using the rules of probability theory \cite{Cox}\cite{Caticha 2009a}, and they are updated by the Maximum Entropy (ME) method \cite{Caticha 2004}-\cite{Giffin}. End of story. The notion of `quantum probabilities' \cite{Von Neumann 1932}\cite{Birkhoff Von Neumann 1936} or `generalized probabilities' \cite{Kirkwood 1933}\cite{Johansen 2007} does not enter. Indeed, it makes no sense in ED to \textit{derive} the `Born rule' from the notion of an `amplitude' $\Psi$ (see e.g., \cite{Caticha 1998}\cite{Goyal Knuth Skilling 2010}) --- our goal is, rather, to derive the notion of an amplitude from the more fundamental concept of probability.

More specifically, ED is a framework for constructing dynamical theories of probability. A challenging aspect of this approach is that a fully epistemic interpretation of the wave function $\Psi $ is not achieved by merely declaring that the square of a wave function $|\Psi |^{2}$ yields a Bayesian probability. Rather, to insure consistency, one must \textit{also} show that the rules for updating those probabilities --- which include both the unitary evolution of the wave function $\Psi $ and its \textquotedblleft collapse\textquotedblright\ during a measurement --- are in strict accord with Bayesian and entropic methods.\footnote{In \cite{Caticha/Johnson}-\cite{Caticha Vanslette 2017} wave function collapse was demonstrated to be an instance of discrete entropic updating \textit{via} Bayes' rule.}

This viewpoint highlights serious differences between ED and other probabilistic reconstructions (see e.g., \cite{Chiribela et al 2011}\cite{Hardy 2001}\cite{Goyal 2008}). To start, it is not clear that these approaches take probability seriously \textit{enough}: if probabilities are being updated, then this should be accomplished within the framework of probability theory and entropic inference. It is not obvious, for instance, that the operational language of `amplitudes', `experiments', `transformations', etc., is entirely adherent to these more fundamental principles. Another key distinction arises in \textit{how} probabilities are utilized. Let us explain. An important task in any probabilistic theory is identifying the propositions, or \textit{microstates}, that are the subject of inquiry.\footnote{Are we uncertain, for instance, of the rise or fall of a stock index price, which side of a coin has landed face up, or perhaps, the location of a particle?} Some approaches, such as quantum Bayesianism \cite{Fuchs 2002} (or `QBism'), for example, assign probabilities to the outcomes of measurements. And certainly, there is nothing strictly speaking `wrong' about this; probabilities are, by design, flexible in this regard. Nevertheless, in doing so, QBists and others (see e.g., \cite{Goyal 2008}) are aligning themselves to many of the philosophical and (indefinite) ontological commitments of complementarity, and of the Copenhagen interpretation, more generally. How much ``physical" insight, after all, can possibly be gained from such a viewpoint?


By contrast, ED commits to a definite ontology --- if we are talking particles, then the positions of those particles are \textit{real}; if we are talking fields, then the values of the fields are \textit{real}! In the language of Kochen and Specker \cite{Kochen Specker 1963}: in ED we commit to a set of (abelian) non-contextual variables, similar to those appearing in the Bohmian approach \cite{Bohm 1952}\cite{Bohm/Hiley 1987}\cite{Bohm/Hiley 1989}. In fact, at first glance there seems to be a significant amount of formal overlap between ED and the de Broglie-Bohm theory.\footnote{It is, of course, quite characteristic of many reconstructions that they converge upon a particular formulation of standard quantum theory --- whether it be Hilbert spaces, Feynman path integrals, etc. Depending on the design of your reconstruction, one or another formulation may be more natural to obtain.} But it is important to stress that they are quite different. For instance, we do not subscribe to the belief, as the Bohmians do, that $\Psi$ has ontic status. The state $\Psi$ is epistemic. Moreover, the Bohmian approach is completely deterministic at the microscopic level: the particles, fields, etc., follow \textit{smooth} trajectories. ED, on the other hand, is completely \textit{indeterministic} from the outset: our goal is only to infer this underlying behavior on the basis of some limited information. (It may even be, in fact, that ED contains the formalism of the Bohmian theory as a special case! See e.g., \cite{Bartolomeo Caticha 2016}.)

Phrased a bit differently, our aim in ED is to obtain the probability $\rho$ that the microstates achieve certain values, and to then update that probability according to the available information. Thus, in an \emph{entropic} dynamics the evolution must be driven by information codified into constraints; it is through these constraints that the \textquotedblleft physics\textquotedblright\ is introduced. And, as one might suspect, significant complications lie in identifying those constraints. 

In recent years, much work has been done \cite{Caticha 2010a}-\cite{Caticha 2019b} to determine the constraints leading to a quantum dynamics, and they appear in stages.\footnote{There is no way around the fact that inference is a messy business, just as science, in general, is a messy business --- we start with some constraints, modify them, and modify them some more until we begin to converge on a theory that is more appealing, in that it better explains, predicts, and controls the observed phenomenon.} A preliminary constraint involves the introduction \cite{Caticha 2013b}\cite{Bartolomeo et al 2014} of an auxiliary variable $\phi$, the \textit{drift potential}, which guides the probabilities; in a more familiar sense, $\phi$ is the mathematical manifestation of the `wave' part of De Broglie's famous `wave-particle' duality, or the pilot wave in Bohmian mechanics.\footnote{Interestingly, the ED approach helps to clarify this duality: the particle aspect is real, the wave aspect is an epistemic property of the ensemble, not an ontological property of the particle.} A secondary constraint emerges once we recognize that it is not just the evolution of $\rho$, but the \textit{coupled} evolution of $\rho$ and $\phi$ that is important --- namely, concerns arise as to how one updates the drift potential $\phi$ in response to the evolving probability $\rho$. Addressing this issue is one of the topics of this thesis.
\paragraph*{Wait, hold on a second!--- }
Before proceeding to develop the specific contents of this work, it is important to preemptively counter some concerns about our methodology. The issue is the following. In the program of quantum reconstruction, we know the answer --- linear, unitary time evolution, Hilbert spaces, Feynman rules, all of that. We know it. No doubt, when the answer is known, it is very easy --- too easy, in fact --- to cook up axioms, postulates, principles, or in our case, constraints, that produce these correct answers. So how do you know that you've succeeded in your reconstruction? Simple: if you've recovered the quantum formalism from your first principles, then clearly it succeeded. The issue of whether it's a \textit{compelling} reconstruction, however, is an entirely different matter; surely what is compelling to one physicist may not be to another, i.e. some subjectivity is involved.

One (mildly objective) criterion for success that we offer here is that any compelling reconstruction should also provide a new perspective on quantum theory.
\begin{center}
\epigraph{\textit{``Therefore psychologically we must keep all the theories in our heads, and every theoretical physicist who is any good knows six or seven theoretical representations for exactly the same physics. He knows that they are equivalent, and nobody is ever going to be able to decide which one is right at that level, but he keeps them in his head, hoping that they will give him different ideas for guessing."}}{--- \textup{Richard Feynman}, The Character of Physical Law \cite{Feynman 1965}}
\end{center}
Put another way, a worthwhile reconstruction is one that equips us with new and \textit{``different ideas for guessing."} But, for guessing what, exactly? Quantum theory, as is, already correctly accounts for much of what is known. (And with a couple new ingenious models, it'll describe the rest, right?) There is, however, still one domain where there is still plenty of room left for guessing: \textit{gravity}. That is, if your reconstruction paves the way for a novel view on a quantum theory of gravity, then this is successful; in our view, at least.
\subsubsection*{Quantum field theory on curved space-time}
The study of quantum field theory on a curved space-time (QFTCS) deals with the interplay of quantum field theory (QFT) together with classical general relativity (GR) (see \cite{Birrell/Davies}\cite{Wald 1994} and the recent review \cite{Hollands Wald 2014}). In the past few decades interest in QFTCS has been driven by a variety of developments. Foremost among these was the recognition by Bekenstein \cite{Bekenstein 1973}, Hawking \cite{Bardeen 1973}\cite{Hawking 1975}\cite{Hawking 1976}, Unruh \cite{Unruh}, and others \cite{Fulling}\cite{Davies} in the 1970's that the intersection of gravitation and quantum theory yields novel features of which \textit{information} and \textit{entropy} are inseparable ingredients.

Since that time, QFTCS has reached a rather mature stage of development; having been accepted as a reasonable effective theory, valid wherever the effects of gravity are not too strong; or more importantly, as a \textit{necessary} step intermediate to a full theory of interacting matter and gravity. Because of this, it is not uncommon to look upon QFTCS as a harbinger of a quantum gravity yet to come --- that is, any such theory must, in some way or another, touch upon QFTCS and its quintessential information-based features.\footnote{A prominent example of this is, of course, the development of the holographic principle \cite{Stephens 1994}\cite{Susskind 1995} and correspondingly, the ADS/CFT correspondence \cite{Maldecena 1999}.} This makes QFTCS a worthy topic for foundational study and reconstruction.
\paragraph*{What's new?--- }
The objective of our thesis is a formulation of QFTCS from the viewpoint of ED. This contribution builds off of a period of sustained effort within the ED program in which several aspects of the quantum formalism have been reconstructed --- including theories of non-relativistic scalar particles \cite{Bartolomeo et al 2014}\cite{Caticha 2019b}; particles with half-integer spin \cite{Carrara Caticha 2020}; and, relativistic scalar fields \cite{Caticha 2013}\cite{Caticha Ipek 2014}. But the singular focus on flat space-time that dominates these developments stands in some contrast to the rather more ambitious goals of the ED program: to ultimately formulate quantum gravity from within the framework of inference and information processing. Here we make inroads on these loftier aims by re-examining and revising the constraints and assumptions of ED in flat space-time in light of the transition to more general, curved space-times.

More to the point: here we propose a model for a manifestly relativistic ED, based on work \cite{Ipek et al 2017}\cite{Ipek et al 2019a} in collaboration with M. Abedi and A. Caticha, that is valid for arbitrarily curved space-times with a globally hyperbolic topology; which, of course, includes Minkowski space as a special case. The scheme requires a couple of conceptual innovations with respect to older versions of ED; one involving the notion of time, the other dealing with the update of  $\phi$, the drift potential. The essential issue is that in a flat space-time with translational symmetry, it is convenient to introduce a \textit{global} notion of time in which all degrees of freedom are updated in a uniform fashion; and in such instances, a \textit{natural} constraint is one where $\phi$ evolves such that a \textit{global} energy functional be \textit{conserved}. However both of these ingredients are inadequate in a generic space-time where it is better to proceed by abandoning such \textit{global} notions in favor of a more \textit{local} description. This takes place in two steps. One is the adoption of a \textit{local} notion of time, in the spirit of Dirac \cite{Dirac 1932}\cite{Dirac 1948}, Tomonaga \cite{Tomonaga 1946}, and Schwinger \cite{Schwinger 1948 I}, where time evolution is allowed to vary across spatial locations. Correspondingly, the update of $\phi$ is determined, not by conserving a global energy, but by introducing a set of \textit{local} Hamiltonian generators, four per spatial point, to perform the updating.

Embedded in this last statement is the central assumption of this reconstruction, which is the adoption of a Hamiltonian framework for evolving $\phi$ --- or more technically, the introduction of a \textit{symplectic} structure on the joint space of $\rho$ and $\phi$. An advantage of this approach is that it allows us to borrow from schemes developed for classical physics in pursuit of a manifestly relativistic ED that is both statistical and quantum. Of particular relevance to us are the covariant Hamiltonian methods pioneered by Dirac \cite{Dirac Lectures}\cite{Dirac 1951}, Kucha\v{r} \cite{Kuchar 1972}, Teitelboim \cite{Teitelboim thesis}\cite{Teitelboim 1972}, and others \cite{Hojman Kuchar Teitelboim 1976}. These approaches are, in short, characterized by expressing the local Lorentz invariance of space-time in terms of a new symmetry --- foliation invariance. In turn, this requirement is implemented by a consistency condition called \textit{path independence}: the evolution of an initial to a final state is independent of the intervening choice of foliation. The result is a collection of Poisson brackets and constraints that are to be satisfied by the aforementioned local Hamiltonian generators.

One feature of these developments is that there is very little friction in transitioning from a fixed background geometry to a dynamical one that reacts in response to the evolving quantum state. We demonstrate that this is essentially what is called ``semi-classical" gravity (see e.g., \cite{Birrell/Davies}\cite{Wald 1994}) in the literature, and it is obtained in our framework in a very natural way. And, although such an approach has long been outright dismissed as a viable theory of quantum gravity (see e.g., \cite{Unruh 1984}), more recent work \cite{Albers et al 2008}\cite{Hall Reginatto 2018} suggests that semi-classical theories of gravity cannot be totally ruled out. Put another way, the gravity theory we put forth blends together some intriguing features --- quantum fields, entropy, and geometry --- we ask whether this could be quantum gravity.
\paragraph*{Layout--- }
The layout of this thesis is as follows. Chapter \ref{Ch EI} proceeds with a short overview of the framework of entropic inference, borrowing heavily from many influential sources. We then develop the ED of non-relativistic particles in chapter \ref{Ch ED NR} as a way of introducing the key features of an ED model. Next, we build off of these developments in chapter \ref{CH ED fields} to address the topic of relativistic fields in Minkowski space. This is followed by our main contributions in chapters \ref{5CH CED} and \ref{CH SCG} where we develop the framework of covariant ED in curved space-time, first in the context of a fixed background in chapter \ref{5CH CED}, then in a dynamical background in chapter \ref{CH SCG}. In chapter \ref{CH conclusion} we make some concluding remarks and analyze the model developed in the previous chapter.

\chapter{Entropic inference}\label{Ch EI}
\epigraph{``But in this world, nothing can be said to be certain, except death and taxes."}{--- \textup{Benjamin Franklin}, 1789}

Uncertainty is a fact of life; it is the rule, not the exception. This is a viewpoint that we feel is so uncontroversial that it serves as the bedrock perspective from which we move forward in this dissertation (and if you don't take this author's humble word for it, perhaps the above words of Benjamin Franklin will be more convincing!).

We understand the notion of uncertainty, colloquially at least, to be related to the fact that we have insufficient information to predict, or know, a particular something (a proposition, perhaps) with complete certainty. This is, indeed, usually the case; we do not posses and, in fact, will often never possess the information necessary to attain certainty. So what do we do then? How do we proceed? What conclusions can be made, what \textit{inferences} can be drawn from the limited information that we do, in fact, possess?

These are, in fact, difficult questions to answer and \textit{a priori} it is not obvious that we can answer any of these questions independent of the context in which they were asked. Indeed, situations of uncertainty constitute almost all the situations we can conceive of --- from estimating the number of different species on earth, to whether there is any species at all on Mars, to whether the Dow Jones will go up or down tomorrow --- all are situations where our information is incomplete. But in each case, the domain knowledge, the data collected, and the general sources of information are extremely disparate. From this one might conclude that the problem is not well posed, that we should give up! Let econometrics be unto the economists  and let biostatistics be unto the biologists and pray that they never need to walk into a bar together!

Luckily, we need not resort to praying in this instance; we can rely instead on the important contributions of many pioneers, such as the early progenitors of probabilistic thought J. Bernoulli \cite{Bernoulli} and P.S. Laplace, as well as the breakthrough 20$^{\text{th}}$ century insights of  R.T. Cox \cite{Cox} and E.T. Jaynes \cite{Jaynes 1957}, and the more recent works of A. Caticha \cite{Caticha 2004}, K. Vanslette \cite{Vanslette 2017}\cite{Vanslette thesis}, as well as many other fine contributions (see e.g., \cite{Giffin}\cite{Caticha Giffin 2006}\cite{Knuth Skilling 2012}\cite{Knuth 2018}).\footnote{See also the monograph of Caticha \cite{Caticha 2012} and the thesis of A. Giffin \cite{Giffin} for a detailed discussion of these issues and review of the history of probability and entropy.}

The framework for handling incomplete information, based on these groundbreaking works, will be called \textit{entropic inference} (for reasons that will become apparent) and can be summarized by two key ingredients. The first is that a partial state of knowledge --- degrees of rational belief --- can be quantified by a set of real numbers. These degrees of rational belief are in turn (subject to a consistency requirement, first due to Cox) manipulated by a set of rules that are in one-to-one correspondence with the rules of probability, thus prompting the controversial\footnote{It is controversial in the sense that there is an active debate surrounding the interpretation of probability with \textit{Bayesians} and \textit{frequentists} on opposing sides. And, even within the Bayesian camp, there are a wide spectrum of views ranging from the personalistic subjective de Finetti approach to the ``logical" objective view espoused by Jaynes. See Jaynes \cite{Jaynes 2003} for a thorough discussion of this topic, albeit coming from a partisan actor.} statement:
\begin{equation*}
\text{Degrees of Rational Belief ~=~ Probabilities~.}
\end{equation*}

Once we have established degrees of rational belief as being quantified by probabilities, the second aspect of entropic inference concerns how one \textit{updates} their beliefs, i.e. updates probabilities, when new information is made available. Indeed, such a procedure is not trivial. To what extent should we change our previously held, or \textit{prior} beliefs on the basis of a single new experimental outcome, for example? Certainly new data should induce some change in our state of knowledge, but how much exactly? How about two, or three outcomes? What if, for instance, we learn that the average expected roll of a die was not 3.5, as is expected for a fair die, but actually 4.0? How does one incorporate such information into their state of knowledge? What are the optimal set of probabilities that satisfy this constraint? Amazingly, the universal solution to this problem is provided by the method of maximum (relative) entropy (ME). The scheme hinges on the basic principle that prior information is valuable. That is, we would like to incorporate any newly available information, but without sacrificing, to the greatest extent possible, the information held in our prior beliefs.

Together, probability theory and entropic updating form the foundation of inductive reasoning. And more importantly, they are \textit{universal} in this regard; implying that these are the default tools for whenever we lack complete information. The universality is key since it justifies the use of these tools in any situation where uncertainty plays a factor. (This, we will find, is also applicable to quantum theory, the subject to which the majority of this thesis is devoted.)

The structure of this short chapter is as follows. Section \ref{Probability theory} introduces probability theory as a set of rules for consistently manipulating degrees of rational belief. Section \ref{Entropic updating} introduces relative entropy as a tool for ranking probability distributions, and thus as a means for updating probabilities, according to a set of desideratum with wide appeal.

The treatment of probability theory and entropic updating that we include here is not meant to be exhaustive, nor is it an extensively detailed account. For further details please consult \cite{Caticha 2012} by A. Caticha as well as the theses of A. Giffin \cite{Giffin} and K. Vanslette \cite{Vanslette thesis}, and the references therein for a list of original works and a more detailed presentation.
\section{Probability theory \label{Probability theory}}
In situations where complete conviction in an outcome is absent, we are left to grapple with the possibility that multiple different outcomes may be realized, and as such, we are left to deliberate on the relative merits of each outcome. That is to say, given limited (or no) knowledge of some assertion, we wish to gauge the truth content of the potential outcomes. This is the problem that we aim to address here. And faced with such a task, we set out to \textit{design} a tool for aiding and facilitating our judgment in this regard.
\subsection*{The design of probability theory}
The key idea in designing a tool for rational inductive reasoning is the following: we wish to formulate a \textit{quantitative} measure of the \textit{degree} to which we may \textit{rationally presume} a particular assertion to be true. There are several components to this idea that require explanation. The first is that the object of interest here is our belief in the truth of a particular proposition; something we will later find can be identified with \textit{probabilities}. Additionally we, by \textit{design}, require that our beliefs be amenable to quantitative analysis; the purpose of any tool, after all, is to be useful in practice, which is exactly what a mathematical description allows for. The third, and most crucial factor, is that the rules for manipulating our beliefs must be \textit{rational}. The difficulty in this requirement, however, is that we cannot easily identify a set of universal guidelines that lead to rational behavior in all circumstances. Instead, we settle for a much more modest goal: we aim to avoid those glaring cases that we can all agree are \textit{irrational}. This implies that we hope to avoid obvious \textit{inconsistencies} in our reasoning; the more we believe, for instance, that $A$ is true, then the less we \textit{ought} to believe that its negation $\tilde{A}$ is true, and anything to the contrary, most would agree, would be irrational.

\paragraph*{Quantifying rational belief--- }
Concerning the mathematical description, the objective is to model the pure \textit{intensity} with which we believe a particular \textit{proposition} to be true. Two words here are operative: intensity and proposition. The latter is important since it makes clear that we are assigning degrees of rational belief to logical \textit{Boolean} propositions; thus, as will be important later, more complex propositions may be built from simpler propositions using the AND and OR operations of Boolean logic. 

The word \textit{intensity} is also key: an intensity is defined simply as being either more or less intense. As such, our mathematical description should allow for intuitive notion that the intensity of rational beliefs is \textit{transitive}: if we believe in $A$ is more than $B$, and believe in $B$ is more than $C$, then it should be that we believe in $A$ more than $C$. These criterion suggest the claim that
\begin{quote}
\textit{degrees of rational belief (which we can later identify with probabilities) can be represented by real numbers.}
\end{quote}
\paragraph*{Notation--- }
We establish here some standard notation which is consistent with that of \cite{Caticha 2012}. We assume some basic familiarity with Boolean logic. For any proposition $a$, it has a negation, denoted $\tilde{a}$, such that if $a$ is true then $\tilde{a}$ is false. Sometimes we are interested in the assertion that $a$ is true subject to some background information $c$. This is denoted $a|c$ and read as ``$a$ given $c$".

From any two propositions $a$ and $b$ we can construct more complicated composite propositions using the conjunction and disjunction operations. The conjunction, or AND operation will be denoted $ab$ and the disjunction of two propositions $a$ and $b$ will be denoted $a\vee b $. The truth value of the composite propositions $a\vee b$ and $ab$ that are conditional on some given background information $c$ will be denoted as $a\vee b|c$ and $ab|c$, respectively.

The degree to which we believe the proposition $a$ to be true is denoted $[a]$, and is such that $[a] \in \mathbb{R}$ is a real number. Eventually, we will find that there is a direct correspondence between $[a]$ and the probability $p(a)$. Finally, degrees of belief will range from some extreme lower value $v_{F}$ reflecting extreme disbelief in a proposition (such as $[a|\tilde{a}] = v_{F}$, for any $a$) to the extreme upper bound $v_{T}$ which reflects complete belief in a proposition (such as $[a|a] = v_{T}$).
\paragraph*{Strategy--- }
The basic element that one has access to is the real number $[a|c]$ associated to any proposition $a$, subject to some background information denoted by $c$. (The truth value of any proposition is, of course, dependent on some context, which for conciseness we denote simply using $c$.) The question of how one assigns a value to $[a|c]$ in the first place is interesting, and is relevant for the subsequent section on entropic updating, but here we assume that such an assignment has already been made.

The problem that we would like to solve, on the other hand, is how to rationally assign values to the degrees of belief $[ab|c]$ and $[a\vee b|c]$ for the composite propositions $ab|c$ and $a\vee b|c$, respectively.
\begin{quote}
\textbf{Assumption:}\textit{ in order to be rational, our beliefs in $a\vee b$ and $ab$ must somehow be related to our separate beliefs in $a$ and $b$.}
\end{quote}
A mathematical representation of this assumption would be that there exist some functions $F$ and $G$ that satisfy
\begin{equation}
[a\vee b|c] = F\left([a|c],[b|c],[a|bc],[b|ac]\right)
\label{Sum rule F}
\end{equation}
and
\begin{equation}
[ab|c] = G\left([a|c],[b|c],[a|bc],[b|ac]\right).
\label{Product rule G}
\end{equation}
Our goal is to determine the unknown functions $F$ and $G$ that must hold for \textit{any} choice of propositions $a$, $b$, $c$.

The essential constraint is that the appropriate functions $F$ and $G$ must respect the associative and distributive properties of the Boolean AND and OR operations. This turns out to be extremely restrictive, resulting in a family of equivalent representations, of which standard probability theory is a member.
\subsection*{Sum rule}
We aim here to outline the derivation of the sum rule of probability theory, which amounts to a determination of the function $F$ in (\ref{Sum rule F}). We proceed in two steps. We first determine the form of $F$ for propositions that are mutually exclusive by imposing an associativity constraint. The general case is easily obtained from there.
\paragraph*{Special case: mutual exclusivity--- }
For the $F$ in eq.(\ref{Sum rule F}), when $a$ and $b$ are mutually exclusive in the context of $d$ then $[a|bd] = [b|ad] = v_{F}$, and so $F$ is essentially a function of two arguments:
\begin{equation}
[a\vee b|d] = F\left([a|d],[b|d],v_{F},v_{F}\right) = F\left([a|d],[b|d]\right).\label{Sum rule F MEx}
\end{equation}

Of course the $F$ cannot be chosen in any haphazard fashion, it must reflect some kernel of rationality. The basic idea is consistency: if a degree of rational can be computed in two different ways, they must agree. To execute this idea it is useful to consider three mutually exclusive propositions $a$, $b$, $c$ in the context of some information $d$. The requirement of consistency follows from the associativity of the Boolean OR,
\begin{equation}
(a\vee b) \vee c = a\vee (b\vee c).
\label{associativity}
\end{equation}

The consequences of imposing such a constraint leads to a functional equation (see e.g., \cite{Caticha 2009a}) whose general solution was first found by Cox (see e.g., \cite{Cox}\cite{Jaynes 1957a}). Unfortunately, for an arbitrary ranking scheme $[\, \cdot \, ]$, the general solution is quite complicated. Cox, however, demonstrated that one could always regraduate (or rescale) the degrees of belief
\begin{equation}
[\, \cdot \, ]\to\xi(\cdot),
\end{equation}
while maintaining the original ordering, the solution takes a very simple form:
\begin{equation}
\xi(a\vee b|d) = \xi(a|d)+\xi(b|d),
 \label{Sum rule xi MEx}
\end{equation}
which is just a sum rule.
\subparagraph*{Utter disbelief:}
In the previous ranking scheme using $[\,\cdot\,]$, the value $v_{T}$ reflected complete certainty, while $v_{F}$ demonstrated complete disbelief. In the modified ranking scheme using $\xi(\cdot)$, complete certainty is given, by definition, as $\xi_{T} = \phi(v_{T})+\beta$; its value cannot be determined exactly. On the other hand, it is a simple matter to show that (\ref{Sum rule xi MEx}) determines the value of complete disbelief to be $\xi_{F} = 0$ in the new scale \cite{Caticha 2012}\cite{Caticha 2009a}.
\paragraph*{General sum rule--- }
The mutual exclusivity assumption can be easily relaxed to obtain a general representation of the OR rule. Note that any pair of \textit{arbitrary} propositions $a$ and $b$ can be written as $a = (ab)\vee(a\tilde{b})$ and $b = (ab)\vee(\tilde{a}b)$, respectively. Thus we have that
\begin{equation}
a\vee b = (a b)\vee (\tilde{a}b) \vee (a\tilde{b}) = a\vee(\tilde{a}b),
\end{equation}
which is composed of two mutually exclusive options. As such, the sum rule in (\ref{Sum rule xi MEx}) applies, and we obtain
\begin{align}
\xi(a\vee b|d) &= \xi(a|d)+\xi(\tilde{a}b|d) + (\xi(ab|d)-\xi(ab|d))\notag\\
               &= \xi(a|d) + \xi(\tilde{a}b\vee ab|d) - \xi(ab|d).
\end{align}
Using the fact that $\tilde{a}b\vee ab = b$ gives us
\begin{equation}
\xi(a\vee b|d) = \xi(a|d)+\xi(b|d) - \xi(ab|d),
\label{Sum rule general}
\end{equation}
which is the general sum rule.
\subsection*{Product rule}
For arbitrary propositions $a$ and $b$ we would like to assign a degree of rational belief $\xi(ab|c)$ to the conjunction $ab$ in the context of some background information $c$. The idea is to identify the function $G$, which in its most general version depends on four inputs
\begin{equation}
\xi(ab|c) = G\left(\xi(a|c),\xi(b|c),\xi(a|bc),\xi(b|ac)\right).
\end{equation}
A lengthy argument due to Caticha \cite{Caticha 2012}\cite{Caticha 2009a}, which is beyond the scope of this thesis, narrows the space of  functions $G$ to those of just \textit{two} variables. In particular, $G = G(\xi(a|c),\xi(b|ac))$.\footnote{Essentially, one considers all possible choices of functions for $G$, ranging from two inputs to all four. By examining certain clever special choices of inputs, one can eliminate all but one of the functions as the others demonstrate clearly irrational results (such as, for example $G = $constant).}
\paragraph*{Consistency constraint--- }
To determine $G(\xi(a|c),\xi(b|ac))$ we impose consistency once again: whatever the function $G$ is, it must be consistent with the sum rule eq.(\ref{Sum rule general}) and the distributive properties of the Boolean AND and OR operations. Our brief derivation follows Caticha \cite{Caticha 2012}.

Consider the propositions $a$, $b$, $c$, where $b$ and $c$ are mutually exclusive, in the context of a fourth assertion $d$. Using the distributivity of AND over OR we have
\begin{equation}
a(b\vee c) = ab\vee ac.
\end{equation}
This implies that we can compute the degree of belief $\xi(a(b\vee c) |d)$ in two ways,
\begin{equation}
\xi(a(b\vee c)|d) = \xi(ab\vee ac|d) = \xi(ab|d)+\xi(ac|d),
\end{equation}
where the second equality follows because $b$ and $c$ are mutually exclusive. Fulfilling this constraint requires that
\begin{equation}
\frac{\xi(ab|d)}{\xi_{T}} = \frac{\xi(a|d)}{\xi_{T}}\frac{\xi(b|ad)}{\xi_{T}}.
\label{product rule xi}
\end{equation}
\paragraph*{The main result--- }
Equation (\ref{product rule xi}) suggests introducing a new, more convenient, set of values, scaled by $\xi_{T}$:
\begin{equation}
p(\cdot) = \frac{\xi(\cdot)}{\xi_{T}}.
\end{equation}

In this \textit{regraduated} scale, the representation of the AND operation of Boolean logic is a simple product rule,
\begin{equation}
p(ab|c) = p(a|c)\, p(b|ac).
\label{Product rule prob}
\end{equation}
On the other hand, in this new scale, the sum rule retains the same form
\begin{equation}
p(a\vee b|c) = p(a|c) + p(b|c) - p(ab|c).
\label{sum rule prob}
\end{equation}
The values $p(\cdot)$ takes values from $p_{F} = 0$ --- complete disbelief --- to $p_{T} = 1$ --- absolute certainty. This justifies calling the values $p(\cdot)$ \textit{probabilities}.
\paragraph*{Discussion--- }
In short we conclude that:
\begin{quote}
\textit{Degrees of rational belief are represented by probabilities and those probabilities are manipulated by the rules of probability theory, i.e. the sum and product rules.}
\end{quote}
While other formal choices are possible, probability theory stands alone as a particularly convenient representation of the rules for manipulating degrees of plausibility. Once one obtains the standard sum and product rules, there is no impediment to obtaining all the standard results of probability theory --- the central limit theorem, marginalization, etc.\footnote{There are many good books on probability theory. This author has gained much insight from the book by Caticha \cite{Caticha 2012} and the book by Jaynes \cite{Jaynes 2003}.}

The implications are also vast: a partial state of knowledge is to be represented by probabilities. Thus probability theory provides the framework for reasoning in the face of uncertainty. (The question, of course, is whether this applies to so-called ``quantum" uncertainties as well.)
\paragraph*{A comment on continuous distributions}
It is worth mentioning that we did not specify the features of the propositions $a$, $b$, etc. They could, for instance, be discrete or continuous (and indeed, our derivation relied on this generality). In the continuous case, however, plenty of subtleties arise owing to an infinite number of possible outcomes. In particular, for a variable $X$ taking on a continuous set of values $x$, the probability of a single point $p(X=x) = 0$.

A well defined notion, on the other hand, is that of the probability of an infinitesimal range of values
\begin{equation}
 p(x\leq X \leq x+\delta x) = \rho(x)dx,
\end{equation} 
where $\rho(x)$ is called the \textit{probability density function}. And, while a true probability is only assigned to the product $\rho(x)dx$, it is common to also call the probability density $\rho(x)$ a probability, or probability distribution; although it is always the case that actual probabilities must come accompanied with some measure $dx$.
\section{Entropic updating \label{Entropic updating}}
That one can assign an interpretation (or any interpretation at all) to the notion of probability is important for the simple fact that knowing \textit{what} a probability is suggests \textit{how} probabilities should be used. Which is to say that, if probabilities are nothing more than a set of rational convictions, then we should treat them as such! And indeed, to this point, another common facet of rational beliefs is that they are not \textit{rigid}: a rational agent changes her beliefs in light of new evidence. This suggests that, yes, probabilities too must be updated when new information becomes available. Here we introduce the universal tool for accomplishing just this, called the maximum entropy (ME) method.

Throughout this section we borrow heavily from \cite{Caticha 2012} in crafting the motivation and rationale for the ME method, however, we have updated the assumptions and design specifications to reflect the work of K. Vanslette in \cite{Vanslette 2017}\cite{Vanslette thesis}, which improves on the treatment in \cite{Caticha 2012} by way of reducing the number of necessary desiderata.
\subsection*{The design of entropic updating}
The idea is the following. We are interested in predicting the values of a variable $x$ about which we are uncertain.\footnote{We assume $x$ is continuous throughout, but the simplification to discrete variables is straightforward.} This state of partial knowledge is encoded, naturally, by a probability distribution $q(x)$. The problem we would like to address is, given that we are presented with some new information about the $x$'s, how should we modify our current beliefs, i.e. $q(x)$, to accommodate this new information. That is, how do we update from a \textit{prior} distribution $q(x)$ to a particular \textit{posterior} distribution $p(x)$ that takes into account the new information that we have learned?
\paragraph*{An aside on information--- }
A chief concern in our task is an identification of what we mean by ``information". To be sure, the meaning of the word ``information" is quite contentious, and it is not our purpose here to present an exhaustive discourse on this. In the simplest sense, we define information operationally: information is whatever \textit{constrains} the beliefs of a rational agent. That is to say, information takes the form of constraints on our state of knowledge, i.e. constraints on the allowed probability distributions. This is the stance that we take throughout this thesis. A more thorough discussion of our perspective is available in \cite{Caticha 2012}\cite{Caticha 2014}.
\paragraph*{The general design criteria--- }
The goal is to \textit{design} a tool for updating probability distributions based on some design criteria that are sufficiently reasonable so as to garner wide appeal.
\subparagraph*{Universality:}
The hope is not to develop a tool that works in this problem or that problem, but a tool that is robust enough to function in \textit{any} problem where inference is needed.
\subparagraph*{Parsimony:}
We take here a conservative viewpoint: prior information is valuable. Thus we should update our beliefs to the minimum extent required by the new information. We call this the \textit{principle of minimum updating} (PMU). This has an appealing practical consequence: we do not proceed by instructing a rational agent how to change their beliefs (there are, of course, many ways to do this), but by detailing when \textit{not} to update our beliefs.\footnote{For example, such a principle guards against a clear form of irrationality: we should not change our beliefs at all when \textit{no} information has been provided. As usual, this idea is more eloquently and lucidly stated in \cite{Caticha 2012}.}
\subparagraph*{Independence:}
When doing science we always tacitly assume that we can identify a set of relevant variables which can be considered \textit{independently} from all others --- if we could not assume, for instance, that the global wind patterns on earth are independent of the behavior of the Andromeda galaxy, then science would be a hopeless endeavor. This principle will be implemented in two manners --- subsystem independence and subdomain independence --- which will we develop more rigorously later on.
\paragraph*{Strategy--- }
We seek to update from a prior distribution $q(x)$ to a candidate posterior distribution $p(x)$
\begin{equation}
q(x) \longrightarrow p(x),
\end{equation}
where $p(x)$  captures some newly obtained information. (The notation convention follows \cite{Caticha 2012}: $q(x)$ denotes priors, while $p(x)$ denotes a possible candidate posterior, and $P(x)$ signifies the optimal posterior.) The strategy we implement for accomplishing this is the following: we \textit{rank} the possible posteriors $p(x)$ according to \textit{preference}. (We will be more specific about what we mean by preference shortly.) That is, to each possible posterior distribution $p_{i}(x)$ we assign a single real number $S[p_{i},q]\equiv S[p_{i}]$ --- the \textit{entropy} functional --- such that if we prefer $p_{1}$ to $p_{2}$ then $S[p_{1}] > S[p_{2}]$.\footnote{The ranking scheme that we utilize for updating probability distributions can traces its origins to the work of Shore and Johnson \cite{Shore Johnson 1980} as well as J. Skilling \cite{Skilling 1988}.} Naturally, such a ranking, if it is to be useful, should also be transitive: if $p_{1}$ is preferred to $p_{2}$ --- $S[p_{1}] > S[p_{2}]$ --- and $p_{2}$ preferred to $p_{3}$ --- $S[p_{2}] > S[p_{3}]$ --- then $p_{1}$ is preferred to $p_{3}$ --- $S[p_{1}] > S[p_{3}]$.

The objective is then, obviously, to choose the most preferred distribution $\hat{p}$; this is the one that \textit{maximizes} the entropy functional $S[p]$. Hence the scheme is called the method of Maximum Entropy (ME).
\subparagraph*{Eliminative induction:}
The strategy for identifying an appropriate entropy functional is called \textit{eliminative induction}. Briefly: once one identifies a set of reasonable criteria to be satisfied, one imposes these conditions on the allowed form of the entropy $S[p]$. As we shall see, the criteria chosen here are sufficiently constraining, such that only a single entropy functional is allowed.

\subsection*{The relative entropy functional}
We consider here two specific design criteria (DC) that our entropy functional must satisfy. We follow the argument of K. Vanslette \cite{Vanslette 2017}\cite{Vanslette thesis} whose method largely follows that of A. Caticha \cite{Caticha 2012}, but improves it by removing one superfluous ingredient. Here we simply state the criteria and their implications; proofs can be found in the associated references.
\paragraph*{DC1: Subdomain independence--- }
The requirement of subdomain independence is an instance of specifying when it is rational to \textit{not} update our beliefs. Consider that the space $\mathcal{X}$ of $x$'s is divided into two domains such that $D \cup \tilde{D} = \mathcal{X}$, where $D$ and $\tilde{D}$ are complements. DC1 states that when a constraint $C$ refers only to the domain $\tilde{D}$, then the \textit{conditional} probabilities $p(x|D)$ should not be revised, i.e. $p(x|D) = q(x|D)$. If one imposes this for every conceivable division $\mathcal{X}$ into sets of subdomains, then this implies that each subdomain contributes additively to the entropy:
\begin{equation}
S[p] = \int dx F\left(p(x),q(x),x\right),
\label{Entropy DC1 F}
\end{equation}
where $F\left(p(x),q(x)\right)$ is a regular function (not functional) of the probabilities $p(x)$ and $q(x)$, and may depend explicitly on the coordinates $x$.\footnote{The proof of this result can be found in \cite{Caticha 2012} under the label of the ``locality" DC.}
\begin{remark*}
A special case of DC1 is that if we do not impose any constraints at all, then we should have that the optimal posterior is just the prior: $P(x) = q(x)$.
\end{remark*}
\paragraph*{DC2: Subsystem independence--- }
Consider the joint space of variables $\mathcal{X} =\mathcal{X}_{1}\cup \mathcal{X}_{2}$ where $x_{1}\in \mathcal{X}_{1}$ and $x_{2}\in \mathcal{X}_{2}$ are their respective coordinates, and moreover, consider in addition that the joint prior $q(x_{1},x_{2}) = q(x_{1})q(x_{2})$ reflects the property of statistical independence. DC2 states that if the new information contained in some constraint $C$ does not explicitly introduce correlations between the variables $x_{1}$ and $x_{2}$, then the optimal posterior $P(x_{1},x_{2}) = P(x_{1})P(x_{2})$ should \textit{also} reflect the statistical independence properties of the prior $q(x_{1},x_{2}) = q(x_{1})q(x_{2})$. This is a clear instance of the PMU: if we believe the systems to be independent \textit{a prior} and the new information does not \textit{explicitly} introduce information to the contrary, then we should maintain the belief \textit{a posteriori} that the system remain independent.

The implications of this DC are particularly constraining, requiring (subject to some caveats of convenience) that the entropy functional be of the form
\begin{equation}
S[p,q] = -A\int dx \, p(x)\log\frac{p(x)}{q(x)},
\label{Entropy DC2}
\end{equation}
which is the well-known relative entropy functional (see e.g., \cite{Jaynes 2003} and references therein). 
\subsection*{The Maximum Entropy method}
Once a suitable entropy functional has been designed, we can proceed to use it for inference. The method can be presented algorithmically as follows:
\begin{itemize}
\item[1.] Establish the prior $q(x)$ that requires updating.
\item[2.] Obtain that information which is deemed relevant for updating: this obliges us to identify the family of posteriors $\left\{p(x)\right\}$ that are consistent with this information
\begin{equation*}
q(x)\quad \overset{\text{new information}}{\xrightarrow{\hspace*{2.5cm}}}\quad p(x).
\end{equation*}
\item[3.] Define a relative entropy $S[p,q]$ for ranking the posteriors $p(x)$ relative to the prior $q(x)$.
\item[4.] Select the optimal $p^{*}(x) = P(x)$; the one that \emph{maximizes} the entropy subject to the constraints, $S^{*} = \text{max}_{p} \, S[p,q]$.
\end{itemize}
\paragraph*{Discussion--- }
There are two types of information that are common enough to warrant specific mention: data and expected value constraints. An interesting aspect of the entropic inference framework is that both types of information can be handled using the ME method. Indeed, the ME method generalizes both Bayesian data analysis as well as the MaxEnt formalism of E.T. Jaynes \cite{Jaynes 1957}.

On one hand, ME easily allows for processing some data that is not readily accommodated in the standard framework; such as if the data is itself a bit uncertain. On the other, the original MaxEnt of Jaynes was not designed for updating from an arbitrary prior to a posterior; its purpose was rather for assigning probabilities when the prior was a uniform distribution.

As it turns out, both types of updating are crucial for developing quantum theory from a perspective that takes entropic inference as the foundation. Since measurement is nothing but the collection of data, it is possible to model the infamous collapse of the wave function as nothing but an application of entropic updating using Bayes' rule. Moreover, entropic updating plays a crucial role in obtaining the Brownian stochastic processes that underlie all quantum systems.\footnote{Another aspect of entropic inference crucial to quantum theory is \textit{information geometry}. While this aspect of entropic inference will be briefly reviewed within the course of this thesis, this does not do the subject justice. See the book of Caticha \cite{Caticha 2012} for an accessible introduction and the book by Amari \cite{Amari 1985} for further details.}

\chapter{Non-relativistic entropic dynamics}\label{Ch ED NR}
The first instance of inferential methods playing a prominent role in fundamental physics occurred with the work of E.T. Jaynes in statistical mechanics \cite{Jaynes 1957}. Up to that point, the development of statistical mechanics --- from Boltzmann to Maxwell to Gibbs and so on --- had been plagued by general confusion, as was evident with Maxwell's demon, the Gibbs paradox, irreversibility, etc.\footnote{See, for example, the writings of Jaynes himself \cite{Jaynes 1967}\cite{Jaynes 1992} as well as the monograph by A. Caticha \cite{Caticha 2012} for an in-depth discussions of these issues and how the inference viewpoint casts light on them.} The beauty of Jaynes' approach was that by placing the notions of probability, entropy, and information at the forefront of the theory, many of these fundamental issues and paradoxes in statistical mechanics were clarified, while many of the enigmatic building blocks of the conventional formulation were left behind. Indeed, once one adopts, as Jaynes did, a Bayesian view of probability, conceptual issues that hindered statistical mechanics --- such as the ergodic hypothesis, etc. ---  were no longer relevant! The fuzzy picture of entropy as a measure of disorder, or chaos or (enter the interpretation \textit{du jour}), of a system's physical state that was inherited from thermodynamics was replaced by entropy as a tool for processing information; a notion that began with the work of C. Shannon \cite{Shannon 1948}.

While attaining clarity in statistical mechanics was a significant achievement in its own right, Jaynes' statistical mechanics is important in another regard as well: it shows that statistical mechanics results from an \textit{application} of inferential methods. From this viewpoint, statistical mechanics is not a theory that describes the mechanics of a physical system --- it is a theory about \textit{processing} the information we possess about a physical system. This begs the question of whether statistical mechanics is a unique theory in this regard, or are other theories of physics amenable to being treated within this framework? Are the laws of nature firm and mechanical, or is it as John Wheeler \cite{Wheeler/Zurek 1983} once said:
\begin{quote}
\textit{``that every law of physics, pushed to the extreme, will be found statistical and approximate, not mathematically perfect and precise"}?
\end{quote}
\subsection*{An entropic view of quantum theory --- why bother?}
\label{An entropic view of quantum theory}
One might look at the state of quantum theory today and think that it is reminiscent of the status of statistical mechanics in the time of Jaynes, almost 70 years ago. Indeed, quantum mechanics is another theory of physics in which uncertainties and probabilities enter in an intrinsic way. And, not coincidentally, quantum theory is itself infamous for the conceptual issues that it poses --- measurement and wavefunction collapse, non-locality and entanglement, and the list goes on.\footnote{That conventional quantum theory is chock-full of paradoxes that challenge the intuition we've gained from classical mechanics is well known and we do not attempt here to compile an exhaustive list. The textbook by L. Ballentine \cite{Ballentine} is more sensitive to these issues than most and the papers by J.S. Bell \cite{Bell} provide penetrating insight.}

It is natural to wonder, in light of Jaynes' work, if we could do for quantum mechanics, what Jaynes did for statistical mechanics? In \cite{Jaynes 1957}, Jaynes said of statistical mechanics:
\begin{quote}
\textit{``it is possible to maintain a sharp distinction between its physical and statistical aspects. The former consists only of the enumeration of the correct states of the system and their properties; the latter is a straightforward example of statistical inference."}
\end{quote}
Can we clarify and understand the foundations of quantum theory in a similar fashion by recognizing, in a clear-eyed manner, the probabilistic and statistical aspects of the theory as distinct from its physical features?



The program of entropic dynamics (ED) is an attempt to do just that --- to derive quantum theory using the methods and tools typical of information processing and inference. This involves understanding the probabilities $\rho=|\Psi|^{2}$ that occur in quantum mechanics to be epistemic \textit{Bayesian} probabilities and, accordingly, that those probabilities be updated by rules that are consistent with entropic and inference methods.

As one might expect, this has significant effects on how the theory is to be interpreted. Indeed, one question that is common when discussing the foundations of quantum mechanics is, what is the interpretation of the quantum state $\Psi$? Is the wave function an element of the ontology? That is to say, is it a ``real" object, like a particle or a field? Or, is the wave function epistemic, related to our state of knowledge of the system? Clearly in ED the answer is the latter, and as a result, interpretations of quantum mechanics where the wave function is given an ontic status are dismissed outright; such interpretations include the Bohmian theory \cite{Bohm 1952}\cite{Holland 1995} and the provocative ``Many-Worlds" interpretations first proposed by Everett \cite{Everett 1957}.

This viewpoint also has utility beyond simply interpretation. Viewing probabilities in quantum theory as Bayesian goes a large way towards solving the measurement problem; perhaps the most infamous open problem in the foundations of quantum mechanics.\footnote{see \cite{Caticha/Johnson}\cite{Caticha Vanslette 2017} for a full discussion of how the problem is resolved in ED.} Once probabilities are Bayesian, one has access to the full toolkit of entropic and Bayesian inference; including the Bayesian and entropic modes of updating probabilities. As a consequence, in ED, an independent collapse postulate is not required --- one can show \cite{Caticha/Johnson} that the infamous collapse of the wave function is nothing but an application of Bayes rule! 


But is ED truly \textit{necessary}? Quantum mechanics is, after all, the most successful scientific theory that humankind is in possession of --- time after time its predictions have been validated with great precision. The conventional viewpoint, it seems, is doing just \textit{fine}! And to be absolutely clear, our goal is not to claim otherwise. However the goals of theoretical physics have expanded beyond describing tabletop and accelerator physics, and now include research into quantum gravity and quantum cosmology --- situations describing the universe at large. It is pertinent to ask, how does one quantify the uncertainty of the conditions of the big bang? What is the significance of the entropy of a black hole? What does it mean to assign a probability to a space-time, as some quantum gravity approaches seem to do? Can we interpret (or even conceive of) the second law of thermodynamics in a cosmological setting?

These questions raise several issues. If the relationship between gravity, entropy, and quantum theory is indeed important, as we expect it is, then perhaps the entropic framework can shed some light on it. Moreover, as Ernst Mach astutely noted: the universe is not twice-given, but only once. Thus it is an open question as to how frequentist views of probability can handle questions at this scale. Is the Bayesian viewpoint not more coherent in this case? And if it is, then consistency suggests that we treat probabilities as Bayesian all the way down to microscopic scales.

This all implies that rethinking physics, in a philosophically sound manner, should be considered a necessary aspect of doing important theoretical work going forward. Feynman, it is said, would often comment that (paraphrasing) ``it is more important for a physicist to be physically rigorous before being mathematically rigorous." It seems, however, that more and more, theorists must be ``philosophically rigorous" as well.  As the ever quotable John Wheeler said \cite{Misner/Thorne/Zurek 2009},
\begin{quote}
\textit{``Philosophy is too important to be left to the philosophers!"}
\end{quote}
\section{Entropic dynamics}
Here we develop the ED approach to formulating a quantum theory. This task consists primarily of determining an evolution for the probability $\rho = |\Psi|^{2}$ and its complement, the drift potential $\phi$, that is in accordance with the rules for processing information. That information is then supplied, in turn, in the form of constraints. Thus ED is a dynamical scheme driven by constraints. A chief aim of the ED program is to identify those constraints that inevitably result in the Schr\"{o}dinger equation and quantum theory. It is only once we have understood these constraints that we can hope to make progress in domains where our understanding is more incomplete, such as gravity. 

ED as a dynamical scheme proceeds in stages; each stage being characterized by the corresponding constraints. Imposing constraints on the microscopic dynamics, for instance, leads to a Brownian motion, and a diffusion equation in the form of a Fokker-Planck equation for $\rho$. While this is a perfectly legitimate form of dynamics, it falls short of the intended aim of a quantum dynamics, which contains two dynamical degrees of freedom --- a magnitude $\rho = |\Psi|^{2}$ and a phase. A natural way to proceed is then to promote the drift potential $\phi$ to a true dynamical variable that guides the probabilities, but which is itself then affected by the probabilities. The natural question that follows is, how does one determine the evolution of $\phi$? For this, we require, naturally, additional constraints.

In ED this choice of additional constraints has been pursued from various directions. An early hint was provided in Nelson's work on stochastic mechanics \cite{Nelson} --- quantum theory can be formulated as a non-dissipative Brownian motion by imposing that $\phi$ evolve such that an ``energy" functional be conserved. This insight was first adopted in ED \cite{Caticha 2010a}\cite{Caticha 2013} in a rather basic fashion: one assumes the existence of an energy functional $\tilde{H}$ and assumes entirely its functional form. This results in a set of coupled dynamical equations for $\rho$ and $\phi$ that are wholly equivalent to the Schr\"{o}dinger equation, but which provides little in the way of clarifying the underlying apparatus of quantum mechanics --- i.e. the Hilbert spaces, operators, Lie algebras, etc.

A similar, but more sophisticated, path was pursued in \cite{Bartolomeo et al 2014}\cite{Caticha Ipek 2014}. There, while the central focus was still on conserving an energy to update $\phi$, the  energy functional $\tilde{H}$ in question was now recognized as a true \textit{Hamiltonian} functional with $\rho$ and $\phi$ playing the role of a canonical pair whose evolution was subject to Hamilton's equations. In other words, $\rho$ and $\phi$ formed a phase space, called the \textit{ensemble} phase space (EPS), or \textit{e}-phase space for short. This represented quite a breakthrough in ED. The canonical framework is extremely rich; coming equipped with Hamiltonian generators, Lie algebras, Noether theorems, and the like. Thus with this insight ED could be seen to be encroaching on the kind of structures --- symmetries, algebras, conserved quantities --- that seem to underpin all of modern theoretical physics, including quantum theory.

As the astute reader may notice, however, this flavor of updating scheme does not work in curved space-time where the notion of a globally conserved energy is not generally available. Thus it is desirable to pursue a more robust criterion for updating that survives, even when energy conservation as an updating principle does not. One such alternative strategy is to pursue a criterion for updating that is founded on geometric principles. For instance, in \cite{Caticha 2019b}, A. Caticha models the e-phase space as a cotangent bundle, which comes naturally equipped with a symplectic structure and thus all the usual accoutrements of the canonical framework, including Hamiltonian generators, Poisson brackets, etc. However, while such an approach provides an interesting way forward in ED, there are still some lingering issues needing to be resolved; not least of which is whether the cotangent bundle is a geometric structure that is rich enough to model all physically relevant quantum states (see e.g., \cite{Molitor 2012}).


To sidestep these issues, we pursue here a manner of updating that encompasses the various methods we have discussed thus far --- energy conservation, cotangent bundles, etc. --- but which is not bound to any one procedure, in particular. The basic idea we pursue here is inspired by the work of M. Hall and M. Reginatto (see e.g., \cite{Hall Reginatto 2002}) and is as follows. Since quantum theory is obtained from ED when the probability $\rho$ and drift potential $\phi$ evolve according to Hamilton's equations, we directly impose this structure on ED. This is accomplished by requiring the e-phase space to, in fact, have the formal structure of a phase space, namely, by imposing the existence of a symplectic structure on our space; a geometrical object that allows for notions of Poisson brackets, Hamiltonian generators, and all the typical tools of the canonical formalism. This approach neither requires us to impose a conserved energy --- which may or may not be reasonable --- nor does it require us to introduce additional assumptions about the e-phase space, which may or may not not be warranted. Thus while this approach is quite versatile and useful in that it allows us to develop various aspects of ED, such as the theory of relativistic fields, it is also silent as to why such structures prove to be so useful in describing physics.

In the remainder of this chapter we deal with a system of non-relativistic quantum particles, the simplest ED model. We develop the theory of relativistic quantum scalar fields in subsequent chapters. In this section we introduce the subject matter of ED, i.e. the elements about which we make inferences, and introduce the information that guides these inferences. In \ref{Entropic time} we introduce entropic time as a scheme for organizing our succession of inferences. In section \ref{Info geo} we use information geometry to assign a metric to the configuration space of particles. Section \ref{Diffusive dynamics} introduces the dynamical equation for the probability $\rho$ while section \ref{Non-dissipative dynamics} discusses how to integrate the drift potential $\phi$ into the dynamics, while in section \ref{Canonical_ED_Particles} we introduce the canonical framework in ED. In section \ref{Quantum ED} we address how a quantum theory is obtained from these developments and in section \ref{Discussion/Conclusion} we conclude with some remarks that are relevant for subsequent chapters.
\paragraph*{On subject matter--- }
Our goal is to build a model of physics from the viewpoint of inference and information processing. The first step in formulating an inference problem consists of a specification of the subject matter: what are the microstates about which we are inquiring? Without this basic (but difficult) step, we cannot proceed.

However, a daunting aspect of formulating quantum mechanics from our viewpoint is exactly the fact that capturing what the ``\text{is}" in quantum mechanics is, has remained elusive. What, after all, are the objective building blocks of the theory, or in the words of J.S. Bell, what are its ``beables"? In the conventional Copenhagen interpretation, no answer is given, or worse yet, no answer can even, in principle, be given. It is not at all uncommon, for example, in conventional quantum mechanics to speak of quantum ``particles" that are neither particulate nor wave-like, and that have neither position nor momentum until measured; such particles are very strange indeed!

To move forward it would seem that such questions must, eventually, find a satisfactory resolution. Here the words of Jaynes \cite{Jaynes 1990} are particularly insightful:
\begin{quote}
\textit{``[Quantum mechanics] is not purely epistemological; it is a peculiar mixture describing in part realities of Nature, in part incomplete human information about Nature, all scrambled up by Heisenberg and Bohr into an omelette that nobody has seen how to unscramble. Yet we think that the unscrambling is a prerequisite for any further advance in basic physical theory. For, if we cannot separate the subjective and objective aspects of the formalism, we cannot know what we are talking about; it is just that simple.”}
\end{quote}

\paragraph*{On microstates---} In ED we proceed by taking a definitive stance on the choice of microstates. Here we consider a system of $N$ point particles in a flat Euclidean space $\mathbf{X}$ with metric $\delta_{ab}$. The particles, by assumption, always have definite positions, but their exact locations are \textit{uncertain}, and so our goal is to infer their unknown positions.

The positions of these particles can be given by their Cartesian coordinates $x_{n}^{a} \in \mathbf{X}$ ( with $n = 1 \cdots N$ denoting particle labels and $a = 1,2,3$ denoting spatial indices). The collective set of these positions $\{x_{n}^{a}\}$ can be denoted efficiently as just $x$, which is a point in the $3N$ dimensional configuration space $\mathbf{X}_{N} = \mathbf{X}\times \mathbf{X} \times \cdots \times \mathbf{X}$. (In the language of probability, the configuration space $\mathbf{X}_{N}$ would be referred to as the space of potential outcomes.) Our goal is to obtain the probability $\rho(x)$ that the system of particles forms a particular configuration $x$.

That this viewpoint, from the very outset, contradicts the Copenhagen interpretation, where observable quantities attain a value only when elicited during a measurement process, is clear. In ED, we may not know where the particles reside, but they do, in fact, reside \textit{somewhere}. The particle positions $x$ are said, in other words, to represent what is called the \textit{ontic} state of the system; they are the physical constituents of the theory. Other quantities that are typical of physics --- such as energy, momentum, angular momentum, etc. --- are not, however, given the same physical status. These quantities, rather, are statistical in nature; they are inferred from the process of measurement, and thus reflect properties of the \textit{epistemic} state, quantified by the probability $\rho(x)$ (and later, the drift potential $\phi(x)$ as well).
\paragraph*{On notation--- }
For our treatment here of the ED of particles we use notation that is in agreement with \cite{Bartolomeo et al 2014} (and \cite{Caticha 2019b}, where applicable). As we've seen so far, particle positions are denoted $x_{n}^{a}$ where $n = 1\cdots N$ denotes the particle label, and lower case Latin letters $a, b,c,$ etc. denote spatial indices, so that $a = 1,2,3$. A point in the configuration space $\mathbf{X}_{N}$ is given by $x$, while configuration space vectors are denoted $V^{A} = V_{n}^{a}$, where capital Latin letters $A, B, C,$ etc. denote the configuration space index, so that $A = (a,n)$.

Unless otherwise specified, we use the Einstein summation convention throughout. For spatial vectors $v^{a}$ and one-forms $u_{b}$ this implies that
\[v^{a} \, u_{a} = \sum_{a = 1}^{3} v^{a} \, u_{a},\quad\text{whereas}\quad V^{A} \, U_{A} = \sum_{n=1}^{N}\sum_{a=1}^{3} V^{a}_{n}\,U_{an}\]
is true for configuration space vectors $V^{A}$ and one-forms $U_{B}$. For convenience, we denote the $3N$-dimensional integration measure $d^{3N}x=dx$. Furthermore, as is standard, we use the notation $f(x): \mathbf{X}_{N}\to \mathbb{R}$ to denote functions, which map points on the configuration space $\mathbf{X}_{N}$ to a real number. On the other hand, square brackets, such as $g[\cdot]:\mathcal{F}\to\mathbb{R}$, denotes a functional dependence, which maps a point on a function space $\mathcal{F}$ to a real number. Additional notation will be introduced as the relevant material presents itself; we do not feel it is convenient to the reader to introduce new notation in a context completely divorced from the pertinent subject matter.

\paragraph*{The inference problem--- }
Having established the microstates of the system, we can proceed to considering the dynamics. ED has two primary assumptions in this regard. The first is that change happens; we do not ask why change happens, but given that it does, we wish to predict what changes are likely to occur. The second assumption is that motion of the particles is continuous such that finite changes result from the accumulation of many infinitesimally small steps. This latter assumption is crucial because it implies that to predict a trajectory we only need knowledge of an infinitesimal step; with finite changes occurring from the iteration of many small steps.

We deal with the following dynamical problem: given some initial state $x\in\mathbf{X}_{N}$ we look to determine the subsequent neighboring state $x^{\prime} = x + \Delta x \in \mathbf{X}_{N}^{\prime}$.\footnote{The question of whether the spaces $\mathbf{X}_{N}^{\prime}$ and $\mathbf{X}_{N}$ are the same dates back to Newton. Here we take the view that while $\mathbf{X}_{N}^{\prime}$ and $\mathbf{X}_{N}$ are necessarily isomorphic to one another, they are not necessarily the same. This is because they are answers to two distinct sets of questions --- where is the particle now? Where is the particle later? We discuss this point further in the section on entropic time.} In the strictest sense, however, this problem is not well posed here because both $x$ and $x'$ are in general uncertain. The distribution that encodes this uncertainty in the initial state $x$ and final state $x'$ is given by the \textit{joint} probability $\rho(x',x)$. Using the product rule we can rewrite this probability as $\rho(x',x) = P(x'|x)\, \rho(x)$. It is the probability of the infinitesimal step, given by the \textit{transition probability} $P(x'|x)$, that determines the information relevant for determining a final state given an arbitrary initial state. Thus our goal at this juncture is to determine an appropriate transition probability $P(x'|x)$.

\paragraph*{Maximum Entropy Method--- }
The procedure for selecting the appropriate transition probability is, of course, provided by the method of Maximum Entropy (ME). (The details of this method have been discussed elsewhere in this thesis; see Ch. 2 for details.) That is, the optimal $P(x'|x)$ is that which maximizes the entropy
\begin{equation}
S[P,Q] = - \int dx' P(x'|x)\ln\frac{P(x'|x)}{Q(x'|x)},
\label{Particles Entropy Transition probability}
\end{equation}
relative to a \textit{prior} distribution $Q(x'|x)$ and subject to the appropriate constraints.
\paragraph*{Prior--- }
A crucial part of the ME method is the choice of prior. Here the prior distribution $Q(x'|x)$ reflects the information that the motion unfolds in infinitesimal steps, but which is otherwise maximally uninformative. Such an extreme state of ignorance is one that ignores possible correlations between degrees of freedom
\begin{equation}
Q(x^{\prime}|x) = \prod_{n} Q(x_{n}^{\prime}|x_{n}).
\end{equation}
Moreover, when confronted with a complete lack of information, it is natural to appeal to symmetry; a principle dating back to the time of Laplace (see e.g. \cite{Jaynes 1968} or \cite{Caticha/Preuss 2004}) and his principle of insufficient reason. Here, for instance, we can take advantage of the rotational and translational symmetries of the three-dimensional Euclidean space $\mathbf{X}$, which the particles reside in, so that
\begin{equation}
Q(x'|x) = \prod_{n} Q(x_{n}'|x_{n}) = \prod_{n} Q(|\Delta x_{n}|).
\end{equation}

The prior $Q(x'|x)$ that reflects this state of affairs can itself be derived as a result of the ME method. Consider the entropy
\begin{equation}
S[Q,\mu] = - \int dx' Q(x'|x)\ln\frac{Q(x'|x)}{\mu(x'|x)}
\label{Particles Relative Entropy Priors}
\end{equation}
relative to the uniform measure $\mu(x'|x)$. Now maximize $S[Q,\mu]$ subject to normalization and the $N$ constraints\footnote{We use a notation here where $ \left\langle A(x') \right\rangle_{Q} = \int dx' Q(x'|x) A(x')$. If the subscript $Q$ is not written, then it means we perform an expectation value with respect to the posterior $P(x'|x)$ such that $ \left\langle A(x') \right\rangle = \int dx' P(x'|x) A(x')$. We suspect the convention will be obvious from context.}
\begin{equation}
 \left\langle \Delta x_{n}^{a}\Delta x_{n'}^{b}\right\rangle_{Q} = \int dx^{\prime} \, Q(x^{\prime}|x) \, \delta_{ab} \Delta x_{n}^{a}\, \Delta x_{n}^{b} = \Delta\kappa_{n},
\label{Particles Continuity Constraint}
\end{equation}
which number one per particle, and which respect the rotational and translational symmetries of the underlying space. Continuity is enforced by requiring the constants $\Delta\kappa_{n}$ to be small, i.e. $\Delta\kappa_{n}\to 0$. The result is a product of Gaussians,
\begin{equation}
 Q(x'|x) \propto \exp - \frac{1}{2}\sum_{n} \alpha_{n} \Delta x_{n}^{a}\Delta x_{n}^{b}\delta_{ab}.
\label{Particles Prior Q}
\end{equation}  
The $N$ Lagrange multipliers $\alpha_{n}$ are a set of particle dependent parameters, which we will later take to be spatially constant. The common practice for determining these multipliers is familiar from statistical mechanics: we compute the constraints (\ref{Particles Continuity Constraint}) using the transition probability (\ref{Particles Prior Q}) and solve for the multipliers $\alpha_{n}$ in terms of the constraints. The same procedure applies here as well.

To determine these multipliers in the standard style we must compute the expected Euclidean distance $ \delta_{ab}\left\langle \Delta x_{n}^{a} \Delta x_{n}^{a} \right\rangle_{Q}$ in (\ref{Particles Continuity Constraint}). To do this we exploit the fact that we deal with a Gaussian distribution which has mean
\begin{equation}
\left\langle \Delta x_{n}^{a}\right\rangle_{Q} = 0 \quad\text{and variance}\quad \left\langle \Delta x_{n}^{a}\Delta x_{n'}^{b}\right\rangle_{Q} = \frac{1}{\alpha_{n}}\delta_{nn'}\delta^{ab}.
\end{equation}
Thus we can easily compute (\ref{Particles Continuity Constraint}) to obtain
\begin{equation}
 \delta_{ab}\left\langle \Delta x_{n}^{a} \Delta x_{n}^{a} \right\rangle_{Q} = \frac{3}{\alpha_{n}} = \Delta\kappa_{n},
\end{equation}
where $\Delta\kappa_{n}$ tends to zero. This implies that $1/\alpha_{n}$ must be small, or rather, that $\alpha_{n}\to \infty$.

Alternatively, inspection of (\ref{Particles Prior Q}) also shows that the probability of finite steps is vanishingly small only in the limit $\alpha_{n}\to \infty$. This, of course, amounts to setting the width of the Gaussian to be infinitely small so that (\ref{Particles Prior Q}) is, for all intents and purposes, a delta function. (The matter of choosing an explicit form for these multipliers will be dealt with later on, within the context of constructing time in ED.) This result is unaltered as we update to the posterior distribution $P(x'|x)$.
\paragraph*{Drift constraint--- }
On their own, the $N$ constraints eqns.(\ref{Particles Continuity Constraint}) lead to a rather simple dynamics; an isotropic diffusion that demonstrates no correlations among the particles. To go further towards a richer dynamics that exhibits such correlations, we require a device that allows for interdependence between the particles. This is achieved by introducing a global scalar function $\phi(x)$ (i.e. a function defined on the configuration space $\mathbf{X}_{N}$, rather than real space $\mathbf{X}$), which we refer to as the \textit{drift potential} and that mediates the inter-particle relationships in our system.\footnote{Alternative formulations of ED (see e.g., \cite{Caticha 2010a}\cite{Caticha 2013}) favor a different strategy, introducing some additional \textit{y}-variables whose entropy $S_{y}(x)$ assumes the role normally reserved for the drift potential. However, these models run into some trouble, since entropy must necessarily be single-valued function, whereas the phase of the quantum state $\Psi(x)$ must, in fact, be multi-valued to accommodate, for example, states of non-zero angular momentum. For this reason we work exclusively with the drift potential $\phi(x)$.}

We include this information into the dynamics of the particles by imposing an additional global constraint on the change of the drift potential $\phi(x)$:
\begin{equation}
\Delta \phi(x) = \phi(x^{\prime}) - \phi(x) = \frac{\partial\phi(x)}{\partial x_{n}^{a}} \Delta x_{n}^{a},
\label{Particles phi taylor expansion}
\end{equation}
so that
\begin{equation}
\left\langle \Delta \phi(x) \right\rangle = \int dx^{\prime} \, P(x^{\prime}|x) \, \frac{\partial\phi(x)}{\partial x_{n}^{a}} \Delta x_{n}^{a} = \Delta\kappa^{\prime},
\label{Particles Drift Constraint}
\end{equation}
where $\Delta\kappa^{\prime}$ is a small number. (Additional terms higher than first order in the Taylor expansion (\ref{Particles phi taylor expansion}) contribute negligibly as $\kappa_{n}\to 0$.) 

\begin{remark*}
The net result of imposing this constraint is that the particles are guided by the gradient of the drift potential, and as such, $\phi(x)$ plays the role of a pilot wave, similar in spirit to that of Bohmian mechanics. Eventually $\phi(x)$ will be responsible for many quintessential quantum phenomenon, such as entanglement and interference. A recent improvement in ED \cite{Caticha 2019b} has been to recognize that the introduction of such a function is not as \emph{ad hoc} as it might, nevertheless, first appear. Note that in the ED we develop here there are two important pairings. The first pertains to the entropic origins of ED, and consists of probabilities and their constraints. Another important pairing occurs when we adopt a Hamiltonian formalism to update the evolving constraints. It is here that we must introduce a canonical coordinate, the probability $\rho(x)$, and its conjugate momentum. In an ED driven by constraints and by Hamiltonians it is therefore exceedingly natural that the two pairings may have some intimate connection. This is accomplished by assigning to the drift potential $\phi(x)$ a dual role, first as encoding the information in the form of constraints, and second, by identifying it as the momentum conjugate to $\rho(x)$.
\end{remark*}

\begin{remark*} Some properties of the drift potential $\phi(x)$ are inherited from its role as the momentum conjugate to $\rho(x)$, while others must be imposed for our model to be sufficiently flexible. To this end, we claim that $\phi(x)$ is a \textit{multi-valued} function, such that $\phi(x)$ and $\phi(x)+2\pi n \beta$ represent the same constraint, where $n$ is an integer and $\beta$ is a dimensionless constant. This will be discussed more thoroughly later in the chapter.\end{remark*}

\paragraph*{The posterior distribution--- }
To select a particular posterior $P(x^{\prime}|x)$ that satisfies the constraint (\ref{Particles Drift Constraint}), we again resort to the ME method. The appropriate criterion for ranking the posteriors $P(x'|x)$, which satisfy (\ref{Particles Drift Constraint}) relative to the prior $Q(x'|x)$ is the relative entropy $S[P,Q]$ in (\ref{Particles Entropy Transition probability}). The most preferred distribution $P(x^{\prime}|x)$ is the one that maximizes $S[P,Q]$. Choosing the optimal $P(x^{\prime}|x)$ is, as usual, implemented by employing the method of Lagrange multipliers. Maximizing the entropy $S[P,Q]$ subject to (\ref{Particles Drift Constraint}), and normalization, we have
\begin{equation}
P(x^{\prime}|x) = \frac{1}{\zeta(\alpha_{n},\alpha^{\prime})} \, \exp -\sum_{n} \left( \frac{\alpha_{n}}{2} \delta_{ab} \Delta x_{n}^{a}\, \Delta x_{n}^{b} - \alpha^{\prime}\,\frac{\partial\phi(x)}{\partial x_{n}^{a}} \Delta x_{n}^{a}\right),
\end{equation}
where $\zeta$ is a normalization constant, and $\alpha^{\prime}$ is a single multiplier for the global constraint (\ref{Particles Drift Constraint}). Since the distribution $P(x^{\prime}|x)$ consists only of terms linear and quadratic in $x^{\prime}$, it is a Gaussian,
\begin{equation}
P(x^{\prime}|x) = \frac{1}{Z(\alpha_{n})} \, \exp -\sum_{n} \, \frac{\alpha_{n}}{2} \left( \Delta x_{n}^{a}- \frac{\alpha^{\prime}}{\alpha_{n}}\,\frac{\partial\phi(x)}{\partial x_{na}}\right)\left( \Delta x_{na}- \frac{\alpha^{\prime}}{\alpha_{n}}\,\frac{\partial\phi(x)}{\partial x_{n}^{a}}\right),
\label{Particles Transition probability}
\end{equation}
where, for convenience, we have raised and lowered indices appropriately using the Euclidean metric $\delta_{ab}$.

Concerning the remaining multiplier $\alpha^{\prime}$, we have no strong restrictions here on what its value must be. Work done by D. Bartolomeo and A. Caticha in \cite{Bartolomeo Caticha 2016} suggests that this freedom in the choice of $\alpha^{\prime}$ reflects a universality class of microscopic models; which includes the smooth trajectories of Bohmian mechanics as a special case. More succinctly: different choices of $\alpha^{\prime}$ lead to different microscopic dynamics, but all of which result in the same quantum theory. For our work here, we continue along the lines of \cite{Bartolomeo et al 2014}\cite{Caticha Ipek 2014} by absorbing, without loss of generality, $\alpha^{\prime}$ into the undetermined $\phi$, such that $\alpha^{\prime} \, \phi = \phi^{\prime}$ and then redefine $\phi^{\prime}\to \phi$; or, equivalently, we can simply set $\alpha^{\prime} = 1$.

\paragraph*{Particle motion--- }
We are now in a position to analyze what microscopic dynamics results from the the transition probability $P(x'|x)$. Since $P(x^{\prime}|x)$ is a Gaussian, it is appropriate to write a typical step as consisting of an expected drift with additional fluctuations, $\Delta x_{n}^{a} = \Delta \bar{x}_{n}^{a} + \Delta w_{n}^{a}$. This allows us to easily compute the moments of $P(x^{\prime}|x)$. For example, the expected step is given by
\begin{equation}
\left\langle \Delta x_{n}^{a} \right\rangle = \frac{1}{\alpha_{n}}\delta^{ab} \frac{\partial\phi(x)}{\partial x_{n}^{b}} = \Delta\bar{x}_{n}^{a}.
\label{Particles Expected Step}
\end{equation}
The fluctuations, on the other hand, are given by,
\begin{equation}
\left\langle \Delta w^{a}_{n} \right\rangle = 0 \quad\text{and}\quad \left\langle \Delta w^{a}_{n} \, w^{b}_{n'} \right\rangle = \frac{1}{
\alpha_{n}}\delta_{ab}\delta_{nn'}.
\label{Particles Fluctutions}
\end{equation}
(Additional moments can be computed using the standard techniques for computing Gaussian integrals.) As anticipated, the expected drift $\Delta \bar{x}_{n}^{a}$ and fluctuations $\Delta w_{n}^{a}$ do, indeed, tend towards small values in the limit $\alpha_{n}\to \infty$; which confirms that the particles in our model do, in fact, travel along continuous paths.

To tease out additional insight, let us inspect the behavior of these trajectories in the small step limit, $\alpha_{n}\to \infty$. From equations (\ref{Particles Expected Step}) and (\ref{Particles Fluctutions}) we see there that, while the fluctuations are of order $O(\alpha_{n}^{-1/2})$, the expected steps are of order $O(\alpha_{n}^{-1})$. Thus, as $\alpha_{n}$ tends to large values, the fluctuations dominate; the resulting motion is thus continuous, but not differentiable. In short: the particles undergo a Brownian motion.
\begin{remark*}
It is worth pointing out that this simple result has quite wide implications. Brownian motions are ubiquitous; with applications in physics, biology, finance, etc. That such models are easily obtained within the ED framework suggests that it may be possible to apply the notions developed here in other fields as well.\footnote{See e.g., the work by M. Abedi and D. Bartolomeo  \cite{Abedi Bartolomeo 2019a}\cite{Abedi Bartolomeo 2019b} in using ED for developing financial models.}
\end{remark*}
\section{Entropic time\label{Entropic time}}
In an inference-based theory, our knowledge of the relevant physics is implemented through constraints. The exact nature of these constraints depends on what information one is trying to incorporate into the model. For ED, a dynamical theory, one place where such considerations arise is at the level of time. While time is intimately related with dynamics and \textit{vice versa}, neither are necessary ingredients of inference; these features must be supplemented to the theory. Here we introduce a scheme for incorporating time into ED that we call \textit{entropic time}.\footnote{Entropic time, as a scheme for introducing time into a theory of inference, was first introduced by A. Caticha \cite{Caticha 2010a}. The current scheme mirrors very closely this original viewpoint by Caticha, but contributes a slight twist in perspective that makes the framework more amenable to future compatibility with relativistic theories.}

\subsubsection*{Entropic instants}
In ED, change happens, and it happens in a way such that motion occurs in a continuous sequence of steps. Or rather, we might say that motion unfolds as a succession of \textit{states}. It is these states that we are uncertain of, but wish to infer, prompting for each such instance the singular question: ``What \textit{is} the state of our system?" In fact, one might turn this on its head and instead claim that ED unfolds as a sequence of questions; a perspective that makes clear the fundamental role of information, or \textit{lack} thereof.

Compare this now with the structure of dynamical theories: dynamics evolves as a series of instants: ED unfolds as a sequence of questions. The language itself is suggestive: in ED, a question is an instant. The construction of this question relies on a few important, but distinct, elements:
\begin{itemize}
\item[1.] A formal tool for labeling the succession of queries;
\item[2.] A specification of the subject matter to which our question pertains; i.e. the propositions $x\in\mathbf{X}_{N}$ about which we are uncertain;
\item[3.] The variables $\rho(x)$ and $\phi(x)$ that captures the informational state.
\end{itemize}
Let us explain. On one hand, a question is itself an admission of uncertainty. Why else, after all, would we ask a question, if not for the fact that we lack complete information? This, in turn, evokes the use of probabilities for quantifying this uncertainty, hence $\rho(x)$. On the other hand, in order to even begin an inquiry, we must supply enough structure to phrase the question in the first place. This amounts to specifying not only which question, in particular, that we are asking, but also what the possible answers to the question could be; what we might call the kinematic structure of the problem.\footnote{The relation between questions, inquiry, and probability is an interesting one that dates back to the work of R.T. Cox \cite{Cox 1979}. Some more recent research that explores this issue along the lines developed by Cox was done by K. Knuth, and can be found in \cite{Knuth 2003}\cite{Knuth 2005}. The viewpoint that a probability distributions and questions are intimately linked is explored by A. Caticha in \cite{Caticha 2004b}. Our colorful exposition on instants in ED was inspired heavily by these considerations.}

The first ingredient is crucial in regards to this latter point as it is through the introduction of just such a tool that we can even keep track of which question we are entertaining. Indeed, ``where are the particles now?" and ``where are the particles later?" are different questions because the information available for answering these questions through $\rho(x)$ and $\phi(x)$, has changed. Thus we require a scheme that allows us maintain these distinctions. In a flat space-time it is reasonable to adopt a notion of time and of instant that is in the Newtonian spirit. This means that a single parameter will be sufficient for labeling the succession of queries and amounts to a \textit{unique} and global notion of simultaneity.

\paragraph*{Comments}
Let us highlight some interesting points. When dealing with a single particle, for example, we desire to know its location, which is labeled by three values $\vec{x} = (x^{1},x^{2},x^{3})$. It comes as no surprise here that the components $(x^{1},x^{2},x^{3})$ of a single particle occur at an instant; this is, after all, the definition of a point particle. Things become more peculiar when we consider multiple particles such that a configuration is defined by $x\in\mathbf{X}_{N}$. The implication is that the point $x = (\vec{x}_{1},\cdots,\vec{x}_{N})$, and consequently all its components --- the locations of each particle --- occur at the very same instant.


However our construction is weirder still --- the set of simultaneous ``events" is the \textit{entire} configuration space $\mathbf{X}_{N}$, not just a single point $x\in\mathbf{X}_{N}$. This latter statement is closely related to the probabilistic nature of our theory: although a particular configuration of particles $x$ occurs at an instant, we are uncertain of \textit{which} $x$ is the correct one. Thus \textit{any} $x$ is, in principle, possible (with some probability $\rho(x)$), and so, all must be treated \textit{equally} at the kinematic level. This justifies calling $\mathbf{X}_{N}$ the set of simultaneous events.

Things are even more interesting in a relativistic theory. In that scenario, there is greater flexibility in the notion of an instant, owing to the relativity of simultaneity. In order to accommodate this situation, we will require an additional construction, that of a hypersurface in space-time. It is this additional kinematic structure that allows us to even phrase the question --- ``what is the state of the system \textit{now}?" --- by identifying a notion of ``now" that is compatible with a relativistic theory. Addressing this issue will be the topic of subsequent chapters.
\subsubsection*{Ordered instants}
The goal is to predict, given what we know now, what will likely occur next. That is, once supplied with some information $\rho(x)$ about where the particles currently reside $x\in\mathbf{X}_{N}$, we wish to determine the probability $\rho^{\prime}(x^{\prime})$ that they are located at some $x'\in\mathbf{X}^{\prime}_{N}$ at a subsequent instant. It helps to be explicit with the notation. The proposition that the particles are found at the state $x_{i}$ at the $i^{\text{th}}$ step will be denoted $s_{i} = x_{i}$; for the $(i+1)^{\text{th}}$ step, $s_{i+1} = x_{i+1}$, and so on, with each instant labeled by the index $i$.

To make predictions about where the particles are at the step $s_{i+1}$ we require information both about where the particles are, and where they will be going. This information is captured by the \textit{joint} probability distribution introduced earlier
\begin{equation}
\rho(s_{i+1} = x_{i+1}, s_{i} = x_{i}) = P(s_{i+1} = x_{i+1}| s_{i} = x_{i})\rho(s_{i} = x_{i}),
\label{Particles joint distribution}
\end{equation}
which gives the probability that the particles begin in a configuration $x_{i}\in\mathbf{X}_{N}^{(i)}$ \textit{and} end at the configuration $x_{i+1}\in\mathbf{X}^{(i+1)}_{N}$. When the configurations $x_{i}$ and $x_{i+1}$ correspond to successive states (as the labels $i$ and $i+1$ imply) the transition probability $P(s_{i+1} = x_{i+1}| s_{i} = x_{i})$ is given precisely by the ME distribution (\ref{Particles Transition probability}).

To determine the probability of this later position $x_{i+1}$ we apply the ``sum rule" of probability theory to the joint distribution (\ref{Particles joint distribution}) to obtain
\begin{equation}
\rho(s_{i+1} = x_{i+1}) = \int dx_{i}\, P(s_{i+1} = x_{i+1}| s_{i} = x_{i})\rho(s_{i} = x_{i}).
\label{Particles stepping equation}
\end{equation}
Just as $\rho(s_{i} = x_{i})$ reflects the informational state at a \textit{prior} instant, the distribution $\rho(s_{i+1} = x_{i+1})$ reflects the \textit{posterior} statistical state corresponding to the next instant.

The temporal ordering can be made even more apparent by introducing a label $t\in\mathbb{R}$, which we call time. This allows us to write $\rho(s_{i} = x_{i})=\rho_{t}(x)$ and $\rho(s_{i+1} = x_{i+1}) = \rho_{t'}(x')$, where $t'$ is a time that is slightly later than $t$. The result is an iterative process: from an instant $\rho_{t}(x)$ we update to a subsequent instant $\rho_{t'}(x')$ using the maximum entropy distribution $P(x'|x)$, and from $\rho_{t'}(x')$ to another later instant $\rho_{t''}(x'')$, etc. The equation (\ref{Particles stepping equation}) can now be rewritten as
\begin{equation}
\rho_{t'}(x') = \int dx\, P(x'| x)\rho_{t}(x),
\label{Particles Chapman kolmogorov}
\end{equation}
which is the dynamical equation we seek. In short: dynamics is composed of a succession of instants. Each \textit{posterior} $\rho_{t'}(x')$ is obtained from a \textit{prior} $\rho_{t}(x)$ by way of maximizing an entropy to obtain $P(x'|x)$; we call this an \textit{entropic updating} scheme.

\paragraph*{The arrow of entropic time--- }
The task of dynamics is, given an arbitrary \textit{initial} state, determine the \textit{final} state that follows. Viewed from the perspective purely of probability, this task is symmetric: we can decompose the joint distribution $\rho(x',x)$ into either
\begin{equation}
\rho(x',x) = P(x'|x)\,\rho_{t}(x)\quad\text{or}\quad\rho(x',x) = \rho_{t'}(x')P(x|x')
\end{equation}
using the product rule. Indeed, there is nothing in the rules of probability that suggests that we update from an initial $\rho_{t}(x)$ to a final $\rho_{t^{\prime}}(x')$ or \textit{vice versa}.

It is the action of obtaining $P(x'|x)$ rather than $P(x|x')$ via the ME method that breaks this symmetry. That is, while $\rho_{t'}(x')$ is obtained from $\rho_{t}(x)$ using (\ref{Particles Chapman kolmogorov}) and the ME distribution $P(x'|x)$, the ``time-reversed" equation
\begin{equation}
\rho_{t}(x) = \int dx' \, P(x|x') \,\rho_{t'}(x')
\label{Particles time reversed CK equation}
\end{equation}
is acquired using $P(x|x')$, which \textit{must} be determined through Bayes' theorem
\begin{equation}
P(x|x') = \frac{\rho_{t}(x)P(x'|x)}{\rho_{t'}(x')}.
\end{equation}
The ``time reversed" dynamics determined by $P(x|x')$ follows only \textit{after} we have computed $\rho_{t'}(x')$ using the forward equation (\ref{Particles Chapman kolmogorov}). 


The consequence of this is that there is a intrinsic directionality to the flow of entropic time: we update from an initial $\rho_{t}(x)$ to a final $\rho_{t'}(x')$ using the ME distribution $P(x'|x)$. Moreover, this asymmetry is inferential in nature; resulting from an asymmetry in what is known, rather than of anything mechanical.
\paragraph*{Duration--- }
The final ingredient to a notion of time, is that of the duration, i.e. the separation $\Delta t$ between instants. The basic criterion here is that of \textit{convenience}: in ED, time is \textit{designed} such that the dynamics is as \textit{simple} as possible. For short steps, eqns.(\ref{Particles Expected Step}) and (\ref{Particles Fluctutions}) reveal that the dynamics is dominated by fluctuations. The dynamics then is most simplified by choosing a measure of duration $\Delta t$ defined in terms of the fluctuations, such that
\begin{equation}
\alpha_{n} = \frac{m_{n}}{\eta}\frac{1}{\Delta t}\quad\text{so that}\quad\left\langle \Delta w^{a}_{n} \, w^{b}_{n'} \right\rangle = \frac{\eta}{
m_{n}}\delta_{ab}\delta_{nn'} \Delta t,
\label{Particles Duration}
\end{equation}
where the proportionality is given by some particle dependent constants $m_{n}$, and a constant $\eta$, which sets the units of length to that of time. While the constants $m_{n}$ will be later identified with the masses of the particles, the constant $\eta$ will be related to Planck's constant $\hbar$.

Regarding the choice of $\Delta t$, a natural option presents itself in the context of flat space and Newtonian-style time: we choose $\Delta t$ to be a spatial constant that is independent of time so that entropic time flows ``equably everywhere and every-when." With this choice of $\alpha_{n}$ in eq.(\ref{Particles Duration}), the transition probability $P(x'|x)$ takes the form of a standard \textit{Wiener} process.

\begin{remark*}
Typically in physics, time is \textit{defined} such that motion is as simple as possible. In Newtonian mechanics, for example, we define time such that the motion of free particles is as simple as possible: free particles travel equal distances in equal times. The free particle provides a clock in ED as well: a particle undergoes equal fluctuations in equal times.
\end{remark*}
\subsubsection*{Formal structure of entropic time}
So far we have eschewed a strictly formal introduction of entropic time in favor of an approach that stresses the subtle interpretational aspects of the formalism. However, it is also, of course, important to be clear about the kind of mathematical structures that this construction implies. It is our goal here to touch on these matters so as to draw contrast to the situation in curved space-time.

We consider here a system of non-relativistic particles. It follows with little surprise then that the notion of time that we adopt here shares some formal similarity with the Newtonian view of time. For instance, it is well known (see e.g., \cite{Schutz}) that a Newtonian theory can be formulated as a fiber bundle. The construction is the following: we can take the time $t\in\mathbb{R}$ to be the base manifold and take each fiber to be a copy of real space $\mathbb{R}^{3}$. (In addition, each fiber is also endowed with a Galilean symmetry group.) Looked at from this angle, it is clear that a single number $t\in\mathbb{R}$ specifies an instant of time to which we associate a copy of three-dimensional Euclidean space; this, in turn, makes it explicit that each point $\vec{x}\in\mathbb{R}^{3}$ is simultaneous with one another in Newtonian physics.

A similar kinematic setup can be attributed to ED as well (although, to be clear, we do \textbf{not} assume Newton's laws of motion). In particular, the design of entropic time too results in a fiber bundle structure, where $t\in\mathbb{R}$ is again the base manifold, but where each fiber is a copy of the configuration space $\mathbf{X}_{N}$ (again, by construction, each $x\in\mathbf{X}_{N}$ is simultaneous with one another, analogous to the role of $\vec{x}\in\mathbf{X}$ in the Newtonian theory).

In addition to this, however, in ED we associate to each instant a probability $\rho(x)$, which is a distribution over the fibers. It is this latter ingredient that is of crucial importance: at the kinematical level, each instant parameterized by $t$ is identical, it is the inclusion of the probability $\rho_{t}(x)$ that truly distinguishes each instant from the next. It is this character of entropic time, for instance, that introduces an asymmetry into the flow of time.
\section{The information geometry of configuration space}
\label{Info geo}
Before proceeding, it is convenient to consider the geometry of the $3N$-dimensional configuration space $\mathbf{X}_{N}$. One might anticipate that $\mathbf{X}_{N} = \mathbf{X}\otimes\cdots\otimes\mathbf{X}$, being $N$ copies of flat Euclidean space $\mathbf{X}$,  that it too might be flat. Considerations from information geometry, in fact, confirm this intuition.

The transition probability $P(x'|x)$ is one probability distribution for every point $x\in\mathbf{X}_{N}$. This endows the configuration space $\mathbf{X}_{N}$ with the structure of a statistical manifold (see e.g. \cite{Caticha 2012}\cite{Amari 1985}). The benefit of this insight is that there is a natural metric associated to a statistical manifold, the \textit{information metric}, which is unique (see e.g. \cite{Campbell 1986} and \cite{Cencov 1981}) up to an overall scale factor.

The information metric that is appropriate to the transition probability $P(x'|x)$ is given by
\begin{equation}
 \gamma_{AB} = C\int dx' \, P(x'|x)\frac{\partial\log P(x'|x)}{\partial x^{A}}\frac{\partial\log P(x'|x)}{\partial x^{B}},
\label{Particles info_metric}
\end{equation} 
where we have used the notation that capital Latin indices $A = (a,n)$ denote both spatial component and particle index, e.g. $x^{A} = x_{n}^{a}$, and where $C$ is an arbitrary positive constant. Using for $P(x'|x)$ the transition probability (\ref{Particles Transition probability}) and making use of equations (\ref{Particles Expected Step}) and (\ref{Particles Fluctutions}), we have for short steps ($\Delta t\to 0$) that
\begin{equation}
\gamma_{AB} = \frac{C m_{n}}{\eta\Delta t}\delta_{nn'}\delta_{ab} \equiv \frac{C m_{n}}{\eta\Delta t}\delta_{AB}.
\label{Particles info_metric config}
\end{equation}
The divergence as $\Delta t\to 0$ occurs because $\gamma_{AB}$ measures the distinguishability between neighboring probability distributions. For small $\Delta t$ the width of the Gaussian in $P(x'|x)$ becomes sharper, and thus it becomes easier to distinguish it from a distribution $P(x'|x+\delta x)$ in its vicinity. Thus $\gamma_{AB}$ diverges.

To define a geometry that remains finite, even for short steps, we can choose the constant $C\propto \Delta t$. Better still, we can define the ``mass" tensor
\begin{equation}
m_{AB} = \gamma_{AB}\frac{\eta\Delta t}{C} = m_{n}\delta_{AB},
\label{Particles mass tensor}
\end{equation}
since it is always this combination that appears in our equations. Moreover, from the inverse metric $\gamma^{AB}$ we can introduce the inverse mass tensor,
\begin{equation}
m^{AB} = \frac{C}{\eta\Delta t}\gamma^{AB} = \frac{1}{m_{n}}\delta^{AB}.
\label{Particles fluctuation tensor}
\end{equation}

Note that it has been common practice in physics (see e.g., \cite{Lanczos}) to attribute to the $N$-particle configuration space a geometry, where the mass tensor plays the role of a metric. It is therefore interesting to see that here that the mass tensor truly is, up to some scale factors, the metric of configuration space, and that this results directly from considerations of information geometry!

The availability of a metric on the configuration space $\mathbf{X}_{N}$ allows us to rewrite our main results thus far. A generic displacement can be written as a drift plus a fluctuation $\Delta x^{A} = b^{A}\Delta t + \Delta w^{A}$. The drift is given by
\begin{equation}
b^{A} = \frac{\left\langle\Delta x^{A}\right\rangle}{\Delta t} = m^{AB}\partial_{B}\phi,
\label{Particles drift velocity}
\end{equation}
where $\partial_{A} = \partial/\partial x_{n}^{a}$, and where the fluctuations yield
\begin{equation}
\left\langle \Delta w^{A} \right\rangle = 0 \quad\text{and}\quad \left\langle \Delta w^{A} \, w^{B} \right\rangle = \eta\, m^{AB}\Delta t.
\label{Particles Fluctution tensor}
\end{equation}
As seen in eq.(\ref{Particles Fluctution tensor}), the expected square fluctuations are thus proportional to the inverse mass tensor, for this reason it is appropriate to also call this the ``fluctuation" tensor.
\section{On multi-valuedness} We previously asserted that $\phi(x)$ must be a multi-valued function. Here we demonstrate this feature by appealing to a special case, one where rotational symmetries are present. Crucially, we do not claim to have derived this property of the drift potential here. Thus ultimately the multi-valuedness of $\phi(x)$ must be regarded as an assumption in this version of ED. While our argument here provides some evidence for this, it does not account for why such a structure is present even in cases where no periodicity is present at all, such as with the case of the free particle. It appears likely that a resolution to this issue will require a deeper analysis of ED and its geometrical structure.





\paragraph*{Uniform circular motion--- }
Consider a single particle whose position, although uncertain, can be described by a point $\vec{x} \in \mathbb{R}^{3}$. The expected infinitesimal displacement of this particle is given by eq.(\ref{Particles Expected Step})
\begin{equation}
\left \langle \Delta \vec{x}\right \rangle = \frac{\eta}{\mu}\vec{\nabla}\phi(\vec{x})\Delta t ~,\label{Particles multi-valued expected step}
\end{equation}
where we have used $\alpha = \mu/\eta \Delta t$, as above, with $\mu$ to be interpreted as the particle mass. The periodic motions we are interested in are most easily expressed in spherical coordinates $\vec{x} = (r,\theta,\varphi)$, where $r$ is the radial distance, $\theta$ is the polar angle, and $\varphi$ is the azimuthal angle. In these coordinates, eq.(\ref{Particles multi-valued expected step}) takes the form
\begin{equation}
\left \langle \Delta \vec{x}\right \rangle  = \frac{\eta}{\mu}\,\vec{\nabla}\phi(\vec{x})\,\Delta t  =  \frac{\eta}{\mu} \left (\frac{\partial \phi}{\partial r}\hat{r}+\frac{1}{r}\frac{\partial \phi}{\partial \theta}\hat{\theta}+\frac{1}{r\sin\theta}\frac{\partial \phi}{\partial \varphi}\hat{\varphi}\right )\Delta t~.
\end{equation}
Furthermore, we consider a situation of extreme symmetry, one in which the expected particle trajectory passes through equal angles in equal times, i.e. uniform circular motion. We choose coordinates so that the motion takes place in the azimuthal plane (i.e. $\theta = \pi )$, leading to
\begin{equation}
\left \langle \Delta \vec{x}\right \rangle  = \frac{\eta}{\mu}\vec{\nabla}\phi(\vec{x})\Delta t = v\, \Delta t \,\hat{\varphi}~,\label{Particles multi-valued circular a}
\end{equation}
where $v$ is a constant parameter with dimensions of length per unit time. Of course, with this the radial and polar angle velocities vanish, which implies that the drift potential is itself independent of the radial length $r$ and polar angle $\theta$, so that $\phi(\vec{x})= \phi(\varphi)$. From this, eq.(\ref{Particles multi-valued circular a}) directly implies that $ \phi(\varphi)$ satisfies
\begin{equation}
\frac{d\phi(\varphi)}{d\varphi} = \frac{\mu\, R \, v}{\eta}~,\label{Particles multi-valued circular b}
\end{equation}
where we have changed from partial to ordinary derivatives, and where $R$ is a fixed length that gives the radius of the orbit. We can easily solve eq.(\ref{Particles multi-valued circular b}) for $\phi(\varphi)$, yielding
\begin{equation}
\phi(\varphi) = \frac{\mu R  v}{\eta}\, \varphi + c~,
\end{equation}
up to an overall integration constant $c$. 
\begin{remark*}
The solution shows that $\phi = \beta \, \varphi$, where $\beta = \mu R v/\eta$. Since $\varphi$ and $\varphi + 2\pi$ represent the exact same angle, then for fixed constant $c$ the drift potentials $\phi$ and $\phi + 2\pi \beta$ correspond to the same constraint. In quantum mechanics we also have that $\beta$ is integer valued, in which case the equivalence between $\phi $ and $ \phi + 2\pi \beta$ is precisely that of an angular variable. In other words, for the drift potential to generate motions consistent with the rotational symmetries of flat space, we must allow it to be a multi-valued function.
\end{remark*}
\begin{remark*}
Of course, the requirement that $\beta$ is integer valued is just a restatement of the \emph{ad hoc} quantization conditions familiar from the old quantum mechanics of Bohr and Sommerfeld. At this stage of the ED approach, there are no such restrictions on $\beta$, nor should they be expected. This, instead, comes from the requirement that the wave function $\Psi$ be single valued. A more complete discussion of this issue in ED is given in \cite{Caticha 2019b}.


\end{remark*}
\begin{remark*}
As discussed above, once it has been established that certain physical solutions of $\phi(x)$ must be multi-valued, it is reasonable to then allow that $\phi(x)$ may be multi-valued in a more general sense. Consider a situation where the motion is not periodic, such as the free particle, which will have $\vec{\nabla}\phi(\vec{x}) = \vec{k}$, where $\vec{k}$ is an arbitrary constant vector.\footnote{One can determine this by requiring the drift velocity be translation invariant, as is appropriate for a free particle, leading to a constant vector.} Therefore up to an additive constant we have that $\phi(\vec{x}) = \vec{k}\cdot\vec{x}$. While in principle the vector $\vec{k}$ is just a collection of constant parameters that determine the (expected) velocity of the particle, once $\phi(x)$ is a multi-valued function the vector $\vec{k}$ must assume an additional interpretation as a wave vector. To see this, consider a shift $\vec{x} \to \vec{x}+\vec{\lambda}$, for some vector $\vec{\lambda}$, so that $\phi(\vec{x}) \to \phi(\vec{x}) + \vec{k}\cdot\vec{\lambda}$. If the drift potential is multi-valued then for any shift $\vec{\lambda}$ there will be a corresponding $\vec{k}$ such that $\phi(\vec{x}) \to \phi(\vec{x}) + 2\pi \, n$, where $n$ is an integer.

\end{remark*}

\section{Diffusive dynamics}
\label{Diffusive dynamics}
The evolution of the probability $\rho_{t}(x)$ is captured by eq.(\ref{Particles Chapman kolmogorov}). A more convenient description of the same dynamics is given by converting that integral equation into a differential equation, known as the Fokker-Planck (FP) equation (see Appendix \ref{appendix FP}):
\begin{equation}
\partial_{t}\rho_{t} = -\partial_{A}\left(\rho_{t}\, b^{A}\right)+\frac{1}{2}\eta\, m^{AB}\partial_{A}\partial_{B}\rho_{t}~.
\label{Particles FP equation}
\end{equation}
The FP equation can, in turn, be cast as a continuity equation,
\begin{equation}
\partial_{t}\rho_{t} = -\partial_{A}\left(\rho_{t}\, v^{A}\right),
\label{Particles Continuity equation}
\end{equation}
where we have introduced $v^{A}$, the \textit{current velocity}, which describes the \textit{flow} of probability.

The current velocity consists of two separate contributions,
\begin{equation}
v^{A} = b^{A} + u^{A},
\end{equation}
where the \textit{osmotic} velocity $u^{A}$ is given by
\begin{equation}
u^{A} = -\eta\, m^{AB}\partial_{B}\log\rho^{1/2}.
\label{Particles osmotic velocity}
\end{equation}
The effect of the osmotic velocity is for $\rho(x)$ to flow down the probability gradient, resulting in a diffusion, which justifies the name ``osmotic".

Owing to the form of the drift velocity and osmotic velocity, eqns.(\ref{Particles drift velocity}) and (\ref{Particles osmotic velocity}), respectively, the current velocity can itself be written as
\begin{equation}
v^{A} = m^{AB}\partial_{B}\Phi\quad\text{where}\quad\Phi = \eta \phi -\eta \log\rho^{1/2}.
\label{Particles def Phi}
\end{equation}
The variable $\Phi$ is particularly interesting. It will eventually be identified as a Hamilton-Jacobi-type function, and later, as the phase of the wave function $\Psi$. (From this point forward, it is more convenient to proceed by dealing directly with $\Phi$, rather than the drift potential $\phi$; although, of course, no results depend on this choice.)

The FP equation (or continuity equation) eq.(\ref{Particles Continuity equation}) can be rewritten as
\begin{equation}
\partial_{t}\rho_{t}(x) = -\partial_{A}\left(\rho_{t}\, v^{A}\right) = \frac{\delta\tilde{H}[\rho,\Phi]}{\delta\Phi(x)}
\label{Particles FP H}
\end{equation}
for a suitably chosen functional $\tilde{H}[\rho,\Phi]$. In fact, one can check that
\begin{equation}
\tilde{H}[\rho,\Phi] = \int dx \rho\frac{1}{2}m^{AB}\partial_{A}\Phi\,\partial_{B}\Phi + \tilde{F}[\rho],
\label{Particles Hamiltonian F}
\end{equation}
where $F[\rho]$ is an ``integration constant", does reproduce the eq.(\ref{Particles FP H}).

The notation in eq.(\ref{Particles FP H}) is, by design, suggestive: the FP equation, written this way, looks very much like one half of Hamilton's equations. And, while this will eventually be the case, it is important to guard against the conclusion that the FP equation was derived from an action principle. It was not. The evolution of $\rho$ has been derived, through and through, from a completely inferential framework; it is only once we consider the full joint dynamics of $\rho$ and $\Phi$ together that $\tilde{H}[\rho,\Phi]$ obtains its interpretation as a true Hamiltonian functional.\footnote{The observant among us will have noticed that there was a bit of subjectivity in choosing the FP equation to be written as in (\ref{Particles FP H}) since it automatically assumes $\rho(x)$ to be the canonical coordinate rather than the conjugate momentum. The choice is not, however, unreasonable. For example, this follows naturally when the e-phase space is modeled as a cotangent bundle, as in \cite{Caticha 2019b}. Alternatively, had we made the opposite choice, the e-functional $\tilde{H}$ would've been negative, and thus we could not have identified the generator $\tilde{H}$ as the ``energy". A convenient convention is then to choose the generator of time translations to coincide with the energy, as the definition eq.(\ref{Particles FP H}) then guarantees.}
\section{Non-dissipative dynamics}
\label{Non-dissipative dynamics}
The dynamical equation, eq.(\ref{Particles FP equation}), describes a standard diffusion process for the probability $\rho(x)$ under the influence of an externally prescribed drift potential $\phi(x)$. The role of $\phi(x)$ is particularly important: different assignments of an external potential $\phi(x)$ imply different flavors of dynamics, but each of which is inevitably a diffusion process. While such a scenario may be desirable for a variety of systems (see e.g., \cite{H Risken 1989}\cite{Gardiner 2009} for some standard applications in physics, finance, and otherwise), it is not accurate for the purposes of predicting the behavior of atoms and subatomic particles. Such microscopic systems display a host of interesting phenomena --- bound states, oscillations, etc. --- which cannot be described by eq.(\ref{Particles FP equation}) alone. In short: a richer dynamics is needed.

An early insight into what such a dynamics could possibly look like was provided by the work of E. Nelson \cite{Nelson} in stochastic mechanics. The idea is a \textit{non-dissipative} diffusion: while the drift potential $\phi(x)$ guides the probability $\rho(x)$, now it too responds, and in turn reacts back on $\phi(x)$, resulting in a coupled dynamics between $\rho(x)$ and $\phi(x)$. This scheme amounts to updating the constraint eq.(\ref{Particles Drift Constraint}) suitably at each instant, and thereby producing a Brownian dynamics capable of more interesting and diverse motions than was previously available.
\paragraph*{A canonical choice--- }
What remains to be done then, is to implement this idea. And here a variety of options are available. As mentioned above, one approach would be to follow Nelson, and earlier iterations of ED \cite{Caticha 2010a}\cite{Bartolomeo et al 2014}, by imposing that $\phi(x)$, or more conveniently $\Phi(x)$, be updated such that a global ``energy" functional be conserved. The resulting dynamics was \textit{Hamiltonian}, conferring the joint space, $\left(\rho(x),\Phi(x)\right)$ of dynamical variables with a \textit{canonical} structure, i.e. a conserved symplectic form, Hamiltonian generators, Poisson brackets, etc. From here, one can go further towards obtaining a quantum dynamics by choosing an appropriate set of potentials in the Hamiltonian; namely, the \textit{quantum potential}, familiar from Bohmian mechanics (see e.g. \cite{Bohm 1952}). But also, as previously mentioned, the reliance on a conserved energy is not conducive to further development of the theory, and thus this criterion fails.

Instead, we follow the strategy outlined above: we take serious the hints provided in \cite{Bartolomeo et al 2014} that the probability $\rho(x)$ and phase function $\Phi(x)$ form a canonical pair. But rather than requiring a conserved energy to accomplish this, we impose this structure on $\rho(x)$ and $\Phi(x)$ outright, as has been considered by Reginatto and Hall \cite{Hall Reginatto 2002}. This is accomplished by imposing that the our space of states be equipped with a symplectic structure; one that is compatible with a notion of Poisson brackets, Hamiltonian generators, and all the tools typical of the canonical framework. A primary advantage of the canonical toolkit, of course, is that it affords us the means to update the constraint eq.(\ref{Particles Drift Constraint}) without reference to energy conservation, thus making the whole formalism more robust.

\section{The canonical formalism in ED}\label{Canonical_ED_Particles}
The goal is to update the phase function $\Phi(x)$ in response to the changing probability $\rho(x)$. That is, we seek a dynamical flow for the pair of variables $\rho(x)$ and $\Phi(x)$. While the dynamics of $\rho(x)$ is fixed by the FP equation, the set of evolutions for $\Phi(x)$ have no restrictions. The goal is to identify a criterion for selecting a desirable subset. In the following we use a notation introduced in the appendix. In particular, the space $\Gamma$ will be called the \textit{ensemble} phase space, or \textit{e}-phase space for short, in which a point $X\in\Gamma$ denotes a pair of functions $\left ( \rho(x),\Phi(x)\right )$. That is, coordinates in $\Gamma$ will be given by $X^{\alpha x} = \left(\rho(x),\Phi(x)\right)$, where $\alpha$ takes on either of two values, $\rho$ or $\Phi$.

\paragraph*{Symplectic form--- } The criterion that we seek is to only consider those flows such that a \textit{symplectic} form $\mathbf{\Omega}\left(\cdot,\cdot\right)$ is preserved by the dynamical evolution. By symplectic form we mean here a two-form that satisfies \cite{Chernoff Marsden 2006}:
\begin{center}
\begin{itemize}
\item[(a)]\hspace{1 cm} $\mathbf{\Omega}$ is closed, i.e. $\tilde{\mathbf{d}}\mathbf{\Omega} = 0$,\label{Particles symplectic form a}\\
\item[(b)]\hspace{1 cm} for every $X\in\Gamma$, $\mathbf{\Omega}_{X}:T_{X}\otimes T_{X}\to\mathbb{R}$ is non-degenerate.\label{Particles symplectic form b}
\end{itemize}
\end{center}
(For a review of notation, see appendix \ref{ch:formalism}.)

The first condition is necessary if our dynamical flows are to leave $\mathbf{\Omega}$ invariant. (More technically, if the Lie derivative of $\mathbf{\Omega}$ along the flow vanishes.) Eventually it is also this condition that allows the Poisson brackets to satisfy the Jacobi identity; this allows for the notion of an algebra. The second condition is also interesting: it says that $\mathbf{\Omega}$ can be viewed as a \textit{mapping} from the tangent to cotangent spaces of $\Gamma$. Consider, for instance, inserting a vector $\bar{\mathbf{V}}$ into $\mathbf{\Omega}$ such that we have $\mathbf{\Omega}\left(\bar{\mathbf{V}},\cdot\right)$. Since $\mathbf{\Omega}$ is a two-form that requires two vectors as input to form a scalar, $\mathbf{\Omega}\left(\bar{\mathbf{V}},\cdot\right)$ now requires us to input an additional vector to produce a real number. That is, this object has all the properties of a one-form. This suggests the notation
\begin{equation}
\tilde{\mathbf{V}} = \mathbf{\Omega}\left(\bar{\mathbf{V}},\cdot\right)\quad\text{where}\quad \tilde{\mathbf{V}}:T_{X}\to\mathbb{R},
\end{equation}
so that $\tilde{\mathbf{V}}$ is the image of the vector $\bar{\mathbf{V}}$ in the cotangent space, hence it is a mapping.

Moreover, a manifold (such as the EPS) that has such a symplectic structure is called a symplectic manifold. And since, in addition, $\mathbf{\Omega}$ is non-degenerate, the mapping induced by $\mathbf{\Omega}$ is one-to-one (wherever it is defined), which implies that it has an inverse $\mathbf{\Pi} = \mathbf{\Omega}^{-1}$, which we call the Poisson tensor, and that we denote (in component form) $\Omega^{\alpha x,\beta x'}$, which satisfies
\begin{equation}
\Omega_{\alpha x,\beta x'}\Omega^{\beta x', \gamma x''} = \delta_{\alpha x}^{\gamma x''}.
\end{equation}
Also, since $\mathbf{\Omega}$, a two-form, is a $(0,2)$ rank tensor that is completely \textit{anti-symmetric} it must be that $\mathbf{\Omega}\left(\bar{\mathbf{V}},\bar{\mathbf{V}}\right)=0$; of course, the inverse $\mathbf{\Omega}^{-1}$ is then also anti-symmetric.\footnote{Technically, when we deal with infinite dimensional spaces, it is often the case that we deal with so-called \textit{weakly} non-degenerate symplectic forms, i.e. there may be some vectors which do not have an image in the cotangent space or \textit{vice versa} (see e.g., \cite{Chernoff Marsden 2006}). That is, the mapping is one-to-one --- read: non-degenerate --- but it may not be \textit{onto}. Interestingly, this subtlety does not appear in quantum theory: the symplectic structure is \textit{strongly} non-degenerate --- it is an isomorphism between the tangent and cotangent spaces of $\Gamma$ at every point. This is a direct result of the fact that the complex Hilbert space is a kind of K\"{a}hler manifold.}

\begin{remark*}
The criterion of a conserved symplectic form seems a bit arbitrary. Why a symplectic structure? Why not something else? To this, no simple answers present themselves. We proceed simply by assuming this as a necessary ingredient for our theory to resemble physics; it is the symplectic structure, after all, that pervades both classical and quantum physics. The hope is, however, that a suitably developed ED will discard with this assumption.
\end{remark*}
\paragraph*{Hamiltonian generators--- }
The flows, or trajectories of interest are those along which the symplectic form $\mathbf{\Omega}$ remains unchanged (such flows are called \textit{symplectomorphisms}). As was previously mentioned, such trajectories can be identified with the integral curves of some set of vector fields $\{\bar{\mathbf{V}}\}$. To show that the symplectic stucture is preserved along these trajectories it is sufficient, locally at least, for any vector field $\bar{\mathbf{V}}$ in this set to satisfy
\begin{equation}
\pounds_{\bar{\mathbf{V}}}\,\mathbf{\Omega} = 0.
\end{equation}
Computing this Lie derivative more explicitly \cite{Schutz}, we have that
\begin{equation}
\pounds_{\bar{\mathbf{V}}}\,\mathbf{\Omega} = \left(\tilde{\mathbf{d}}\mathbf{\Omega}\right)\left(\bar{\mathbf{V}}\right)+\tilde{\mathbf{d}}\left(\mathbf{\Omega}\left(\bar{\mathbf{V}}\right)\right) = \tilde{\mathbf{d}}\left(\mathbf{\Omega}\left(\bar{\mathbf{V}}\right)\right)=0,
\end{equation}
where the second equality follows from the requirement (a) that $\mathbf{\Omega}$ be closed. This condition implies that the one-form $\tilde{\mathbf{V}} = \mathbf{\Omega}\left(\bar{\mathbf{V}}\right)$ is also closed, i.e. $\tilde{\mathbf{d}}\tilde{\mathbf{V}} = 0$. Clearly this is satisfied if $\tilde{\mathbf{V}}$ is exact, that is, $\tilde{\mathbf{V}} = \tilde{\mathbf{d}}\tilde{G}[X]$, for some e-functional $\tilde{G}[X]$ since $\tilde{\mathbf{d}}^{2}\tilde{G}=0$ identically. As is known, however, a closed form need not be exact; this is always true \textit{locally}, but the global statement depends on the topology of $\Gamma$. Here we go further and assume that the topology of $\Gamma$ is such that it allows the identification of $\tilde{\mathbf{V}}=\tilde{\mathbf{d}}\tilde{G}[X]$ always.\footnote{This is consistent, for example, with the fact that the space of pure states in quantum theory is a simply connected K\"{a}hler manifold. See e.g., \cite{Cirelli I 1990}.} As one might expect, the e-functionals $\tilde{G}$ are special, they are the \textit{Hamiltonian} generators.

\begin{definition}
A vector field $\bar{\mathbf{V}}_{\tilde{G}}\in T_{X}$ is a \textit{Hamiltonian} vector field (HVF) if
\begin{equation}
\mathbf{\Omega}\left(\bar{\mathbf{V}}_{\tilde{G}}\right) = \tilde{\mathbf{d}}\tilde{G}.
\label{Particles Definition HVF}
\end{equation}
\end{definition}

\noindent In the coordinates defined by $\rho(x)$ and $\Phi(x)$ the HVFs have the explicit form
\begin{equation}
V^{\alpha x}_{\tilde{G}} = \Omega^{\alpha x\beta x'}\frac{\delta\tilde{G}}{\delta X^{\beta x'}} = \left(\frac{\delta\tilde{G}}{\delta\Phi(x)},-\frac{\delta\tilde{G}}{\delta\rho(x)}\right)
\label{Particles Definition HVF Components}
\end{equation}
so that
\begin{equation}
\bar{\mathbf{V}}_{\tilde{G}} = \int dx\left(\frac{\delta\tilde{G}}{\delta\Phi(x)}\frac{\delta}{\delta\rho(x)}-\frac{\delta\tilde{G}}{\delta\rho(x)}\frac{\delta}{\delta\Phi(x)}\right).
\label{Particles Definition HVF Vector}
\end{equation}
By definition, since such flows leave $\mathbf{\Omega}$ invariant, it is these flows that we are interested in. The problem of finding a joint evolution for the variables $\rho(x)$ and $\Phi(x)$ thus reduces to finding an appropriate $\bar{\mathbf{V}}_{\tilde{H}}$ that generates a dynamical flow, i.e. the flow whose integral curves are labeled by the distinguished entropic time parameter $t$. And, since these HVFs are in one-to-one correspondence with the Hamiltonian generators, this task amounts to just finding an appropriate Hamiltonian e-functional $\tilde{H}[\rho,\Phi]$ .

The natural question that follows is, what e-functional $\tilde{H}$ is suitable for updating the variables $\rho(x)$ and $\Phi(x)$? Indeed, at first the problem seems quite severe since \textit{any} e-functional generates a flow that leaves $\mathbf{\Omega}$ invariant. The situation is, however, not quite so dire since $\tilde{H}[\rho,\Phi]$ cannot be completely arbitrary: it must reproduce the FP equation, eq.(\ref{Particles FP equation}). Recalling the rewritten FP equation in (\ref{Particles FP H}), this implies taking as our dynamical generator the e-functional $\tilde{H}[\rho,\Phi]$, introduced first in (\ref{Particles Hamiltonian F}).
\paragraph*{On Lagrangians--- }
The developments described above put ED in an interesting position. While in classical physics it is often difficult to construct a Hamiltonian directly, without first appealing to the notion of a Lagrangian,\footnote{Indeed, in \cite{Teitelboim thesis}\cite{Teitelboim 1972}, for example, C. Teitelboim sets out to determine a set of covariant dynamical theories without first appealing to a Lagrangian. His method is successful, but his method of constructing the Hamiltonian is much more tedious than if he had just began with a Lagrangian. And Dirac, who was a major proponent of the Hamiltonian viewpoint, in \cite{Dirac Lectures} himself advocated beginning from a Lagrangian.} in ED much of the Hamiltonian is fixed simply by requiring that it reproduce the FP equation. That is, the FP equation, which results directly from the rules of inference, fixes a good portion of the eventual quantum dynamics. 

Interestingly, ED therefore has a problem that is somewhat inverted: how does one construct a Lagrangian formulation given a Hamiltonian? This, it seems is, in fact, not straightforward, or potentially, not possible at all. For example, Reginatto and Hall \cite{Hall/Reginatto 2005} suggest that a Lagrangian formulation for the variables $\rho(x)$ and $\Phi(x)$ would require the introduction of two additional variables to play the role of ``velocities" corresponding to the ``coordinates" $\rho(x)$ and $\Phi(x)$.

But why should we need such an indirect route to a Lagrangian formulation? Recall that in classical mechanics the Hamiltonian and Lagrangian viewpoints are often used interchangeably, largely because it is often easy to perform the Legendre transform which transitions between the Hamiltonian picture based on the cotangent bundle and the Lagrangian picture based on the tangent bundle (see e.g., \cite{Arnold 2013}). In \cite{Caticha 2019b}, however, A. Caticha does, in fact, model the e-phase space as a cotangent bundle. Should this not suggest a corresponding Lagrangian formulation of ED? In short, yes. Nevertheless, while this route may be available in principle, in practice obtaining such a Lagrangian picture would require us to invert the local differential FP equation to obtain the equivalent of $p = \partial L /\partial \dot{q}$ in classical mechanics, which would here be a non-local integral equation. Thus suggests that the Hamiltonian viewpoint, being more simple and compact, should actually be considered the more fundamental one.



%
\paragraph*{Poisson brackets--- }
As we mentioned in passing above, the symplectic form $\mathbf{\Omega}$ is a skew-symmetric bilinear product. The fact that it is skew-symmetric means that it is completely anti-symmetric on its components $\Omega_{\alpha x,\beta x'}$, while it is bilinear because it requires two vectors as inputs.

Having introduced the notion of a HVF $\bar{\mathbf{V}}_{\tilde{G}}$, it is interesting to consider supplying two such vectors to $\mathbf{\Omega}$. The result is, of course, a real number (an e-functional), but owing to the anti-symmetry of $\mathbf{\Omega}$, the resulting object also retains an anti-symmetric character. It is, in fact, nothing but the \textit{Poisson} bracket between two e-functionals
\begin{equation}
\mathbf{\Omega}\left(\bar{\mathbf{V}}_{\tilde{F}},\bar{\mathbf{V}}_{\tilde{G}}\right) = \Omega_{\alpha x,\beta x'}V_{\tilde{F}}^{\alpha x}V_{\tilde{G}}^{\beta x'}\equiv \left\{\tilde{F},\tilde{G}\right\}.
\end{equation}
Using the inverse symplectic form $\mathbf{\Omega}^{-1}$ the Poisson bracket can be written as
\begin{equation}
\left\{\tilde{F},\tilde{G}\right\} = \Omega^{\alpha x,\beta x'}\frac{\delta\tilde{F}}{\delta X^{\alpha x}}\frac{\delta\tilde{G}}{\delta X^{\beta x'}},
\end{equation}
or more familiarly, it can be put (locally at least) in the canonical form
\begin{equation}
\left\{\tilde{F},\tilde{G}\right\} = \int dx\left(\frac{\delta\tilde{F}}{\delta\rho(x)}\frac{\delta\tilde{G}}{\delta\Phi(x)}-\frac{\delta\tilde{G}}{\delta\rho(x)}\frac{\delta\tilde{F}}{\delta\Phi(x)}\right).\label{Particles Poisson bracket}
\end{equation}

The Poisson bracket has a few interesting properties.
\begin{center}
\begin{itemize}
\item[(a)]It is skew-symmetric:
\begin{equation}
\left\{\tilde{F},\tilde{G}\right\}=-\left\{\tilde{G},\tilde{F}\right\},
\end{equation}
\item[(b)] It satisfies the Jacobi identity:
\begin{equation}
\left\{\tilde{F},\left\{\tilde{G},\tilde{H}\right\}\right\}+\left\{\tilde{G},\left\{\tilde{H},\tilde{F}\right\}\right\}+\left\{\tilde{H},\left\{\tilde{F},\tilde{G}\right\}\right\}=0,\label{Particles Jacobi Identity}
\end{equation}
\item[(c)] It also satisfies the Leibniz property:
\begin{equation}
\left\{\tilde{F}\tilde{G},\tilde{H}\right\} = \tilde{F}\left\{\tilde{G},\tilde{H}\right\} + \tilde{G}\left\{\tilde{F},\tilde{H}\right\}.
\end{equation}
\end{itemize}
\end{center}

The first two properties endow the Poisson bracket with the properties of a Lie bracket, such that it allows for a notion of an \textit{algebra} between e-functionals. The last property allows for a notion of derivation, i.e. $\left\{\tilde{F},\tilde{G}\right\} $ can be viewed as the derivative of $\tilde{F}$ with respect to the flow generated by $\tilde{G}$. This is made explicit by the relation
\begin{equation}
\left\{\tilde{F},\tilde{G}\right\} = \bar{\mathbf{V}}_{\tilde{G}}\left(\tilde{F}\right),
\end{equation}
which uses the form of $\bar{\mathbf{V}}_{\tilde{G}}$ as a ``differential operator", which is explicit in (\ref{Particles Definition HVF Vector}).

Pertaining to the notion of algebra, consider two HVFs, $\bar{\mathbf{V}}_{\tilde{F}}$ and $\bar{\mathbf{V}}_{\tilde{G}}$. Clearly the flows generated by $\tilde{F}$ and $\tilde{G}$ both preserve the symplectic form, but what of the flow generated by their Poisson bracket $\left\{\tilde{F},\tilde{G}\right\} $? In the generic case this too leaves $\mathbf{\Omega}$ unchanged
\begin{equation}
\pounds_{\bar{\mathbf{V}}_{\{\tilde{F},\tilde{G}\}}}\mathbf{\Omega} = 0.
\end{equation}
And so, the set of e-functionals form a Lie algebra since for each $\tilde{F}$ and $\tilde{G}$, which preserves the symplectic structure, so too does their Poisson bracket $\{\tilde{F},\tilde{G}\}$. On the other hand, there is a natural Lie bracket for HVFs,
\begin{equation}
\left[\bar{\mathbf{V}}_{\tilde{F}},\bar{\mathbf{V}}_{\tilde{G}}\right] \equiv \bar{\mathbf{V}}_{\tilde{F}}\bar{\mathbf{V}}_{\tilde{G}}-\bar{\mathbf{V}}_{\tilde{G}}\bar{\mathbf{V}}_{\tilde{F}}.
\end{equation}
Is there a relation between the two brackets? As it turns out, there is: the Lie bracket of two HVFs produce a third which is also a HVF, and in fact, it is nothing but the HVF generated by $\{\tilde{F},\tilde{G}\}$. That is to say,
\begin{equation}
\left[\bar{\mathbf{V}}_{\tilde{F}},\bar{\mathbf{V}}_{\tilde{G}}\right] = \bar{\mathbf{V}}_{\{\tilde{F},\tilde{G}\}}.
\end{equation}
Thus the algebra of HVFs is also a Lie algebra.\footnote{The set of all vectors in a tangent space form a Lie algebra \cite{Schutz} since, by definition, the Lie bracket of two vectors must produce a third which is in this set. The algebra of HVFs is then technically a Lie sub-algebra, since its elements are only those vectors whose flow preserve $\mathbf{\Omega}$.}

\paragraph*{Normalization constraint--- }
There is a subtlety that occurs when one of the coordinates in the phase space is a probability which owes to the fact that probabilities must, of course, be normalized
\begin{equation}
\int dx \, \rho(x) = 1~.\label{Particles Normalization}
\end{equation}
Following \cite{Caticha 2019b} (see also, \cite{Schilling thesis}\cite{Schilling Ashtekar 1999}), we define the e-phase space $\Gamma$ to include all smooth probabilities, including those that are unnormalized. We then proceed to remedy this situation by imposing an additional constraint
\begin{equation}
\tilde{N} = \int dx \, \rho(x) - 1~,\label{Particles Normalization Constraint}
\end{equation}
which serves as a restriction on the allowed states.

In the Dirac canonical theory \cite{Dirac Lectures} constraints may be distinguished as either first or second class. Briefly stated, a constraint is first class when its Poisson bracket with all other constraints vanishes weakly.\footnote{More specifically, for a set of $N$ constraints $\tilde{C}_{i}$ ($i = 1, \cdots, N$), each of which weakly vanish, i.e. $\tilde{C}_{i}\approx 0$, a constraint is first class when its Poisson bracket with all other constraints vanishes weakly
\begin{equation}
\left \{\tilde{C}_{i},\tilde{C}_{j} \right \}\approx 0~.
\end{equation}
This therefore implies that the Poisson bracket of two first class constraints must itself be strongly equal to a linear combination of constraints such that
\begin{equation}
\left \{\tilde{C}_{i},\tilde{C}_{j} \right \} = \sum_{k}\alpha_{k}\tilde{C}_{k}~.
\end{equation}
} Here we have just one constraint
\begin{equation}
\left \{\tilde{N},\tilde{N} \right \} = 0~,
\end{equation}
and thus it is trivially first class.

The Hamiltonian flows generated by such first class constraints are known to correspond to \emph{gauge transformations}, i.e. a change of variables that correspond to the same physical state. Here the HVF corresponding to $\tilde{N}$ and parameterized by $\xi$ is given by
\begin{equation}
V_{\tilde{N}}^{\alpha x} = \left (\frac{\delta \tilde{N}}{\delta\Phi(x)},-\frac{\delta \tilde{N}}{\delta\rho(x)} \right ) = \left (0,-1\right )~,
\end{equation}
which yields the transformation
\begin{equation}
\rho_{\xi}(x) = \rho_{\xi = 0}(x)\quad \text{where}\quad \Phi_{\xi}(x) = \Phi_{\xi = 0}(x) + \xi~.
\end{equation}
This shows that the symmetry generated by $\tilde{N}$ is to shift the phase $\Phi$ by a constant $\alpha$, but otherwise leaves the dynamics unchanged.\footnote{Note, whether the $\xi$ is added or subtracted to the phase is a matter of convention.}

\subsubsection*{Hamiltonian dynamics in ED}
Our purpose is to construct a joint dynamics for $\rho(x)$ and $\Phi(x)$, with the criterion being that we update $\Phi(x)$ such that a symplectic form $\mathbf{\Omega}$ is preserved by the time evolution. The key development in this regard was to recognize that this task amounts to making an appropriate choice of Hamiltonian generator, $\tilde{H}$. We ask here what evolutions for $\Phi(x)$ are consistent with these assumptions.

First, note that since $\rho(x)$ and $\Phi(x)$ are, by definition, a canonical pair, we require that
\begin{equation}
\{\rho(x),\rho(x')\} = \{\Phi(x),\Phi(x')\} = 0\quad\text{and}\quad\{\rho(x),\Phi(x')\} = \delta(x,x'),
\end{equation}
using the Poisson brackets introduced above. Next, we point out that the FP equation, written as in eq.(\ref{Particles FP H}) can also be expressed as
\begin{equation}
 \partial_{t}\rho(x) = \{\rho(x),\tilde{H}\} = \frac{\delta\tilde{H}}{\delta\Phi(x)},\label{Particles Hamilton's equation rho}
 \end{equation} 
for the $\tilde{H}$ given in (\ref{Particles Hamiltonian F}). This is one half of Hamilton's equations for the canonical coordinate $\rho(x)$. The other half of Hamilton's equations is given by
\begin{equation}
\partial_{t}\Phi(x) = \{\Phi(x),\tilde{H}\} = -\frac{\delta\tilde{H}}{\delta\rho(x)}.\label{Particles Hamilton's equation Phi}
\end{equation}
Computing this derivative explicitly yields
\begin{equation}
-\partial_{t}\Phi(x) = \frac{1}{2}m^{AB}\partial_{A}\Phi(x)\partial_{B}\Phi(x) + \frac{\delta\tilde{F}[\rho]}{\delta\rho(x)},
\label{Particles Hamilton Jacobi}
\end{equation}
which we identify as an equation of the Hamilton-Jacobi (HJ) type, familiar from classical mechanics (see e.g. \cite{Lanczos}), or most notably recognized in the quantum setting from Bohmian mechanics (see e.g. \cite{Bohm 1952}\cite{Holland 1995}).

What is most apparent from eq.(\ref{Particles Hamilton Jacobi}) is that the family of evolutions for $\Phi(x)$ amount to the various choices of the undetermined functional $\tilde{F}[\rho]$ --- \textit{any} choice of $\tilde{F}[\rho]$ is consistent with our assumptions of a conserved symplectic form. This is remarkably similar to the situation in classical mechanics where, often times, the ``kinetic energy" --- here the quadratic piece $\left(\partial\Phi(x)\right)^{2}$ --- is determined, but the choice of ``potential" is left arbitrary.
\paragraph*{Hybrid ED--- }
The simplest possible choice for $\tilde{F}[\rho]$ is linear in $\rho(x)$ and takes the form of the expected value of a ``potential energy" $V(x)$
\begin{equation}
\tilde{F}[\rho] = \int dx \rho(x) \, V(x).
\end{equation}
The resulting HJ equation has the form
\begin{equation}
-\partial_{t}\Phi(x) = \frac{1}{2}m^{AB}\partial_{A}\Phi(x)\partial_{B}\Phi(x) + V(x),
\label{Particles Hamilton Jacobi Classical}
\end{equation}
which is identical to the HJ equation of classical mechanics. The theory thus contains features particular to both classical and quantum theories. For example, the particles undergo an indeterministic Brownian motion and are subject to a set of uncertainty relations (see e.g. \cite{Bartolomeo Caticha 2016}), such as in the quantum theory, but particles are guided by a completely classical Hamilton-Jacobi function. For this reason this model has been dubbed a ``hybrid" model of ED.

It is possible that such models may be useful for analyzing ``mesoscopic" systems that lie between the quantum and classical regimes. Indeed, a promising application of such a theory lie in developing models of interacting quantum and non-quantum systems, such as has been considered by Reginatto and Hall (see e.g. \cite{Hall/Reginatto 2005}). Although the possibilities for this hybrid theory are interesting, we will not develop it any further.
\subsubsection*{Symmetries and invariance in ED}
An important concept in any theory of inference is that of symmetry and invariance, i.e. the identification of a set of transformations which leave the problem at hand unchanged. While certain symmetries may, for example, be imposed (or exploited) as a means to simplify an otherwise difficult problem, other symmetries are more privileged; they express certain ``fundamental" immutable properties of the system. These latter symmetries are of particular interest to us here, as they express important invariance properties of the assumed underlying structure of space and time; for this reason, one might call these \textit{kinematical} symmetries.

The chief correspondence that demonstrates this relationship between symmetries and invariance is, of course, Noether's theorem (see e.g. \cite{Greiner Field Quant}). The theorem states that if a quantity is conserved along a flow, then that quantity must itself be related to the generator of a symmetry. These ideas are conveniently expressed using HVFs and Poisson brackets.

The statement that a particular quantity $\tilde{G}$ is conserved along the flow generated by $\tilde{F}$ is given by
\begin{equation}
\pounds_{\bar{\mathbf{V}}_{\tilde{F}}}\tilde{G} = \bar{\mathbf{V}}_{\tilde{F}}(\tilde{G}) = \{\tilde{G},\tilde{F}\} = 0.
\end{equation}
On the other hand, this also implies that
\begin{equation}
\{\tilde{F},\tilde{G}\} =-\{\tilde{G},\tilde{F}\} =\bar{\mathbf{V}}_{\tilde{G}}(\tilde{F}) = 0.
\end{equation}
Thus there is a kind of reciprocity between the generators of the two flows: the flow of $\tilde{G}$ conserves $\tilde{F}$ and \textit{vice versa}.
\paragraph*{Conservation of energy--- }
A principal ingredient to previous iterations of ED, such as that developed in \cite{Bartolomeo et al 2014}, for example, was the requirement that energy be conserved. Thus far we have not explicitly assumed that energy is conserved, but of course it has been implicit in our assumptions thus far. 

Consider the dynamical flow generated by the Hamiltonian $\tilde{H}$, and parameterized by the time $t$. While it may seem a trivial thing, we ask, how does $\tilde{H}$ change along the flow that it itself generates? To answer this we compute the Lie derivative of $\tilde{H}$ along $\bar{\mathbf{V}}_{\tilde{H}}$
\begin{equation}
\pounds_{\bar{\mathbf{V}}_{\tilde{H}}}\tilde{H} = \bar{\mathbf{V}}_{\tilde{H}}(\tilde{H}) = \{\tilde{H},\tilde{H}\} = 0,
\end{equation}
which vanishes due to the skew-symmetry of the Poisson bracket.

Clearly, based on our previous discussion, if $\tilde{H}$ is conserved, then it must be that it corresponds to some symmetry; the symmetry is called \textit{time translation symmetry}, and the corresponding conserved quantity is called the \textit{energy}. Thus we say energy is conserved.\footnote{Note that we have implicitly assumed, as we do throughout, that the Hamiltonian does not depend explicitly on time. That is, any potentials that may appear are independent of time.}
\paragraph*{Generators of translations and rotations--- }
While a general discourse on symmetries and conserved quantities in ED takes us outside the scope of this thesis, there are two more symmetries (and thus conserved quantities) that are worth mentioning: rotations and translations. Indeed, we deal here with a set of $N$ particles in a flat space $\mathbf{X}$ with metric $\delta_{ab}$. It is well known that the isometries of flat space are the set of global rotations and translations. These transformations also have a representation in terms of Hamiltonian generators, which are related to the total linear and angular momentum.
\subparagraph*{Linear momentum:}
A generator of translations induces in some quantity a change $x_{n}^{a}\to x_{n}^{a}+\epsilon^{a}$, where $\epsilon^{a}$ is a small parameter that is the same for all particles $n$ (see e.g. \cite{Bartolomeo et al 2014}\cite{Ipek Caticha 2015}). It is not difficult to verify that
\begin{equation}
\tilde{P}_{a} = \int dx \rho(x)\frac{\partial\Phi(x)}{\partial X^{a}}~,
\label{Particles Generator Translations}
\end{equation}
where $X^{a}$ are center of mass coordinates, generates global spatial translations and thus deserves the name ``total momentum". Indeed, just as in classical mechanics, when the Hamiltonian $\tilde{H}$ is translation invariant,
\begin{equation}
 \left\{\tilde{H},\tilde{P}_{a}\right\} = 0,
 \end{equation} 
then we say that momentum is \textit{conserved}.
\subparagraph*{Angular momentum:}
It is also possible to identify a generator of rotations
\begin{equation}
\tilde{L}_{a} = \int dx \, \rho(x)\sum_{n}\epsilon_{abc}x_{n}^{b}\frac{\partial\Phi(x)}{\partial x_{n}^{c}}.\label{Particles Generator Rotations}
\end{equation}
The effect of $\tilde{L}_{a}$ is to generate an infinitesimal rotation. For example, it is easy to show that, for some small rotation angle $\theta^{a}$ that
\begin{equation}
\delta\rho(x) = \{\rho(x),\tilde{L}_{a}\}\theta^{a}=-\sum_{n}\epsilon_{abc}\theta^{a}x_{n}^{b}\frac{\partial\rho(x)}{\partial x_{n}^{c}},
\end{equation}
which is exactly what one expects from performing a Lie derivative. And, just as in the case of translations, when the Hamiltonian $\tilde{H}$ is invariant under rotations
\begin{equation}
 \left\{\tilde{H},\tilde{L}_{a}\right\} = 0,
\end{equation} 
then we claim that \textit{angular} momentum is conserved.
\section{Quantum ED}
\label{Quantum ED}
The purpose of deriving quantum mechanics is to identify those ingredients that appear most integral to the description of quintessentially quantum effects. Thus far, however, we have only modeled a generic \textit{non-dissipative} ED, one that is not necessarily quantum. Our task here is therefore to discuss what additional elements are required to make this transition in ED and to note how we recover many of the expressions familiar from ordinary quantum mechanics.


\subsubsection*{The quantum Hamiltonian}
In many standard approaches a quantum theory is obtained from a classical one through an \emph{ad hoc} procedure known as quantization. While this process may often be put on a mathematically sound footing, its guiding principles are notoriously unclear. By contrast the transition to a quantum theory in ED is more natural. The goal in ED is simply to update probabilities subject to the appropriate constraints. Thus the process of constructing a quantum ED amounts to an identification of the appropriate constraints.

One class of constraints that have already proven quite useful are those based on symmetry and invariance. This could be seen, for instance, in our identification of the relevant dynamical flows as those that conserve a symplectic structure. Moreover, additional notions of symmetry, such as the invariance under the space-time symmetries of the Galilean group, can similarly be evoked to identify more physically relevant classes of Hamiltonians. 
\paragraph*{The quantum potential--- }
The specific symmetries above are not on their own sufficient to deliver a quantum ED. For this one requires a special \emph{quantum} potential, one which takes the form
\begin{equation}
\tilde{F}_{Q} = \lambda\int dx \rho\,m^{AB}\partial_{A}\rho^{1/2}\,\partial_{B}\rho^{1/2},
\label{Particles Quantum Potential}
\end{equation}
where $\lambda$ is a constant that determines the strength of the quantum potential. 

However, as has been discussed by Caticha in \cite{Caticha 2019b}, as well as, for instance, by Hall and Reginatto \cite{Reginatto Hall 2011} in a slightly different context, this potential can too be \emph{derived} from symmetry principles. While such approaches greatly improve upon previous heuristic arguments for the quantum potential, such as in \cite{Bartolomeo et al 2014}, they have yet to be tested in the regime of relativistic quantum fields, which is our primary focus here. Thus for our purposes we simply include such a potential, as further study into the origin of the quantum potential falls outside the scope of this thesis.

\paragraph*{The quantum Hamilton-Jacobi equation--- }
The minimal model that contains such a term such as $\tilde{F}_{Q}$ is one that has no ``direct" interactions among the particles, i.e. $V(x) = 0$. The Hamiltonian $\tilde{H}_{0}$ for this minimal model reads
\begin{equation}
\tilde{H}_{0} = \int dx \rho\left(\frac{1}{2}m^{AB}\partial_{A}\Phi\partial_{B}\Phi +\lambda \, m^{AB}\partial_{A}\rho^{1/2}\,\partial_{B}\rho^{1/2} \right).
\end{equation}
The HJ equation that results from this particular choice of Hamiltonian is given by,
\begin{equation}
\partial_{t}\Phi = \left \{\Phi,\tilde{H}_{0}\right \} = -\left(\frac{1}{2}m^{AB}\partial_{A}\Phi\partial_{B}\Phi -4\,\lambda \, m^{AB}\frac{\partial_{A}\partial_{B}\rho^{1/2}}{\rho^{1/2}} \right).\label{Particles Quantum HJ}
\end{equation}
\begin{remark*}
A quick note on terminology. In the literature on Bohmian mechanics the second term in the HJ equation is often called the ``quantum potential". This is perhaps confusing because in ED we call $\tilde{F}_{Q}$ the quantum potential. The rationale for this in ED is that $\tilde{H}$ is a true Hamiltonian, and thus $\tilde{F}_{Q}$ is a true potential. The use of quantum potential in the Bohmian theory is related to the fact that in classical mechanics the HJ equation reads $-\partial_{t}\Phi = H$, where $H$ is the classical Hamiltonian of the system. This seemingly justifies referring to the second term in (\ref{Particles Quantum HJ}) as a ``potential". Here both terms may be used, but it should be obvious from context which one is meant.
\end{remark*}
\paragraph*{The ``free" Schr\"{o}dinger equation--- }
Althought it is not obvious yet, this is the Hamiltonian necessary for a collection of ``free" quantum particles. To make the connection more apparent, consider a canonical transformation to the pair of \textit{complex} variables,\footnote{This is also called a ``Madelung transformation". See e.g., \cite{Madelung}.}
\begin{equation}
\Psi_{k} = \rho^{1/2}\exp ik\Phi/\eta\quad\text{and}\quad \Psi_{k}^{*} = \rho^{1/2}\exp -ik\Phi/\eta~.
\end{equation}
In terms of these new variables the equations of motion read as
\begin{equation}
i\frac{\eta}{k}\partial_{t}\Psi_{k} = -\frac{\eta^{2}}{2k^{2}}m^{AB}\partial_{A}\partial_{B}\Psi_{k} + \left(\frac{\eta^{2}}{2k^{2}} - 4\lambda\right)m^{AB}\frac{\partial_{A}\partial_{B}|\Psi_{k}|}{|\Psi_{k}|}\Psi_{k},
\label{Particles nonlinear SE}
\end{equation}
where $|\Psi_{k}| = \rho^{1/2}$, and where a similar equation holds for $\Psi_{k}^{*}$. This is a family of equivalent equations, one for each distinct $k$.

In principle, nothing requires a particular choice of $k$, but by the same token, nothing prevents us from making a convenient choice $\hat{k}$ that makes it particularly simple to integrate the equations of motion.\footnote{This is not unlike, after all, the rationale one uses in classical mechanics. There, to integrate the equations of motion one chooses a particularly convenient set of variables that makes the dynamics as simple as possible, i.e. choose the canonical variables such that the new Hamiltonian vanishes. This view of classical mechanics leads to the Hamilton-Jacobi equation. Here we pursue a similar logic, the result is a linear Schr\"{o}dinger equation.} To this end, one notices that the equations of motion take a particularly simple, linear form when in eq.(\ref{Particles nonlinear SE}) we choose
\begin{equation}
 \frac{\eta^{2}}{2\hat{k}^{2}} - 4\lambda = 0~.
 \end{equation} 
The optimal $\hat{k}$ then becomes
\begin{equation}
 \hat{k} = \sqrt{\frac{\eta^{2}}{8\lambda}}~,
\end{equation}
which suggests a rescaling of parameters $\eta/\hat{k} = \hbar$. With this choice eq.(\ref{Particles nonlinear SE}) becomes linear
\begin{equation}
i\hbar\partial_{t}\Psi = -\frac{\hbar^{2}}{2}m^{AB}\partial_{A}\partial_{B}\Psi~,
\label{Particles linear free SE}
\end{equation}
which we recognize as a Schr\"{o}dinger equation. This allows us to identify $\hbar$ with Planck's constant. The Hamiltonian takes the form
\begin{equation}
\tilde{H}_{0} = \int dx \frac{\hbar^{2}}{2}m^{AB}\partial_{A}\Psi^{*}\partial_{B}\Psi.
\label{Particles ensemble Hamiltonian free}
\end{equation}

\paragraph*{The Schr\"{o}dinger equation with interactions--- }
The Hamiltonian (\ref{Particles ensemble Hamiltonian free}) describes a model in which no interactions are present. To add such interactions one introduces a ``classical" potential $V(x)$ so that the Hamiltonian becomes
\begin{equation}
\tilde{H} = \int dx \left(\frac{\hbar^{2}}{2}m^{AB}\partial_{A}\Psi^{*}\partial_{B}\Psi +\Psi^{*} \,V\,\Psi\right)~.
\label{Particles ensemble Hamiltonian quantum}
\end{equation}
The Schr\"{o}dinger equation then reads
\begin{equation}
i\hbar\partial_{t}\Psi = -\frac{\hbar^{2}}{2}m^{AB}\partial_{A}\partial_{B}\Psi + V\Psi~.
\label{Particles linear SE}
\end{equation}
This is the linear quantum dynamics we seek. And, being linear, it is possible to identify the linear differential operator
\begin{equation}
\hat{H} =-\frac{\hbar^{2}}{2}m^{AB}\partial_{A}\partial_{B}+ V~,
\end{equation}
as the Hamiltonian operator. It is also straightforward to show that the ensemble Hamiltonian $\tilde{H}$ is nothing but the expectation value of this operator,
\begin{equation}
\tilde{H} = \int dx \Psi^{*}\hat{H}\Psi.
\end{equation}
\subsubsection*{Relation to the conventional quantum theory}
In ED one has access to a symplectic structure, and thus Hamiltonians, and one can argue for (or derive, see e.g., \cite{Caticha 2019b}) a linear Schr\"{o}dinger dynamics. These pieces on their own, while composing important aspects of quantum theory, are not equivalent to it --- the conventional quantum theory is built upon a complex Hilbert space with many additional constructions: inner products, metric structures, unitary operators, etc. Introducing this additional architecture from a natural inferential viewpoint is a task worthy of pursuit, but one that is largely outside the purview of this thesis in particular.

Nonetheless, it is useful to outline here some of the simple relations between the ED developed thus far, and the conventional quantum theory. These relationships have been noticed independently, and frequently, by many authors in the past few decades (see e.g., \cite{Schilling thesis}\cite{Schilling Ashtekar 1999}\cite{Kibble 1979}\cite{Cirelli Lanzavecchia 1984}\cite{Cirelli II 1990}\cite{Heslot 1985}); the treatment given by T.A. Schilling and A. Ashtekar \cite{Schilling thesis}\cite{Schilling Ashtekar 1999} is particularly clear and thorough. Here we draw on these works to highlight how the generators of ED -- the Hamiltonian functionals -- and those of quantum theory -- the unitary Hermitian operators -- are related, as well as to compare the algebras between those generators, which in the former case are implemented using Poisson brackets, while in the latter case by commutators of operators.
\paragraph*{Symmetry generators--- } As we saw above, the Hamiltonian generator of time translations $\tilde{H}$ was simply related to the standard quantum Hamiltonian operator by expectation: $\tilde{H} = \left\langle \hat{H}\right\rangle$. As it turns out, other familiar quantum operators satisfy similar relations. For instance, the linear momentum Hamiltonian generator $\tilde{P}_{a}$ in eq.(\ref{Particles Generator Translations}) can be written as
\begin{equation}
\tilde{P}_{a} = \int dx \Psi^{*}\hat{P}_{a}\Psi\quad\text{where}\quad\hat{P}_{a} = -i\hbar \sum_{n}\nabla_{na}
\end{equation}
is the quantum linear momentum operator. Similarly, for $\tilde{L}_{a}$, the generator of rotations given in eq.(\ref{Particles Generator Rotations}), we have that
\begin{equation}
\tilde{L}_{a} = \int dx \Psi^{*}\hat{L}_{a}\Psi\quad\text{where}\quad\hat{L}_{a} = -i\hbar \sum_{n}\epsilon_{abc}\, x_{n}^{b}\nabla_{n}^{c}
\end{equation}
is the familiar angular momentum operator from quantum mechanics.

\paragraph*{On generators in ED and quantum theory--- }
It is worth noting the more general relationship between generators in ED and those in quantum theory. In ED generators are Hamiltonian functionals defined on the e-phase space, whereas generators in quantum theory are unitary Hermitian operators acting on a complex Hilbert space. However, while the former constitute a set of infinite-dimension, the latter set of generators in quantum theory are much more limited. The relationship between the two is that, if $\tilde{A}$ is a Hamiltonian whose flow leaves both the symplectic and metric structures of the e-phase space invariant, then it can be written as the expectation value of a Hermitian operator $\hat{A}$, such that
\begin{equation}
\tilde{A} = \int dx \Psi^{*}\hat{A}\Psi~.
\label{Particles Generators Hamiltonian-Hermitian}
\end{equation} 
Clearly the generators of translations and rotations are instances of this. The Hamiltonian itself is another example. Additional examples for scalar particles, i.e. those without spin, are expectation values of products of $\hat{P}_{i}$, $\hat{L}_{i}$, and $\hat{H}$.
\paragraph*{Lie algebras--- }
This relationship can, in fact, be pushed further. An important aspect of Hamiltonian generators is that they form a Lie algebra with the Poisson bracket playing the role of the skew-symmetric product. That is, for two Hamiltonian generators $\tilde{Y}$ and $\tilde{Z}$, their Poisson bracket $\{\tilde{Y},\tilde{Z}\}$ is itself a Hamiltonian generator.

When the Hamiltonian generators $\tilde{Y}$ and $\tilde{Z}$ in question are of the form given in eq.(\ref{Particles Generators Hamiltonian-Hermitian}), then the an interesting relation arises:
\begin{equation}
\left\{\tilde{Y},\tilde{Z}\right\} = \frac{1}{i\hbar}\int dx \Psi^{*}\left[\hat{Y},\hat{Z}\right]\Psi =\frac{1}{i\hbar} \left\langle \left[\hat{Y},\hat{Z}\right] \right\rangle,
\label{Particles Generator Poisson bracket}
\end{equation}
so that the Poisson brackets of ED are directly related by expectation value to the commutators of operators in quantum theory.
\paragraph*{Comments--- }
Ever since the advent of the ``new" quantum theory of Heisenberg, Dirac, Schr\"{o}dinger, etc., in 1925, it has been recognized that there is some, at the very least, qualitative similarities between quantum mechanics and classical mechanics, in its Hamiltonian formulation. Each has notions of symmetries, Noether theorems, generators, and so on, but the formal realization of these ideas are worlds apart --- while the former uses functions on phase space, the latter utilizes operators on a Hilbert space. Where the former is based on a geometrical picture --- manifolds, symplectic structures, metric tensors, etc. --- the latter is algebraic in nature --- vector spaces, linear operators, etc. The beauty of this so-called \textit{geometric} quantum theory is that it allows for the quantum formalism to be understood in terms of the concepts familiar from classical mechanics: unitary Hermitian operators \textit{are} Hamitonian generators, commutators \textit{are} Poisson brackets. Quantum mechanics is not just \textit{like} a Hamiltonian theory, it is a Hamiltonian theory. This is such an elegant rewriting of quantum theory that we believe that every physicist should be acquainted with it at one time or another.

From the viewpoint of ED, the reason for studying this alternative formulation is largely practical: since ED deals with the dynamical evolution of $\rho$ and $\Phi$, it already largely resembles a ``classical" dynamical theory. Thus a fully quantum ED will not resemble the algebraic formulation of quantum theory based on Hilbert spaces, but rather, the alternative, geometrical formulation described in \cite{Schilling thesis}-\cite{Heslot 1985}. As mentioned above, the first steps towards accomplishing this were set out in \cite{Caticha 2019b}, but more work remains to be done.
\section{Discussion and conclusion\label{Discussion/Conclusion}}
ED is a reconstruction of quantum theory that is grounded in the principles of inference. The basic template for a generic ED goes as follows.
\begin{itemize}
\item[1.] Specify the ontic microstates about which we are uncertain;
\item[2.] Identify the constraints relevant for analyzing a small change;
\item[3.] Construct a notion of time and scheme for updating probabilities;
\item[4.] Identify the constraints needed for updating the constraints themselves.
\end{itemize}
Perhaps the most crucial aspect of this reconstruction occurs in the first step wherein we designate a set of ontic microstates. This allows for a satisfactory resolution to the problem of measurement \cite{Caticha/Johnson}\cite{Dave Johnson thesis}\cite{Caticha Vanslette 2017}, which is completely settled within the framework of Bayesian inference. The special focus placed on time in ED is also crucial. As we will later see, the notion of entropic time plays a central role in our later development of covariant ED models. And to this latter point, the identification of a conserved symplectic structure as a means for updating the evolving constraints is particularly important as it remains a viable criterion even in curved space-times.

An important issue in the ED reconstruction of quantum is whether ED is, in fact, actually equivalent to quantum theory. While this issue was first raised by T.C. Wallstrom \cite{Wallstrom 1994} in the context of Nelson's stochastic mechanics, it poses a significant challenge to any reconstruction of quantum theory whose principal variables are expressed in terms of probabilities and phases, rather than wave functions; these include, of course, Nelson's stochastic mechanics \cite{Nelson 1964}, but also, for instance, the hydrodynamical formulations of Takabayasi \cite{Takabayasi} and Madelung \cite{Madelung}. In short, Wallstrom's objection is as follows. Conventional quantum theory requires the wave function $\Psi$ to be single valued, all while allowing its phase $\Phi$ to be a multi-valued function. The difficulty faced by ED, and those theories formally similar to ED, is that the phase function $\Phi$ is either too restricted --- it is singled valued --- or when $\Phi$ is allowed to be multi-valued, there is no plausible rationale for requiring $\Psi$ to then be single valued. The former case leads to a rather stunted quantum theory where crucial states of non-zero angular momentum, for example, are excluded, whereas the latter case is too flexible; the total angular momentum is not quantized, which disagrees with the empirical evidence. In either case, the theories do not reproduce the physical Hilbert space of solutions from the standard theory.

Here, as mentioned above, we allow for the drift potential $\phi$, and thus $\Phi$, to be multi-valued. The argument then that $\Psi$ be single-valued is addressed in \cite{Caticha 2019b}. In short, once one has a linear Schr\"{o}dinger equation, and therefore, a superposition principle, we must require that these additional ingredients remain consistent with the probabilistic structure of ED. This, it turns out, is sufficiently constraining --- for the probability $\rho = | \Psi|^{2}$ to remain single valued, the wave function $\Psi$ must be either single-valued or double-valued.

Relevant to subsequent chapters is the related issue of whether a similar objection can be raised at the level of quantum fields. Indeed, much of the discussion surrounding Wallstrom's objection is centered on the regime of non-relativistic particles, but as we will see in the next chapter, there are important differences brought about by the introduction of fields.

\let\cleardoublepage\clearpage

\chapter{Entropic dynamics of fields}\label{CH ED fields}
A common feature of both physics and inference is the importance of variable choice. Indeed, proposing a set of variables for the purposes of describing a set of phenomena is a problem typical of model building. And, in some cases, this is an impossibly difficult task --- is, for instance, the microscopic structure of gravity comprised of loops? Or perhaps strings? Or, is space-time discrete, like in causal set theory? Other times, however, a model based on certain variables so readily succeeds at explaining many desirable features that one cannot help but regard this model as capturing more objective ``truth" than the alternatives; and in this sense we claim such a model is ``better".

In the previous chapter we highlighted how the ED framework could fruitfully be applied to a system of non-relativistic quantum particles. Among the most important aspects of this development was that the particles were assumed to have \textit{definite} positions. Questions that appear paradoxical and ambiguous in the standard Copenhagen appear reasonable from this change in perspective. Are quantum objects, for example, truly particles, or are they waves? From this viewpoint quantum particles are just \textit{particles}; the wave-like properties of matter can be \textit{explained} simply as features of the epistemic state, not of the particles themselves. Similarly, whereas the quantum measurement problem is typically treated in an \textit{ad hoc} manner, here it is handled completely within the framework of Bayesian and entropic updating. 

Still, this model with particle positions taking center stage is not completely satisfactory, either. For one, the theory deals only with distinguishable particles; and the transition to a theory of indistinguishable particles is not straightforward (see e.g., \cite{Laidlaw DeWitt-Morette 1971}\cite{Leinaas Myrheim 1977}). Another drawback is that the theory is inherently \textit{non-relativistic} and deals only with a fixed number of particles. This runs counter to the fact that relativistic processes allow for an indefinite number of particles through absorbtion and emission of.\footnote{On a side note, despite its success, it is certainly possible that the principles of relativity might break down. In recent years this possibility has inspired some to develop models where the usual Lorentz invariance is violated (see e.g., \cite{Gambini Pullin 1999}\cite{Jacobson Mattingly 2001}\cite{Ameline-Camelia 2002}\cite{Barbour et al 2005}). To date, however, there does not seem to be any empirical support for these claims. (For an older review of Lorentz invariance violations, see e.g., \cite{D Mattingly 2005}). Thus for the time the principle of relativity appears safe as a fundamental symmetry.} Moreover, there is strong evidence that the notion of a particle might itself be a fraught concept (see e.g., \cite{Birrell/Davies}\cite{Unruh}\cite{Fulling}\cite{Davies}), and thus a weak starting point for a model. A better way forward, one that settles these criticisms, is afforded by a model that treats the \textit{field} concept as primary. And although, for instance, some attempts have been made in the past to model relativistic theories of particles \cite{Dirac 1932}\cite{Weiss 1938}, it is by now well agreed upon that the field concept is the natural setting for relativistic theories.\footnote{The book by S. Schweber \cite{Schweber} has an interesting discussion on the ``field versus particle" debate that occurred in the early days of quantum field theory.}

Here we develop a relativistic ED by way of analyzing the dynamics of a scalar field $\chi(x)$. The primary result that we present is a derivation of the functional Fokker-Planck and functional Hamilton-Jacobi equations. By combining these two equations we produce a third equation, a functional Schr\"{o}dinger equation, which is equivalent to the standard quantum theory of relativistic scalar fields (see e.g., \cite{Jackiw 1989}\cite{Long Shore 1998}). The procedure for accomplishing this is largely adapted from Ipek and Caticha \cite{Caticha Ipek 2014}, but deviates slightly from their method. In particular, we follow the developments of the previous chapter by outright imposing the use of a Hamiltonian for updating the (functional) drift potential $\phi[\chi]$, as opposed to imposing energy conservation to meet these same ends, such as in \cite{Caticha Ipek 2014}.

Among the insights gained by this reconstruction is that the state $\Psi$ in quantum field theory is also epistemic, just as in the usual quantum mechanics. This is particularly important in the context of quantum fields as this viewpoint sheds new light on the divergences that typically plague quantum field theories. Indeed, we find that these divergences appear as \textit{epistemic}, not physical effects; arising in the computation of variances, expected values, and other quantities pertaining to our \textit{knowledge} of the system. Moreover, a virtue of our current method is that the quantum field theory presented here automatically satisfies the Bose-Einstein statistics expected of scalar field theory. Put another way, here we can claim to have derived the relationship between spin and statistics for a scalar field, all without recourse to the \emph{ad hoc} notion of commutation (or anti-commutation) relations.

Additional work that is novel to this thesis pertains to the study of the phase $\Phi$ in quantum field theory. As discussed in the previous chapter, T. C. Wallstrom \cite{Wallstrom 1994} objected to the claim that Nelson's stochastic mechanics was, in fact, equivalent to conventional quantum mechanics. The argument, which we will discuss in more detail below, levied that reconstructions of quantum theory based on the real-valued variables $\rho$ and $\Phi$ tend to be problematic in that either the phase $\Phi$ is single valued, or if it is not, then the quantum state $\Psi$ is multi-valued, unless one postulates an additional \emph{ad hoc} quantization condition. Below we discuss how this criticism is evaded in the ED approach to quantum scalar fields. It should also be noted that such an analysis appears to be entirely unique, as most consideration of this issue appears confined to the domain of particles in quantum mechanics.

This chapter is laid out as follows. In section \ref{dynamics/fields} we introduce the field concept and consider the short step dynamics of the scalar field $\chi(x)$. In section \ref{Fields Entropic_time} we briefly recap the entropic time construct, as it is largely unchanged from the previous chapter. Section \ref{Fields Non_DissipativeED} introduces the model we use for updating our constraints; as before, we use the notion of a conserved symplectic form, while section \ref{RQFT} discusses relativistic quantum scalar fields coming from the ED approach. In section \ref{Ch 5 Wallstrom} we discuss how the ED approach to quantum scalar fields avoids the so-called Wallstrom objection. Finally, we conclude in section \ref{Fields Conclusion} with some remarks about our results and discuss the shortcomings with our model.

\section{The infinitesimal dynamics of fields\label{dynamics/fields}}
Our goal here is to predict and understand the dynamical behavior of a scalar field. We proceed by making two main dynamical assumptions. The first is that change happens; we do not ask why, but given that it does, what can we rationally infer from this knowledge. Our second assumption is that, in ED, motion is continuous, such that large changes follow from the accumulation of many infinitesimally small steps. The ED framework provides a methodology for estimating this infinitesimal dynamics.
\paragraph{Subject matter--- }
In an inference scheme the physics is introduced through
the choice of variables and of constraints. Here we deal with a real scalar field $\chi \left( x^{i}\right)\equiv \chi _{x}$. The field $\chi_{x} $ lives on a three dimensional flat Euclidean space with metric $\delta_{ij}$, and is labeled by Cartesian coordinates $x^{i}$ ($i=1,2,3$). The field $\chi_{x}$ is a scalar in the sense that it is invariant with respect to the isometries of Euclidean space: global translations and rotations.
\paragraph*{Some notation--- }
A single scalar field $\chi_{x}$ associates one real degree of freedom to each spatial point. Thus an entire field configuration $\chi = \{\chi_{x_{1}}, \chi_{x_{2}},\cdots\}$, which denotes the simultaneous values of the field at all spatial points, can be represented by a point in an infinite dimensional configuration space $\mathcal{C}$.\footnote{$\infty$-infinite dimensional spaces are complicated objects. We make no claim of mathematical rigor and follow the standard assumptions, notation, and practices of the subject. To be definite we can, for example, assume that the fields are initially defined on a discrete lattice (which makes the dimension of $\mathcal{C}$ infinite but countable) and that the continuum is eventually reached in the limit of vanishing lattice constant.} The probability that the true values of the field $\hat{\chi}$ lie in some infinitesimal region of $\mathcal{C}$ is given, in the usual way, by
\begin{equation}
\text{Pr}\left [\chi < \hat{\chi} < \chi +\delta\chi\right ] = \rho\left [ \chi \right ]D\chi~,
\end{equation}
where
\begin{equation}
D\chi \sim \prod_{x} d\chi_{x}~
\end{equation}
is an integration measure over the configuration space $\mathcal{C}$. The probability \emph{density} $\rho[\chi]$ assigns a real number to every field configuration, or point in configuration space, i.e. $\rho[\chi]: \mathcal{C}\longrightarrow \mathbb{R}$, which makes it a \emph{functional} defined over $\mathcal{C}$. We follow standard practice in denoting such functional dependence with bracket notation, such as $\rho[\chi]$.


\begin{remark*}
In the ED model that we set forth, it is the field variables $\chi_{x}$ which have values that are definite, but uncertain; it is these values that we wish to infer. Interestingly, taking the field concept as primary presents a new challenge. Whereas in the ED of particles the problem was to explain the wave-like properties of matter, here the situation is reversed: while the wave-like features of matter are somewhat more natural with a field picture, it is the particulate nature of matter that requires explaining.
\end{remark*}



\paragraph{Maximum Entropy--- }
The dynamical problem that we wish to make progress on is, given some initial information about our system, what future behavior do we expect?  The distribution that quantifies this informational state is the transition probability $P[\chi^{\prime}|\chi]$, which gives the probability of the field jumping from $\chi\in \mathcal{C}$ to an unknown $\chi^{\prime}\in \mathcal{C}$. We obtain this distribution using the ME method by maximizing the entropy
\begin{equation}
S[P,Q] = - \int D\chi^{\prime}P[\chi^{\prime}|\chi]\log\frac{P[\chi^{\prime}|\chi]}{Q[\chi^{\prime}|\chi]}
\label{Fields entropy a}
\end{equation}
subject to the appropriate constraints and relative to the \textit{prior} distribution $Q[\chi^{\prime}|\chi]$.
\paragraph*{The prior--- }
We adopt here a prior $Q\left[ \chi ^{\prime }|\chi \right] $ that incorporates the information that the fields move in a continuous fashion, but is otherwise maximally uninformative. Such a prior state reflects the knowledge that the field changes in infinitesimally small increments, but is completely silent on the correlations between the degrees of freedom. The prior distribution $Q\left[ \chi ^{\prime }|\chi \right] $ that expresses this is given by a product of Gaussians
\begin{equation}
Q\left[ \chi ^{\prime }|\chi \right] \propto \,\exp -\frac{1}{2}\int
dx\,\alpha _{x}\left( \Delta \chi _{x}\right) ^{2}~. \label{Fields prior}
\end{equation}
Such a distribution, however, can itself be derived from the principle of maximum entropy. Indeed, maximize 
\begin{equation}
S[Q,\mu ]=-\int d\chi ^{\prime }\,Q\left[ \chi ^{\prime }|\chi \right] \log 
\frac{Q\left[ \chi ^{\prime }|\chi \right] }{\mu (\chi ^{\prime })}~,
\label{Fields entropy b}
\end{equation}%
relative to the measure $\mu (\chi ^{\prime })$, which we take to be
uniform, and subject to the relevant constraints. Here these constraints contain the information that the motion of the field is continuous, and is implemented by imposing an infinite number of independent constraints
\begin{equation}
\left\langle \left( \Delta \chi _{x}\right) ^{2}\right\rangle \equiv \int
D\chi ^{\prime }\,P\left[ \chi ^{\prime }|\chi
\right] \left( \Delta \chi _{x}\right) ^{2}=\Delta\kappa _{x}~,
\label{Fields Constraint 1}
\end{equation}%
one per spatial point $x\in\mathbb{R}^{3}$. This yields the prior $Q[\chi^{\prime}|\chi]$ given in (\ref{Fields prior}). Here the $\Delta\kappa _{x}$ are required to be small quantities, which can be enforced by taking the limit $\alpha_{x}\to\infty$ for the associated Lagrange multipliers.

\paragraph*{The drift constraint--- }
The prior $Q[\chi^{\prime}|\chi]$ alone depicts a rather simple dynamics where the field variables change independently of one another and exhibit no correlations. A more interesting dynamical process can be obtained by imposing a single additional constraint that couples the degrees of freedom together. This is implemented through the introduction of the functional $\phi[\chi]$ which we call the \textit{drift potential}. More explicitly, we impose that
\begin{equation}
\left\langle \Delta \phi\right\rangle =\int D\chi ^{\prime }\,\int dx\,\,P\left[ \chi ^{\prime }|\chi
\right]  \,\Delta \chi _{x}%
\frac{\delta \phi \left[ \chi \right] }{\delta \chi _{x}}=\Delta\kappa ^{\prime },
\label{Fields Constraint 2}
\end{equation}
where we also take the limit $\Delta\kappa ^{\prime }\to 0$.

\begin{remark*}
The introduction of the drift potential can be justified, in part, by two complementary principles. From one viewpoint, a defining feature of quantum theory is the presence of non-local correlations, thus the presence of a single global constraint serves to incorporate this crucial information. But this is not quite enough for a quantum dynamics. Another important insight comes from recognizing that we deal with the dynamics of probability. Thus if we take the probabilities as a dynamical coordinate, we must also introduce its conjugate momentum. The constraint (\ref{Fields Constraint 2}) lets us to unify all of these ingredients by allowing us to eventually identify the drift potential $\phi[\chi]$ with this conjugate momentum.


\end{remark*}

\begin{remark*}
As in the previous chapter, we must supplement the inclusion of the drift potential with another \emph{assumption}: for ED to reproduce quantum mechanics the drift potential must have the topological properties of an angle, such that $\phi$ and $\phi + 2\pi$ can be identified with the same physical state.\footnote{In the approach taken by Caticha in \cite{Caticha 2019b} one assumes a spherical geometry for the e-phase space so that the angular property of $\phi$ is more natural.}
\end{remark*}
\paragraph*{The transition probability --- }
To determine the transition probability $P\left[ \chi ^{\prime }|\chi
\right]$ we maximize the entropy, eq.(\ref{Fields entropy a}) subject to the constraint eq.(\ref{Fields Constraint 2}) and normalization. The result is a Gaussian transition probability distribution%
\begin{equation}
P\left[ \chi ^{\prime }|\chi\right] =\frac{1}{Z%
\left[ \alpha _{x}\right] }\,\exp -\frac{1}{2}\int dx\,\alpha _{x}\left( \Delta \chi _{x}-\frac{\alpha^{\prime}}{\alpha _{x}}\frac{\delta \phi \left[ \chi \right] }{\delta \chi
_{x}}\right) ^{2},  \label{Fields Trans Prob}
\end{equation}%
where $Z\left[ \alpha _{x}\right] $ is a normalization constant and $\alpha^{\prime}$ is the multiplier conjugate to the constraint (\ref{Fields Constraint 2}).

The multiplier $\alpha^{\prime}$ regulates the strength of the constraint eq.(\ref{Fields Constraint 2}); and, as was mentioned in the previous chapter, and in more detail in \cite{Bartolomeo Caticha 2016}, we can set $\alpha^{\prime} = 1$. With regards to the set of multipliers $\alpha_{x}$, we choose them here to be a spatial constant $\alpha_{x} = \alpha$ to reflect the translational symmetry of flat Euclidean space. In subsequent chapters, in the context of curved space-time, we relax this assumption.
\paragraph*{Motion--- }
The Gaussian form of (\ref{Fields Trans Prob}) allows us to present a generic
change $\Delta \chi _{x}=\left\langle \Delta \chi _{x}\right\rangle +\Delta
w_{x}$ as resulting from an expected drift $\left\langle \Delta \chi
_{x}\right\rangle $ plus fluctuations $\Delta w_{x}$. The expected step is 
\begin{equation}
\left\langle \Delta \chi _{x}\right\rangle =\frac{1}{\alpha}\frac{\delta \phi \left[ \chi \right] }{\delta \chi_{x}}~,  \label{Fields Exp Step 1}
\end{equation}%
while the fluctuations $\Delta w_{x}$ satisfy,%
\begin{equation}
\left\langle \Delta w_{x}\right\rangle =\left\langle \Delta \chi
_{x}-\left\langle \Delta \chi_{x}\right\rangle \right\rangle =0,\hspace{.5 cm}\textrm{as well as,}\hspace{.5 cm}
\left\langle \Delta w_{x}\Delta w_{x^{\prime }}\right\rangle
=\frac{1}{\alpha}\delta _{xx^{\prime }}~.
\label{Fluctuations}
\end{equation}%
We see that even as $\langle \Delta \chi_{x}\rangle\sim 1/\alpha$, the fluctuations $\Delta w_{x}\sim 1/\alpha^{1/2}$, therefore in the limit $\alpha\rightarrow \infty $ the fluctuations dominate the motion. This leads to a continuous, but non-differentiable dynamics for the field variables $\chi_{x}$.
\section{Entropic time and dissipative dynamics\label{Fields Entropic_time}}
With knowledge of what infinitesimal steps we should expect in place, our goal here is to devise a scheme for iterating and organizing the sequence of steps. The scheme that we implement here is identical to the scheme that was implemented for non-relativistic particles in the previous chapter, with the result being a dynamical evolution for the probability $\rho[\chi]$ in the form of a Fokker-Planck equation. Surprisingly, the ED based on this version of entropic time still leads to a relativistic field theory, but one that is not manifestly relativistic, which will be addressed in the next chapter.
\paragraph*{Relativistic considerations--- }
In the previous chapter, an entropic instant was defined as a question, containing three ingredients: (1) a tool for labeling the succession of queries. (2) A specification of the microstates under investigation. (3) An identification of the relevant information. In the case of particles, these three ingredients could be distilled into the pair of variables $\rho_{t}(x)$ and $\phi_{t}(x)$, defined on a given spatial surface.

The situation for relativistic fields can be framed in somewhat analogous terms. We slice Minkowski space into a succession of level hyperplanes labeled by a time parameter. On each such hyperplane we define our ontic field variables $\chi_{x}$ and construct a corresponding configuration space which describes the possible values of the field at each spatial slice, after which it is appropriate to assign an epistemic state, given by $\rho_{t}[\chi]$ and $\phi_{t}[\chi]$.
\begin{remark*}
While the procedure we have just outlined is a perfectly valid approach, it suffers from an important defect. Namely, by virtue of a particular slicing of Minkowski space, it commits to one specific notion of simultaneity, seemingly in violation of the principle of relativity. Despite this apparent violation of relativity, it will nevertheless be possible to proceed with this scheme and show that the final results are independent of the initial choice of reference frame.
\end{remark*}
\paragraph*{Ordered instants--- }
The goal is to update from one instant to the next. That is, given the information available at one instant, we wish to construct the next instant. For this we look to the joint distribution $\rho[\chi,\chi^{\prime}] = \rho[\chi]P[\chi^{\prime}|\chi]$, where $P[\chi^{\prime}|\chi]$ is the infinitesimal transition probability given in eq.(\ref{Fields Trans Prob}). 

As outlined in the previous chapter, if an initial instant is defined by the probability $\rho[\chi]$, then the subsequent instant $\rho^{\prime}[\chi^{\prime}]$ is obtained using the sum rule:
\begin{equation}
\rho^{\prime}[\chi^{\prime}] = \int D\chi \, \rho[\chi,\chi^{\prime}] = \int D\chi \, \rho[\chi]P[\chi^{\prime}|\chi].
\label{Fields ED update}
\end{equation}
Accordingly, if $\rho[\chi]$ represents the informational state an initial instant $t$, then $\rho^{\prime}[\chi^{\prime}]$ represents the informational state at the next instant $t^{\prime}$, thus suggesting the notation $\rho[\chi] = \rho_{t}[\chi]$ and $\rho^{\prime}[\chi^{\prime}] = \rho_{t^{\prime}}[\chi^{\prime}]$, so that
\begin{equation}
\rho_{t'}[\chi^{\prime}] = \int D\chi \, \rho_{t}[\chi]P[\chi^{\prime}|\chi].
\label{Fields Chapman-Kolmogorov eqn}
\end{equation}
The scheme being iterative by nature --- $\rho_{t}[\chi]$ begets a $\rho_{t'}[\chi^{\prime}]$, which itself begets a $\rho_{t^{\prime\prime}}[\chi^{\prime\prime}]$, and so on --- determines an evolution for the probability $\rho_{t}[\chi]$.
\paragraph*{Duration--- }
What remains to be addressed in our construction of entropic time is the specification of the time interval $\Delta t = t'-t$. In the context of non-relativistic particles it was natural to choose a notion of time that flowed ``\textit{equably, everywhere and everywhen}," in the Newtonian fashion. Here we attempt to develop a theory of relativistic scalar fields in a background Minkowski space. As such, we take advantage of the homogeneity that is intrinsic to Minkowski space and declare that the interval $\Delta t$ be a constant that is independent of position or time --- consistent with the isometries of flat space-time. (In the next chapter, our implementation of a fully covariant ED will require us to relax the assumption of a uniform interval $\Delta t$ in favor of a local notion of duration.)

This is implemented by choosing the Lagrange multiplier $\alpha$ to be
\begin{equation}
\alpha = \frac{1}{\Delta t}\quad\text{so that}\quad\left\langle\Delta w_{x}\Delta w_{x'}\right\rangle = \Delta t \delta_{xx'}~.
\end{equation}
With this choice, the transition probability eq.(\ref{Fields Trans Prob}) describes a \textit{Wiener} process in a background Minkowski space.\footnote{Note that this choice of $\alpha$ eventually implies a system of units where $\hbar = 1$.}
\subsubsection*{The functional Fokker-Planck equation\label{Fields DissipativeED}}
The integral equation, eq.(\ref{Fields Chapman-Kolmogorov eqn}), can be written more conveniently as a differential equation (see Appendix \ref{appendix FP}), called the Fokker-Planck (FP) equation
\begin{equation}
\partial_{t}\rho_{t} =-\int dx\left[\frac{\delta}{\delta\chi_{x}}\left(\rho_{t}\frac{\delta\phi_{t}}{\delta\chi_{x}}\right)-\frac{1}{2}\frac{\delta^{2}\rho_{t}}{\delta\chi^{2}_{x}}\right]~.
\label{Fields FP equation}
\end{equation}
The FP equation can then, in turn, be written more conveniently as a continuity equation
\begin{equation}
\partial_{t}\rho_{t} = -\int dx \frac{\delta}{\delta\chi_{x}}\left(\rho_{t}\,\frac{\delta\Phi_{t}}{\delta\chi_{x}}\right)
\label{Fields continuity equation}
\end{equation} 
with current velocity
\begin{equation}
V_{x}[\chi] = \frac{\delta\Phi_{t}[\chi]}{\delta\chi_{x}}\quad\text{where}\quad \Phi_{t}[\chi] = \phi_{t}[\chi] - \log\rho^{1/2}.\label{Fields current velocity}
\end{equation}

\begin{remark*}
Similar to the previous chapter, the functional $\Phi$ will eventually be identified as the phase of the quantum state $\Psi$. Moreover, the phase $\Phi$ is also the momentum conjugate to $\rho$, as the relationship between $\Phi$ and the drift potential $\phi$, given in (\ref{Fields current velocity}), is just a canonical transformation. Finally, the phase $\Phi$ also has the global topology of an angle, a property that it inherits from the drift potential $\phi$.
\end{remark*}
\paragraph*{Prelude to non-dissipative ED--- }
The FP equation accounts fully for one half of the major dynamical equations in the ED of scalar fields. It can be written in a more compact and suggestive form by the introduction of the concept of an \textit{ensemble}-functional, or e-functional for short. Just as a regular functional such as $\rho_{t} \left[ \chi \right] $ maps a field $\chi $ into a real number (a probability in this case), an e-functional maps a functional, such as $\rho_{t} \left[ \chi \right] $ or $\Phi_{t} \left[ \chi \right] $, into a real number. Then, just as one can define functional derivatives, one can also define e-functional derivatives.\footnote{An excellent brief review of the ensemble calculus is given in the appendix of \cite{Hall et al 2003}.}

Introduce an e-functional $\tilde{H}\left[ \rho _t,\Phi _t\right] $ such that, 
\begin{equation}
\partial_{t}\rho_{t}[\chi] =
\frac{\tilde{\delta}\tilde{H}\left[ \rho _t,\Phi _t%
\right] }{\tilde{\delta}\Phi _t\left[ \chi \right] }
\label{Fields FP equation H}
\end{equation}%
reproduces (\ref{Fields FP equation}). (In what follows we denote all ensemble quantities such as $\tilde{H}$ with a tilde: $\tilde{\delta}/\tilde{\delta}\Phi \lbrack \chi ]$ is the e-functional derivative with
respect to $\Phi \left[ \chi \right] $.) Writing the FP equation in this way does not constitute an independent additional assumption, a suitable $\tilde{H}$ can always be found. In fact, a short calculation shows that any $\tilde{H}$ of the form
\begin{equation}
\tilde{H}[\rho,\Phi] = \int D\chi\int dx\frac{1}{2}\rho\left(\frac{\delta\Phi}{\delta\chi_{x}}\right)^{2} + \tilde{F}[\rho]
\label{Fields Hamiltonian with F}
\end{equation}
would be consistent with eq.(\ref{Fields continuity equation}). As is suggested by the notation, the e-functional $\tilde{H}$ will be later identified with the Hamiltonian.
\section{Non-dissipative dynamics --- the functional Hamilton-Jacobi equation\label{Fields Non_DissipativeED}}
The Fokker-Planck equation, eq.(\ref{Fields FP equation}), describes a standard diffusion process --- it does not describe quantum systems. As discussed in \cite{Caticha 2013}\thinspace\cite{Caticha Ipek 2014}, and the previous chapter, the solution to this problem is to evolve the constraint eq.(\ref{Fields Constraint 2}) in response to the evolving $\rho$. That is: instead of the drift potential $\phi$ being externally prescribed, we allow it to become a dynamical degree of freedom. The appropriate criterion for updating the constraints consists of demanding that the probability $\rho$ and the drift potential $\phi$ (or equivalently, the phase functional $\Phi$) form a phase space, the ensemble phase space (EPS). Which is to say that, the canonical variables $\rho$ and $\Phi$ are updated by a Hamiltonian generator, to be identified as $\tilde{H}$ introduced in eq.(\ref{Fields Hamiltonian with F}). Much of the formalism of the previous chapter can be carried over exactly, with the primary difference being the configuration space variables. 
\subsubsection*{Ensemble Phase space}
Phase spaces are introduced in physics as tools designed to help analyze the dynamics for a system of interest. Such constructions find utility in ED since we are interested in obtaining the time evolution of the probability $\rho_{t}[\chi]$ and its counterpart, the phase functional $\Phi_{t}[\chi]$. Thus our goal here is to identify a notion of phase space that is useful for our purposes. In particular, we would like to be able to construct vectors, curves, tensors, etc., for the express purpose of studying the evolution of $\rho_{t}[\chi]$ together with $\Phi_{t}[\chi]$.

\paragraph*{Some notation--- } Following the previous chapter, we construct a phase space $\Gamma$ whose points $X\in \Gamma$ are labelled by the functionals $X^{\alpha \chi} = (\rho[\chi],\Phi[\chi])$ ($\alpha = \rho, \Phi$ and $\chi \in \mathcal{C}$). Additionally, we endow the space with a symplectic structure by introducing a skew-symmetric, closed, non-degenerate two-form called the symplectic form $\mathbf{\Omega}$, thus satisfying the conditions given in section \ref{Canonical_ED_Particles}. Since $\mathbf{\Omega}$ is non-degenerate, it also has an inverse $\mathbf{\Omega}^{-1}$, which is a $(2,0)$ tensor field called the Poisson tensor.

An ensemble functional $\tilde{F}$, or e-functional, is a $(0,0)$ tensor defined over $\Gamma$ and thus an exact one-form, or $(0,1)$ tensor over $\Gamma$, can be obtained by applying the so-called exterior derivative $\tilde{\mathbf{d}}$ to such an e-functional, resulting in $\tilde{\mathbf{V}}_{\tilde{F}} = \tilde{\mathbf{d}}\tilde{F}$. Moreover, a vector field is a $(1,0)$ tensor defined at every point $X\in\Gamma$ and any vector field $\bar{V}$ that leaves the symplectic form $\mathbf{\Omega}$ invariant, i.e. $\pounds _{\bar{V}}\mathbf{\Omega} = 0$, is called a Hamiltonian vector field (HVF). Up to an additive constant, there is a one-to-one correspondence between exact one-forms and HVFs obtained by application of the symplectic form or the Poisson tensor, respectively,
\begin{equation}
\tilde{\mathbf{V}}_{\tilde{F}} = \mathbf{\Omega}(\bar{\mathbf{V}}_{\tilde{F}},\,\cdot\,) = \tilde{\mathbf{d}}\tilde{F}\quad \text{and}\quad \bar{\mathbf{V}} = \mathbf{\Omega}^{-1}(\tilde{\mathbf{V}}_{\tilde{F}},\,\cdot\,) = \mathbf{\Omega}^{-1}(\tilde{\mathbf{d}}\tilde{F},\,\cdot\,) ~.
\end{equation}
Thus the symplectic form $\mathbf{\Omega}$ provides a mapping at each point $X\in \Gamma$ from elements of the tangent space $T\Gamma_{X}$ to elements of the cotangent space $T^{*}\Gamma_{X}$.

\paragraph*{Ensemble Poisson brackets}
The geometry of the e-phase space defined above allows us to introduce some useful tools. Chief among these is the notion of a Poisson bracket
\begin{equation}
\left \{\tilde{F},\tilde{G} \right \} \equiv \mathbf{\Omega}\left (\bar{\mathbf{V}}_{\tilde{F}},   \bar{\mathbf{V}}_{\tilde{G}} \right )~.
\end{equation}
As mentioned in the previous chapter, the Poisson bracket satisfies a few important properties (a) skew-symmetry, (b) the Leibniz property, (c) the Jacobi identity. 

These properties allow for us to view the Poisson bracket $\left\{\tilde{F},\tilde{G}\right\}$ as the Lie derivative of the generator $\tilde{F}$ along the HVF generated by $\tilde{G}$:
\begin{equation}
\left\{\tilde{F},\tilde{G}\right\} = \pounds_{\bar{\mathbf{V}}_{\tilde{G}}}\tilde{F}.
\end{equation}
Moreover, since $\{\tilde{F},\tilde{G}\}:\Gamma\to\mathbb{R}$ is itself a function on $\Gamma$, then the infinite-dimensional set of all such e-functionals themselves form a group, a Lie group, under the Poisson bracket operation --- provide two e-Hamiltonians and their bi-linear product, i.e. the Poisson bracket, is also an e-Hamiltonian. The notion of Lie groups, algebras, etc., will prove to be useful in constructing relativistic theories of ED.

For any ensemble Hamiltonian, or e-Hamiltonian generator $\tilde{F},\tilde{G}:\Gamma\to\mathbb{R}$ the Poisson bracket, in local coordinates $(\rho,\Phi)$ takes the form
\begin{equation}
\left\{\tilde{F},\tilde{G}\right\} = \int D\chi\left(\frac{\tilde{\delta}\tilde{F}}{\tilde{\delta}\rho[\chi]}\frac{\tilde{\delta}\tilde{G}}{\tilde{\delta}\Phi[\chi]}-\frac{\tilde{\delta}\tilde{G}}{\tilde{\delta}\rho[\chi]}\frac{\tilde{\delta}\tilde{F}}{\tilde{\delta}\Phi[\chi]}\right).
\label{Field Poisson bracket}
\end{equation}
From this one can directly observe that the canonical Poisson bracket relations
\begin{equation}
\left \{\rho[\chi],\rho[\chi^{\prime}]\right \} = \left \{\Phi[\chi],\Phi[\chi^{\prime}]\right \} = 0\quad\text{and}\quad\left \{\rho[\chi],\Phi[\chi^{\prime}]\right \}=\delta[\chi,\chi^{\prime}]~,
\label{Fields Canonical Coordinate PB relations}
\end{equation}
are satisfied automatically, where $\delta[\chi,\chi^{\prime}]$ is a delta \textit{functional} which is the natural extension of the notion of a Dirac delta distribution to infinite dimensional configuration space $\mathcal{C}$.
\subsubsection*{Equations of motion}
The goal of introducing this formalism is, of course, to discuss dynamics. The evolution of the probability $\rho_{t}[\chi]$ is tracked by the entropic time parameter $t$, which leads to the evolution equation for $\rho_{t}[\chi]$, the FP equation, eq.(\ref{Fields continuity equation}). Determining the evolution of $\Phi_{t}[\chi]$ amounts to identifying the e-Hamiltonian that generates a curve in $\Gamma$ parameterized by that very same $t$.

Whatever form this e-Hamiltonian, say $\tilde{H}$, may take, it must reproduce the FP equation. Thus we have that
\begin{equation}
\partial_{t}\rho_{t}[\chi] = \pounds_{\bar{\mathbf{V}}_{\tilde{H}}}\rho_{t}[\chi] = \{\rho_{t}[\chi],\tilde{H}[\rho,\Phi]\}~.
\end{equation}
Using the form of the Poisson brackets given in (\ref{Field Poisson bracket}), this gives us
\begin{equation}
\partial_{t}\rho_{t}[\chi]  = \frac{\tilde{\delta}\tilde{H}[\rho,\Phi]}{\tilde{\delta}\Phi_{t}[\chi]}~,
\end{equation}
which is identical to eq.(\ref{Fields FP equation H}). Thus $\tilde{H}$ must be of the form given in eq.(\ref{Fields Hamiltonian with F}), with $\tilde{F}[\rho]$ an undetermined e-functional. Using this e-Hamiltonian, we determine that the dynamical equation for the phase functional $\Phi_{t}[\chi]$ must take the form
\begin{equation}
-\partial_{t}\Phi_{t}[\chi] = - \left \{\Phi_{t}[\chi],\tilde{H}[\rho,\Phi]\right \} =\int dx\frac{1}{2}\left(\frac{\delta\Phi_{t}}{\delta\chi_{x}}\right)^{2}+\frac{\tilde{\delta}\tilde{F}[\rho]}{\tilde{\delta}\rho_{t}[\chi]}~,
\label{Fields Functional HJ equation}
\end{equation}
which is reminiscent of a functional Hamilton-Jacobi equation, with a yet unspecified choice of ``potential" given by the choice of the e-functional $\tilde{F}[\rho]$.
\subsubsection*{Relativistic Invariance}
The ED that has been developed thus far consists of a family of theories, one for each choice of ``potential" $\tilde{F}[\rho]$. To choose a physically relevant subset of such potentials, it is natural to look to the symmetries of the problem for guidance. For instance, in a Newtonian theory, one might impose that the potentials express the rotational and translational symmetries intrinsic to flat Euclidean space, such as the Newtonian and Coulomb potentials do. Better yet, in fact, is the requirement that the Hamiltonian generators of the theory carry a \textit{representation} of the underlying symmetry group; in a Newtonian theory this is, of course, the Galilean group.

In a relativistic theory the underlying logic is the same. For a theory built upon an underlying Minkowski space, it is natural for the corresponding generators of the theory to carry a representation of the Poincar\`{e} group, in this case. This is the path that we follow here.
\paragraph*{Poincar\'{e} group and algebra--- }
Minkowski space is a flat four-dimensional (pseudo)-Riemannian space-time that, as is implied by its flatness, is both homogeneous and isotropic. First the homogeneity. In space-time homogeneity implies a translation symmetry in three space-like directions, as well as in the time-like direction. Isotropy in Minkowski space implies a symmetry of the standard three-dimensional rotations, as well as symmetry under Lorentz boosts, which ``rotate" time-like and space-like directions.

Homogeneity implies the existence of four symmetry generators --- three space-like $P_{i}$ and one time-like $P_{0}$. Isotropy implies the usual three generators of spatial rotations $J_{i}$ --- where $i = 1,2,3$ is the space-like axis of rotation --- as well as three generators of Lorentz boosts $K_{i}$ --- where $i$ is the space-like direction of the boost --- that mix the time-like direction with each spatial direction. Altogether these ten generators correspond to the isometries of Minkowksi space; that is, the flows drawn by these generators in Minkowski space leave the Minkowski metric $\eta_{\mu\nu}$ $(\mu,\nu = 0,1,2,3)$ invariant.

The Lie algebra of the Poincar\'{e} group is called the \textit{Poincar\'{e} algebra} and is given by \cite{Weinberg I 1995}
\begin{subequations}
\begin{align}
&\left[P_{i},P_{j}\right] = 0\label{Poincare algebra Pi Pj}\\
&\left[P_{i},P_{0}\right] = 0 \label{Poincare algebra Pi H}\\
&\left[J_{i},P_{j}\right] = \epsilon_{ijk}P_{k}\label{Poincare algebra Ji Pj}\\
&\left[J_{i},P_{0}\right] = 0\label{Poincare algebra Ji H}\\
&\left[J_{i},J_{j}\right] = \epsilon_{ijk}J_{k}\label{Poincare algebra Ji Jj}\\
&\left[K_{i},P_{j}\right] = \eta_{ij}P_{0}\label{Poincare algebra Ki Pj}\\
&\left[K_{i},P_{0}\right] = P_{i}\label{Poincare algebra Ki H}\\
&\left[K_{i},J_{j}\right] = -\epsilon_{ijk}K_{k}\label{Poincare algebra Ki Jj}\\
&\left[K_{i},K_{j}\right] = -\epsilon_{ijk}J_{k}.
\label{Poincare algebra Ki Kj}
\end{align}
\end{subequations}
The typical procedure for constructing models of relativistic physics is to require that the theory carries a representation of this algebra. 

Note that the splitting up of generators in this fashion is not covariant, but nevertheless, any generators that satisfy these relations will be in accordance with the special theory of relativity since satisfying this algebra guarantees that we can construct a Poincar\'{e} covariant stress-energy tensor for the dynamical variables. Furthermore, also note that the Poincar\'{e} algebra depends crucially on the fact that Minkowski space is homogeneous and isotropic, which is not the case in a general space-time. For this more general case we require a new scheme for implementing (locally) Einstein's principle of relativity. This is the subject of the next chapter.
\paragraph*{Poincar\'{e} algebra in ED--- }
We construct here a relativistic ED by imposing that our theory carries a representation of the Poincar\'{e} group, and corresponding Poincar\'{e} algebra. Since we are interested only in flows that leave $\mathbf{\Omega}$ invariant, it is natural to look for a representation of the Poincar\'{e} group in terms of e-Hamiltonian generators and to formulate their corresponding Lie algebra in terms of Poisson brackets. This means we must identify a set of 10 e-Hamiltonian generators that perform certain functions; translations in space $\tilde{P}_{i}$ and time $\tilde{P}_{0} = \tilde{H}$; spatial rotations $\tilde{J}_{i}$; and Lorentz Boosts $\tilde{K}_{i}$. These 10 e-Hamiltonians must close in exactly the same way as the Poincar\'{e} algebra, eqns.(\ref{Poincare algebra Pi Pj})-(\ref{Poincare algebra Ki Kj}), above. For completeness, we provide the Poisson bracket algebra here
\begin{subequations}
\begin{align}
&\left\{\tilde{P}_{i},\tilde{P}_{j}\right\} = 0,\label{Poincare PB algebra Pi Pj}\\
&\left\{\tilde{P}_{i},\tilde{H}\right\}=0,\label{Poincare PB algebra Pi H}\\
&\left\{\tilde{J}_{i},\tilde{P}_{j}\right\} = -\epsilon_{ijk}\tilde{P}_{k}\label{Poincare PB algebra Ji Pj}\\
&\left\{\tilde{J}_{i},\tilde{H}\right\} = 0\label{Poincare PB algebra Ji H}\\
&\left\{\tilde{J}_{i},\tilde{J}_{j}\right\} = -\epsilon_{ijk}\tilde{J}_{k}\label{Poincare PB algebra Ji Jj}\\
&\left\{\tilde{K}_{i},\tilde{P}_{j}\right\}= -\eta_{ij}\tilde{H}\label{Poincare algebra PB Ki Pj}\\
&\left\{\tilde{K}_{i},\tilde{H}\right\} = -\tilde{P}_{i}\label{Poincare algebra PB Ki H}\\
&\left\{\tilde{K}_{i},\tilde{J}_{j}\right\} = \epsilon_{ijk}\tilde{K}_{k}\label{Poincare algebra PB Ki Jj}\\
&\left\{\tilde{K}_{i},\tilde{K}_{j}\right\} = \epsilon_{ijk}\tilde{K}_{k}~.
\label{Poincare algebra PB Ki Kj}
\end{align}
\end{subequations}
(The overall minus sign discrepancy arises from the way that generic Lie brackets are defined relative to Poisson brackets. This is mentioned, for instance, in \cite{Isham/Kuchar I}.)
\paragraph*{The e-Hamiltonian--- }
As discussed above, the e-generator of time translations must be given by the $\tilde{H}$ in eq.(\ref{Fields Hamiltonian with F}). However, not just \textit{any} $\tilde{H}$ of this form is acceptable, since not just \textit{any} $\tilde{H}$ will satisfy eqns.(\ref{Poincare PB algebra Pi Pj})-(\ref{Poincare algebra PB Ki Kj}). Essentially, one can view the Poincar\'{e} algebra as a set of conditions that constrain the allowed forms of $\tilde{H}$.

To this end, consider e-Hamiltonians of the form
\begin{equation}
\tilde{H}[\rho,\Phi] = \int dx \,\tilde{\mathcal{H}}_{x}\quad\text{where}\quad \tilde{\mathcal{H}}_{x} = \int D\chi\,\rho\left(\frac{1}{2}\left(\frac{\delta\Phi}{\delta\chi_{x}}\right)^{2}+\mathcal{V}_{x}(\chi,\rho)\right)
\label{Fields Generator time Translations (local)}
\end{equation}
is a spatially local e-Hamiltonian density. More specifically, in terms of the $\tilde{H}$ in eq.(\ref{Fields Hamiltonian with F}), we have restricted ourselves to $\tilde{F}[\rho]$ of the form
\begin{equation}
\tilde{F}[\rho] = \int D\chi \rho \int dx \mathcal{V}_{x}(\chi,\rho),
\end{equation}
where the undetermined ``potential" $\mathcal{V}_{x}$ can depend on $\chi_{x}$ and $\rho[\chi]$, as well as their (respective) derivatives. The Poincar\'{e} algebra serves then to isolate a physically relevant set of potentials $\tilde{\mathcal{V}}_{x}$ that are consistent with the symmetries of Minkowski space. Of course, by virtue of our assumption in (\ref{Fields Generator time Translations (local)}), our treatment cannot be completely general, but for our purposes it is sufficient to simply obtain some solutions to the Poincar\'{e} algebra, not all of them.
 
\paragraph*{Translation and rotation e-generators--- } 
Next, notice that the e-generators $\tilde{P}_{i}$ and $\tilde{J}_{i}$ of spatial translations and rotations, respectively, can themselves be deduced by requiring that they generate the correct variations of the dynamical variables $\rho[\chi]$ and $\Phi[\chi]$ when the underlying space is translated or rotated. These can be shown to take the form
\begin{equation}
\tilde{P}_{i} = \int D\chi \rho[\chi]\int dx \, \frac{\delta\Phi[\chi]}{\delta\chi_{x}}\partial_{ix}\chi_{x}
\label{Fields Generator Translations}
\end{equation}
and
\begin{equation}
\tilde{L}_{i} = \int D\chi \rho[\chi]\int dx \, \frac{\delta\Phi[\chi]}{\delta\chi_{x}}\epsilon_{ijk}\,x^{j}\,\partial^{kx}\chi_{x}
\label{Fields Generator Rotations}
\end{equation}
for spatial translations and rotations, respectively.

\paragraph*{Lorentz boost e-generator--- }
Obtaining the boost generator $\tilde{K}_{i}$, on the other hand, is not quite as obvious. Since a boost mixes time and space, it is inextricably linked to the Hamiltonian $\tilde{H}$ itself --- as can be seen, for example, from equations (\ref{Poincare algebra PB Ki Pj}) and (\ref{Poincare algebra PB Ki H}). The Hamiltonian is itself, however, still to be determined. Nevertheless, given a Hamiltonian of the form given in eq.(\ref{Fields Generator time Translations (local)}) we can supply an ansatz for $\tilde{K}_{i}$, the generator of Lorentz boosts, to be
\begin{equation}
\tilde{K}_{i} = \int dx \, x_{i}\tilde{\mathcal{H}}_{x} - t\tilde{P}_{i}.\label{Fields Generator Lorentz Boosts}
\end{equation}
Of course, this is only a correct choice of boost e-generator provided we choose $\mathcal{V}_{x}$ appropriately.
\paragraph*{Hybrid solutions}
It is not difficult to show that potentials of the form
\begin{equation}
\mathcal{V}_{x}(\chi) = \frac{1}{2}\delta^{ij}\partial_{ix}\chi_{x}\partial_{jx}\chi_{x} + \sum_{n}\lambda_{n}\chi_{x}^{n},
\label{Fields Poincare potentials}
\end{equation}
which depend only on $\chi_{x}$ (but not on higher derivatives of $\chi_{x}$, nor $\rho$ or its derivatives) satisfy the algebra, eqns.(\ref{Poincare PB algebra Pi Pj})-(\ref{Poincare algebra PB Ki Kj}) above. While simple substitution of (\ref{Fields Poincare potentials}) into the $\tilde{H}$ and $\tilde{K}_{i}$ demonstrates that the first term in (\ref{Fields Poincare potentials}) is absolutely \textit{necessary} for the Poincar\'{e} algebra to be satisfied, the other terms do work too, but are merely optional.

Potentials that are of this type are regarded in ED as being \textit{hybrid}, analogous to the hybrid ED discussed in the previous chapter. To see this, consider the potential
\begin{equation}
\mathcal{V}_{x}^{\text{KG}}(\chi) = \frac{1}{2}\delta^{ij}\partial_{ix}\chi_{x}\partial_{jx}\chi_{x} +\frac{1}{2}m^{2}\chi_{x}^{2}.
\end{equation}
The dynamical equation for $\Phi[\chi]$ that results from this choice is
\begin{equation}
-\partial_{t}\Phi = \frac{\tilde{\delta}\tilde{H}}{\tilde{\delta}\rho} = \int dx \left(\frac{1}{2}\left(\frac{\delta\Phi}{\delta\chi_{x}}\right)^{2} + \frac{1}{2}\delta^{ij}\partial_{ix}\chi_{x}\partial_{jx}\chi_{x} +\frac{1}{2}m^{2}\chi_{x}^{2}\right),
\label{Hybrid HJ KG}
\end{equation}
which is identical to a functional HJ equation for a Klein-Gordon scalar field theory with mass $m$. Thus the Hybrid ED models describe a scalar field theory that is both fully Brownian, but which is guided by a HJ functional $\Phi[\chi]$ whose dynamics is completely ``classical". And, although the formalism singles out a particular time parameter $t$ for use, the Poincar\'{e} algebra guarantees that the resulting dynamical evolution is fully relativistic, even if it is not manifestly so.
\section{Relativistic quantum scalar fields\label{RQFT}}
The ED of fields that we have encountered thus far is already quite rich, including notions of symplectic structures, Hamiltonians, the Poincar\'{e} algebra, and so on, but it is not a quantum theory. Therefore here we pursue the steps needed to obtain a fully quantum ED and in the process we highlight many conceptual aspects of quantum field theory that are not typically discussed, such as the nature of divergences, the status of particles, the single-valuedness of the quantum state, etc.

\subsubsection*{The quantum equations of motion}
The chief dynamical equations in ED are the FP and HJ equations, the latter of which is determined by an explicit choice for the Hamiltonian $\tilde{H}$. As in the previous chapter, one we have restricted ourselves to a special class of Hamiltonians, it is convenient, although not at all necessary, to perform a change of variables which leave the equations of motion linear. In analogy with the situation in quantum mechanics, we call this a functional Schr\"{o}dinger equation. We discuss below.


\paragraph*{The quantum potential--- }
Although the quantum potential is often discussed in the context of non-relativistic particles \cite{Bohm 1952}, it is similarly crucial for quantum field theory as well (see e.g., \cite{Bohm/Hiley 1987}\cite{Bohm/Hiley 1989}\cite{Hall et al 2003}). In the ED approach to the quantum theory of scalar fields \cite{Caticha 2013}\cite{Caticha Ipek 2014} there have thus been a couple of arguments used to suggest the inclusion of the quantum potential; either based on the notion of ``osmotic" velocities \cite{Caticha 2013}, or information geometry \cite{Caticha Ipek 2014}. Unfortunately, however, in both of these cases the quantum potential has not so much been derived, but motivated based on heuristic reasoning. A more desirable direction, of course, is to extend the rigorous geometric approach of Caticha \cite{Caticha 2019b} to the domain of fields, something which will be the focus of forthcoming work, but is not included here.



At any rate, regardless of exactly how one arrives at the quantum potential, for a single quantum scalar field its form is
\begin{equation}
\tilde{Q}[\rho] = \frac{\hbar^{2}}{8}\int D\chi \, \rho \int dx \left(\frac{\delta\log\rho}{\delta\chi_{x}}\right)^{2}~,
\label{Fields Quantum potential}
\end{equation}
where the coupling constant $\hbar^{2}/8$ has been chosen for future convenience, with $\hbar$ later being identified as Planck's constant.\footnote{With this choice of units we have $\alpha = 1/\hbar\Delta t$ for the Lagrange multiplier in the transition probability.} For more sophisticated models that include mass terms and relativistic interactions one may also include potentials
\begin{equation}
\tilde{{V}}[\rho] = \int D\chi \rho \int dx \, \mathcal{V}_{x}~,
\end{equation}
where $\mathcal{V}_{x}$ are the potentials in (\ref{Fields Poincare potentials}) allowed by the Poincar\'{e} algebra. The full quantum Hamiltonian $\tilde{H}$ thus has the form
\begin{equation}
\tilde{H}_{Q} = \int D\chi \, \rho \int dx\left(\frac{1}{2}\left(\frac{\delta\Phi}{\delta\chi_{x}}\right)+\frac{1}{2}\partial_{ix}\chi_{x}\partial_{x}^{i}\chi_{x}+\mathcal{V}_{x}+\frac{\hbar^{2}}{8}\left(\frac{\delta\log\rho}{\delta\chi_{x}}\right)^{2}\right)~.
\label{Fields Quantum Hamiltonian}
\end{equation}

\begin{remark*}
Although we do not present it here, it is, nonetheless, a relatively straightforward exercise to show that the $\tilde{H}_{Q}$ in (\ref{Fields Quantum Hamiltonian}) satisfies the Poincar\'{e} algebra above. Indeed, simply insert this potential into eqns.(\ref{Poincare PB algebra Pi Pj})-(\ref{Poincare algebra PB Ki Kj}) and the result follows.
\end{remark*}

\paragraph*{Functional Schr\"{o}dinger equation--- }
The e-Hamiltonian $\tilde{H}_{Q}$ in (\ref{Fields Quantum Hamiltonian}) can be put into a particularly simple form by making a canonical transformation from the two real functionals $\rho$ and $\Phi$ to the complex variables
\begin{equation}
\Psi = \rho^{1/2} \, \exp i\Phi/\hbar\quad\text{and}\quad \Psi^{*} = \rho^{1/2}\, \exp -i\Phi/\hbar~.\label{Fields Quantum Wave Functionals}
\end{equation}
In these new coordinates, the e-Hamiltonian becomes
\begin{equation}
\tilde{H}_{Q} = \int D\chi \,\Psi^{*}\int dx\left(-\frac{\hbar^{2}}{2}\frac{\delta^{2}}{\delta\chi_{x}^{2}}+\frac{1}{2}\partial_{ix}\chi_{x}\partial_{x}^{i}\chi_{x}+\mathcal{V}_{x}\right)\Psi,\label{Fields Quantum Hamiltonian Psi}
\end{equation}
and the corresponding dynamical equation for $\Psi_{t}[\chi]$ becomes
\begin{equation}
i\hbar\partial_{t}\Psi_{t}[\chi] = i\hbar\left  \{\Psi_{t}[\chi],\tilde{H}_{Q}\right \} = \hat{H}\Psi_{t}[\chi]\label{Fields Functional Schrodinger Equation}
\end{equation}
where
\begin{equation}
\hat{H} = \int dx\left(-\frac{\hbar^{2}}{2}\frac{\delta^{2}}{\delta\chi_{x}^{2}}+\frac{1}{2}\partial_{ix}\chi_{x}\partial_{x}^{i}\chi_{x}+\mathcal{V}_{x}\right)\label{Fields Quantum Hamiltonian Operator}
\end{equation}
is a linear differential operator.

\subsubsection*{Relation to the usual QFT}
Equation (\ref{Fields Functional Schrodinger Equation}) is a functional Schr\"{o}dinger equation, and thus this is quantum field theory in the Schr\"{o}dinger representation; which is entirely equivalent to the standard approaches (see e.g., \cite{Jackiw 1989}\cite{Long Shore 1998}). In ED, however, one takes a more indirect route to recovering the conventional quantum formalism, together with the Hilbert spaces, Hermitian operators, and the rest of the usual quantum formalism.

\paragraph*{Generators and Lie brackets--- }
As discussed in the previous chapter (see e.g., \cite{Schilling thesis}\cite{Schilling Ashtekar 1999}), the relationship between Hermitian operators $\hat{A}$ that act on a complex Hilbert space and e-functionals $\tilde{A}$ defined over the e-phase space $\Gamma$ is given by
\begin{equation}
\tilde{A}[\rho,\Phi] = \int D\chi \Psi^{*}\hat{A}\Psi = \left \langle \hat{A} \right \rangle~.\label{Fields Quantum e-functional}
\end{equation}
Although not all e-generators will have an interpretation in terms of quantum operators --- they must be quadratic in $\Psi$, $\Psi^{*}$ --- for those that do, the corresponding Poisson brackets are simply related to the commutators of the respective operators. That is, for $\tilde{A}_{Q} = \left\langle \hat{A}\right\rangle$ and $\tilde{B}_{Q} = \left\langle \hat{B}\right\rangle$, we have that
\begin{equation}
\{\tilde{A}_{Q},\tilde{B}_{Q}\} = \frac{1}{i\hbar}\left\langle [\hat{A},\hat{B}]\right\rangle~.\label{Fields Quantum PB/commutator}
\end{equation}

\paragraph*{Quantization without commutators--- }
In many standard approaches to QFT, a classical field theory is ``quantized" by an \emph{ad hoc} prescription whereby classical observables are promoted to quantum operators, or more technically, operator valued distributions, acting on an abstract Hilbert space. Central to this procedure for Bosonic fields, in particular, are the canonical Poisson bracket relations between the classical phase space variables, say, a scalar field $\chi_{x}$ and its conjugate momentum $\pi_{x}$
\begin{equation}
\left \{ \chi_{x}, \pi_{x^{\prime}}\right \} = \delta(x,x^{\prime})~.
\end{equation}
The usual quantization algorithm then rests crucially on the assumption that an analogous relationship holds in the quantum theory, as well,
\begin{equation}
\left [\hat{\chi}_{x},\hat{\pi}_{x^{\prime}} \right ] = i\hbar \delta(x,x^{\prime})~,
\end{equation}
but with $\hat{\chi}_{x}$ and $\hat{\pi}_{x^{\prime}}$ now understood as operators with the Lie bracket $\left [\,\cdot\, , \,\cdot\,  \right ]$ being the commutator. Again, this step cannot be explained, at best it can be motivated.\footnote{For instance, one might argue that the spin-statistics theorem gives support for the use of commutation versus anti-commutation relations for Boson fields. But such an argument still cedes the point that observables be represented as operators in the first place --- that part must be postulated.}

A major benefit of the ED approach, on the other hand, is that this relationship can be derived. Indeed, consider the expected value of the field $\chi_{x}$
\begin{equation}
\tilde{\chi}_{x} = \int D\chi \, \rho\, \chi_{x}~,
\end{equation}
as well as the e-functional
\begin{equation}
\tilde{\Pi}_{x} = \int D\chi \, \rho \,\frac{\delta\Phi}{\delta\chi_{x}}~.
\end{equation}
Introducing such objects is entirely natural in ED and requires no assumptions. To interpret the e-functional $\tilde{\Pi}_{x}$, notice that it is the expected value of the current velocity introduced in (\ref{Fields current velocity}). Computing the Poisson bracket of these two e-functionals we obtain
\begin{equation}
\left \{ \tilde{\chi}_{x}, \tilde{\Pi}_{x^{\prime}}\right \} = \delta(x,x^{\prime})~,
\end{equation}
which is reminiscent of the canonical Poisson bracket relations one has in the classical theory. This suggests that we can also interpret $\tilde{\Pi}_{x}$ as something like an expected momentum,\footnote{Indeed, in a flat space-time, the current velocity and current momentum are both given by the gradient of $\Phi$.} so that $\tilde{\chi}_{x}$ and $\Pi_{x}$ resemble a canonical pair.\footnote{Note, however, that the resemblance is purely formal. In ED the canonical variables are $\rho$ and $\Phi$, or perhaps even $\Psi$ and $i\hbar\Psi^{*}$, but never are the field and its ``momentum" a true canonical pair.}

Alternatively, consider a transformation to the complex variables $\Psi$ and $\Psi^{*}$ so that
\begin{equation}
\tilde{\chi}_{x} = \int D\chi \Psi^{*}\, \chi_{x}\, \Psi
\end{equation}
and
\begin{equation}
\tilde{\Pi}_{x} = \int D\chi \, \Psi^{*}\hat{\Pi}_{x}\Psi~,\quad\text{where}\quad \hat{\Pi}_{x} = -i\hbar \frac{\delta}{\delta\chi_{x}}~.
\end{equation}
The benefit of this variable change is that $\tilde{\chi}_{x}$ and $\tilde{\Pi}_{x}$ are exactly the types of e-functionals introduced in (\ref{Fields Quantum e-functional}). Thus it follows from (\ref{Fields Quantum PB/commutator}) that the Poisson bracket between them has the form
\begin{equation}
\left \{ \tilde{\chi}_{x}, \tilde{\Pi}_{x^{\prime}}\right \} = \frac{1}{i\hbar}\int D\chi \Psi^{*}\left [\chi_{x},\hat{\Pi}_{x^{\prime}}\right ]\Psi = \delta(x,x^{\prime})~,
\end{equation}
implying for arbitrary choices of $\Psi$ that
\begin{equation}
\left [\chi_{x},\hat{\Pi}_{x^{\prime}}\right ] = i\hbar\delta(x,x^{\prime})~,\label{Fields CCR}
\end{equation}
which is exactly the canonical commutator relation usually found in standard approaches. However, while this typically serves as a basic postulate in the usual quantum field theory, in ED this relationship has been derived.

\begin{remark*}
An advantage of this approach is that it sidesteps the issue of representing observables as operators and therefore of constructing the Hilbert spaces on which they presumably act. Indeed, in QFT, where the number of field degrees of freedom are (uncountably) infinite, the Stone-von Neumann theorem does not hold (see e.g., \cite{Wald 1994}) and thus there are an infinite number of representations of the canonical commutation relations, all of which are unitarily inequivalent. While the issue is less problematic in a flat space-time, where a unique Poincar\'{e} invariant vacuum state can be constructed, it presents a significant problem in curved space-time where there is no Poincar\'{e} invariance, and consequently, no unique vacuum state, either. In ED, however, a quantum theory is obtained without encountering these issues which are emblematic of an emphasis on Hilbert space methods. Thus, as we will see in the next chapter, the fact that ED proceeds without appealing to the notion of Hilbert spaces is thus a decided advantage for constructing a quantum theory in the context of a curved space-time.
\end{remark*}

\paragraph*{Quantum Poincar\'{e} algebra}
As mentioned above, with the inclusion of the quantum potential in the Hamiltonian there is a simple relation between $\tilde{H}_{Q}$  and the operator $\tilde{H}$, namely, $\tilde{H}_{Q} = \left\langle \hat{H}\right\rangle$. Such a relationship holds for the other symmetry generators, as well: $\tilde{P}_{i} = \left\langle \hat{P}_{i}\right\rangle$ for translations, $\tilde{L}_{i} = \left\langle \hat{L}_{i}\right\rangle$ for rotations, and $\tilde{K}_{i} = \left\langle \hat{K}_{i}\right\rangle$ for boosts.

Owing to (\ref{Fields Quantum PB/commutator}), it is therefore possible to formulate the ensemble Poincar\'{e} algebra eqns.(\ref{Poincare PB algebra Pi Pj})-(\ref{Poincare algebra PB Ki Kj}) in terms of quantum operators and their commutators:
\begin{subequations}
\begin{align}
&\left[\hat{P}_{i},\hat{P}_{j}\right] = 0\label{Poincare commutator algebra Pi Pj}\\
&\left[\hat{P}_{i},\hat{H}\right] = 0 \label{Poincare commutator algebra Pi H}\\
&\left[\hat{J}_{i},\hat{P}_{j}\right] = i\hbar\epsilon_{ijk}\hat{P}_{k}\label{Poincare commutator algebra Ji Pj}\\
&\left[\hat{J}_{i},\hat{H}\right] = 0\label{Poincare commutator algebra Ji H}\\
&\left[\hat{J}_{i},\hat{J}_{j}\right] = i\hbar\epsilon_{ijk}\hat{J}_{k}\label{Poincare commutator algebra Ji Jj}\\
&\left[\hat{K}_{i},\hat{P}_{j}\right] = i\hbar\eta_{ij}\hat{H}\label{Poincare commutator algebra Ki Pj}\\
&\left[\hat{K}_{i},\hat{H}\right] = i\hbar \hat{P}_{i}\label{Poincare commutator algebra Ki H}\\
&\left[\hat{K}_{i},\hat{J}_{j}\right] = -i\hbar \epsilon_{ijk}K_{k}\label{Poincare commutator algebra Ki Jj}\\
&\left[\hat{K}_{i},\hat{K}_{j}\right] = -i\hbar\epsilon_{ijk}J_{k}~.
\label{Poincare commutator algebra Ki Kj}
\end{align}
\end{subequations}
Thus a formal equivalence between ED and the usual relativistic QFT can be made.

\subsubsection*{Free field theory}
Here we discuss free fields in ED, which is a context that is appropriate for drawing conceptual distinctions between ED and the standard QFT. Indeed, setting $\mathcal{V}_{x} = \frac{1}{2}m^{2}\chi_{x}^{2}$ (in units $\hbar = c = 1$) the functional Schr\"{o}dinger equation reads
\begin{equation}
i\partial_{t}\Psi=\hat{H}_{\text{KG}}  \Psi
\label{Fields Functional Schrodinger Equation KG}
\end{equation}
with
\begin{equation}
\hat{H}_{\text{KG}} = \frac{1}{2}\int dx\left[  -\frac
{\delta^{2}}{\delta\chi_{x}^{2}} + \chi_{x}(-\nabla^{2}+m^{2})\chi_{x}\right]~,
\label{Fields KG Hamiltonian}
\end{equation}
which exactly reproduces the quantum theory of free real scalar fields \cite{Jackiw 1989}. Alternatively, we can write this Hamiltonian using the notion of a real symmetric kernel operator 
\begin{equation}
\omega^{2}(\vec{x},\vec{y}) =  (-\nabla^{2}+m^{2})\delta^{3}(\vec{x}-\vec{y})
\label{Fields KG kernel operator}
\end{equation}
such that
\begin{equation}
\hat{H}_{\text{KG}} = \frac{1}{2}\int dx\left[  -\frac
{\delta^{2}}{\delta\chi_{x}^{2}} +\int dy \,\chi_{x}\,\omega^{2}(\vec{x},\vec{y})\,\chi_{y}\right].
\label{Fields KG Hamiltonian kernel}
\end{equation}
\paragraph*{Ground state--- }
It is well-known (see e.g., \cite{Long Shore 1998}) that a theory of this type describes a set of harmonic oscillators. As such, we might make an ansatz for the ground state that is a general Gaussian functional
\begin{equation}
\Psi_{0}\left[  \chi\right]  =N_{0}(t)\exp\left[-\frac{1}{2}\int d^{3}x\int d^{3}y\,\,\chi\left(  \vec{x}\right)  G\left(\vec{x},\vec{y}\right)  \chi\left(  \vec{y}\right)  \right] = N_{0}(t)\psi_{0}[\chi]
\label{Fields KG Ground state functional}
\end{equation}
with a generic time-independent kernel $G\left(\vec{x},\vec{y}\right)$. Inserting this ansatz into the Schr\"{o}dinger equation, eq.(\ref{Fields Functional Schrodinger Equation KG}), above we deduce that
\begin{equation}
N_{0}(t) = \frac{1}{Z_{0}^{1/2}}e^{-iE_{0}t}\quad\text{where}\quad E_{0} = \frac{1}{2}\int d^{3}x \, G\left(\vec{x},\vec{x}\right)
\label{Fields KG Ground state energy}
\end{equation}
will later be identified as the (divergent) ground state energy. Furthermore, the kernel $G\left(\vec{x},\vec{y}\right)$ must also satisfy the condition
\begin{equation}
\int d^{3}z \, G\left(\vec{x},\vec{z}\right)G\left(\vec{z},\vec{y}\right) = \left(-\nabla^{2}+m^{2}\right)\delta^{3}(\vec{x}-\vec{y}) = \omega^{2}\left(\vec{x},\vec{y}\right).
\label{Fields KG kernel}
\end{equation}

A standard technique to solve this is to expand eq.(\ref{Fields KG kernel}) in terms of a set of basis functions that diagonalize the kernel $\omega^{2}\left(\vec{x},\vec{y}\right)$. This is, of course, a purely pragmatic strategy for the purposes of solving eq.(\ref{Fields KG kernel}); indeed, any set of functions that accomplish this task will do. However, notice also that whatever basis functions we use to diagonalize the kernel $\omega^{2}\left(\vec{x},\vec{y}\right)$ also diagonalizes the Hamiltonian $\hat{H}_{\text{KG}}$, as can be seen in eq.(\ref{Fields KG Hamiltonian kernel}). Moreover, among the various basis functions one can use for this, on account of the requirement $[\hat{P}_{i},\hat{H}] = 0$ from the Poincar\'{e} algebra, it is possible to also choose a set of basis functions that, in addition, diagonalizes $\hat{P}_{i}$ together with $\hat{H}_{\text{KG}}$.\footnote{Because of $[\hat{L}_{i},\hat{H}] = 0$ it is also possible to choose a different set of basis functions --- the spherical wave functions --- that diagonalize both $\hat{L}_{i}$ and $\hat{H}$, but not all three since $[\hat{P}_{i},\hat{L}_{j}]\neq 0$. We consider this case later in the context of analyzing Wallstrom's objection in QFT.}

All in all, this suggests a Fourier expansion of the kernel
\begin{equation}
G(\vec{x},\vec{y})=\int\frac{dk^{3}}{(2\pi)^{3}}\,e^{i\vec{k}\cdot(\vec{x}-\vec{y})}\, \tilde{G}(\vec{k}).
\label{Fourier expansion kernel}
\end{equation}
Inserting this expansion into eq.(\ref{Fields KG kernel}), and using the expansion of the Dirac delta into plane waves
\begin{equation*}
\delta^{3}(\vec{x}-\vec{y})=\int\frac{dk^{3}}{(2\pi)^{3}}\,e^{i\vec{k}\cdot(\vec{x}-\vec{y})},
\label{Fourier expansion Dirac delta}
\end{equation*}
we obtain
\begin{equation}
\tilde{G}(\vec{k})^{2} = \omega_{k}^{2}\quad\text{where}\quad\omega_{k} = \sqrt{\vec{k}^{2}+m^{2}}
\end{equation}
so that
\begin{equation}
G(\vec{x},\vec{y})=\int\frac{dk^{3}}{(2\pi)^{3}}\omega_{k}\,e^{i\vec{k}\cdot(\vec{x}-\vec{y})}.
\label{KG kernel function}
\end{equation}

The ground state energy $E_{0}$ from eq.(\ref{Fields KG Ground state energy}) can now be computed explicitly and is
\begin{equation}
E_{0}=\left\langle H\right\rangle _{0}=\frac{1}{2}\int d^{3}x\,G\left(
\vec{x},\vec{x}\right)  =\int d^{3}x\int\frac{d^{3}k}{\left(  2\pi\right)^{3}}\frac{1}{2}\sqrt{\vec{k}^{2}+m^{2}},
\label{Ground state energy - divergent}
\end{equation}
which is divergent both due to the infinite volume of space --- i.e. infrared divergent --- and due to the infinite energy contained in the high frequency, small wavelength modes --- i.e. ultraviolet divergent. This is, of course, the case because the ground state energy is merely the sum of the zero-point energies of each oscillator, and we have an infinite number of such oscillators, i.e. one for each $\vec{k}\in\mathbb{R}^{3}$.
\paragraph*{Fourier mode representation--- }
As mentioned above, the use of Fourier modes greatly simplifies other aspects of the theory as well. For instance, using the Fourier expansion of the field we obtain
\begin{equation}
\chi_{x} = \int \frac{d^{3}k}{(2\pi^{3})}\, e^{i\vec{k}\cdot\vec{x}}\, \tilde{\chi}_{k}~,
\label{Fourier expansion chi}
\end{equation}
where $\tilde{\chi}_{k}$ is a complex Fourier coefficient. Using the conventions that
\begin{equation}
\frac{\delta\chi_{x}}{\delta\chi_{x^{\prime}}} = \delta^{3}(x,x^{\prime})\quad\text{and}\quad \frac{\delta\tilde{\chi}_{k}}{\delta\tilde{\chi}_{k^{\prime}}} = (2\pi)^{3}\delta^{3}(k,k^{\prime})
\end{equation}
the functional derivative can be expanded as
\begin{equation}
\frac{\delta}{\delta\chi_{x}} = \int \frac{d^{3}k}{(2\pi)^{3}}\, e^{-i\vec{k}\cdot\vec{x}}\,\frac{\delta}{\delta\tilde{\chi}_{k}}~.
\end{equation}
We can now rewrite the Hamiltonian operator $\hat{H}_{\text{KG}}$ in Fourier modes, yielding
\begin{equation}
\hat{H}_{\text{KG}} = \frac{1}{2}\int \frac{d^{3}k}{(2\pi)^{3}}\left(-\frac{\delta^{2}}{\delta\tilde{\chi}_{k}\delta\tilde{\chi}^{*}_{k}}+\omega_{k}^{2}\,\tilde{\chi}_{k}\,\tilde{\chi}^{*}_{k}\right)~,
\label{KG Hamiltonian Fourier}
\end{equation}
which is reminiscent of a set of Harmonic oscillators. From this we can show that the ground state functional $\Psi_{0}[\chi]$ takes the simplified form
\begin{equation}
\Psi_{0}[\chi] = \frac{1}{Z_{0}^{1/2}}e^{-iE_{0}t}\exp-\frac{1}{2}\int\frac{d^{3}k}{(2\pi)^{3}}\omega_{k}\,\tilde{\chi}_{k}\,\tilde{\chi}^{*}_{k}~,
\label{Ground state Fourier}
\end{equation}
which is a product of Gaussians, one for each oscillator mode $\vec{k}\in\mathbb{R}^{3}$.
\paragraph*{Excited states--- }
Let us now look beyond the vacuum to the excited states. In analogy with the simple harmonic oscillator we introduce the notion of \textit{annihilation} and \textit{creation} operators
\begin{equation}
\hat{a}_{k} = \frac{1}{\sqrt{2}}\int d^{3}x\left[\int d^{3}y \, G^{1/2}(\vec{x},\vec{y})\chi_{y}+G^{-1/2}(\vec{x},\vec{y})\frac{\delta}{\delta\chi_{x}}\right]f_{k}^{*}(\vec{x})
\label{annihilation operator}
\end{equation}
and
\begin{equation}
\hat{a}_{k}^{\dagger} = \frac{1}{\sqrt{2}}\int d^{3}x\left[\int d^{3}y \, G^{1/2}(\vec{x},\vec{y})\chi_{y}-G^{-1/2}(\vec{x},\vec{y})\frac{\delta}{\delta\chi_{x}}\right]f_{k}(\vec{x})
\label{creation operator}
\end{equation}
that create and destroy an excitation of the $\vec{k}^{\text{th}}$ oscillator, respectively. 
Here the kernel $G(\vec{x},\vec{y})$ in eqns.(\ref{annihilation operator}) and (\ref{creation operator}) is the same as that in eq.(\ref{KG kernel function}) and the $f_{k}(\vec{x})$ are a set of basis functions. If we choose $f_{k}(\vec{x})$ to be plane waves then the operators $\hat{a}_{k}$ and $\hat{a}_{k}^{\dagger}$ take a simplified form 
\begin{equation}
\hat{a}_{k} = \sqrt{\frac{\omega_{k}}{2}}\tilde{\chi}_{k} + \frac{1}{\sqrt{2\omega_{k}}}\frac{\delta}{\delta\tilde{\chi}_{k}^{*}}\quad\text{and}\quad\hat{a}_{k}^{\dagger} =  \sqrt{\frac{\omega_{k}}{2}}\tilde{\chi}_{k}^{*} - \frac{1}{\sqrt{2\omega_{k}}}\frac{\delta}{\delta\tilde{\chi}_{k}}~.
\label{ladder operators k Fourier}
\end{equation}
It is easy to check that these operators satisfy the commutation relations
\begin{equation}
[\hat{a}_{k},\hat{a}_{p}^{\dagger}] = (2\pi)^{3}\delta^{3}(\vec{k}-\vec{p})\quad\text{and}\quad [\hat{a}_{k},\hat{a}_{p}]=[\hat{a}_{k}^{\dagger},\hat{a}_{p}^{\dagger}]=0~.
\label{ladder commutation}
\end{equation}
We can now rewrite $\hat{H}$ in terms of these operators, which reads
\begin{equation}
\hat{H}_{\text{KG}} =\frac{1}{2}\int \frac{d^{3}k}{(2\pi)^{3}}\,\omega_{k}\left[\hat{a}_{k}^{\dagger}\hat{a}_{k}+\hat{a}_{k}\hat{a}_{k}^{\dagger}\right].
\label{KG Hamiltonian ladder I}
\end{equation}
Equivalently, using the commutation relations eq.(\ref{ladder commutation}), we have
\begin{equation}
\hat{H}_{\text{KG}} =\int \frac{d^{3}k}{(2\pi)^{3}}\,\omega_{k}\,\hat{a}_{k}^{\dagger}\hat{a}_{k}+E_{0}~,
\label{KG Hamiltonian ladder II}
\end{equation}
where
\begin{equation}
E_{0} = \frac{1}{2}\int \frac{d^{3}k}{(2\pi)^{3}}\omega_{k} (2\pi)^{3}\delta^{3}(0) = \frac{1}{2}\int dx \int \frac{d^{3}k}{(2\pi)^{3}} \omega_{k}
\end{equation}
is interpreted as the divergent zero-point energy.

To interpret the action of the operators $\hat{a}_{k}$ and $\hat{a}^{\dagger}_{k}$, consider first the number operator
\begin{equation}
\hat{N} = \int \frac{d^{3}k}{(2\pi)^{3}}\hat{a}_{k}^{\dagger}\hat{a}_{k},
\label{Number operator}
\end{equation}
which can be interpreted as ``counting" the number of excitations in a particular quantum state. Indeed, a standard calculation shows that $\hat{a}_{k}$ acting on $\Psi_{0}$ gives $\hat{a}_{k}\Psi_{0} = 0$, which implies that $\hat{N}\Psi_{0} = 0$. In other words, the state $\Psi_{0}$ is said to have no excitations, and thus we can be justified in calling $\Psi_{0}$ the ground state or \emph{vacuum} state.

On the other hand, the action of $\hat{a}_{k}^{\dagger}$ is\footnote{As is standard (see e.g., \cite{Long Shore 1998}), we are being rather sloppy with the normalization factors. This is largely because the full Hilbert space on which these our operators act includes the space of probability distributions not normalized to unity.}
\begin{equation}
\Psi_{k} \equiv \hat{a}_{k}^{\dagger}\Psi_{0} =  2\omega_{k}\tilde{\chi}^{*}_{k}\Psi_{0}~.
\end{equation}
Applying now the number operator to $\Psi_{k}$ yields $\hat{N}\Psi_{k} = \Psi_{k}$. Furthermore, acting the Hamiltonian operator on $\Psi_{k}$ gives us
\begin{equation}
\hat{H}_{\text{KG}}\Psi_{k} = (E_{0}+\Delta E_{k})\Psi_{k}\quad\text{where}\quad\Delta E_{k} = \sqrt{\vec{k}^{2}+m^{2}}~.
\end{equation}
This implies that $\hat{a}_{k}^{\dagger}$ creates a state with one additional excitation that contributes an additional energy $\Delta E_{k}$ that is exactly the energy-momentum dispersion that we expect for a relativistic particle. Thus we say that $\hat{a}_{k}^{\dagger}$ creates a state with an additional excitation and conversely that $\hat{a}_{k}$ creates a state with one fewer excitation.

States with additional excitations can be generated iteratively by successive application of $\hat{a}_{k}^{\dagger}$ so that
\begin{equation}
\Psi_{n; k_{1},\cdots,k_{n}} = \hat{a}_{k_{n}}^{\dagger}\cdots \hat{a}_{k_{1}}^{\dagger}\Psi_{0}, 
\end{equation}
which has energy eigenvalue $E_{n} =E_{0} + \sum_{j}\Delta E_{k_{j}}$ and is an eigenstate of the number operator: $\hat{N}\Psi_{n; k_{1},\cdots,k_{n}} = n\Psi_{n; k_{1},\cdots,k_{n}}$. The full set of such states is, of course, known as Fock space, which constitutes the Hilbert space of QFT in a flat space-time.
\paragraph*{On particles--- }
It is tempting to say at this stage that the operator $\hat{a}_{k}^{\dagger}$ \textit{creates} a particle with momentum $\vec{k}$ while $\hat{a}_{k}$ \textit{annihilates} a particle of momentum $\vec{k}$. But this requires that we clarify what is meant by a ``particle". Indeed, in the relativistic quantum ED that we have developed here the physical microstates consist of the field degrees of freedom $\chi_{x}$. Period. The notion of a particle, on the other hand, is a rather derived concept from this perspective. Indeed, it might be more appropriate to view the notion of particles as a convenient way to label certain quantum states $\Psi$, but that is all that particles are in this description, merely a label. No more, no less.


Although such a viewpoint may seem controversial, the concept of the particle is rather tenuous even in the standard view of QFT. C. Rovelli and D. Colosi \cite{Colosi Rovelli 2008}, for instance, correctly note that localized particle detectors do not measure Fock states, which are eigenstates of the global operators $\hat{N}$ and $\hat{H}$, but instead measure the eigenstates of local operators. Moreover, investigations of QFT in curved space-time in the 1970's \cite{Unruh}\cite{Fulling}\cite{Davies} give strong hints that the particle concept itself is highly problematic to begin with, leading some physicists, such as Davies \cite{Davies 1984}, to claim that \textit{``particles do not exist."} 

Nevertheless, despite these issues, from a practical viewpoint, any physicist who holds these views must contest with the fact that particle detectors do, in fact, go ``\textit{click}". Thus from the ED viewpoint our task is to explain how it is that although the fields are ``real", why it is that detectors seem to detect particles. Such a challenge would require an analysis of the measurement process, an issue which has already been tackled with some success in ED (see e.g.,  \cite{Caticha/Johnson}\cite{Caticha Vanslette 2017}). While this gives us some confidence that the matter can be handled by ED, further investigation is outside the scope of our current work.
\subsubsection*{Some comments on divergences}
As is well known, QFT is often plagued by infinities and ED is no different. The energy of the ground state $E_{0} = \langle\hat{H}\rangle_{0}$, as we saw, is both infrared and ultraviolet divergent. And, while the vacuum expectation value of the field $\chi_{x}$ at any point $\vec{x}$ vanishes, its variance diverges:
\begin{equation}
\left\langle \chi\left(  \vec{x}\right)  \right\rangle =0\quad\text{and}\quad\text{Var}\left[  \chi\left(  \vec{x}\right)  \right]  =\langle\chi^{2}\left(  \vec{x}\right)  \rangle_{0}=\int\frac{d^{3}k}{\left(  2\pi\right)^{3}}\frac{1}{2\omega_{k}}~.
\end{equation}
Note, however, that what diverges here are not the physical fields or quantities, but the uncertainty in our predictions of those quantities. ED recognizes the role of incomplete information: the theory is completely unable to predict the field value at a sharply localized point.

The theory does, however, offer meaningful predictions for other quantities. For example, the equal time correlations between two field variables $\chi\left(  \vec{x}\right)  $ and $\chi\left(\vec{y}\right)  $ are \cite{Long Shore 1998},
\begin{equation}
\left\langle \chi\left(  \vec{x}\right)  \chi\left(  \vec{y}\right)
\right\rangle _{0}=\int\frac{d^{3}k}{\left(  2\pi\right)  ^{3}}\frac
{e^{i\vec{k}\cdot\left(  \vec{x}-\vec{y}\right)  }}{2\omega_{k}}=\frac{m}{4\pi^{2}\left\vert \vec{x}-\vec{y}\right\vert }K_{1}\left(  m\left\vert\vec{x}-\vec{y}\right\vert \right)
\end{equation}
where $K_{1}$ is a modified Bessel function.
\begin{remark*}
Since infinities occur in the ED approach to quantum fields, just as in the standard approaches, one must introduce schemes to regularize and renormalize the theory to obtain finite results. With the apparatus of quantum theory now in place, one way to do this is to make a unitary transformation to the Heisenberg representation and use the standard methods (see e.g., \cite{Peskin/Schroeder 1995}). More relevant to us is the work by Symanzik \cite{Symanzik 1981}, where he shows that the Schr\"{o}dinger representation itself exists as a renormalized QFT to all orders in the perturbation expansion; nothing prevents us at this point from borrowing these standard methods wholesale and applying them to ED. On the other hand, a comprehensive construction of QFT from the ED perspective should, in addition, require an inferential understanding of renormalization. To this end, work has been undertaken by A. Caticha and P. Pessoa \cite{Pessoa/Caticha 2018} to develop a theory of renormalization group (RG) techniques using entropic methods, but in the context of equilibrium statistical mechanics; the entropic RG methods that would be relevant here are still a work in progress.
\end{remark*}
\section{Multi-valued phases in the ED of quantum fields}\label{Ch 5 Wallstrom}
At the end of the previous chapter we drew attention to the work of T.C. Wallstrom \cite{Wallstrom 1994} who objected to the equivalence of Nelson's stochastic mechanics with standard quantum theory. Essentially, the issue there was that to recover the full complex Hilbert space of physical states from the real-valued $\rho$ and $\Phi$, one must simultaneous require the phase $\Phi$ to be multi-valued all while the complex $\Psi$ remain single valued. The argument, ultimately, was an empirical one: deny multi-valued $\Phi$ and you deny the experimentally verified, non-vanishing values of angular momentum.

The question arises as to whether a similar argument also holds in the context of quantum fields. While there has been some discussion in the literature of what has been called ``Wallstrom objection" (see e.g., \cite{Smolin 2006}), the focus seems exclusively on non-relativistic quantum particles. Thus our investigation into this issue for relativistic quantum scalar fields appears entirely novel.

Here we perform an analysis similar to Wallstrom's, but in the context of a quantum scalar field. To begin, we must determine whether the phase $\Phi$ of the quantum state $\Psi$ must be multi-valued, as it must be in the case of quantum particles. With non-relativistic particles, it was sufficient to consider eigenstates of angular momentum, and thus that will be our focus here as well. We accomplish this by expanding the field $\chi_{x}$ with a set of spherical basis functions, rather than the usual plane wave basis. This allows us to construct a Fock space with states labeled by particle number $n$, radial quantum number $k$, as well as angular momentum quantum numbers $l$ and $m$. Our analysis proceeds by considering the rotational properties of the eigenstates of angular momentum.


\subsubsection*{Some formalism}
A key aspect of our argument is the choice of basis functions to express the fields, kernels, and so on, that are present in our theory. In what preceded our work here, we used a basis of plane waves $f_{\vec{k}} = \exp i\vec{k}\cdot \vec{x}$ for this purpose. The benefit of this approach is that the kernel operator $\omega^{2}$ in eq.(\ref{Fields KG kernel operator}) is diagonalized in this basis, and consequently, so too is the Hamiltonian as well as the momentum operator $\hat{P}_{i}$. Thus the ladder operators $\hat{a}_{k}$ and $\hat{a}_{k}^{\dagger}$ that result from this approach excite the corresponding Fourier modes, as evidenced from eq.(\ref{ladder operators k Fourier}).

However, such a procedure is not unique. As was mentioned above, nothing prevents us from choosing a different set of basis functions, say, $f_{klm}$ that diagonalizes $\omega^{2}$, and similarly, the Hamiltonian $\hat{H}$. However, rather than working with a basis that simultaneously diagonalizes $\hat{H}$ and $\hat{P}_{i}$, we know from $\left[\hat{L}_{i},\hat{H} \right ] = 0 $ in the Poincar\'{e} algebra that it is must also be possible to simultaneously diagonalize $\hat{H}$ and the angular momentum generator $\hat{L}^{2}$ as well as the angular momentum about one axis, say, $\hat{L}_{z}$.
\paragraph*{Spherical basis representation--- }
To diagonalize the angular momentum operator $\hat{L}_{z}$ we introduce the spherical basis functions\footnote{The spherical wave representation of quantum fields is not very standard in QFT textbooks since scattering experiments typically are interested in states with definite linear momentum. Nevertheless, this representation is worked out in the book by W. Greiner \cite{Greiner Field Quant}, albeit in the Heisenberg rather than Schr\"{o}dinger representation.}
\begin{equation}
f_{klm}(r,\theta,\varphi) = N_{klm}\,j_{l}(kr)\, Y_{lm}(\Omega)~.
\end{equation}
These basis functions are typically introduced as solutions to the Laplace equation in spherical coordinates $(r,\Omega = \theta,\varphi)$, where $j_{l}(kr)$ is the so-called spherical Bessel function which is a solution to the radial part of the equation, while $Y_{lm}(\Omega)$ are the spherical harmonics that solve the angular part of the Laplacian, and here $N_{k}$ is a normalization introduced for convenience.

We look to express the field $\chi_{x} = \chi(r,\Omega)$ as a linear combination of these basis functions. To this point, a rather natural requirement for physical fields is that they be single-valued upon a complete $2\pi$ rotation of the field. As is standard (see e.g., \cite{Ballentine}), this forces the parameter $l$ to take discrete positive integer values while $m$ then must take integer values from $-l$ to $+l$. In terms of these basis functions the field $ \chi(r,\Omega)$ then takes the form
\begin{eqnarray}
\chi(r,\Omega) = \int dk \sum_{lm} N_{klm} j_{l}(kr)Y_{lm}(\Omega)\tilde{\chi}_{klm}~,\label{Fields Chi Spherical basis expansion}
\end{eqnarray}
where we have used an abbreviated notation $\sum_{lm} = \sum_{l = 0}^{\infty}\sum_{m = -l}^{+l} $, and where $\tilde{\chi}_{klm}$ are the spherical wave components of the field.\footnote{As we can see in (\ref{Fields Chi Spherical basis expansion}), the choice of normalization $N_{klm}$ simply amounts to a redefinition of the modes $\tilde{\chi}_{klm}$.}

\paragraph*{Orthogonality and completeness--- }
The basis functions $f_{klm}$, of course, satisfy the standard orthogonality relation
\begin{equation}
\int d^{3}x \, f^{*}_{klm}\, f_{k'l'm'} = \frac{\pi}{2k^{2}} N_{klm}^{2}\delta(k-k')\delta_{ll'}\delta_{mm'}~.
\label{Spherical orthogonality}
\end{equation}
A typical convention is to normalize (\ref{Spherical orthogonality}) to a delta function, implying
\begin{equation}
\frac{\pi}{2k^{2}}N_{klm}^{2} = 1\quad \text{so that}\quad N_{klm} =\sqrt{\frac{2}{\pi}}k~.
\end{equation}
It is then standard to define the normalized radial function $u_{kl}(r) = N_{klm}j_{l}(kr)$. The basis functions also satisfy a completeness relation\footnote{To derive this expression, note first that
\begin{equation}
\int \frac{d^{3}k}{(2\pi)^{3}}\exp i\vec{k}\cdot (\vec{x}-\vec{x}^{\prime}) = \delta^{3}(\vec{x}-\vec{x}^{\prime})~.
\end{equation}
Then, insert the so-called Rayleigh equation \cite{Ugincius 1972}
\begin{equation}
e^{i\vec{k}\cdot\vec{x}} = 4\pi\sum_{lm} i^{l}j_{l}(kr)Y_{lm}(\Omega)Y^{*}_{lm}(\Omega_{k})\label{Fields Rayleigh}
\end{equation}
in for the plane waves above. After using the orthogonality of the spherical basis functions the completeness relation (\ref{Spherical completeness}) follows.}
\begin{equation}
\int_{0}^{\infty} dk\sum_{lm}u_{kl}(r)Y_{lm}^{*}(\Omega_{r})u_{kl}(r')Y_{lm}(\Omega_{r'}) = \delta^{3}(\vec{r}-\vec{r}^{\prime})~.
\label{Spherical completeness}
\end{equation}
It is possible to now use the completeness relation to derive the spherical basis expansion of the functional derivative of $\chi(r,\Omega)$. Indeed, the expressions\footnote{We use a convention where there are no $2\pi$s present, in contrast to the analogous relationship with Fourier modes.}
\begin{equation}
\frac{\delta\chi(\vec{x})}{\delta\chi(\vec{x}^{\,\prime})} = \delta^{3}(\vec{x}-\vec{x}^{\,\prime})\quad \text{and}\quad \frac{\delta\tilde{\chi}_{klm}}{\delta\tilde{\chi}_{k'l'm'}} = \delta(k-k^{\prime})\delta_{ll'}\delta_{mm'}
\end{equation}
together with the expansion in (\ref{Fields Chi Spherical basis expansion}) implies that
\begin{equation}
\frac{\delta}{\delta\chi(\vec{x})} = \int dk \sum_{lm} u_{kl}(r)Y^{*}_{lm}(\Omega)\frac{\delta}{\delta\tilde{\chi}_{klm}}~.\label{Fields Chi func-deriv expansion}
\end{equation}

\paragraph*{Relationship to Fourier modes--- }
To compare with the usual plane wave expansion, it is useful to derive the relationship between the Fourier modes $\tilde{\chi}_{\vec{k}}$ and the spherical basis modes. To do this, expand $\chi(r,\Omega)$ using both plane waves and spherical waves and set the two expressions equal to one another
\begin{equation}
\chi(\vec{x}) = \int \frac{d^{3}k}{(2\pi)^{2}} e^{i\vec{k}\cdot \vec{x}}\tilde{\chi}_{\vec{k}} = \int dk \sum_{lm} \sqrt{\frac{2}{\pi}}kj_{l}(kr)Y_{lm}(\Omega)\tilde{\chi}_{klm}~.
\end{equation}
Continuing on transform to ``spherical" coordinates $\vec{k} = (k, \Omega_{k} = \theta_{k},\varphi_{k})$ so that
\begin{equation}
\int \frac{dk k^{2}d\Omega_{k}}{(2\pi)^{3}}e^{i\vec{k}\cdot\vec{x}}\tilde{\chi}_{\vec{k}} = \int dk \sum_{lm} \sqrt{\frac{2}{\pi}}kj_{l}(kr)Y_{lm}(\Omega)\tilde{\chi}_{klm}~,
\end{equation}
where we have used that $d^{3}k = dk k^{2}d\Omega$. Recalling the Rayleigh expansion (\ref{Fields Rayleigh}) of plane waves into spherical waves, leaves us with
\begin{equation}
\int \frac{dk k^{2}d\Omega_{k}}{(2\pi)^{3}}\left [4\pi\sum_{lm} i^{l} j_{l}(kr)Y_{lm}(\Omega)Y^{*}_{lm}(\Omega_{k}) \right ]\tilde{\chi}_{\vec{k}} = \int dk \sum_{lm} \sqrt{\frac{2}{\pi}}kj_{l}(kr)Y_{lm}(\Omega)\tilde{\chi}_{klm}~.
\end{equation}
Finally, comparing each side and using the orthogonality of the basis functions gives us that
\begin{equation}
\tilde{\chi}_{klm} = \frac{ i^{l}k}{(2\pi)^{3/2}}\int d\Omega_{k}Y_{lm}^{*}(\Omega_{k})\tilde{\chi}_{\vec{k}}~.\label{Fields chi spherical-Fourier}
\end{equation}
A similar procedure can be employed for the functional derivatives, yielding
\begin{equation}
\frac{\delta}{\delta\tilde{\chi}_{klm}} = \frac{(-i)^{l}k}{(2\pi)^{3/2}}\int d\Omega_{k}Y_{lm}(\Omega_{k})\frac{\delta}{\delta\tilde{\chi}_{\vec{k}}}~.\label{Fields Func-deriv spherical-Fourier}
\end{equation}

\subsubsection*{Free quantum fields in the spherical basis}
Having established the expansion of the fields and related quantities in terms of the spherical basis functions, we can now proceed to rewriting the Hamiltonian $\hat{H}_{KG}$, introducing the appropriate creation and annihilation operators, and constructing the Fock space from angular momentum eigenstates.
\paragraph*{The Hamiltonian--- }
To determine the new form of the Hamiltonian, take the expression for $\hat{H}_{KG}$ in (\ref{Fields KG Hamiltonian}) and insert the spherical basis expansions in (\ref{Fields Chi Spherical basis expansion}) and (\ref{Fields Chi func-deriv expansion}). After making use of the orthogonality of the basis functions, we obtain
\begin{equation}
\hat{H}_{KG} = \frac{1}{2}\int dk \sum_{lm} (-1)^{l} \left (\frac{\delta^{2}}{\delta\tilde{\chi}_{klm}\delta\tilde{\chi}_{kl-m}} +\omega_{k}^{2}\,\tilde{\chi}_{klm}\tilde{\chi}_{kl-m}  \right )~,
\end{equation}
where, $\omega_{k}^{2} = k^{2}+m^{2}$, as before. It follows from the fact that $\chi(\vec{x}) = \left ( \chi(\vec{x})\right )^{*}$ is a real scalar field that
\begin{equation}
\tilde{\chi}_{kl-m}= \tilde{\chi}^{*}_{klm}\quad\text{and}\quad\frac{\delta}{\delta\tilde{\chi}_{kl-m}} = (-1)^{l}\frac{\delta}{\delta\tilde{\chi}^{*}_{klm}},
\end{equation}
thus $\hat{H}_{KG}$ simplifies to
\begin{equation}
\hat{H}_{KG} = \frac{1}{2}\int dk \sum_{lm}  \left (\frac{\delta^{2}}{\delta\tilde{\chi}_{klm}\delta\tilde{\chi}^{*}_{klm}} +\omega_{k}^{2}\,\tilde{\chi}_{klm}\tilde{\chi}^{*}_{klm}  \right )~.\label{Fields Ham KG spherical}
\end{equation}

\paragraph*{Creation and annihilation operators--- }
The quadratic form of the Hamiltonian operator suggests the use of ladder operators. Indeed, either from inspection of (\ref{Fields Ham KG spherical}) or by analogy with eqns.(\ref{annihilation operator}) and (\ref{creation operator}), where we choose $f(\vec{x}) = f_{klm}(\vec{x})$, it is possible to introduce the operators
\begin{equation}
\hat{a}_{klm} = \sqrt{\frac{\omega_{k}}{2}}\tilde{\chi}_{klm} + \frac{1}{\sqrt{2\omega_{k}}}\frac{\delta}{\delta\tilde{\chi}_{klm}^{*}}\quad\text{and}\quad\hat{a}_{klm}^{\dagger} =  \sqrt{\frac{\omega_{k}}{2}}\tilde{\chi}_{klm}^{*} - \frac{1}{\sqrt{2\omega_{k}}}\frac{\delta}{\delta\tilde{\chi}_{klm}}~.
\label{ladder operators k spherical}
\end{equation}
These operators satisfy the commutation relations
\begin{equation}
[\hat{a}_{klm},\hat{a}_{k'l'm'}^{\dagger}] = \delta(k-k')\delta_{ll'}\delta_{mm'}\quad\text{and}\quad [\hat{a}_{klm},\hat{a}_{k'l'm'}]=[\hat{a}_{klm}^{\dagger},\hat{a}_{k'l'm'}^{\dagger}]=0~.
\label{ladder commutation spherical}
\end{equation}
To see this, note that $\hat{a}_{klm}$ and $\hat{a}_{klm}^{\dagger}$ can be written in terms of the earlier $\hat{a}_{\vec{k}}$ and $\hat{a}_{\vec{k}}^{\dagger}$ using the relations (\ref{Fields chi spherical-Fourier}) and (\ref{Fields Func-deriv spherical-Fourier}), so that
\begin{eqnarray}
\hat{a}_{klm} &=& \frac{i^{l}k}{(2\pi^{3/2}}\int d\Omega_{k}Y_{lm}^{*}(\Omega_{k})a_{\vec{k}}\label{annihilation spher-Fourier}\\
\hat{a}_{klm}^{\dagger} &=& \frac{(-i)^{l}k}{(2\pi^{3/2}}\int d\Omega_{k}Y_{lm}(\Omega_{k})a^{\dagger}_{\vec{k}}\label{creation spher-Fourier}~.
\end{eqnarray}
Thus the commutator of $\hat{a}_{klm}$ and $\hat{a}_{klm}^{\dagger}$ reads
\begin{equation}
[\hat{a}_{klm},\hat{a}_{k'l'm'}^{\dagger}] = \frac{i^{l}(-i)^{l'}kk'}{(2\pi^{3}}\int d\Omega_{k}\int d\Omega_{k'} Y_{lm}^{*}(\Omega_{k})Y_{l'm'}(\Omega_{k'})[\hat{a}_{\vec{k}},\hat{a}_{\vec{k}^{\,\prime}}^{\dagger}]~.
\end{equation}
Recalling from (\ref{ladder commutation}) the commutator relations
\begin{equation}
[\hat{a}_{\vec{k}},\hat{a}_{\vec{k}^{\,\prime}}^{\dagger}] = (2\pi)^{3}\delta^{3}(\vec{k}-\vec{k}^{\,\prime}) =  (2\pi)^{3}\frac{\delta(k-k')}{k^{2}}\delta^{2}(\Omega_{k}-\Omega_{k'})~,
\end{equation}
we now have
\begin{equation}
[\hat{a}_{klm},\hat{a}_{k'l'm'}^{\dagger}] = i^{l}(-i)^{l'}\delta(k-k')\int d\Omega_{k} Y_{lm}^{*}(\Omega_{k})Y_{l'm'}(\Omega_{k})~,
\end{equation}
from which the orthogonality of the spherical harmonics yields
\begin{equation}
[\hat{a}_{klm},\hat{a}_{k'l'm'}^{\dagger}] = \delta(k-k')\delta_{ll'}\delta_{mm'}~,
\end{equation}
as desired.

\paragraph*{Energy and angular momentum--- }
The introduction of the creation and annihilation operators put the Hamiltonian into a particularly special form
\begin{equation}
\hat{H}_{\text{KG}} =\frac{1}{2}\int dk \sum_{lm}\omega_{k}\left (\hat{a}_{klm}^{\dagger}\hat{a}_{klm} +\hat{a}_{klm}\hat{a}_{klm}^{\dagger} \right )~.
\label{KG Hamiltonian spherical}
\end{equation}
Using this commutation relations (\ref{ladder commutation spherical}) above, the Hamiltonian then becomes
\begin{equation}
\hat{H}_{\text{KG}} = E_{0}+\frac{1}{2}\int dk \sum_{lm}\omega_{k}\hat{a}_{klm}^{\dagger}\hat{a}_{klm},
\label{KG Hamiltonian spherical}
\end{equation}
where $E_{0}$ is the same ground state energy as in (\ref{Fields KG Ground state energy}). Similarly, the angular momentum operator $\hat{L}_{z}$, for example, takes a particularly simple form
\begin{equation}
\hat{L}_{z} = \int dk \sum_{lm} m \, \hat{a}_{klm}^{\dagger}\hat{a}_{klm} \label{Angular Momentum Z operator}
\end{equation}
where $m$ can now be easily identified as the quantum number that represents the projection of the angular momentum to the $z$-axis.

\paragraph*{Fock space--- }
Borrowing the conventional interpretation of these operators, we might say that $\hat{a}_{klm}^{\dagger}$ and $\hat{a}_{klm}$ create and destroy, respectively, particles with quantum numbers $k$, $l$, $m$. Indeed, if the operators $\hat{a}_{k}$ and $\hat{a}_{k}^{\dagger}$ acting on the ground state $\Psi_{0}$ generate the full Fock space of states, then so too do $\hat{a}_{klm}^{\dagger}$ and $\hat{a}_{klm}$ because the relationship in eqns.(\ref{annihilation spher-Fourier}) and (\ref{creation spher-Fourier}) amounts to a unitary transformation.

A single particle state, for instance, can be generated through action of $\hat{a}_{klm}^{\dagger}$ on the ground state
\begin{equation}
\Psi_{klm} = \hat{a}^{\dagger}_{klm}\Psi_{0} = 2\omega_{k}\tilde{\chi}_{klm}^{*}\Psi_{0}~.
\label{Single particle Ang Mom Eigenstate}
\end{equation}
Similarly, a successive application of $\hat{a}_{klm}^{\dagger}$ on the vacuum state generates a complete set of orthogonal basis functionals, just as we have in the standard setup (see e.g., \cite{Long Shore 1998}), yielding a general $n$-particle state
\begin{equation}
\Psi_{n, k_{1}l_{1}m_{1},\cdots k_{n}l_{n}m_{n}} = \hat{a}^{\dagger}_{k_{1}l_{1}m_{1}}\cdots \hat{a}^{\dagger}_{k_{n}l_{n}m_{n}}\Psi_{0}~.
\label{Multiple particle Ang Mom Eigenstate}
\end{equation}
\subsubsection*{Multi-valuedness}
The basic criterion for whether $\Phi$ behaves as an angle or not is given by
\begin{equation}
\oint_{\gamma}\delta\Phi = 2\pi j~,\label{Wallstrom condition}
\end{equation}
which states that the integral of $\Phi$ around a closed loop $\gamma$ in configuration space must be an integer multiple of $2\pi$. In the following we consider just such a loop, one obtained by completely rotating the field $\chi$ about the $z$-axis in real space. As such, it is convenient to restrict our focus to eigenstates of angular momentum, and more specifically, to those eigenstates that contain just a single excitation, such as those given in eq.(\ref{Single particle Ang Mom Eigenstate}) above.
\begin{remark*}
Limiting ourselves to quantum states $\Psi_{klm}$ that are the eigenstates of angular momentum with a single excitation is more than just a matter of convenience, it is a matter of strategy. That is, once we make the conceptual leap that such states can be said to describe the properties of a single particle then it will be possible to directly compare our analysis with that of Wallstrom \cite{Wallstrom 1994}, whose focus was directed towards particles in quantum mechanics. As we will see in the next section, the conclusions drawn by Wallstrom in this instance cannot be extended to quantum fields.
\end{remark*}
\paragraph*{Rotation of the field--- }
To begin, let us inspect how the one-particle eigenstate of angular momentum $\psi_{klm}$ changes under rotation. In equation (\ref{Single particle Ang Mom Eigenstate}) the factor $\omega_{k}$ is, of course, a scalar under rotations, as is the ground state $\Psi_{0}$ since it respects the translational and rotational symmetries of flat Euclidean space. Thus the key rotational aspects of $\Psi_{klm}$ are built into the spherical wave components $\tilde{\chi}_{klm}$.

The condition for a field to be a scalar is that $\chi^{\prime}(\vec{x}^{\,\prime}) = \chi(\vec{x})$ so that the transformed field evaluated in the rotated frame has the same value as the original field in the original frame. Alternatively, since we deal with a simple rotation about the $z$-axis, this is equivalent in this case to
\begin{equation}
\chi^{\,\prime}(r,\theta,\varphi+\alpha) = \chi(r,\theta,\varphi).\label{Scalar condition}
\end{equation}
The full spherical wave expansion of $\chi(\vec{r})$ is
\begin{equation}
\chi(\vec{x}) = \int dk \sum_{lm} u_{kl}(r)P_{lm}(\theta)e^{im\varphi}\tilde{\chi}_{klm},\label{Spherical expansion chi full}
\end{equation}
where we have used that the spherical harmonics $Y_{lm}(\theta,\varphi) = P_{lm}(\theta)e^{im\varphi}$, with $P_{lm}(\theta)$ being the associated Legendre functions (see e.g., \cite{Ballentine}). When we rotate about the $z$-axis we have that $\varphi \to \varphi + \alpha$, but we do not yet know how $\tilde{\chi}_{klm}$ transforms under this rotation. To determine this, note that the scalar condition (\ref{Scalar condition}) in terms of spherical wave components becomes
\begin{equation}
\int dk\sum_{lm}u_{kl}(r)P_{lm}(\theta)e^{im\varphi}e^{im\alpha}\tilde{\chi}_{klm}^{\prime}= \int dk\sum_{lm}u_{kl}(r)P_{lm}(\theta)e^{im\varphi}\tilde{\chi}_{klm}~.
\end{equation}
Using the orthogonality of the basis functions we find that
\begin{equation}
 \tilde{\chi}_{klm}^{\prime} = e^{-im\alpha}\tilde{\chi}_{klm}.
\end{equation}

\paragraph*{Rotation of the quantum state--- }
Looking back to the $\Psi_{klm}$ from eq.(\ref{Single particle Ang Mom Eigenstate}) we see that when we rotate about the $z$-axis by an angle $\alpha$ that this state transforms as
\begin{equation}
\Psi^{\prime}_{klm} = e^{-im\alpha}\Psi_{klm}~,
\end{equation}
therefore this implies that
\begin{equation}
\Psi_{klm}^{\prime} =\left ( \rho_{klm}^{\prime}\right )^{1/2}e^{i\Phi^{\prime}_{klm}} = \rho_{klm}^{1/2}e^{i\Phi_{klm}}e^{-im\alpha}~.
\end{equation}
Clearly a rotation by $\alpha$ around the $z$-axis yields a phase change in $\Psi_{klm}$ by $\delta\Phi = -m\alpha$, which is, in fact, proportional to $\alpha$. Thus we conclude just from this simple case that the phase $\Phi$ must, indeed, have the multi-valued properties of an angle. A full rotation of $\alpha = 2\pi$ then obviously leads to
\begin{equation}
\oint_{\gamma}\delta\Phi_{klm}= -2\pi m~,\label{Fields Single valued}
\end{equation}
exactly as desired since $m$ is an integer.
\paragraph*{Comments--- }
There are a couple of important takeaways from this analysis. The first is simply that the phase $\Phi$ must be multi-valued, even in the context of relativistic fields.\footnote{Although simple scalar fields such as this are, of course, not found in nature, the fact that this property holds even in this relatively stunted model seems to imply that we should certainly expect this structure to persist with other Bosonic fields, and certainly for Fermion fields, as well.} Simply put, single-valued phases simply do not recover states of physical importance. This has some important implications for the ED approach. For instance, in previous iterations of ED (see e.g., \cite{Caticha 2010a}\cite{Caticha 2013}) it was posited that the drift potential $\phi$ could possibly be an entropy. Clearly this explanation is not sufficient since entropy must be single-valued. Instead the drift potential and its attendant angular properties must have a deeper geometrical interpretation. This is made rather manifest in \cite{Caticha 2019b} wherein the geometry of the e-phase space is assumed to be spherical. Thus if probabilities in the e-phase space play the role of a ``polar" coordinate, then it is natural that the drift potential satisfies the role of its corresponding phase angle.

Furthermore, of particularly significant interest is the relation (\ref{Fields Single valued}), which shows that, although $\Phi$ is multi-valued, that $\Psi$ itself remains \emph{singe-valued}. This is in stark contrast to the case of non-relativistic quantum particles where the single-valuedness of the wave function is often introduced as an additional postulate.\footnote{The approach of Caticha \cite{Caticha 2019b} is the exception that proves the rule.} Indeed, the crux of Wallstrom's argument against Nelson's stochastic dynamics was that, once one accepts $\Phi$ as multi-valued, the condition that $\Psi$ be single-valued often appears rather \emph{ad hoc}. But here our brief investigation into the matter reveals that the marriage between multi-valued $\Phi$ and single-valued $\Psi$ can, in fact, be rather natural --- it follows because the azimuthal quantum number $m$ is already quantized by virtue of the single-valuedness of the physical field $\chi(\vec{x})$. Thus the ED approach to the theory of quantum scalar fields avoids the criticism levied by Wallstrom.

From a broader perspective our analysis demonstrates that significant qualitative differences are present between the quantization of fields versus the quantization of particles. And this, in turn, casts some doubt on the mission of understanding quantum foundations through an often express focus on non-relativistic quantum mechanics.  This is not to say that such endeavors are not without value, but only that the program of quantum reconstruction must always be cognizant that some issues which may seem pressing in the non-relativistic domain, are instead just artifacts of a model that is not sufficiently rich.

\section{Concluding remarks\label{Fields Conclusion}}
We have demonstrated here that the ED template is a robust framework for constructing quantum theories; sufficient for describing either quantum particles or fields. Central to our approach is the update of the probability $\rho$, which is guided by the phase $\Phi$. Here notions of Hilbert spaces and the usual quantum structure are introduced \textit{a posteriori}, for convenience. 

One advantage of this approach is that it completely avoids the operator ordering ambiguities that are typically encountered during quantization. Another is that a reduced emphasis on Hilbert spaces implies that the canonical commutation relations cannot be fundamental in the ED approach. Indeed, here such relations are derived, and thus so too are the resulting Bose-Einstein statistics.

Many of the chief benefits of the ED framework are conceptual, however. A strict adherence to the viewpoint that fields are fundamental in ED allows for a possible resolution of the measurement problem, especially in light of the successes in the regime of non-realtivistic particles \cite{Caticha/Johnson}\cite{Caticha Vanslette 2017}. Such tools will prove extremely valuable in how one conceives of the particle concept, which is more ambiguous in curved space-time \cite{Unruh}\cite{Fulling}\cite{Davies}.

Finally, a clear disadvantage of the formalism developed here is that, although the ED developed here carries a representation of the Poincar\'{e} group, and is thus fully relativistic, it is not manifestly so. In the next chapter we remedy this issue by establishing a formalism wherein we develop a theory of quantum scalar fields on curved space-time. The framework has the added advantage that it allows for the mutual interaction of quantum fields with dynamical geometry, which we discuss in the final chapter.

\chapter{The covariant entropic dynamics of fields}\label{5CH CED}
The primary task in ED is to update the probability $\rho$ in response to the evolving constraints, codified by the dynamics of the drift potential $\phi$. However, while the update of the probabilities is completely determined by the Bayesian and entropic methods of inference, the criterion for evolving $\phi$ is not quite as rigid. The question is, of all the possible dynamical evolutions for the drift potential, which of these reproduces quantum theory? As discussed in the preceding chapters, a major insight in this regard was to recognize that the desirable trajectories, or flows, were those that could be traced to symmetry considerations.

The scheme can be distilled into two pieces. First is a dynamical element wherein we isolate a subset of flows, the Hamiltonian flows, that conserve a symplectic structure in the phase space of ED. This equips ED with all of the tools of the canonical formalism. However, for many cases in physics this is not sufficient; largely because many cases of physical interest are subject to additional symmetries of a \emph{kinematical} variety, as well. In such cases we must impose additional constraints on the allowed Hamiltonian flows so that these symmetries may be respected. An prime example of this was made explicit in the previous chapter, in the context of a flat space-time, where the choice of Hamiltonian was restricted so as to complete a canonical representation of the Poincar\'{e} group.\footnote{This principle is not, obviously, exclusive to ED. Whether we deal with classical canonical Hamiltonians, or Hermitian operators in quantum theory, it is a basic physical principle that the generators satisfy a representation of the applicable symmetry group.}

In the context of a curved background, considerations of symmetry present themselves yet again as an important ingredient for ED. The issue is that of what should replace the role of the Poincar\'{e} group in the transition to a curved space-time where such global symmetries are not, in general, available.


Related to this is the role of time. In particular, the relativistic ED models considered previously in ch. \ref{CH ED fields}, as well as in \cite{Caticha 2013}\cite{Caticha Ipek 2014}, all commit to a particular foliation, and in doing so, break manifest Lorentz invariance. This situation is somewhat reminiscent of the status of quantum field theory in the early 1930's, shortly after the advent of the new quantum mechanics by Heisenberg, Schr\"{o}dinger, and Dirac.\footnote{The biographical book by Silvan Schweber \cite{Schweber} provides, in chapter one, an excellent overview of the status of quantum field theory at that time. For instance, it was common to construct the canonical commutation relations  with respect to a particular time variable; as was done, for example, by Pauli and Heisenberg \cite{Heisenberg Pauli 1929}\cite{Heisenberg Pauli 1930}.} There, although such theories could be relativistic in content --- i.e. the generators satisfy the Poincar\'{e} algebra, etc. --- they were not manifestly so. In particular, such models failed to make explicit the notion of relative simultaneity, which is surely a hallmark of the special and general theories of relativity.

Work done to overcome this problem in the standard quantum theory began with the early \textit{many-time} efforts of P. Weiss \cite{Weiss 1938} and Dirac \cite{Dirac 1932}. These ideas were later further developed by Dirac \cite{Dirac 1948}, and were adopted by others, such as S. Tomonaga \cite{Tomonaga 1946} and J. Schwinger \cite{Schwinger 1948 I} in their approaches to a fully covariant quantum theory. Collectively, these initial works could be characterized as precursors for the subsequent covariant Hamiltonian methods developed, again, by Dirac \cite{Dirac Lectures}\cite{Dirac 1951}, but which were later refined by K. Kucha\v{r} \cite{Kuchar 1972} and C. Teitelboim \cite{Teitelboim thesis}\cite{Teitelboim 1972} in their pursuit of a covariant approach to geometrodynamics.

A central idea is the notion of an instant of time, which contains two main ingredients. First, in a properly relativistic theory, an instant is associated with an arbitrary space-like hypersurface $\sigma$ in space-time, called simply a \emph{surface}. Second, to each such surface one must attach an epistemic state, expressed by $\rho_{\sigma}$ and $\phi_{\sigma}$, which captures the information available for making predictions. Put another way, it is these variables that take the role of Cauchy data: they contain the information necessary to generate the very next instant, and its corresponding epistemic state. Thus, In ED, time is generated instant by instant.



Following the scheme established in \cite{Ipek et al 2017}\cite{Ipek et al 2019a}, which was itself inspired by the covariant Hamiltonian methods of Dirac, Hojman, Kucha\v{r}, and Teitelboim (DHKT), we construct a manifestly covariant dynamics. Since the formalism deals equally well with a flat or curved space-time, being more general, we deal with the latter case. Following DHKT, we slice, or foliate, the background space-time into a succession of surfaces. While this procedure obscures the original local four-dimensional Lorentz symmetry of space-time, it can be recovered by the implementation of a new symmetry: foliation invariance. This foliation symmetry is then, in turn, implemented by the kinematic requirement of embeddability \textit{\`{a} la } Teitelboim and Kucha\v{r}: if a final surface can be reached from an initial surface by a number of intermediate paths, then all such paths should agree. The result of such considerations is that surface deformations form a ``group" with a corresponding ``algebra".\footnote{The use of quotations is a reminder that surface deformations do not, strictly speaking, form a group. We discuss in more detail below.}

A major insight coming from DHKT was that imposing the structure of surface deformations on the dynamical generators leads to a manifestly covariant dynamics. This is implemented through a principle of consistency, referred to as \emph{path independence}: the evolution from an initial surface to a final surface should be independent of the intervening path. The result is that the local Hamiltonian generators, four per spatial point, which are responsible for updating the evolving constraints, must satisfy an ``algebra" identical to that of surface deformations and a set of initial value constraints needed to make the scheme consistent.

Here we implement this approach for a single scalar field $\chi(x)$ in a fixed curved background. The material is based on work done in \cite{Ipek et al 2017}\cite{Ipek et al 2019a} in collaboration with M. Abedi and A. Caticha. In section \ref{55S Entropic Dynamics} we outline our model for inferring the dynamics of the scalar field. In section \ref{55S Entropic time} we construct a notion of time to assist in organizing our inferences. Section \ref{55S Embeddability} introduces the kinematic notion of embeddability and section \ref{55S Path independence} familiarizes the dynamical principle of path independence \textit{\`{a} la } Teitelboim and Kucha\v{r}. In section \ref{55S Canonical representation} we build a canonical representation of the ``algebra" of surface deformations, which we then use in section \ref{55S Evolution equations} to determine the dynamical equations of motion for ED. In section \ref{55S General solution} we discuss a broader class of dynamical models resulting from the Poisson bracket relations. Next we discuss the Ehrenfest equations for fields in a curved background in section \ref{55S Ehrenfest}. We then summarize our work and discuss future applications in section \ref{55S Conclusion}.

\section{The Entropic Dynamics of infinitesimal steps}
\label{55S Entropic Dynamics}
\paragraph{The microstates---}

Here we consider the dynamics of a scalar field $\chi \left( x\right) $. For notational convenience we will again write the $x$-dependence as a subscript, $\chi \left( x\right) =\chi _{x}$. This field has definite values at all times, however, they are unknown and their dynamics is indeterministic.

The field $\chi $ lives on a $3$-dimensional curved space $\sigma $, the points of which are labeled by coordinates $x^{i}$ ($i=1,2,3$). The space $\sigma $ is endowed with a metric $g_{ij}$ induced by the fixed space-time in which it is embedded. Thus $\sigma $ is an embedded hypersurface, or just \emph{surface} for short. The field $\chi _{x}$ is a scalar with respect to
3-diffeomorphisms on the surface $\sigma $. The $\infty $-dimensional space
of all possible field configurations is the configuration space $\mathcal{C}$%
. A single field configuration, labelled $\chi $, is represented as a point
in $\mathcal{C}$, and the uncertainty in the field is described by a
probability distribution $\rho \lbrack \chi ]$ over $\mathcal{C}$.

\paragraph{Maximum Entropy---}

Our goal is the standard one: to predict the evolution of the scalar field $\chi $. We make one major assumption in this regard: in ED, motion is continuous, the fields follow continuous trajectories in $\mathcal{C}$ such that finite change can be analyzed as the accumulation of many infinitesimally short steps. Therefore our first goal is to calculate the probability $P\left[ \chi ^{\prime }|\chi \right] $ that the field undergoes a small change from an initial configuration $\chi $ to a neighboring $\chi^{\prime }=\chi +\Delta \chi $ and later we calculate the probability of the finite change that results from a sequence of short steps. The transition
probability $P\left[ \chi ^{\prime }|\chi \right] $ is found by maximizing the entropy functional, 
\begin{equation}
S\left[ P,Q\right] =-\int D\chi ^{\prime }P\left[ \chi ^{\prime }|\chi %
\right] \log \frac{P\left[ \chi ^{\prime }|\chi \right] }{Q\left[ \chi
^{\prime }|\chi \right] },  \label{5entropy a}
\end{equation}%
relative to a prior $Q\left[ \chi ^{\prime }|\chi \right] $ and subject to appropriate constraints. 
\paragraph*{The prior ---}

We adopt a prior $Q\left[ \chi ^{\prime }|\chi \right] $ that incorporates
the information that the fields change by infinitesimally small amounts, but
is otherwise maximally uninformative. In particular, before the constraints are taken into
account, knowledge of the dynamics at one point does not convey information about the dynamics at another point, i.e. the degrees of freedom are \emph{a priori} uncorrelated. 

Such a prior can itself be derived from the principle of maximum entropy. Indeed, maximize
\begin{eqnarray}
S[Q,\mu ]=-\int D\chi ^{\prime }\,Q\left[ \chi ^{\prime }|\chi \right] \log 
\frac{Q\left[ \chi ^{\prime }|\chi \right] }{\mu (\chi ^{\prime })}~,
\label{5entropy b}
\end{eqnarray}%
relative to the measure $\mu (\chi ^{\prime })$, which we assume to be
uniform, and subject to appropriate constraints.\footnote{Since we deal with infinitesimally short steps, the prior $Q$ turns out to be quite independent of the background measure $\mu$.} The requirement that the
field undergoes changes that are small and uncorrelated is implemented by
imposing an infinite number of independent constraints, one per spatial point $x$,\\
\begin{eqnarray}
\langle \Delta \chi _{x}^{2}\rangle =\int D\chi ^{\prime }\,Q\left[ \chi
^{\prime }|\chi \right] (\Delta \chi _{x})^{2}=\kappa _{x}\,,
\label{5Constraint 1}
\end{eqnarray}%
where $\Delta \chi _{x}=\chi _{x}^{\prime }-\chi _{x}$ and the $\kappa _{x}$
are small quantities. The result of maximizing (\ref{5entropy b}) subject to (%
\ref{5Constraint 1}) and normalization is a product of Gaussians
\begin{eqnarray}
Q\left[ \chi ^{\prime }|\chi \right] \propto \,\exp -\frac{1}{2}\int
dx\,g_{x}^{1/2}\alpha _{x}\left( \Delta \chi _{x}\right) ^{2}~,  \label{5prior}
\end{eqnarray}%
where $\alpha _{x}$ are the Lagrange multipliers associated to each
constraint (\ref{5Constraint 1}); the scalar density $g_{x}^{1/2}=\,\left(
\det \,g_{ij}\right) ^{1/2}$ is introduced so that $\alpha _{x}$ is a scalar
field. For notational simplicity we write $dx^{\prime }$ instead of $%
d^{3}x^{\prime }$. To enforce the continuity of the motion we will eventually
take the limit $\kappa _{x}\rightarrow 0$ which amounts to taking $\alpha
_{x}\rightarrow \infty $.

\paragraph*{The global constraint--- }

The motion induced by the prior (\ref{5prior}) leads to a rather simple diffusion process in the probabilities, in which the field variables evolve independently of each other. To model a dynamics that exhibits correlations and is capable of demonstrating the full suite of quantum effects, such as the superposition of states, interference, and entanglement, however, we require additional structure. This is accomplished by imposing a \emph{single} additional constraint that is \emph{non-local} in space but local in configuration space, which involves the introduction of a \emph{drift} potential $\phi \lbrack \chi ]$ which is a scalar-valued functional defined over the configuration space $\mathcal{C}$. More explicitly, we impose
\begin{eqnarray}
\langle \Delta \phi \rangle = \int D\chi ^{\prime }\,P\left[ \chi ^{\prime }|\chi \right] \int
dx\,\,\Delta \chi _{x}\frac{\delta \phi \left[ \chi \right] }{\delta \chi
_{x}}=\kappa ^{\prime },  \label{5Constraint 2}
\end{eqnarray}%
where we require $\kappa^{\prime}\to 0$. (Note that since $\chi _{x}$ and $\Delta \chi _{x}$ are scalars, in order that (\ref{5Constraint 2}) be invariant under coordinate transformations of the surface, the derivative $\delta /\delta \chi _{x}$ must transform as a scalar density.)


\paragraph*{The transition probability---}

Next we maximize (\ref{5entropy a}) subject to (\ref{5Constraint 2}) and
normalization.\footnote{As discussed in \cite{Bartolomeo Caticha 2016} the multiplier $\alpha ^{\prime }$ associated to the global constraint (\ref{5Constraint 2}) turns out to have no influence on the dynamics: it can be absorbed into the drift potential $\alpha ^{\prime }\phi \rightarrow \phi $ \ which means we can effectively set $\alpha ^{\prime }=1$.} The resulting transition probability is a Gaussian distribution, 
\begin{equation}
P\left[ \chi ^{\prime }|\chi \right] =\frac{1}{Z\left[ \alpha _{x},g_{x}\right] }\,\exp -\frac{1}{2}\int dx\,g_{x}^{1/2}\alpha _{x}\left( \Delta\chi _{x}-\frac{1}{g_{x}^{1/2}\alpha _{x}}\frac{\delta \phi \left[ \chi \right] }{\delta \chi _{x}}\right) ^{2},  \label{5Trans Prob}
\end{equation}
where $Z\left[ \alpha _{x},g_{x}\right] $ is the normalization constant. In
previous work \cite{Caticha 2013}\cite{Caticha Ipek 2014} $\alpha _{x}$ was
chosen to be a spatial constant $\alpha $ to reflect the translational
symmetry of flat space. Here we make no such restriction and instead relax
the global constant $\alpha $ in favor of a non-uniform spatial scalar $%
\alpha _{x}$ which will be a key element in implementing our scheme for a
local entropic time.

The Gaussian form of (\ref{5Trans Prob}) allows us to present a generic
change, 
\begin{equation}
\Delta \chi _{x}=\left\langle \Delta \chi _{x}\right\rangle +\Delta w_{x}~,
\end{equation}
as resulting from an expected drift $\left\langle \Delta \chi
_{x}\right\rangle $ plus fluctuations $\Delta w_{x}$. At each $x$ the
expected short step is 
\begin{equation}
\left\langle \Delta \chi _{x}\right\rangle =\frac{1}{g_{x}^{1/2}\,\alpha _{x}%
}\frac{\delta \phi \left[ \chi \right] }{\delta \chi _{x}}~,  \label{5Exp Step 1}
\end{equation}%
while the fluctuations $\Delta w_{x}$ satisfy,%
\begin{equation}
\left\langle \Delta w_{x}\right\rangle =0\,,\quad \text{and}\hspace{0.4cm}%
\left\langle \Delta w_{x}\Delta w_{x^{\prime }}\right\rangle =\frac{1}{%
g_{x}^{1/2}\alpha _{x}}\delta _{xx^{\prime }}.  \label{5Fluctuations}
\end{equation}%
Thus we see that $\Delta \bar{\chi}_{x}\sim 1/\alpha _{x}$ and $\Delta
w_{x}\sim 1/\alpha _{x}^{1/2}$, so that for short steps, $\alpha
_{x}\rightarrow \infty $, the fluctuations dominate the motion. The
resulting trajectory is continuous but non-differentiable --- a Brownian
motion.

\section{Some space-time kinematics}

We deal with a curved space-time. Events are labeled by coordinates $X^{\mu} $, and the metric is $g_{\mu \nu }\left( X^{\lambda }\right) $.\footnote{We use Greek indices ($\mu ,\nu ,...$ $=0,1,2,3$) for space-time coordinates  $X^{\mu }$ and latin indices ($a,b,...i,j,...$ $=1,2,3$) for coordinates $x^{i}$ on the surface $\sigma $. The space-time metric has signature $(-+++)$.} Space-time is foliated by a sequence of space-like surfaces $\left\{\sigma \right\} $. Points on the surface $\sigma $ are labeled by coordinates $x^{i}$ and the embedding of the surface within space-time is defined by four functions $X^{\mu }=X^{\mu }\left( x^{i}\right) $. The metric induced on the surface is
\begin{equation}
g_{ij}\left( x\right) =X_{i}^{\mu }X_{j}^{\nu }g_{\mu \nu }\quad \text{where}%
\quad X_{i}^{\mu }=\frac{\partial X^{\mu }}{\partial x^{i}}\ .
\label{5induced metric}
\end{equation}%
gives the space-time coordinates of three-vectors along the surface. Although the metric $g_{ij}$ will in general depend on the particular surface, it is not a genuine dynamical variable in its own right.\footnote{The case of a dynamically evolving geometry will be considered in the next chapter.}

Following Teitelboim and Kucha\u{r}, we consider an infinitesimal
deformation of the surface $\sigma $ to a neighboring surface $\sigma
^{\prime }$. This is specified by the deformation vector $\delta \xi ^{\mu }$
which connects the point in $\sigma $ with coordinates $x^{i}$ to the point
in $\sigma ^{\prime }$ with the same coordinates $x^{i}$, 
\begin{equation}
\delta \xi ^{\mu }=\delta \xi ^{\bot }n^{\mu }+\delta \xi ^{i}X_{i}^{\mu }~,
\label{5deformation vector}
\end{equation}%
where $n^{\mu }$ is the unit normal to the surface ($n_{\mu }n^{\mu }=-1$, $%
n_{\mu }X_{i}^{\mu }=0$). The normal and tangential components are given by%
\begin{equation}
\delta \xi _{x}^{\bot }=-n_{\mu x}\delta \xi _{x}^{\mu }\quad \text{and}%
\quad \delta \xi _{x}^{i}=X_{\mu x}^{i}\delta \xi _{x}^{\mu }~,
\end{equation}%
where $X_{\mu x}^{i}=g^{ij}g_{\mu \nu }X_{jx}^{\nu }$. They are known as the infinitesimal
lapse and shift, respectively, and are collectively denoted ${}(\delta \xi
^{\bot },\delta \xi ^{i})=\delta \xi ^{A}$. As a matter of convention a
deformation is identified by its normal $\delta \xi ^{\bot }$ and tangential 
$\delta \xi ^{i}$ components independently of the surface upon which it acts
(\emph{i.e.}, independently of the normal $n^{\mu }$). This allows us to
speak about applying \emph{the same deformation} to different surfaces; a
useful concept for our discussion of path independence.

\section{Entropic time}\label{55S Entropic time}
In ED time is introduced as a device to figure out how the accumulation of many infinitesimal changes builds up into a finite change. Questions such as, \textquotedblleft What is an instant?\textquotedblright\ \textquotedblleft How are they ordered?\textquotedblright\ and \textquotedblleft To what extent are they separated?\textquotedblright\ are central to constructing any dynamical theory and ED is no exception.

\paragraph{Ordered instants---}

Of particular importance is the notion of an instant, which in ED involves a couple key ingredients: (1) A spacelike surface $\sigma $ that provides a criterion of simultaneity and codifies spatial relations among the degrees of freedom. (2) The \textquotedblleft epistemic contents\textquotedblright\ of the surfaces. This is a specification of a statistical state that is sufficient for the prediction of future states. It is given by a probability distribution $\rho_{\sigma} \lbrack \chi ]$ and a drift potential $\phi_{\sigma} \lbrack \chi ]$.

\paragraph{The statistical state and its evolution---}

Entropic dynamics is generated by the short-step transition probability $P%
\left[ \chi ^{\prime }|\chi \right] $. In a generic short step both the
initial $\chi $ and the final $\chi ^{\prime }$ are unknown. Integrating the
joint probability, $P\left[ \chi ^{\prime },\chi \right] $, over $\chi $
gives 
\begin{equation}
P[\chi ^{\prime }]=\int d\chi \,P\left[ \chi ^{\prime },\chi \right] =\int
d\chi P\left[ \chi ^{\prime }|\chi \right] P\left[ \chi \right] ~.
\label{5CK a}
\end{equation}%
These equations are true by virtue of the laws of probability; they involve
no assumptions. However, if $P\left[ \chi \right] $ happens to be the
probability of $\chi $ \emph{at an \textquotedblleft
instant\textquotedblright\ labelled }$\sigma $, then we can interpret $P%
\left[ \chi ^{\prime }\right] $ as the probability of $\chi ^{\prime }$ 
\emph{at the \textquotedblleft next instant,\textquotedblright } which we
will label $\sigma ^{\prime }$. Accordingly, we write $P\left[ \chi \right]
=\rho _{\sigma }\left[ \chi \right] $ and $P\left[ \chi ^{\prime }\right]
=\rho _{\sigma ^{\prime }}\left[ \chi ^{\prime }\right] $ so that%
\begin{equation}
\rho _{\sigma ^{\prime }}\left[ \chi ^{\prime }\right] =\int D\chi \,P\left[
\chi ^{\prime }|\chi \right] \,\rho _{\sigma }\left[ \chi \right] ~.
\label{5Evolution equation}
\end{equation}%
This is the basic dynamical equation; it allows one to update the
statistical state $\rho _{\sigma }\left[ \chi \right] $ from one instant to
the next. Note that since $P\left[ \chi ^{\prime }|\chi \right] $ is found
by maximizing entropy not only are these instants ordered but there is a
natural \emph{entropic} arrow of time: $\sigma ^{\prime }$ occurs \emph{after%
} $\sigma $.

\paragraph{Duration---}

In ED time is defined so that motion looks simple. Since for short steps the
dynamics is dominated by fluctuations, eq.(\ref{5Fluctuations}), the
specification of the time interval between two successive instants is
achieved through an appropriate choice of the multipliers $\alpha _{x}$. So
far the present development of ED has followed closely along the lines of
the non-covariant models discussed in \cite{Caticha Ipek 2014}\cite{Bartolomeo et al 2014}.

The important point of departure is that here we are
concerned with instants defined on the curved embedded surfaces $\sigma $
and $\sigma ^{\prime }$. It is then natural to define a local notion of
duration in terms of proper time. The idea is the familiar one: at the point 
$x$ in $\sigma $ draw a normal segment reaching out to $\sigma ^{\prime }$. 
\emph{The proper time }$\delta \xi _{x}^{\bot }$\emph{\ along this normal
segment provides us with the local measure of duration between }$\sigma $%
\emph{\ and }$\sigma ^{\prime }$\emph{\ at }$x$\emph{.} More specifically,
let 
\begin{equation}
\alpha _{x}=\frac{1}{ \delta \xi _{x}^{\bot }}\quad \text{so that}\quad
\left\langle \Delta w_{x}\Delta w_{x^{\prime }}\right\rangle =\frac{
\,\delta \xi _{x}^{\bot }}{g_{x}^{1/2}}\delta _{xx^{\prime }}~.
\label{5Duration}
\end{equation}%

\begin{remark*}
But we are not done yet. With the definition (\ref{5Duration})
of duration, the dynamics given by (\ref{5Evolution equation}) and (\ref%
{5Trans Prob}) describes a Wiener process evolving along a given foliation of
space-time. To obtain a fully covariant dynamics we require that the
evolution of any dynamical quantity such as $\rho _{\sigma }[\chi ]$ from an
initial $\sigma _{i}$ to a final $\sigma _{f}$ must be independent of the
intermediate choice of surfaces. This \textquotedblleft foliation
invariance\textquotedblright\ or \textquotedblleft path
independence\textquotedblright is a consistency requirement: if there are different ways to
evolve from a given initial instant into a given final instant, then all
these ways must agree. The conditions to implement this consistency are the
subject of the next section.
\end{remark*}

\paragraph{The local-time diffusion equations---}

The dynamics expressed in integral form by (\ref{5Evolution equation}) with (%
\ref{5Trans Prob}) and (\ref{5Duration}) can be rewritten in differential form\footnote{Note that this means we restrict our attention only to probability distributions that are both continuous and differentiable. These distributions form a dense subset of the full statistical manifold.}
as an infinite set of local equations, one for each spatial point, 
\begin{equation}
\frac{\delta \rho _{\sigma }}{\delta \xi _{x}^{\bot }}=-\frac{1}{g_{x}^{1/2}}\frac{%
\delta }{\delta \chi _{x}}\left( \rho _{\sigma }\,\frac{\delta \Phi _{\sigma
}}{\delta \chi _{x}}\right) \hspace{0.5cm}\text{with}\hspace{0.5cm}\Phi
_{\sigma }\left[ \chi \right] = \,\phi _{\sigma }\left[ \chi \right]
- \log \rho _{\sigma }^{1/2}\left[ \chi \right] ~.  \label{5FP equation}
\end{equation}%
(The derivation is given in Appendix \ref{appendix FP}.) This set of
equations describes the flow of the probability $\rho _{\sigma }\left[ \chi %
\right] $ in the configuration space $\mathcal{C}$ as the surface $\sigma $
is deformed. More explicitly, the actual change in $\rho \lbrack \chi ]$ as $%
\sigma $ is infinitesimally deformed to $\sigma ^{\prime }$ is 
\begin{equation}
\delta \rho _{\sigma }\left[ \chi \right] =\int dx\frac{\delta \rho _{\sigma
}\left[ \chi \right] }{\delta \xi _{x}^{\bot }}\delta \xi _{x}^{\bot }=-\int
dx\frac{\delta \xi _{x}^{\bot }}{g_{x}^{1/2}}\frac{\delta }{\delta \chi _{x}}%
\left( \rho _{\sigma }\left[ \chi \right] \frac{\delta \Phi _{\sigma }\left[
\chi \right] }{\delta \chi _{x}}\right) \,.  \label{5FP b}
\end{equation}%
In the special case when both surfaces $\sigma $ and $\sigma ^{\prime }$
happen to be flat then $g_{x}^{1/2}=1$ and $\delta \xi _{x}^{\bot }=dt$ are
constants and eq.(\ref{5FP b}) becomes 
\begin{equation}
\frac{\partial \rho _{t}\left[ \chi \right] }{\partial t}=-\int dx\frac{%
\delta }{\delta \chi _{x}}\left( \rho _{t}\left[ \chi \right] \frac{\delta
\Phi _{t}\left[ \chi \right] }{\delta \chi _{x}}\right) \,,  \label{5FP c}
\end{equation}%
which we recognize as a diffusion or Fokker-Planck equation written as a
continuity equation for the flow of probability in configuration space $%
\mathcal{C}$. Accordingly we will refer to (\ref{5FP equation}) as the
\textquotedblleft local-time Fokker-Planck\textquotedblright\ equations
(LTFP). These equations describe the flow of probability with a current
velocity $v_{x}\left[ \chi \right] =\delta \Phi /\delta \chi _{x}$\thinspace . As we've seen, the functional $\Phi $ will be identified as the Hamilton-Jacobi functional, or the phase of the wave functional in the quantum theory.
\section{Kinematics of surface deformations and their generators}\label{55S Embeddability}
Dynamics in local time must reflect the kinematics of surface deformations,
and this kinematics can be studied independently of the particular dynamics
being considered. Consider a generic functional $T\left[ X(x)\right] $ that assigns a
number to every surface $X^{\mu }(x)$. The change in the functional $\delta
T $ resulting from an arbitrary deformation $\delta \xi _{x}^{A}$ has the
form 
\begin{equation}
\delta T=\int dx\,\delta \xi _{x}^{\mu }\frac{\delta T}{\delta \xi _{x}^{\mu
}}=\int dx\,\left( \delta \xi _{x}^{\bot }G_{\bot x}+\delta \xi
_{x}^{i}G_{ix}\right) T~\,,  \label{5delta T}
\end{equation}%
where 
\begin{equation}
G_{\bot x}=\frac{\delta }{\delta \xi _{x}^{\bot }}=n_{x}^{\mu }\frac{\delta 
}{\delta X_{x}^{\mu }}\quad \text{and}\quad G_{ix}=\frac{\delta }{\delta \xi
_{x}^{i}}=X_{ix}^{\mu }\frac{\delta }{\delta X_{x}^{\mu }}
\end{equation}%
are the generators of normal and tangential deformations respectively.
Unlike the vectors, $\delta /\delta X_{x}^{\mu }$, which form a coordinate
basis in the space of hypersurfaces and therefore commute, the generators of
deformations $\delta /\delta \xi _{x}^{A}$ form a non-holonomic basis. Their
non-vanishing commutator is 
\begin{equation}
\frac{\delta }{\delta \xi _{x}^{A}}\frac{\delta }{\delta \xi _{x^{\prime
}}^{B}}-\frac{\delta }{\delta \xi _{x^{\prime }}^{B}}\frac{\delta }{\delta
\xi _{x}^{A}}=\int dx^{\prime \prime }\,\kappa ^{C}{}_{BA}(x^{\prime \prime
};x^{\prime },x)\frac{\delta }{\delta \xi _{x^{\prime \prime }}^{C}}
\label{5commutator}
\end{equation}%
where $\kappa ^{C}{}_{BA}$ are the \textquotedblleft structure
constants\textquotedblright\ of the \textquotedblleft
group\textquotedblright\ of deformations.

The calculation of $\kappa ^{C}{}_{BA}$ is given in \cite{Teitelboim 1972}\cite{Kuchar 1972}. The basic idea is \emph{embeddability}: When we perform
two successive infinitesimal deformations $\delta \xi ^{A}$ followed by $%
\delta \eta ^{B}$, $\sigma \overset{\delta \xi }{\rightarrow }\sigma _{1}%
\overset{\delta \eta }{\rightarrow }\sigma ^{\prime }$, the three surfaces
are all embedded in the same space-time. The same happens when we execute
them in the opposite order, $\sigma \overset{\delta \eta }{\rightarrow }%
\sigma _{2}\overset{\delta \xi }{\rightarrow }\sigma ^{\prime \prime }$. The
key point is that the since the surfaces $\sigma ^{\prime }$ and $\sigma
^{\prime \prime }$ are embedded in the same space-time there must exist a
third deformation $\delta \zeta ^{A }$ that takes $\sigma ^{\prime }$
to $\sigma ^{\prime \prime }$: $\sigma ^{\prime }\overset{\delta \zeta }{%
\rightarrow }\sigma ^{\prime \prime }$. Thus the deformation from $\sigma $
to $\sigma ^{\prime \prime }$ can be attained by following two different
paths: either we follow the direct path $\sigma \overset{\delta \eta }{%
\rightarrow }\sigma _{2}\overset{\delta \xi }{\rightarrow }\sigma ^{\prime
\prime }$ or we follow the indirect path $\sigma \overset{\delta \xi }{%
\rightarrow }\sigma _{1}\overset{\delta \eta }{\rightarrow }\sigma ^{\prime }%
\overset{\delta \zeta }{\rightarrow }\sigma ^{\prime \prime }$. Then, as
shown in \cite{Teitelboim 1972}\cite{Kuchar 1972}, eq.(\ref{5commutator})
leads to the \textquotedblleft algebra\textquotedblright , 
\begin{eqnarray}
\lbrack G_{\bot x},G_{\bot x^{\prime }}] &=&-(g_{x}^{ij}G_{jx}+g_{x^{\prime
}}^{ij}G_{jx^{\prime }})\partial _{ix}\delta (x,x^{\prime })~,  \label{5LB1}
\\
\lbrack G_{ix},G_{\bot x^{\prime }}] &=&-G_{\bot x}\partial _{ix}\delta
(x,x^{\prime })~,  \label{5LB2} \\
\lbrack G_{ix},G_{jx^{\prime }}] &=&-G_{ix^{\prime }}\,\partial _{jx}\delta
(x,x^{\prime })-G_{jx}\,\partial _{ix}\delta (x,x^{\prime })~.  \label{5LB3}
\end{eqnarray}%
The previous quotes in \textquotedblleft group\textquotedblright\ and
\textquotedblleft algebra\textquotedblright\ are a reminder that strictly,
the set of deformations do not form a group. The composition of two
successive deformations is itself a deformation, of course, but it also
depends on the surface to which the first deformation is applied. Thus, the
\textquotedblleft structure constants\textquotedblright\ on the right hand
sides of (\ref{5LB1})-(\ref{5LB3}) are not constant, they depend on the surface $%
\sigma $ through its metric $g_{ij}$ which appears explicitly on the right
hand side of (\ref{5LB1}).

\section{Consistent entropic dynamics: path independence}
\label{55S Path independence}
In ED the relevant physical information --- supplied through the prior (\ref{5prior}) and the constraint (\ref{5Constraint 2}) --- have led us to a diffusive\ dynamics in which the probability $\rho _{\sigma }[\chi ]$ evolves under the action of the externally prescribed drift potential $\phi_{\sigma}\lbrack \chi ]$. It is a curious diffusion in a curved background space-time, but it is a diffusion nonetheless. The dynamics can be made richer by introducing $\phi_{\sigma}$, or more conveniently $\Phi_{\sigma}$, as an additional dynamical variable. 

However, it is not enough simply to have a dynamical $\rho_{\sigma}$ and $\Phi_{\sigma}$. We must also ensure that their joint time evolution is \emph{independent} of any particular foliation chosen. This translates into a consistency requirement, referred to by Kucha\v{r} \cite{Kuchar 1972} and Teitelboim \cite{Teitelboim 1972} as \emph{path independence}: if there are different paths to evolve from an initial instant into a final instant, then all these paths must lead to the same final values for all quantities.

The major insight of DHKT was to recognize that this principle amounted to representing the brackets in eqns.(\ref{5LB1})-(\ref{5LB3}) as relations among the generators of the dynamical variables. What remains to be done at present is therefore to choose a setting in which this may be enacted. Given the developments in this dissertation, in which we have established the vital role played by the canonical formalism in ED, it is \emph{sufficient}, and it is \emph{reasonable}, to adopt a Hamiltonian framework for this purpose. That is, following the previous chapters, we take $\rho_{\sigma}$ and $\Phi_{\sigma}$ as a canonical pair.

An additional \emph{pragmatic} benefit to this approach is that it affords us access to a roster of techniques designed for covariant classical field theories, but which can be used for an ED that is indeterministic and quantum. Indeed, once a Hamiltonian framework is adopted, we can borrow the techniques of Dirac \cite{Dirac Lectures}\cite{Dirac 1951} in his development of parameterized field theories. That is, for the task of depicting the changing surface geometry we treat the surface variables $X^{\mu}_{x}$ as if they were dynamical variables too. This is accomplished by formally introducing the auxiliary variables $\pi_{\mu }(x)=\pi _{\mu x}$ that will play the role of momenta conjugate to $X_{x}^{\mu }$. These $\pi $'s are defined through the Poisson bracket (PB)\
relations, 
\begin{equation}
\left \{ X_{x}^{\mu },X_{x^{\prime }}^{\nu }\right \}=0\,,\quad \left \{ \pi _{\mu
x},\pi _{\nu x^{\prime }}\right \}=0\,,\quad \left \{ X_{x}^{\mu },\pi _{\nu
x^{\prime }}\right \}=\delta _{\nu }^{\mu }\delta (x,x^{\prime })~.  \label{5PB HS}
\end{equation}%
It is important to stress that the technique of promoting the $X^{\mu}_{x}$'s to canonical variables is merely a \emph{trick}. The surface geometry is \emph{not} dynamical, rather, we deal here with a space-time that is \emph{externally} prescribed. The benefit of this approach, however, is that it allows us to treat genuine dynamics together with the kinematics of surfaces in unified way.


Having established this, the change of a generic functional $T[X,\pi ,\rho ,\Phi ]$ resulting from an arbitrary deformation $\delta \xi _{x}^{A}=(\delta \xi _{x}^{\bot },\delta \xi _{x}^{i})$ is expressed in terms of PBs, 
\begin{equation}
\delta T=\int dx\,\delta \xi _{x}^{\mu }\left\{ T,H_{\mu x}\right\} =\int dx\,\left( \delta \xi _{x}^{\bot }\left\{ T,H_{\bot x}\right\} +\delta \xi_{x}^{i}\left\{ T,H_{ix}\right\} \right) ~,
\end{equation}%
where $H_{\bot x}[X,\pi ,\rho ,\Phi ]$ and $H_{ix}[X,\pi ,\rho ,\Phi ]$ are the generators of normal and tangential deformations respectively, and the generic PB of two arbitrary functionals $U$ and $V$ is
\begin{equation}
\left\{ U,V\right\} =\int dx\,\left( \frac{\delta U}{\delta X_{x}^{\mu }}\frac{\delta V}{\delta \pi _{\mu x}}-\frac{\delta U}{\delta \pi _{\mu x}}\frac{\delta V}{\delta X_{x}^{\mu }}\right) +\int D\chi \,\left( \frac{\tilde{\delta}U}{\tilde{\delta}\rho }\frac{\tilde{\delta}V}{\tilde{\delta}\Phi }-\frac{\tilde{\delta}U}{\tilde{\delta}\Phi }\frac{\tilde{\delta}V}{\tilde{\delta}\rho }\right) .
\label{5General PB}
\end{equation}%
Thus, the PBs perform a double duty: on one hand they reflect the kinematics of deformations of surfaces embedded in a background space-time, and on the other hand they express the genuine entropic dynamics of $\rho $ and $\Phi $.

To comply with the requirement of path independence we follow Teitelboim and Kucha\u{r} \cite{Kuchar 1972}\cite{Teitelboim thesis}\cite{Teitelboim 1972} to seek generators $H_{\bot x}$ and $H_{ix}$ that provide a \emph{canonical} \emph{representation} of the DHKT \textquotedblleft algebra\textquotedblright\ of surface deformations. Unlike DHKT who developed a classical formalism based on choosing the field $\chi (x)$ and its momentum as canonical variables, here we develop a quantum formalism. We choose the functionals $\rho \lbrack \chi]$ and $\Phi \lbrack \chi ]$ as the pair of canonical variables.

The idea then is that in order for the dynamics to be consistent with the kinematics of surface deformations the PBs of $H_{\bot x}$ and $H_{ix}$ must close in the same way as the \textquotedblleft group\textquotedblright\ of deformations --- that is, they must provide a \textquotedblleft representation\textquotedblright\ involving the same \textquotedblleft structure constants\textquotedblright , 
\begin{eqnarray}
\left \{ H_{\bot x},H_{\bot x^{\prime }}\right \} &=&(g_{x}^{ij}H_{jx}+g_{x^{\prime
}}^{ij}H_{jx^{\prime }})\partial _{ix}\delta (x,x^{\prime })~,  \label{5PB 1}
\\
\left \{ H_{ix},H_{\bot x^{\prime }}\right \} &=&H_{\bot x}\partial _{ix}\delta
(x,x^{\prime })~,  \label{5PB 2} \\
\left \{ H_{ix},H_{jx^{\prime }}\right \} &=&H_{ix^{\prime }}\,\partial _{jx}\delta
(x,x^{\prime })+H_{jx}\,\partial _{ix}\delta (x,x^{\prime })~.  \label{5PB 3}
\end{eqnarray}%
It may be worth noting that these equations have not been derived; it is more appropriate to say that imposing (\ref{5PB 1})-(\ref{5PB 3}) as strong constraints constitutes our \emph{definition} of what we mean by a \textquotedblleft representation\textquotedblright . To complete the definition, we add that, as shown in \cite{Teitelboim thesis}\cite{Teitelboim 1972}, the requirement that the evolution of an arbitrary functional $T[X,\pi ,\rho ,\Phi ]$ satisfy path independence implies that the initial values of the canonical variables must be restricted to obey the weak constraints 
\begin{equation}
H_{\bot x}\approx 0\quad \text{and}\quad H_{ix}\approx 0~.
\label{5Hp Hi constr}
\end{equation}%
Furthermore, once satisfied on an initial surface $\sigma $ the dynamics
will be such as to preserve (\ref{5Hp Hi constr}) for all subsequent surfaces
of the foliation.

\section{The canonical representation}
\label{55S Canonical representation}
Next we seek explicit expressions for $H_{\bot x}$ and $H_{ix}$. A surface
deformation is described by (\ref{5deformation vector}), 
\begin{equation}
\delta X_{x}^{\mu }=\delta \xi _{x}^{\bot }n_{x}^{\mu }+\delta \xi
_{x}^{i}X_{ix}^{\mu }~.  \label{5HS deform a}
\end{equation}%
On the other hand, we can evaluate $\delta X_{x}^{\mu }$ using (\ref{5General
PB}), 
\begin{equation}
\delta X_{x}^{\mu }=\int dx^{\prime }\left( \left \{X_{x}^{\mu },H_{\bot x^{\prime
}}\right \}\delta \xi _{x^{\prime }}^{\bot }+\left \{X_{x}^{\mu },H_{ix^{\prime }}\right \}\delta
\xi _{x^{\prime }}^{i}\right) ~.  \label{5HS deform b}
\end{equation}%
Since 
\begin{equation}
\left \{ X_{x}^{\mu },H_{\bot x^{\prime }}\right \}=\frac{\delta H_{\bot x^{\prime }}%
}{\delta \pi _{\mu x}}\quad \text{and}\quad \left \{ X_{x}^{\mu
},H_{ix^{\prime }}\right \}=\frac{\delta H_{ix^{\prime }}}{\delta \pi _{\mu x}}~,
\end{equation}%
comparing (\ref{5HS deform a}) and (\ref{5HS deform b}) leads to 
\begin{equation}
\frac{\delta H_{\bot x^{\prime }}}{\delta \pi _{\mu x}}=n_{x}^{\mu }\,\delta
(x,x^{\prime })\quad \text{and}\quad \frac{\delta H_{ix^{\prime }}}{\delta
\pi _{\mu x}}=X_{ix}^{\mu }\,\delta (x,x^{\prime })~.
\end{equation}%
These equations can be integrated to give, 
\begin{equation}
H_{\bot x}=\pi _{\bot x}+\tilde{H}_{\bot }\quad \text{and}\quad H_{ix}=\pi
_{ix}+\tilde{H}_{i}~,  \label{5gen a}
\end{equation}%
where 
\begin{equation}
\pi _{\bot x}=n_{x}^{\mu }\,\pi _{\mu x}\quad \text{and}\quad \pi
_{ix}=X_{ix}^{\mu }\pi _{\mu x}~.  \label{5gen b}
\end{equation}%
Here $\tilde{H}_{\bot }$ and $\tilde{H}_{i}$ are ``constants" of integration
that are independent of the surface momenta $\pi _{\mu }$ but can, in
principle, depend on the remaining canonical variables.

Thus, the generators $H_{\perp x}$ and $H_{ix}$ separate into two
sectors: one pair, $\pi _{\bot x}$ and $\pi _{ix}$, that acts only on the
geometry and another pair, $\tilde{H}_{\bot x}$ and $\tilde{H}_{ix}$,\ that
acts on the epistemic variables. The
latter, $\tilde{H}_{\bot x}$ and $\tilde{H}_{ix}$, are called the ensemble
Hamiltonian and the ensemble momentum. In what follows we refer to them as the e-Hamiltonian and e-momentum, respectively.

\begin{remark*}
Given the canonical PBs and the expressions for $\pi_{\perp x}$ and $\pi _{ix}$ in (\ref{5gen b}), it is straightforward to check that $\pi _{\bot x}$ and $\pi _{ix}$ automatically satisfy the DHKT \textquotedblleft algebra\textquotedblright ,
\begin{subequations}
\begin{eqnarray}
\left \{ \pi _{\bot x},\pi _{\bot x^{\prime }}\right \} &=&(g_{x}^{ij}\pi
_{jx}+g_{x^{\prime }}^{ij}\pi _{jx^{\prime }})\partial _{ix}\delta
(x,x^{\prime })~,  \label{5pi PB 1} \\
\left \{ \pi _{ix},\pi _{\bot x^{\prime }}\right \} &=&\pi _{\bot x}\partial
_{ix}\delta (x,x^{\prime })~,  \label{5pi PB 2} \\
\left \{ \pi _{ix},\pi _{jx^{\prime }}\right \} &=&\pi _{ix^{\prime }}\,\partial
_{jx}\delta (x,x^{\prime })+\pi _{jx}\,\partial _{ix}\delta (x,x^{\prime })~.
\label{5pi PB 3}
\end{eqnarray}%
\end{subequations}
\end{remark*}

\subsubsection{The e-momentum generators}

The generators of tangential deformations are the simpler ones: they induce
translations of the dynamical variables parallel to the surface. The change
in $\rho $ and $\Phi $ (and functionals thereof) under a tangential
deformation $\delta \xi _{x}^{a}$ is
\begin{equation}
\frac{\delta \rho_{\sigma} }{\delta \xi _{x}^{i}}=\frac{\delta \rho_{\sigma} }{\delta \chi
_{x}}(\partial _{ix}\chi _{x})\quad \text{and}\quad \frac{\delta \Phi_{\sigma} }{%
\delta \xi _{x}^{i}}=\frac{\delta \Phi_{\sigma} }{\delta \chi _{x}}(\partial
_{ix}\chi _{x})~.\label{5Tangential variation}
\end{equation}
This change is generated by the e-momentum $\tilde{H}_{ix}$ according to

\begin{eqnarray}
\frac{\delta \rho_{\sigma} }{\delta \xi _{x}^{i}} &=&\left \{\rho_{\sigma} ,\tilde{H}_{ix}\right \}=\frac{%
\tilde{\delta}\tilde{H}_{ix}}{\tilde{\delta}\Phi_{\sigma}}~,  \notag \\
\frac{\delta \Phi_{\sigma} }{\delta \xi _{x}^{i}} &=&\left \{\Phi_{\sigma} ,\tilde{H}_{ix}\right \}=-\frac{%
\tilde{\delta}\tilde{H}_{ix}}{\tilde{\delta}\rho_{\sigma} }~.\label{5Tangential variation PB}
\end{eqnarray}%
These equations can be integrated to yield the e-momentum 
\begin{equation}
\tilde{H}_{ix}[\rho ,\Phi ]=-\int D\chi \,\rho_{\sigma} \lbrack \chi ]\frac{\delta
\Phi_{\sigma} \lbrack \chi ]}{\delta \chi _{x}}\,\partial _{ix}\chi _{x}~.
\label{5e-Momentum}
\end{equation}%
Note that $\tilde{H}_{ix}$ depends only on the epistemic variables, any dependence on the geometrical variables is absent. It is also straightforward to check that $\tilde{H}_{ix}$
satisfies the condition (\ref{5PB 3}),%
\begin{equation}
\left\{ \tilde{H}_{ix},\tilde{H}_{jx^{\prime }}\right \}=\tilde{H}_{ix^{\prime
}}\,\partial _{jx}\delta (x,x^{\prime })+\tilde{H}_{jx}\,\partial
_{ix}\delta (x,x^{\prime })~,  \label{5e-H PB 3}
\end{equation}%
so that the tangential generators $\pi _{ix}$ and $\tilde{H}_{ix}$ satisfy (%
\ref{5PB 3}) separately.

\subsubsection{The e-Hamiltonian generators}

The mixed Poisson bracket relations (\ref{5PB 2}) are the easiest to satisfy and
therefore the least informative. They merely tell us that $H_{\perp x}$, and therefore $\tilde{H}_{\bot x} $, must be a scalar density. In contrast, the normal-normal Poisson bracket in (\ref{5PB 1}) is crucial; it provides a necessary criterion for identifying the appropriate $\tilde{H}_{\perp x}$. Our goal here is to determine an acceptable family of such generators.

\paragraph*{Constraints from ED--- }
One criterion for choosing a suitable $\tilde{H}_{\perp x}$ comes from ED itself. Indeed, $\tilde{H}_{\perp x}$ is not just \emph{any} generator, it is a generator that \emph{defined} so as to reproduce the LTFP equations (\ref{5FP equation}). Namely, $\tilde{H}_{\perp x}$ must be such that
\begin{equation}
\frac{\delta \rho _{\sigma }\left[ \chi \right] }{\delta \xi _{x}^{\bot }}=%
\frac{\tilde{\delta}\tilde{H}_{\perp x}}{\tilde{\delta}\Phi _{\sigma }\left[ \chi \right] }~.
\label{5FP equation H}
\end{equation}%
The desired generator is therefore required to be of the form
\begin{equation}
\tilde{H}_{\perp x} =\int D\chi \,\rho \frac{1}{%
2g_{x}^{1/2}}\left( \frac{\delta \Phi }{\delta \chi _{x}}\right) ^{2}+F_{x}%
\left[ \rho; X^{\mu} \right] \,,  \label{5e-Hp a}
\end{equation}%
where $F_{x}$ is an undetermined integration ``constant" which is a \emph{functional} of just $\rho $ and $X^{\mu}_{x}$.

\paragraph*{Constraints from path independence--- }
Determining an appropriate $\tilde{H}_{\perp x}$ amounts to making a suitable choice for $F_{x}$. To determine this, insert $H_{\perp x} = \pi_{\perp x} + \tilde{H}_{\perp x}$ into the Poisson bracket (\ref{5PB 1}) and use eq.(\ref{5pi PB 1}) for $\pi_{\perp x}$, which yields
\begin{eqnarray}
\left \{ \tilde{H}_{\bot x},\tilde{H}_{\bot x^{\prime }}\right \} + \left \{\pi_{\perp x},\tilde{H}_{\perp x^{\prime}}   \right \} - \left \{\pi_{\perp x^{\prime}},\tilde{H}_{\perp x}   \right \} &=&(g_{x}^{ij}\tilde{H}_{jx}+g_{x^{\prime
}}^{ij}\tilde{H}_{jx^{\prime }})\partial _{ix}\delta (x,x^{\prime })~.\label{5PB 1 split a}
\end{eqnarray}%
Next we rewrite $\tilde{H}_{\perp x}$ as 
\begin{equation}
\tilde{H}_{\perp x}=\tilde{H}_{\perp x}^{0}+F_{x}^{0}  \label{5e-Hp b}
\end{equation}%
where%
\begin{equation}
\tilde{H}_{\perp x}^{0}=\int D\chi \rho\left[ \frac{1 }{%
2g_{x}^{1/2}}\left( \frac{\delta \Phi }{\delta \chi _{x}}\right) ^{2}+
\frac{g_{x}^{1/2}}{2}g^{ij}\partial _{i}\chi _{x}\partial _{j}\chi
_{x}\right] ~.  \label{5e-Hp c}
\end{equation}%
There is no loss of generality here because both the new $F_{x}^{0}$ and old $%
F_{x}$ are entirely arbitrary. The advantage of this new definition, however, is that the newly introduced $\tilde{H}_{0}$ satisfies
\begin{equation}
\left\{ \tilde{H}_{\bot x}^{0},\tilde{H}_{\bot x^{\prime }}^{0}\right \}=(g_{x}^{ij}%
\tilde{H}_{jx}+g_{x^{\prime }}^{ij}\tilde{H}_{jx^{\prime }})\partial
_{ix}\delta (x,x^{\prime })~.
\end{equation}%
Thus (\ref{5PB 1 split a}) reduces to
\begin{eqnarray}
\left\{ \tilde{H}_{\bot x}^{0},F_{x^{\prime }}^{0}\right \} - \left \{\tilde{H}_{\bot
x^{\prime }}^{0},F_{x}^{0}\right \}+ \left \{\pi_{\perp x},\tilde{H}_{\perp x^{\prime}}   \right \} - \left \{\pi_{\perp x^{\prime}},\tilde{H}_{\perp x}   \right \}  = 0~,\label{5PB 1 split b}
\end{eqnarray}%
where we have used that
\begin{equation}
\left\{ F_{x}^{0},F_{x^{\prime }}^{0}\right \}=0~.
\end{equation}%

\paragraph*{Some assumptions--- }
Equation (\ref{5PB 1 split b}) is the condition for the undetermined $F_{x}$. However, a general solution to this relation is outside the scope of this work, therefore we make the following \emph{assumption}: We restrict our search to those $F_{x}$ where the surface variables $X^{\mu}_{x}$ appear through undifferentiated functions of the metric components $g_{ij}$, i.e.,\footnote{Such a family of solutions appears rather naturally in a dynamical background when one assumes a ``non-derivative" coupling between ``matter" and geometry (see e.g., \cite{Teitelboim thesis}). This is discussed in the next chapter.}
\begin{equation}
F_{x}[\rho; X^{\mu}] = F_{x}[\rho; g_{ij}]\quad \text{where}\quad g_{ij} = g_{\mu\nu}X^{\mu}_{i}X^{\nu}_{j}~.
\end{equation}
To see the consequence of this assumption, recall first that \cite{Hojman Kuchar Teitelboim 1976}
\begin{equation}
\frac{\delta g_{ijx^{\prime}}}{\delta \xi^{\perp}_{x}} = n^{\mu}_{x}\frac{\delta g_{ijx^{\prime}}}{\delta X^{\mu}_{x}} = 2 K_{ijx}\delta(x^{\prime},x)~,
\end{equation}
where the symmetric tensor $K_{ijx}$ is the extrinsic curvature, from which it follows that
\begin{equation}
 \left \{\pi_{\perp x},\tilde{H}_{\perp x^{\prime}}   \right \} = 2K_{ijx}\frac{\partial \tilde{H}_{\perp x}}{\partial g_{ijx}}\delta(x^{\prime},x)~,
\end{equation}
which is an expression that is symmetric upon exchange of $x$ and $x^{\prime}$. This implies that
\begin{equation}
 \left \{\pi_{\perp x},\tilde{H}_{\perp x^{\prime}}   \right \} =  \left \{\pi_{\perp x^{\prime}},\tilde{H}_{\perp x}   \right \} ~,
\end{equation}
simplifying (\ref{5PB 1 split b}) to
\begin{equation}
\left\{ \tilde{H}_{\bot x}^{0},F_{x^{\prime }}^{0}\right \}=\left \{\tilde{H}_{\bot
x^{\prime }}^{0},F_{x}^{0}\right \}~,  \label{55 PB of HF}
\end{equation}%
which is a homogeneous and linear equation for $F_{x}^{0}$. Thus, the
condition for a functional $F_{x}^{0}$ to be acceptable is that the Poisson bracket $\{%
\tilde{H}_{\bot x}^{0},F_{x^{\prime }}^{0}\}$ be symmetric under exchange of $x$ and $x^{\prime }$.

\paragraph*{The modificed DHKT ``algebra"--- }
Before moving forward, note that the assumption made above simplifies the Poisson brackets (\ref{5PB 1})-(\ref{5PB 3}) so that the surface and ensemble sectors essentially decouple. To see this, recall first that the e-momentum $\tilde{H}_{ix}$ is independent of both $X^{\mu}$ and $\pi_{\mu}$, so that $\tilde{H}_{ix} = \tilde{H}_{ix}[\rho,\Phi]$. Next, we have also shown above that $\tilde{H}_{\perp}$ is itself independent of $\pi_{\mu}$. Once we also take into account the assumption made above that the surface variables appear in $\tilde{H}_{\perp}$ only as simple functions of $g_{ij}$ without derivatives, then it is possible to show that the Poisson brackets (\ref{5PB 1})-(\ref{5PB 3}) split into the two pieces. One is a set of PBs for the surface generators, given in eqns.(\ref{5pi PB 1})-(\ref{5pi PB 3}), that mirrors exactly the DHKT ``algebra". The remaining piece then satisfies
\begin{subequations}
\begin{eqnarray}
\left \{ \tilde{H}_{\bot x},\tilde{H}_{\bot x^{\prime }}\right \} &=&(g_{x}^{ij}\tilde{H}_{jx}+g_{x^{\prime
}}^{ij}\tilde{H}_{jx^{\prime }})\partial _{ix}\delta (x,x^{\prime })~,  \label{5PB 1 e}
\\
\left \{ H_{ix},\tilde{H}_{\bot x^{\prime }}\right \} &=&\tilde{H}_{\bot x}\partial _{ix}\delta
(x,x^{\prime })~,  \label{5PB 2 e} \\
\left \{ \tilde{H}_{ix}, \tilde{H}_{jx^{\prime }}\right \} &=& \tilde{H}_{ix^{\prime }}\,\partial _{jx}\delta
(x,x^{\prime })+ \tilde{H}_{jx}\,\partial _{ix}\delta (x,x^{\prime })~.  \label{5PB 3 e}
\end{eqnarray}%
\end{subequations}
Notice that these relations are almost identical to those of (\ref{5PB 1})-(\ref{5PB 3}) apart from an important distinction in (\ref{5PB 2 e}) where the Poisson bracket is between $\tilde{H}_{\perp x}$ and $H_{ix}$, not just $\tilde{H}_{ix}$. This discrepancy arises because $\tilde{H}_{\perp x}$ contains the surface and epistemic variables both, therefore we require the full generator $H_{ix}$.

\paragraph*{Some solutions---}
A slightly more general treatment of (\ref{55 PB of HF}) is offered in subsequent sections. However, for now, we identify a class of solutions that, nonetheless, prove to be of physical interest, given by $F_{x}^{0}$ of the form
\begin{equation}
F_{x}^{0}[X,\rho ;\chi ]=\int D\chi \,f_{x}(X_{x},\rho ,\frac{\delta \rho }{%
\delta \chi _{x}};\chi _{x},\partial \chi _{x})~,\label{5Fx local}
\end{equation}%
where $f_{x}$ is a \emph{function} (not a functional) of its arguments. Of course, any $F_{x}^{0}$ must be a scalar density, which implies that $f_{x}$ must be a scalar density too. Since the available scalar densities are $g_{x}^{1/2}$ and $\delta /\delta \chi_{x}$, some natural proposals are\footnote{Since each $\delta /\delta \chi_{x}$ introduces a scalar density, the second term in (\ref{5PB HF e}) introduces two scalar densities, and thus must be multiplied by an inverse density.}
\begin{equation}
f_{x}\sim g_{x}^{1/2}\rho \chi _{x}^{n}\quad \text{(integer }n\text{)}\quad \text{and}\quad f_{x}\sim \frac{1}{g_{x}^{1/2}\rho }\left( \frac{\delta \rho }{\delta \chi _{x}}\right) ^{2}~. \label{5PB HF e}
\end{equation}%
Moreover, because (\ref{55 PB of HF}) is linear in $F_{x}^{0}$ we can also consider linear combinations of these trial forms, resulting in a complicated dependence that we simply denote
\begin{equation}
f_{x}\sim g_{x}^{1/2}\rho V_{x}(\chi_{x},X_{x}^{\mu}).\label{5local potential V}
\end{equation}
A simple calculation of eq.(\ref{55 PB of HF}) using the $f_{x}$ given above confirms the suitability of these trials. What is perhaps not apparent, however, is that for an $F_{x}^{0}$ of the type given in (\ref{5Fx local}), these are the \textit{only} choices that we can make. We detail this in the next section.
\paragraph*{The quantum potential--- }
The set of acceptable $F_{x}^{0}$'s imply a family of Hamiltonians given by
\begin{equation}
\tilde{H}_{\perp x} =\int D\chi \rho \left[ \frac{1}{2g_{x}^{1/2}}
\left( \frac{\delta \Phi }{\delta \chi _{x}}\right) ^{2}+\frac{%
g_{x}^{1/2}}{2}g^{ij}\partial _{i}\chi _{x}\partial _{j}\chi _{x}+ g_{x}^{1/2} V(\chi _{x})+\frac{\lambda }{g_{x}^{1/2} }%
\left( \frac{\delta \log\rho }{\delta \chi _{x}}\right) ^{2}\right].
\label{5e-Hp}
\end{equation}%
The first two terms of this expression are \emph{required} by the principles of ED and path independence, the latter terms are merely optional. The potential $V_{x}(\chi _{x},X_{x}^{\mu})$, as mentioned above, involves self-interactions and interactions with the background geometry.\footnote{Although we do not address this here, we might expect arguments from the renormalization group to demonstrate which types of potentials are \textit{relevant}, in the technical sense.} To interpret the last term in (\ref{5e-Hp}) we recall that in flat space-time the quantum potential is given by \cite{Caticha Ipek 2014} 
\begin{equation}
Q=\lambda\int d^{3}x\int D\chi \rho \left( \frac{\delta \log\rho }{\delta \chi _{x}}\right) ^{2}~.\,
\end{equation}%
The transition to curved coordinates and to curved space-time is achieved by setting 
\begin{equation}
d^{3}x\rightarrow g_{x}^{1/2}d^{3}x\quad \text{and}\quad \frac{\delta }{%
\delta \chi _{x}}\rightarrow \frac{1}{g_{x}^{1/2}}\frac{\delta }{\delta \chi _{x}}
\end{equation}%
which gives 
\begin{equation}
Q_{\sigma }=\lambda\int d^{3}x\int D\chi \rho\left( 
\frac{\delta \log\rho }{\delta \chi _{x}}\right) ^{2}~.\,  \label{5Q}
\end{equation}%
Therefore the last term in (\ref{5e-Hp}) may be called the \textquotedblleft
local quantum potential\textquotedblright\ or the \textquotedblleft local
osmotic potential.\textquotedblright\ Its contribution to the energy is such
that those states that are more smoothly spread out in configuration space
tend to have lower energy. The corresponding coupling constant $\lambda >0$
controls the relative importance of the quantum potential; the case $\lambda
<0$ is excluded because it leads to instabilities.
\section{Evolution equations}\label{55S Evolution equations}
We will now summarize the main results of the previous sections by writing down the equations that describe how the dynamical variables $\rho_{\sigma}$ and $\Phi_{\sigma}$ evolve in a curved space-time.

\subsubsection*{Entropic Dynamics in a curved space-time}

Given a space-time with metric $g_{\mu \nu }(X)$ we start by specifying a foliation of surfaces $\sigma _{t}$ labeled by a time parameter $t$, $X^{\mu}=X^{\mu }(x,t)$, where $x^{i}$ are coordinates on the surface. The metric induced on the surface is given by (\ref{5induced metric}). The deformation of $\sigma _{t}$ to $\sigma _{t+dt}$ is given by (\ref{5deformation vector}), 
\begin{equation}
\delta X^{\mu }=\delta \xi ^{\bot }n^{\mu }+\delta \xi ^{i}X_{i}^{\mu
}=[N_{xt}n^{\mu }+N^{i}_{xt}X_{i}^{\mu }]dt~,
\end{equation}%
where
\begin{equation}
N_{xt} = \frac{\delta\xi^{\perp}_{x}}{dt}\quad\text{and}\quad N_{xt}^{i} = \frac{\delta\xi^{i}_{x}}{dt}
\end{equation}
are the scalar \emph{lapse} and vector \emph{shift}, respectively, which we denote together as $N^{A}_{xt} = (N_{xt}, N_{xt}^{i})$.

Similarly, consider a functional of the canonical variables $T_{t} = T_{t}[\rho,\Phi;X,\pi]$ defined on a surface $\sigma_{t}$ labeled by parameter $t$. A generic variation of $T_{t}$ from one surface to the next is given by
\begin{equation}
\delta T_{t} = \int d^{3}x \frac{\delta T_{t}}{\delta\xi_{x}^{A}}\delta\xi_{x}^{A} =\int d^{3}x \{T_{t},H_{Ax}\}\delta\xi_{x}^{A}.
\end{equation}
For a particular foliation $\sigma_{t}$ with the lapse $N_{xt}$ and shift $N^{i}_{xt}$ prescribed above, this can be written instead as
\begin{equation}
\delta T_{t} = \int d^{3}x N^{A}_{xt}\{T_{t},H_{Ax}\}dt~.
\end{equation}

The evolution of the functionals $\rho_{t}[\chi]$ and $\Phi_{t}[\chi]$ can now be determined in an analogous fashion. In fact, we have that
\begin{align}
&\partial_{t}\rho_{t} = \int d^{3}x N^{A}_{xt}\frac{\delta \rho_{t}}{\delta\xi_{x}^{A}} =\int d^{3}x N^{A}_{xt}\{\rho_{t},H_{Ax}\}\\
& \partial_{t}\Phi_{t} = \int d^{3}x N^{A}_{xt}\frac{\delta \Phi_{t}}{\delta\xi_{x}^{A}} =\int d^{3}x N^{A}_{xt}\{\Phi_{t},H_{Ax}\}~.
\end{align}
The tangential pieces\[ \frac{\delta \rho_{t}}{\delta\xi_{x}^{i}} =\frac{\tilde{\delta}\tilde{H}_{i x}}{\tilde{\delta}\Phi_{t}}\quad\text{and}\quad \frac{\delta \Phi_{t}}{\delta\xi_{x}^{i}} =-\frac{\tilde{\delta}\tilde{H}_{i x}}{\tilde{\delta}\rho_{t}}\] are given in eqns.(\ref{5Tangential variation}) and (\ref{5Tangential variation PB}). For the normal components we have
\begin{equation}
\frac{\delta \rho_{t}}{\delta\xi_{x}^{\perp}} = -\frac{1}{g_{x}^{1/2}} \frac{\delta}{\delta\chi_{x}}\left(\rho \frac{\delta\Phi}{\delta\chi_{x}}\right) = \frac{\tilde{\delta}\tilde{H}_{\perp x}}{\tilde{\delta}\Phi_{t}}
\end{equation}
which are the LTFP equations, given by (\ref{5FP equation}) and (\ref{5FP equation H}). To complete this descriptionwe can use the $\tilde{H}_{\perp x}$ given in eq.(\ref{5e-Hp}) to determine
\begin{equation}
\frac{\delta \Phi _{t}}{\delta \xi _{x}^{\bot }} = - \frac{\tilde{\delta}\tilde{H}_{\perp x}}{\tilde{\delta}\rho_{t}},
\end{equation}
where
\begin{align}
-\frac{\delta \Phi _{t}}{\delta \xi _{x}^{\bot }}=\frac{1}{2g_{x}^{1/2}}\left( \frac{\delta \Phi _{t}}{\delta \chi _{x}}\right) ^{2}+\frac{g_{x}^{1/2}}{2}g^{ij}\partial _{i}\chi _{x}\partial _{j}\chi
_{x}+g_{x}^{1/2}V_{x}-\frac{4\lambda }{g_{x}^{1/2}\rho ^{1/2}}\frac{%
\delta ^{2}\rho _{t}^{1/2}}{\delta \chi _{x}^{2}}~.  \label{5LT HJ}
\end{align}%
Owing to its resemblance to the Hamilton-Jacobi equation in (\ref{Fields Functional HJ equation}), we call (\ref{5LT HJ}) the local time Hamilton-Jacobi (LTHJ) equation. With this, the formulation of the ED of fields in curved space-time is complete. While it may not yet be obvious, this is a quantum theory; to make it explicit is the next task. 
\subsubsection*{The Local time Schr\"{o}dinger functional equations}
The relation of the ED formalism to quantum theory can made explicit by choosing $\lambda = 1/8$ for the coupling constant in eq.(\ref{5e-Hp}) and making a canonical transformation (often called a \textit{Madelung}
transformation) from the dynamical variables $\rho $ and $\Phi $ into a pair of complex variables,\footnote{Based on this choice, we are working in coordinates where $\hbar = c = 1$.} 
\begin{equation}
\Psi_{t}=\rho ^{1/2}_{t}e^{i\Phi _{t} }\quad \text{and}\quad \Psi_{t}^{*}=\rho ^{1/2}_{t}e^{-i\Phi _{t} }~.
\end{equation}%
The equation of evolution for the new variable $\Psi _{t}$ is then given by
\begin{equation}
\partial_{t}\Psi_{t} = \int d^{3}x N^{A}_{xt}\frac{\delta \Psi_{t}}{\delta\xi_{x}^{A}} = \int d^{3}x N^{A}_{xt}\{\Psi_{t},H_{Ax}\}\label{5SchEq gen a}
\end{equation}
where
\begin{equation}
i\frac{\delta\Psi_{t}}{\delta\xi_{x}^{\perp}} = \hat{H}_{\perp x}\Psi_{t} 
\end{equation}
with
\begin{equation}
\hat{H}_{\perp x} = -\frac{1}{2 g_{x}^{1/2}}\frac{\delta^{2}}{\delta\chi_{x}^{2}}+\frac{g_{x}^{1/2}}{2}g_{ij}\partial_{i}\chi_{x}\partial_{j}\chi_{x}+g_{x}^{1/2}V_{x}
\end{equation}
is the local time version of the functional Schr\"{o}dinger equation. The tangential deformations are similarly given by
\begin{equation}
i\frac{\delta\Psi_{t}}{\delta\xi_{x}^{i}} = \hat{H}_{ix}\Psi \quad\text{where}\quad \hat{H}_{ix} = -i \partial_{ix}\chi_{x}\frac{\delta}{\delta\chi_{x}}~.
\end{equation}
Thus a general evolution for the complex coordinate $\Psi_{t}$ is given by
\begin{equation}
i\partial_{t}\Psi_{t} = \int d^{3}x \left(N_{xt}\hat{H}_{\perp x}+N^{i}_{xt}\hat{H}_{ix}\right)\Psi_{t},\label{5SchEq a}
\end{equation}
which is a functional Schr\"{o}dinger equation in curved space-time; a result that is recognizable, for instance, from \cite{Long Shore 1998}.

The limit of flat space-time is obtained setting $g_{x}^{1/2}=1$, $\delta\xi _{x}^{\bot }=dt$, $N=1$, and $N^{i} = 0$. Then eq.(\ref{5SchEq a}) becomes the Schr\"{o}dinger functional equation, 
\begin{equation}
i \frac{\partial \Psi _{t}}{\partial t}=\int d^{3}x\,\left( -\frac{1}{2}\frac{\delta ^{2}}{\delta \chi _{x}^{2}}+\frac{1}{2}g^{ij}\partial _{i}\chi _{x}\partial _{j}\chi _{x}+V_{x}\right) \Psi _{t}~.  \label{5SchEq c}
\end{equation}%
This recovers the standard quantum field theory in the Schr\"{o}dinger functional representation, which we recognize from the previous chapter (see also, \cite{Jackiw 1989}) and thus justifies calling this a \textit{quantum} theory.
\paragraph*{Discussion--- }
Our result is to be interpreted as follows: given an initial state $\Psi_{\sigma}$, eq.(\ref{5SchEq a}) determines the subsequent state $\Psi_{\sigma^{\prime}}$; by iteration, one can then construct an entire trajectory for the state $\Psi_{\sigma}$. Thus, the ED we have developed here determines the evolution of the quantum state $\Psi_{\sigma}$, in a generic (globally hyperbolic) background space-time. A decided advantage of this framework is that no notion of Fock space (or otherwise) is introduced, and none is needed. 

This is especially relevant for constructing quantum field theory on curved space-time (QFTCS). Canonical approaches to quantum theory require a Hilbert space on which to represent the canonical commutation relations, field equations, operator algebra, etc. In QFTCS, the problem is that there are an infinite number of such representations, each of which are unitarily inequivalent, and with no way to choose among them.\footnote{As a general rule, the issue of unitarily inequivalent Hilbert space representations has to do with the infinite dimensional configuration space of field theories; in such cases, the Stone-von Neumann theorem (see e.g., \cite{Reed Simon 1972} for a reference), which holds for finite-dimensional configuration spaces, fails. Consequently, issues can also crop up in flat space-time; for instance, with gauge theories \cite{Kibble 1967}.} In recent years, the development of QFTCS has thus favored approaches that do not commit to a particular Hilbert space representation. A popular example of this is the local algebraic quantum field theory (AQFT) approach initiated by Haag \cite{Haag 2012}, and more recently pursued by, for example, by Wald and Hollands (see e.g., \cite{Wald 1994}\cite{Hollands Wald 2014}).

The ED framework can be seen as an alternative construction of QFTCS that shares many of its appealing features with AQFT. They are both agnostic pertaining to the choice of Fock space. Moreover, both place a special focus on a local algebras of observables --- in ED, the Dirac-Kucha\v{r}-Teitelboim ``algebra" --- which can be defined independent of a chosen Hilbert space.
\section{Some solutions to the Poisson brackets}\label{55S General solution}\label{5General Solution}
The covariant ED that we have developed here can be regarded as an entire family of dynamical models, with each member characterized by a particular choice of the e-functional $F_{x}^{0}$. As was noted above, however, such choices are not entirely arbitrary, but subject to a constraint, given by the Poisson bracket relation eq.(\ref{55 PB of HF}). In the previous section we enumerated some solutions to eq.(\ref{55 PB of HF}) for a particular form of $F_{x}^{0}$. Here, we generalize this result by considering a larger class of e-functionals from which $F_{x}^{0}$ could be selected from. From an alternative viewpoint, this process can instead be viewed as suggesting potential `non-linear' models of dynamics that extend the linear functional Schr\"{o}dinger equation --- but that are also relativistically covariant.
\subsubsection*{Background}
Beginning in the 1970's, there began a serious interest in generalizations of quantum theory. The motivation was the following. Quantum theory was a linear theory; consisting of Hilbert spaces, linear Schr\"{o}dinger evolution, and so on. In physics, however, we tend to view such linear theories as approximations to some complicated `non-linear' theory, rather than as fundamental theories in their own right. This begets the question of whether such a generalization might be appropriate for quantum theory as well. Thus the program of generalized quantum theory (GQT) was conceived.\footnote{The origins of this program might be traced to the work of B. Mielnik \cite{Mielnik 1974} aptly titled ``Generalized Quantum Mechanics". See references therein for additional background.} The chief aim was of finding some suitable non-linear `quantum theory' that might supersede the standard formulation, in a spirit similar to how general Riemannian space-times had come to supersede that of Minkowski in the treatment of gravity.

A particularly robust framework for tackling such a problem was provided by the geometric quantum theory approach, developed over the years by many authors (e.g., \cite{Kibble 1979}\cite{Schilling thesis}) with the basic premise of reformulating the complex Hilbert space as a genuine manifold of states with additional structure --- i.e. symplectic, Riemannian (i.e. metric), and complex. The modifications proposed within this framework typically boiled down to two chief maneuvers: either (a) keeping the symplectic structure intact, but changing the dynamics, mainly by changing the Hamiltonians, or (b) modifying the geometric structures themselves. The latter is certainly a more difficult endeavor; and has been discussed, for example, by Cirelli \textit{et al.} \cite{Cirelli I 1990}\cite{Cirelli II 1990} as well as by Schilling \cite{Schilling thesis} and company \cite{Schilling Ashtekar 1999}. The former is simpler to implement, and many of the proposed extensions, for example by Weinberg \cite{Weinberg 1989} or Kibble \cite{Kibble 1979}\cite{Kibble 1978}, are exactly of this type.

The work of Kibble in \cite{Kibble 1978} is of particular interest. While many proposals for alternative non-linear `Schr\"{o}dinger equations' focused on the non-relativistic regime \cite{Weinberg 1989}\cite{BB Mycielski 1976}, Kibble's work examined possible non-linear extensions for \textit{relativistic} quantum dynamics. His approach was remarkably similar to our own --- requiring a local Hamiltonian density to fulfill an algebraic relation analogous to the normal-normal Poisson bracket, eq.(\ref{5PB 3}). The treatment was not exhaustive, however; leaving it unclear what the full scope of these generalizations were. Our task in this section is thus to flesh out a wider range of potentials by explicitly solving the Poisson bracket relations for a desirable, but somewhat general, class of potentials.
\subsubsection*{Setup}
We consider solutions to the Poisson bracket in (\ref{55 PB of HF}). By virtue of the assumption we made in section \ref{55S Canonical representation}, this means we have already restricted ourselves to solutions where the surface variables $X^{\mu}_{x}$ appear through functions of the metric $g_{ij}$.

The allowed solutions for the bracket eq.(\ref{55 PB of HF}) are most conveniently examined by considering instead the Jacobi identity
\begin{equation}
\left\{\left\{\rho ,\tilde{H}_{\perp x}\right\},\tilde{H}_{\perp y}\right\}-\left\{\left\{\rho ,\tilde{H}_{\perp y}\right\},\tilde{H}_{\perp x}\right\}=\left\{\rho,\left\{\tilde{H}_{\perp y} ,\tilde{H}_{\perp x}\right\}\right\}~.
\end{equation}
Using the Poisson bracket relation (\ref{5PB 1}) we have
\begin{equation}
\left\{\left\{\rho ,\tilde{H}_{\perp x}\right\},\tilde{H}_{\perp y}\right\}-\left\{\left\{\rho ,\tilde{H}_{\perp y}\right\},\tilde{H}_{\perp x}\right\}=\left\{\rho,\left(\tilde{H}_{x}^{i}+\tilde{H}_{y}^{i}\right)\partial_{ix}\delta\left(x,y\right)\right\},
\label{5Necessary condition}
\end{equation}
where
\begin{equation}
\tilde{H}_{x}^{i}=g_{x}^{ij}\tilde{H}_{jx}\hspace{.75 cm}\text{and}\hspace{.75 cm}\tilde{H}_{y}^{i}=g_{y}^{ij}\tilde{H}_{jy}
\end{equation}
are the contravariant tangential generators that are given by eq.(\ref{5e-Momentum}). If we split $\tilde{H}_{x}$ as in (\ref{5e-Hp b}) then eq.(\ref{5Necessary condition}) implies that $F_{x}^{0}[X,\rho;\chi]$ must satisfy
\begin{equation}
\frac{1}{g_{x'}^{1/2}}\frac{\delta}{\delta\chi_{x'}}\left(\rho\,\frac{\delta}{\delta\chi_{x'}} \, \frac{\tilde{\delta}F_{x}^{0}}{\tilde{\delta}\rho}\right)=\frac{1}{g_{x}^{1/2}}\frac{\delta}{\delta\chi_{x}}\left(\rho\,\frac{\delta}{\delta\chi_{x}} \, \frac{\tilde{\delta}F_{x^{\prime}}^{0}}{\tilde{\delta}\rho}\right),
\label{5Necessary condition b}
\end{equation}
which, again, is an equation linear in $F_{x}^{0}$. This implies that a general solution is simply the sum of individual solutions. For brevity and conciseness we denote the linear functional differential operator in eq.(\ref{5Necessary condition b}) to be
\begin{equation}
\hat{O}_{x} = \frac{1}{g_{x}^{1/2}}\frac{\delta}{\delta\chi_{x}}\left(\rho\,\frac{\delta}{\delta\chi_{x}} \right) = \frac{1}{g_{x}^{1/2}}\left(\frac{\delta\rho}{\delta\chi_{x}}\frac{\delta}{\delta\chi_{x}} +\rho \frac{\delta^{2}}{\delta\chi_{x}^{2}} \right).
\label{5Necessary condition Op}
\end{equation}
\paragraph*{Ansatz--- }
We consider here a family of possible e-functionals for the unknown $F_{x}^{0}$. The subsequent solutions will not therefore be the most \emph{general} solution possible, but hopefully the family comprises a physically reasonable subset of solutions. In particular, consider $F_{x}^{0}$ of the form
\begin{equation}
F_{x}^{0}[\rho, X;\chi] = \prod_{n = 1}^{N} \mathcal{F}^{(n)}_{x}[\rho, X;\chi],
\label{5Fx ansatz}
\end{equation}
which is an $N$-fold product ($N$ is as large as desired) of e-functionals that are integrals over \textit{functions} $f^{(n)}_{x}$ of the arguments $\rho$, $\chi$, $X^{\mu}$, and their derivatives:
\begin{equation}
\mathcal{F}_{x}^{(n)}[\rho, X;\chi] = \int D\chi_{n} f_{x}^{(n)}\left(\rho,\frac{\delta^{(m)}\rho}{\delta\chi_{nx}^{m}}; X^{\mu}_{x},\chi_{nx}\right).
\label{5Fx Local Definition}
\end{equation}
The dependence of these functions $f_{x}^{(n)}$ can be as complicated as desired, so long as $F_{x}^{0}$ is a scalar density; in particular, dependence on $\rho$ could include derivatives up to any order, the field $\chi_{x}$ can appear as a polynomial to arbitrary order, and the geometry can be any complicated scalar function of the metric components.\footnote{What does not appear, for instance, are the analogue of kernel operators inside the integrals or non polynomial functions of $\chi_{x}$.}

\subsubsection{Solutions}
The solution of the condition (\ref{5Necessary condition Op}) with ansatz given by (\ref{5Fx ansatz}) can be split into two pieces consisting of solutions where $N = 1$, and those for $N \geq 2$. For a proof of these results, we refer the reader to Appendix C.
\paragraph*{Results: $N\geq 2$: }
As shown in the appendix, for the case $N\geq 2$ we have that
\begin{equation}
F_{x}^{0}[\rho; X^{\mu}] = \prod_{n = 1}^{N} \mathcal{F}_{x}^{(n)}[\rho; X^{\mu}]
\end{equation}
where we have
\begin{equation}
 \mathcal{F}_{x}^{(n)}[\rho; X^{\mu}] = \int D\chi\, \rho[\chi] \, V_{x}^{(n)}(\chi_{x}, X^{\mu}_{x}) \quad\text{and}\quad V_{x}^{(n)}(\chi_{x}, X^{\mu}_{x}) = \sum_{m} c_{m}(X_{x}^{\mu})\,\chi_{x}^{n}~.\label{5 Fx Solution N >= 2}
\end{equation}
Thus to include nonlinear potentials that are $N$-fold products with $N\geq 2$ it must be that each product is particularly simple: they can be interpreted as the expected value of functions of the field. In particular, according to our assumptions, $V_{x}^{(n)}$ must be polynomial in powers of the field, while the surface variables must, of course, appear as the metric. Moreover $\mathcal{F}_{x}^{(n)}$ overall must be a scalar density.

\paragraph*{Results: $N = 1$: } It turns out that the proof used to determine $F_{x}$ for $N \geq 2$ does not apply to the case $N = 1$. The end result of this is that there is a single additional solution allowed for the function $f_{x}$ in this case. Thus in this instance we have
\begin{equation}
F_{x}^{0} = g_{x}^{1/2}\int D\chi \, \rho \, V_{x}(\chi_{x},X_{x}^{\mu})+Q_{x}\label{5 Fx Solution N = 1}
\end{equation}
where
\begin{equation}
Q_{x} = \frac{\lambda}{g_{x}^{1/2}}\int D\chi \frac{1}{\rho}\left(\frac{\delta\rho}{\delta\chi_{x}}\right)^{2}~.
\end{equation}
The first term in (\ref{5 Fx Solution N = 1}) is identical in form to the $\mathcal{F}_{x}$ contained in (\ref{5 Fx Solution N >= 2}). However, the remaining term is new and is exactly the ``local" quantum potential introduced in (\ref{5PB HF e}) and (\ref{5Q}). Thus the full class of acceptable relativistic potentials, including all terms, with $N = 1$ and $N\geq 2$, and linear combinations thereof, can thus be written as
\begin{equation}
F_{x}^{0} = g_{x}^{1/2}\int D\chi \, \rho V_{x}(\chi_{x},X_{x}^{\mu})+Q_{x} + g_{x}^{1/2} \sum_{m=1}^{M}\,\prod_{n=1}^{N}\int D\chi\,\rho \, V_{x}^{(mn)}~,
\label{55PB Full solution}
\end{equation}
where $M$ and $N$ are finite numbers and $V_{x}^{(mn)} = V_{x}^{(mn)}(\chi_{x},X)$ is a distinct function, one for each $m$, $n$.

\paragraph*{Comparison to Kibble}
It is not difficult to show that this agrees with Kibble's suggestion from \cite{Kibble 1978}. For reference, the non-linear equation put forth by Kibble in a flat space-time takes the form
\begin{equation}
i\partial_{t}\Psi_{t} = \int d^{3}x\left(\hat{H}^{0}_{x}+V_{x}^{NL}\right)\Psi_{t}
\end{equation}
where
\begin{equation}
\hat{H}^{0}_{x} = -\frac{1}{2g_{x}^{1/2}}\frac{\delta^{2}}{\delta\chi_{x}^{2}}+\frac{g_{x}^{1/2}}{2}g^{ij}\partial_{i}\chi_{x}\partial_{j}\chi_{x}
\end{equation}
and
\begin{equation}
V_{x}^{NL} = \left\langle f^{1}_{x}(\chi_{x})\right\rangle+\left\langle f^{2}_{x}(\chi_{x})\right\rangle\chi_{x}^{2}+\left\langle f^{3}_{x}(\chi_{x})\right\rangle \chi_{x}^{4}
\end{equation}
with $\left\langle f^{i}_{x}(\chi_{x})\right\rangle = \int D\chi \rho f^{i}_{x}$ just the expected value of the function $f^{i}_{x}$. It should be mentioned that these potentials are non-linear, in the sense that $V^{NL}\Psi\sim \Psi^{3}$ since $\rho = |\Psi|^{2}$.

Clearly by an appropriate choice of $F_{x}^{0}$ in eq.(\ref{55PB Full solution}), these nonlinear potentials can be reproduced and are, therefore, in agreement. On the other hand, our approach goes much further than Kibble in \emph{deriving} that these are the \emph{only} nonlinear generalizations in the relativistic regime, within a reasonable family of potentials. Kibble's work seems to imply that these potentials were sufficient, but did not discuss how general they were, or were not.
\paragraph*{Why non-linear, anyway?} 
From the viewpoint of orthodox quantum theory the search for additional, nonlinear potentials seems, at best, a waste of time, and perhaps even downright foolish. But the calculus is somewhat different in ED. Here linearity is not a sacrosanct principle of our derivation, thus it is completely reasonable to characterize a family of relativistic ED models; whether linear or not. Indeed, in the prolonged search for a theory of quantum gravity, are we so sure that a fundamental theory of gravity will come with the full complement of linear, quantum features? Is it necessarily undesirable that ED affords such flexibility? Such questions will be considered more carefully in the next chapter when we deal with a dynamical background.
\paragraph*{Actually, why linear?}
Quite apart from these considerations, the challenge faced by ED is not ``why nonlinear?" Rather, the question is, why should the dynamics even be linear in the first place? In the context of non-relativistic physics, Caticha \cite{Caticha 2019b} has supplied an answer that appeals to the symmetries of the e-phase space, but this approach has not been attempted in the case of relativistic fields.

Absent such a rigorous derivation here, one must supply a heuristic argument for including the quantum potential, and thus obtaining linearity. One possibility is to argue on the basis of simplicity. In ED the generators of significant importance are $\tilde{H}_{\perp x}$ and $\tilde{H}_{i x}$. The latter has a simple statistical interpretation (up to an overall minus sign) as the expected value of a local momentum density
\begin{equation}
 \tilde{H}_{ix} = -\int D\chi \rho\left(\frac{\delta\Phi}{\delta\chi_{x}}\partial_{ix}\chi_{x}\right).
 \end{equation}
Furthermore, as we have discussed above, the normal generator $\tilde{H}_{\perp x} = \tilde{H}^{0}_{\perp x}+F_{x}$ can be split up into two pieces. The first term is particularly special in that its presence is strictly required; first by ED and then by path independence. But an interesting aspect of $\tilde{H}^{0}_{\perp x}$ is that it is quite similar in form to $\tilde{H}_{ix}$, meaning that it too can be interpreted as the expected value of a density, but here an energy density rather than a momentum density.

This suggests the following possibility: Let us \emph{insist} that $F_{x}$ be of the same functional form as both $\tilde{H}_{\perp x}^{0}$ and $\tilde{H}_{ix}$. Pushing this analogy leads us to exclude the nonlinear terms in (\ref{55PB Full solution}), leaving only
\begin{align}
\tilde{H}_{\perp x} &= \int D\chi \rho\left[\frac{1}{g_{x}^{1/2}}\left(\left(\frac{\delta\Phi}{\delta\chi_{x}}\right)^{2}+\lambda\left(\frac{\delta\log \rho}{\delta\chi_{x}}\right)^{2}\right)+ g_{x}^{1/2}\left(\frac{1}{2}g^{ij}\partial_{ix}\chi_{x}\partial_{jx}\chi_{x}+V_{x}\right)\right]~.
\end{align}
This answer is almost uniquely quantum theory.

Of course, this argument is not entirely convincing. First it is not obvious why we should even push this analogy in the first place. But besides that, the argument does not suggest why we must include the quantum potential at all. Indeed, not choosing the quantum potential, i.e. setting $\lambda = 0$, is also entirely consistent with this line of reasoning. Thus there are clearly some additional ingredients that appear to be missing in our reconstruction. Such is the subject of future investigation.

That being said, the ED approach here does have some advantages over the GQTs considered in the non-relativistic regime. As demonstrated by Bialynicki-Birula and Mycielski \cite{BB Mycielski 1976} in the context non-relativistic quantum mechanics, for instance, one possible nonlinear generalization to the usual Schr\"{o}dinger equation, which satisfies certain requirements, is given (up to an overall additive constant) by
\begin{equation}
 F[\rho]  = \int dx \rho \log\rho~,
 \end{equation} 
which we can identify as the entropy of $\rho$ relative to a uniform prior. Such alternatives are, however, ruled out in our scheme. This is not insignificant because the model of Bialynicki-Birula and Mycielski \cite{BB Mycielski 1976} is still being used for the purposes of testing quantum theory (see e.g., \cite{Georg Rosenzweig 2020}), even despite its apparent conflict with relativity. Moreover, other choices available in the non-relativistic regime (see e.g.,  \cite{Kibble 1979}) also fail to be viable here. Thus the combination of ED and covariance is much more restrictive than many previous approaches to the issue of GQTs.
\section{An application: the Ehrenfest equations}\label{55S Ehrenfest}
As laid out above, the ED that we have developed can be made formally identical to the standard quantum field theory in the Schr\"{o}dinger functional representation. This means that predictions made on the basis of equations (\ref{5SchEq gen a})-(\ref{5SchEq a}) are identical to those obtained using the standard methods. And indeed, there have been a wide range of topics pursued within this formalism; including, studies of vacuum states in curved space-time \cite{Long Shore 1998}, research on the Hawking effect \cite{Freese et al 1985}, applications in cosmology \cite{Guven et al 1989}, and investigations into symmetries \cite{Floreanini et al 1987}\cite{Halliwell 1991}, just to name a few. In other words: the ED developed here, while being fully \emph{consistent} with all of these developments, it does not go beyond them. And, indeed, the purpose of the ED developed here is not to generate better techniques for calculation, but to put QFTCS on a firm conceptual foundation.

The ED formulation of QFTCS, nonetheless, is sufficiently different from the usual approaches that a demonstration of the framework is, in fact, warranted. This is highlighted, for instance, in the derivation of the Ehrenfest relations for a quantum scalar field in ED, which we pursue here.
\paragraph{Some background}
In ED, the focus of our inquiries are the field variables $\chi_{x}$ which have definite, but unknown values. Thus, in the absence of such definite information, our goal is to obtain an estimate of these values. One such estimate is provided by the \emph{expected value} of the field variables
\begin{equation}
\tilde{\chi}_{x} = \int D\chi \, \rho \, \chi_{x}~.\label{5Expected chi}
\end{equation}
This, in turn, defines a functional $\tilde{\chi}_{x}$ on the ensemble phase space, which makes it amenable to treatment through the canonical formalism.

However, this expected value is not static and we wish to know how it changes in time. To determine this evolution, we carry over the formalism of the previous section, including the lapse $N_{xt}$, shift $N^{i}_{xt}$, and foliation parameter $t$. Moreover, we are concerned with a quantum dynamics so we choose our generators $H_{\perp x}$ and $H_{ix}$ to correspond with those given in eqns.(\ref{5e-Momentum}) and (\ref{5e-Hp}).

\paragraph*{Evolution of the expected values}
Since $\tilde{\chi}_{x} = \tilde{\chi}_{x}[\rho]$ is a Hamiltonian functional, its update can be obtained by taking the appropriate Poisson brackets. Indeed, the velocity of $\tilde{\chi}_{x}$ is given by
\begin{equation}
\partial_{t}\tilde{\chi}_{x} = \int dx^{\prime} \left (N_{x^{\prime}t}\left \{\tilde{\chi}_{x},H_{\perp x^{\prime}}\right \}+N^{i}_{x^{\prime}t}\left \{\tilde{\chi}_{x},H_{i x^{\prime}}\right \}\right )\label{5Expected chi velocity a}
\end{equation}
with
\begin{equation}
\left\{\tilde{\chi}_{x},H_{\perp x^{\prime}}\right \} = \delta(x,x^{\prime})\int D\chi \, \rho \frac{1}{g_{x^{\prime}}^{1/2}}\frac{\delta\Phi}{\delta\chi_{x^{\prime}}} \label{5Expected chi velocity perp}
\end{equation}
and
\begin{equation}
\left\{\tilde{\chi}_{x},H_{i x^{\prime}}\right \} =- \delta(x,x^{\prime})\int D\chi \, \rho\, \partial_{ix^{\prime}}\chi_{x^{\prime}}~.\label{5Expected chi velocity tang}
\end{equation}
Taken together, eqns.(\ref{5Expected chi velocity a})-(\ref{5Expected chi velocity tang}) result in
\begin{equation}
\partial_{t}\tilde{\chi}_{x} =  \int D\chi \, \rho \left ( \frac{N}{g^{1/2}_{x}}\frac{\delta\Phi}{\delta\chi_{x}}-N^{i}\partial_{ix}\chi_{x}\right )~,\label{5Expected chi velocity}
\end{equation}
which contains two contributions. The latter contribution in eq.(\ref{5Expected chi velocity}) is just due to the shift, while the former is due to the flow of probability, which is characterized by the appearance of the current velocity \[v_{x} = \frac{1}{g^{1/2}_{x}}\frac{\delta\Phi}{\delta\chi_{x}}~, \] first introduced in the context of the LTFP equations (\ref{5FP equation}).

Consider here a related quantity, the \emph{current momentum} and its expectation, the \emph{ensemble} current momentum,\footnote{Note that while the current velocity is a scalar valued quantity, the current \emph{momentum} is a scalar \emph{density} of weight one.}
\begin{equation}
P_{x} \equiv \frac{\delta\Phi}{\delta\chi_{x}} = g_{x}^{1/2} \, v_{x}\quad \text{and}\quad \tilde{P}_{x} = \int D\chi \, \rho\, P_{x}~,\label{5Expected current momentum}
\end{equation}
respectively.\footnote{Alternatively, introduce the differential operator $\hat{P}_{x} = -i\delta/\delta\chi_{x}$. Then the ensemble current momentum translates in the conventional language to the expected value of this operator. That is, $\tilde{P}_{x} = \int \Psi^{*}\hat{P}_{x}\Psi$, using the complex functionals $(\Psi^{*},\Psi)$ introduced above.} We can now conveniently rewrite the velocity $\partial_{t}\tilde{\chi}_{x}$ in terms of the current momentum, yielding
\begin{equation}
\partial_{t}\tilde{\chi}_{x} = \frac{N}{g^{1/2}_{x}}\tilde{P}_{x}-N^{i}\left \langle \partial_{ix}\chi_{x} \right \rangle~,\label{5Expected chi velocity b}
\end{equation}
where we have used the notation $\left\langle A \right \rangle = \int \rho A$ to denote expectation.

The advantage of introducing the current momentum (as opposed to the current velocity) is that the expected current momentum $\tilde{P}_{x}$, together with the expected field value $\tilde{\chi}_{x}$, satisfy the canonical Poisson bracket relations
\begin{equation}
 \left \{ \tilde{\chi}_{x}, \tilde{P}_{x^{\prime}}\right \} = \delta(x,x^{\prime})~. 
 \end{equation}
This shows that $\tilde{P}_{x}$ plays the role of a momentum \emph{conjugate} to $\tilde{\chi}_{x}$.

Indeed, let us take this hint seriously and compute the corresponding Hamilton's equations for this canonical pair. The time derivative of $\tilde{\chi}_{x}$ was provided earlier, in eq.(\ref{5Expected chi velocity b}). The velocity of $\tilde{P}_{x}$, on the other hand, is given by
\begin{equation}
\partial_{t}\tilde{P}_{x} = \int dx^{\prime} \left (N_{x^{\prime}t}\left \{\tilde{P}_{x},H_{\perp x^{\prime}}\right \}+N^{i}_{x^{\prime}t}\left \{\tilde{P}_{x},H_{i x^{\prime}}\right \}\right )~.\label{5Expected current momentum velocity a}
\end{equation}
To compute this we need the two Poisson brackets in eq.(\ref{5Expected current momentum velocity a}). A quick calculation gives
\begin{equation}
\left\{\tilde{P}_{x},H_{\perp x^{\prime}}\right \} =-\int D\chi \, \rho\, g_{x^{\prime}}^{1/2}\left (\delta(x,x^{\prime})\frac{\partial V_{x^{\prime}}}{\partial \chi_{x^{\prime}}}+g^{ij}_{x^{\prime}}\partial_{ix^{\prime}}\chi_{x^{\prime}}\partial_{jx^{\prime}}\delta(x^{\prime},x)\right ) \label{5Expected current momentum velocity perp}
\end{equation}
and
\begin{equation}
\left\{\tilde{P}_{x},H_{i x^{\prime}}\right \} =\int D\chi \, \rho \frac{\delta\Phi}{\delta\chi_{x^{\prime}}}\partial_{ix^{\prime}}\delta(x^{\prime},x)\label{5Expected current momentum velocity tang}
\end{equation}
so that from eq.(\ref{5Expected current momentum velocity a}) we obtain
\begin{equation}
\partial_{t}\tilde{P}_{x} =  \partial_{i}\left ( N\, g_{x}^{1/2} \, g^{ij} \partial_{j}\tilde{\chi}_{x}  \right ) - \partial_{i}\left (N^{i}\tilde{P}_{x}\right ) - N\, g_{x}^{1/2} \left\langle \frac{\partial V}{\partial\chi_{x}} \right \rangle~.\label{5Expected current momentum velocity}
\end{equation}
Thus an initial assignment of $\tilde{\chi}_{x}$, $\tilde{P}_{x}$, and higher statistical moments of $\chi_{x}$, will be sufficient to determine the evolution of $\tilde{\chi}_{x}$, i.e. no further derivatives are required.
\paragraph*{Ehrenfest equations}
The equations (\ref{5Expected chi velocity}) and (\ref{5Expected current momentum velocity}) taken together have the character of classical field equations for some \textquotedblleft classical\textquotedblright\ field variables $\tilde{\chi}_{x}$ and $\tilde{P}_{x}$. In fact, it is not difficult to show that these are nothing but the Ehrenfest equations (see e.g.,\cite{Ballentine}).

To see this, note that the velocity $\partial_{t}\tilde{\chi}_{x}$ is linear in the current momentum $\tilde{P}_{x}$. Now, invert this relation for $\tilde{P}_{x}$ in terms of the velocity $\partial_{t}\tilde{\chi}_{x}$ and substitute into eq.(\ref{5Expected current momentum velocity}). The result is that $\tilde{\chi}_{x}$ evolves according to the equation
\begin{equation}
\hat{\Box}\tilde{\chi}_{x} =  \left\langle \frac{\partial V(\chi_{x})}{\partial\chi_{x}} \right \rangle~,\label{5Ehrenfest relations}
\end{equation}
where
\begin{align}
\hat{\Box} = -\frac{1}{N\, g^{1/2}}\bar{\partial}_{t}\left [ \left ( \frac{g^{1/2}}{N}\right )\bar{\partial}_{t}\right ]-\frac{\partial_{i}N^{i}}{N^{2}}\bar{\partial}_{t}+\frac{1}{N\, g^{1/2}}\partial_{i}\left [  N\, g^{1/2} g^{ij}\partial_{j}  \right] \label{5Wave operator}
\end{align}
is the wave operator in curved space-time in foliation-adapted coordinates \cite{Long Shore 1998}\cite{Gourgoulhon 2007}, and where we have introduced the operator $\bar{\partial}_{t} = \partial_{t}+N^{i}\partial_{i}$.


Equation (\ref{5Ehrenfest relations}) comprises the Ehrenfest relations that we seek. An appealing feature of such relations is that they are \emph{exact}, and thus contain complete information about the underlying quantum dynamics. This makes the Ehrenfest relations ideal for probing the behavior of quantum fields and their deviations from classical behavior. For instance, it is not difficult to see that $\tilde{\chi}_{x}$ follows a classical evolution only when
\begin{equation}
\left\langle \frac{\partial V(\chi_{x})}{\partial\chi_{x}} \right \rangle =  \frac{\partial V(\left\langle\chi_{x}\right \rangle)}{\partial\chi_{x}}~.
\end{equation}
But this, of course, only occurs when the potential is itself quadratic in the field.

Indeed, choose for $V_{x}$ the potential $V_{x} = \frac{1}{2}m^{2}\chi_{x}^{2}$ so that $\partial V/\partial \chi_{x} = m^{2}\chi_{x}$. Equation (\ref{5Ehrenfest relations}) then reduces to
\begin{equation}
\left (\hat{\Box}-m^{2}\right )\tilde{\chi}_{x} = 0~,\label{5Ehrenfest equation KG}
\end{equation}
which is a classical Klein-Gordon equation in curved space-time (see e.g., \cite{Hollands Wald 2014}) for the expected field configuration. Thus $\tilde{\chi}_{x}$ follows --- exactly --- the classical equations of motion, which is precisely the content of Ehrenfest's theorem, familiar from non-relativistic quantum mechanics (see e.g., \cite{Ballentine}). For potentials that are \emph{not} quadratic, however, we can expect deviations from classical behavior and it is legitimate to obtain quantum corrections via an approximation scheme.
\paragraph*{Comments}
As opposed to standard formulations of equations of this type, our derivation of the Ehrenfest relations was performed entirely within the framework of ED, using Hamiltonians, Poisson brackets, etc., rather than commutators and the standard quantum machinery. These methods, which are geometric in nature, are thus of more general applicability than the standard techniques, which rely heavily on the \emph{linearity} of quantum theory. Indeed, while the assumption of linearity has thus far proved quite robust, it is not immediately obvious that quantum gravity needs to follow suit (see e.g., \cite{Albers et al 2008}\cite{Callender Huggett 2001}-\cite{Carlip 2008}). In such cases the ED approach might provide a viable alternative framework.
\section{Conclusion}
\label{55S Conclusion}
The covariant ED framework we've developed here can be distilled down to three main ingredients:
\begin{itemize}
\item entropic updating in local time;
\item adoption of a canonical formalism;
\item a canonical representation of the DHKT ``algebra".
\end{itemize}
While these ingredients were primarily introduced in \cite{Ipek et al 2017}\cite{Ipek et al 2019a}, an entirely new contribution to this program was presented in section \ref{55S General solution}, where we demonstrated a much wider range of covariant dynamical models.

From a broad perspective, the covariant quantum ED that we present can be viewed as an alternative to the standard Dirac method of quantization. The key feature is that we were able to obtain a theory that is quantum and statistical by implementing methods that were developed for classical theories. Thus our approach does not require any of the ad hoc rules of standard quantization and is free of the operator ordering ambiguities that plague the conventional quantization methods. Moreover, the theory sidesteps many of the tricky issues pertaining to Fock space representation that complicates other approaches.

The framework is also rather flexible; forthcoming work suggests that extensions to other ontic fields, such as charged scalar fields, gauge fields, and so on, are possible.\footnote{The extension to Fermion fields, on the other hand, are not quite on the horizon. Stay tuned.} Additionally, our methods have a certain ``classical" flavor, including notions of phase space, Hamiltonians, and so on. This therefore lends itself rather naturally to the study of interacting classical-quantum systems. In particular, one could consider the coupling of classical Einsteinian gravity to a quantum scalar field. We explore this in the next chapter.

\chapter{The entropic dynamics of quantum scalar fields coupled to gravity}\label{CH SCG}
\section{Introduction}
Without any empirical matter of fact, and none likely on the near horizon, quantum gravity (QG) research has largely split off into distinct channels, each reflecting a different set of attitudes, and yes, \emph{philosophies} directed towards the problem at hand. (See e.g., \cite{Carney et al 2018} for a recent overview of some feasible experimental proposals.) But this state of affairs should not be entirely surprising. Solving the problem of QG should, after all, entail first addressing the issue of what QG even \emph{is} --- indeed, what is meant by \emph{quantum}? What is meant by \emph{gravity}? Which, if any, elements of Einstein's gravity, or the standard quantum formalism, should be abandoned in the transition to QG?

Pursuant to this, a common view is that the gravitational field itself should, in some way, be quantized. There are several routes to accomplishing this.\footnote{Indeed, highlighting the interpretational difficulties of QG, is the way that notions of \emph{quantum} can vary from model to model. A popular class of theories postulates that the combination of quantum theory with gravity should give way to fundamentally discrete models of space and time, such as in the causal set \cite{Surya 2019} and causal dynamical triangulation \cite{Loll 2019} programs.} Most typical, however, is through some manner of quantization algorithm: extending the existing quantum formalism to the gravitational domain by application of the standard \emph{ad hoc} quantization rules to the \emph{appropriate} gravitational degrees of freedom. Such is the tack taken by the principal QG candidates --- string theory (ST) and loop quantum gravity (LQG).\footnote{To be sure, although there are vast differences between canonical approaches to QG, including LQG, and ST (see e.g., \cite{Butterfield Isham 2001}), the adoption of a quantization algorithm is possibly one of the few points of intersection. For some historical overviews in the development of ST and LQG, see the reviews by S. Mukhi \cite{Mukhi 2011} and C. Rovelli \cite{Rovelli 2011}, respectively.}

But such approaches should be met with some suspicion in the context of QG. One issue is the quantization procedure itself. Take models with constraints, such as general relativity (GR), for instance. Should we solve the constraints then quantize, or quantize then solve the constraints? The two methods are not, in general, the same (see e.g., \cite{Henneaux Teitelboim 1994}). Another issue is choosing which degrees of freedom to quantize in the first place. Historically this process has been guided by tight coordination between theory and experiment, allowing trial and error to supplement an incomplete understanding of the quantization process itself. But absent such help from experiment, a more fundamental understanding of quantization may be necessary to make progress.

The lack of clarity around quantization is made particularly acute in the case of the gravitational field, which plays a dual role in GR as an object with genuine dynamical modes, but one that also serves to establish spatial and temporal relationships. This calls into question the precise nature of the gravitational field. Is it just another field to be quantized, along the lines of the gauge fields of Yang-Mills theories, or is it something else entirely? Canonical approaches to QG, such as LQG, have wagered, with varying degrees of success, that the former is true. But there are various clues (see e.g., \cite{Jacobson 1995}-\cite{Caticha 2019c}) suggesting that gravity is an emergent, statistical phenomenon, not unlike temperature and pressure in statistical physics. Would any physicist quantize a temperature field? Perhaps not, but this is exactly what we might be doing when we quantize the gravitational field.

All of this suggests that a different attitude with regards to QG may be in order. In the past couple of decades there has been increased interest (see e.g., \cite{Hall Reginatto 2005}-\cite{Doring Isham 2008b}) in developing approaches to QG that seek to treat gravitational considerations hand-in-hand with a more robust understanding of quantum theory (QT). The goal in these cases, however, is not \emph{just} the development of specific models (although such models must necessarily follow), but the construction of entire frameworks that can readily integrate salient features of gravitational and quantum physics alike. One such approach is afforded by entropic dynamics (ED), which is a general framework for constructing indeterministic dynamical models based on the principles of Bayesian probability and entropic inference.

The previous chapter detailed the ED of of quantum scalar fields on a \emph{fixed} curved background based in part on the efforts of Ipek, Abedi, and Caticha (IAC) in \cite{Ipek et al 2017}\cite{Ipek et al 2019a}, which was itself inspired by the seminal works of Dirac \cite{Dirac Lectures}\cite{Dirac 1951} as well as Hojman \emph{et al.} \cite{Hojman Kuchar Teitelboim 1976}, Kucha\v{r} \cite{Kuchar 1972}, and Teitelboim \cite{Teitelboim thesis} \cite{Teitelboim 1972}(DHKT) in their development of covariant canonical methods in classical field theory. To recap, the key insight of DHKT was that a dynamics that unfolds in space-time must mirror a certain pattern, and that this pattern is itself reflective of the structure of surface deformations. As stressed by DHKT, the geometrical character of these conditions implies that they are of a rather \emph{universal} nature and apply to any such dynamics. Thus, as shown earlier, ED can be made manifestly covariant by imposing just such a structure on the generators of ED. From this perspective, the covariant ED developed by IAC generalizes to the quantum domain the results of DHKT for classical fields evolving on a fixed background.

Here we model the indeterministic dynamics of a scalar field $\chi(x)$. The material laid out here is based on work done in \cite{Ipek Caticha 2019}\cite{Ipek Caticha 2020}, in conjunction with A. Caticha. Our current presentation extends the efforts of IAC and the previous chapter in one crucial way: we allow the background geometry to become \emph{dynamical}. Following DHKT in their development of a manifestly covariant classical geometrodynamics, the transition to a dynamical background is not accomplished by any modification to the ``embeddability" conditions of DHKT, but by an appropriate choice of variables for describing the evolving geometry. The result is a hybrid ED model that approaches quantum field theory in a fixed background in one limit and approaches classical general relativity in another limit, but is not fully described by either. In particular, the model shares some formal similarities with the so-called Semi-classical Einstein equations (SCEE) (see e.g., \cite{Wald 1994}), but here we model the fluctuations of the quantum fields and \emph{derive} their coupling to gravity from first principles without the \emph{ad hoc} arguments typically used to justify the SCEE.

The outline of the chapter is the following.\footnote{The current chapter has a fair deal of overlap with the previous one dealing with a fixed background. Indeed, as mentioned earlier, one virtue of our approach here is that the cases of fixed and dynamical backgrounds can be handled within the same general framework. The chief distinction is in the choice of variables to describe the geometry. The interested reader may skip ahead to section \ref{ND_ED}.} In section \ref{ED_Steps}, we review the ED of short steps. Following this, in section \ref{ST_Notation} we introduce some key notation, which is useful in our development of entropic time in section \ref{ED_time}. Key concepts of surface deformations and ``embeddability" are introduced in section \ref{Embeddability}. In section \ref{ND_ED} we introduce the canonical formalism in ED, which is a necessary step before we review the condition of path independence developed by DHKT in section \ref{Path Independence}. Section \ref{Canonical Rep} outlines the construction of the local generators. In section \ref{ED_DyEqns} we describe the resulting dynamical equations, while in section \ref{ED_QT} we apply these results to obtain an ostensibly quantum theory. We discuss our results in section \ref{conclusion}.

\section{Statistical Model for Short Steps}\label{ED_Steps}
We present a short review of the ED of infinitesimal steps in curved space-time, adopting the notations and conventions of the previous chapter. Here the object of analysis is single a scalar field $\chi \left( x\right)\equiv \chi_{x} $ that populates space and whose values are posited to be definite, but unknown. An entire field configuration, which we denote $\chi $, lives on a $3$-dimensional space $\sigma $, the points of which are labeled by coordinates $x^{i}$ ($i=1,2,3$). The space $\sigma$ is itself curved and comes equipped with a metric $g_{ij}$ that is currently fixed, but that will later become dynamical. A single field configuration $\chi$ is a point in an $\infty $-dimensional configuration space $\mathcal{C}$. Our uncertainty in the values of this field is then quantified by a probability distribution $\rho \lbrack \chi ]$ over $\mathcal{C}$, so that the probability that the field attains a value $\hat{\chi}$ in an infinitesimal region of $\mathcal{C}$ is $\text{Prob}[\chi < \hat{\chi} < \chi+\delta\chi]=\rho[\chi]\, D\chi$, where $D\chi$ is an integration measure over $\mathcal{C}$.

\paragraph*{On microstates--- }
In ED the field distributions $\chi_{x}$ play a singularly special role: they define the ontic state of the system. This ontological commitment is in direct contrast with the usual Copenhagen interpretation in which such microscopic values become actualized only through the process of measurement. The Bohmian interpretation shares with ED that fact that in both the fields are ontic but the resemblance ends there; Bohmian wave functions are ontic while ED wave functions are fully epistemic \cite{Bartolomeo Caticha 2016}\cite{Caticha 2019b}. The metric $g_{ij}$, on the other hand, in our approach is a \textit{tool} whose purpose is to measure distances, areas, etc., and to characterize the spatial relations between the physical degrees of freedom, the $\chi_{x}$. While the geometry may later become dynamical, we do \textit{not} interpret this to mean that $g_{ij}$ is itself an ontic variable; it is not.\footnote{A model in which the metric tensor is itself of statistical origin is proposed in \cite{Caticha 2016}.} Put another way, the geometric variables enter much like parameters in a typical statistical model. The value of those parameters are important in guiding the distribution of outcomes. However, unlike the ontic variables, their values are not detected directly, but inferred from an ensemble of measurements.

\paragraph*{Maximum Entropy---}
Our goal is to predict the indeterministic dynamics of the scalar field $\chi $ whose statistical features are captured by a probabilistic model. To this end, we make one major assumption: in ED, the fields follow continuous trajectories such that finite changes can be analyzed as an accumulation of many infinitesimally small ones. Such an assumption allows us to focus our interest on obtaining the probability $P\left[ \chi ^{\prime }|\chi \right] $ of a transition from an initial configuration $\chi $ to a neighboring $\chi ^{\prime }=\chi +\Delta \chi $. This is accomplished via the Maximum Entropy (ME) method by maximizing the entropy functional, 
\begin{eqnarray}
S\left[ P,Q\right] =-\int D\chi ^{\prime }P\left[ \chi ^{\prime }|\chi %
\right] \log \frac{P\left[ \chi ^{\prime }|\chi \right] }{Q\left[ \chi
^{\prime }|\chi \right] },  \label{6 entropy a}
\end{eqnarray}%
relative to a prior $Q\left[ \chi ^{\prime }|\chi \right] $ and subject to appropriate constraints.

\paragraph*{The prior ---}

We adopt a prior $Q\left[ \chi ^{\prime }|\chi \right] $ that incorporates
the information that the fields change by infinitesimally small amounts, but
is otherwise maximally uninformative. In particular, before the constraints are taken into
account, knowledge of the dynamics at one point $x_{1}$ does not convey information about the dynamics at another point $x_{2}$, in other words, the degrees of freedom are \emph{a priori} uncorrelated. 

Such a prior can itself be derived from the principle of maximum entropy. Indeed, maximize
\begin{eqnarray}
S[Q,\mu ]=-\int D\chi ^{\prime }\,Q\left[ \chi ^{\prime }|\chi \right] \log 
\frac{Q\left[ \chi ^{\prime }|\chi \right] }{\mu (\chi ^{\prime })}~,
\label{entropy b}
\end{eqnarray}%
relative to the measure $\mu (\chi ^{\prime })$, which we assume to be
uniform, and subject to appropriate constraints.\footnote{Since we deal with infinitesimally short steps, the prior $Q$ turns out to be quite independent of the background measure $\mu$.} The requirement that the
field undergoes changes that are small and uncorrelated is implemented by
imposing an infinite number of independent constraints, one per spatial point $x$,\\
\begin{eqnarray}
\langle \Delta \chi _{x}^{2}\rangle =\int D\chi ^{\prime }\,Q\left[ \chi
^{\prime }|\chi \right] (\Delta \chi _{x})^{2}=\kappa _{x}\,,
\label{Constraint 1}
\end{eqnarray}%
where $\Delta \chi _{x}=\chi _{x}^{\prime }-\chi _{x}$ and the $\kappa _{x}$
are small quantities. The result of maximizing (\ref{entropy b}) subject to (%
\ref{Constraint 1}) and normalization is a product of Gaussians
\begin{eqnarray}
Q\left[ \chi ^{\prime }|\chi \right] \propto \,\exp -\frac{1}{2}\int
dx\,g_{x}^{1/2}\alpha _{x}\left( \Delta \chi _{x}\right) ^{2}~,  \label{prior}
\end{eqnarray}%
where $\alpha _{x}$ are the Lagrange multipliers associated to each
constraint (\ref{Constraint 1}); the scalar density $g_{x}^{1/2}=\,\left(
\det \,g_{ij}\right) ^{1/2}$ is introduced so that $\alpha _{x}$ is a scalar
field. For notational simplicity we write $dx^{\prime }$ instead of $%
d^{3}x^{\prime }$. To enforce the continuity of the motion we will eventually
take the limit $\kappa _{x}\rightarrow 0$ which amounts to taking $\alpha
_{x}\rightarrow \infty $.

\paragraph*{The global constraint--- }

The motion induced by the prior (\ref{prior}) leads to a rather simple diffusion process in the probabilities, in which the field variables evolve independently of each other. To model a dynamics that exhibits correlations and is capable of demonstrating the full suite of quantum effects, such as the superposition of states, interference, and entanglement, however, we require additional structure. This is accomplished by imposing a \emph{single} additional constraint that is \emph{non-local} in space but local in configuration space, which involves the introduction of a \emph{drift} potential $\phi \lbrack \chi ]$ which is a scalar-valued functional defined over the configuration space $\mathcal{C}$. More explicitly, we impose
\begin{eqnarray}
\langle \Delta \phi \rangle = \int D\chi ^{\prime }\,P\left[ \chi ^{\prime }|\chi \right] \int
dx\,\,\Delta \chi _{x}\frac{\delta \phi \left[ \chi \right] }{\delta \chi
_{x}}=\kappa ^{\prime },  \label{Constraint 2}
\end{eqnarray}%
where we require $\kappa^{\prime}\to 0$. (Note that since $\chi _{x}$ and $\Delta \chi _{x}$ are scalars, in order that (\ref{Constraint 2}) be invariant under coordinate transformations of the surface, the derivative $\delta /\delta \chi _{x}$ must transform as a scalar density.)

Before moving on we remark that the drift potential plays two central roles. On one hand, the dynamics of probabilities consists in their entropic updating which must necessarily involve constraints. On the other hand, as mentioned earlier, a key criterion for the choice of constraints will be the preservation of a symplectic structure which amounts to adopting a Hamiltonian formalism. Therefore, if what we seek is a dynamics in which probabilities are to be treated as generalized coordinates, then one must also introduce the corresponding canonical momenta. It is remarkable that in the ED framework one can, with maximum economy, perform the two functions with a single variable — the drift potential defines the constraint and is the conjugate momentum.


\paragraph*{The transition probability ---}

Next we maximize (\ref{6 entropy a}) subject to (\ref{Constraint 2}) and
normalization. The multiplier $\alpha ^{\prime }$ associated to the global constraint (\ref{Constraint 2}) turns out to have no influence on the dynamics: it can be absorbed into the yet undetermined drift potential $\alpha ^{\prime }\phi \rightarrow \phi $, effectively setting $\alpha ^{\prime }=1$.

The result is a Gaussian transition probability distribution, 
\begin{eqnarray}
P\left[ \chi ^{\prime }|\chi \right] \propto \exp -\frac{1}{2}\int dx\,g_{x}^{1/2}\alpha _{x}\left( \Delta
\chi _{x}-\frac{1}{g_{x}^{1/2}\alpha _{x}}\frac{\delta \phi \left[ \chi %
\right] }{\delta \chi _{x}}\right) ^{2}.  \label{Trans Prob}
\end{eqnarray}%
In previous work \cite{Caticha 2013}\cite{Caticha Ipek 2014}, $\alpha _{x}$ was
chosen to be a spatial constant $\alpha $ to reflect the translational
symmetry of flat space. Such a requirement, however, turns out to be inappropriate in the context of curved space-time. Instead, we follow \cite{Ipek et al 2017}\cite{Ipek et al 2019a} in allowing $\alpha _{x}$ to remain a non-uniform spatial scalar. This will be a key element in developing our scheme for a local entropic time.

The form of (\ref{Trans Prob}) allows us to present a generic
change, 
\begin{eqnarray}
\Delta \chi _{x}=\left\langle \Delta \chi _{x}\right\rangle +\Delta w_{x}~,\notag
\end{eqnarray}
as resulting from an expected drift $\left\langle \Delta \chi
_{x}\right\rangle $ plus Gaussian fluctuations $\Delta w_{x}$. Computing the expected short step for $\chi_{x}$ gives
\begin{eqnarray}
\left\langle \Delta \chi _{x}\right\rangle =\frac{1}{g_{x}^{1/2}\,\alpha _{x}%
}\frac{\delta \phi \left[ \chi \right] }{\delta \chi _{x}}~,  \label{Exp Step 1}
\end{eqnarray}%
while the fluctuations $\Delta w_{x}$ satisfy,%
\begin{eqnarray}
\left\langle \Delta w_{x}\right\rangle =0\,,\quad \text{and}\hspace{0.4cm}%
\left\langle \Delta w_{x}\Delta w_{x^{\prime }}\right\rangle =\frac{1}{%
g_{x}^{1/2}\alpha _{x}}\delta _{xx^{\prime }}.  \label{Fluctuations}
\end{eqnarray}%
Thus we see that while the expected step size is of order $\Delta \bar{\chi}_{x}\sim 1/\alpha _{x}$, the fluctuations go as $\Delta w_{x}\sim 1/\alpha _{x}^{1/2}$. Thus, for short steps, i.e. $\alpha_{x}\rightarrow \infty $, the fluctuations overwhelm the drift, resulting in a trajectory that is continuous but not, in general, differentiable. Such a model describes a Brownian motion in the field variables $\chi_{x}$.\footnote{As discussed in \cite{Caticha 2019b}\cite{Bartolomeo Caticha 2016}, with appropriate choices of the multipliers $\alpha$ and $\alpha^{\prime}$ the fields or the particles, as the case might be, can be made to follow paths typical of a Brownian motion or alternatively paths that are smooth and resemble a Bohmian motion. It is remarkable that these different paths at the sub-quantum level all lead to the same Schr\"{o}dinger equation.}

\section{Some notation}\label{ST_Notation}
The ED developed here deals with the coupled evolution of a quantum scalar field together with a classical dynamical background. The class of theories that allow such evolving geometries are often called instances of \emph{geometrodynamics}. In a typical geometrodynamics (see e.g., \cite{Dirac 1958}\cite{ADM 1960}), the primary object of interest is the evolving three-metric $g_{ij}(x)$, whose dynamics must be suitably constrained so that the time evolution sweeps a four-dimensional space-time with metric $^{4}g_{\mu\nu}$ ($\mu ,\nu ,...$ $=0,1,2,3$). Thus, despite the intrinsic dynamics at play, it is nonetheless appropriate to make reference to the enveloping space-time, if only as formal scaffolding that can be later removed.\footnote{Remarkably, as discussed by Teitelboim \cite{Teitelboim thesis}, although certain aspects of the formalism, indeed, rely crucially on space-time itself, such notions are ultimately absent from the operative equations of geometrodynamics. In view of this, we can rightfully regard three-dimensional space as taking a primary role, with space-time being taken as a secondary construct.}

It is therefore possible to assign coordinates $X^{\mu}$ to the space-time manifold. Furthermore, we deal with a space-time with globally \emph{hyperbolic} topology, admitting a foliation by space-like surfaces $\left\{ \sigma \right\} $. The embedding of such surfaces in space-time is defined by four scalar embedding functions $X^{\mu }\left(x^{i}\right) = X^{\mu}_{x}$. An infinitesimal deformation of the surface $\sigma $ to a neighboring surface $\sigma^{\prime }$ is determined by the deformation vector,
\begin{eqnarray}
\delta \xi ^{\mu }=\delta \xi ^{\bot }n^{\mu }+\delta \xi ^{i}X_{i}^{\mu }~.
\label{6 deformation vector}
\end{eqnarray}%
Here we have introduced $n^{\mu }$, which is the unit normal to the surface that is determined by the conditions $n_{\mu }n^{\mu }=-1$ and $n_{\mu }X_{ix}^{\mu }=0$), and where we have introduced $X_{ix}^{\mu} =\partial_{ix}X^{\mu}_{x} $, which are the space-time components of three-vectors tangent to $\sigma$. The normal and tangential components of $\delta \xi ^{\mu }$, also known as the infinitesimal lapse and shift, are collectively denoted $\delta \xi ^{A}_{x}=(\delta \xi
^{\bot }_{x},\delta \xi ^{i}_{x})$ and are given by%
\begin{eqnarray}
\delta \xi _{x}^{\bot }=-n_{\mu x}\delta \xi _{x}^{\mu }\quad \text{and}%
\quad \delta \xi _{x}^{i}=X_{\mu x}^{i}\delta \xi _{x}^{\mu }~,\label{6 deformation components}
\end{eqnarray}
where $X_{\mu x}^{i}=\,^{4}g_{\mu \nu }g^{ij}X_{jx}^{\nu }$. Additionally, a particular deformation is defined by its components $\xi^{A}_{x}$. This allows us to speak unambiguously about applying the same deformations to different surfaces.\footnote{Note, however, that differing surfaces will, in general have distinct normal vectors $n^{\mu}_{x}$. Thus, even as the very same deformation $\xi^{A}_{x}$ is being applied, the actual deformation vector $\xi^{\mu}_{x}$ on each distinct surface will not generally be the same, as per eq.(\ref{6 deformation vector}).}

\section{Entropic time}\label{ED_time}
In ED, entropic time is introduced as a tool for keeping track of the accumulation of many short steps. (For additional details on entropic time, see e.g., \cite{Caticha 2012}.) Here we introduce a manifestly covariant notion of entropic time, along the lines of that in \cite{Ipek et al 2017}\cite{Ipek et al 2019a}.

\paragraph*{An instant---}
Central to the formulation of entropic time is the notion of an instant which includes two main components. One is kinematic the other informational. The former amounts to specifying a particular space-like surface. The latter consists of specifying the contents of the instant, namely, the information — the relevant probability distributions, drift potentials, geometries, etc.— that are necessary to generate the next instant.

\paragraph*{Ordered instants--- }
Establishing the notion of an instant proves crucial because it supplies the structure necessary to inquire about the field $\chi_{x}$ at a \emph{moment} of time. Equivalently, such a notion allows us to consider a probability distribution $\rho_{\sigma}[\chi]$ corresponding to the informational state of the field $\chi_{x}$ at an instant labeled by the surface $\sigma$.

Dynamics in ED, on the other hand, is constructed step-by-step, as a sequence of instants. Thus it is appropriate to turn our attention to the issue of updating from some distribution $\rho_{\sigma}[\chi]$ at some initial instant, to another distribution $\rho_{\sigma^{\prime}}[\chi]$ at a subsequent instant. Such dynamical information is encoded in the short-step transition probability from eq.(\ref{Trans Prob}), or better yet, the joint probability $ P\left[ \chi ^{\prime },\chi \right]= P\left[ \chi ^{\prime }|\chi \right]\rho_{\sigma}[\chi] $.

Application of the ``sum rule" of probability theory to the joint distribution yields
\begin{eqnarray}
\rho_{\sigma^{\prime}}[\chi^{\prime}] = \int D\chi \, P\left[ \chi ^{\prime }|\chi \right]\rho_{\sigma}[\chi]~.\label{6 CK equation}
\end{eqnarray}
The structure of eq.(\ref{6 CK equation}) is highly suggestive: if we interpret $\rho_{\sigma}[\chi]$ as being an \emph{initial} state, then can interpret $\rho_{\sigma^{\prime}}[\chi^{\prime}] $ as being \emph{posterior} to it in the sense that it has taken into account the new information captured by the transition probability $P\left[ \chi ^{\prime }|\chi \right]$. Here we take this hint seriously and adopt eq.(\ref{6 CK equation}) as the equation that we seek for updating probabilities.

\paragraph*{Duration--- }
A final aspect of time to be addressed is the duration between instants. In doing so, one must distinguish between two separate issues. On one hand there is a natural notion of time that can be inherited from space-time itself; this being the local proper time $\delta\xi_{x}^{\perp}$ experienced by an observer at the point $x$ (see e.g., \cite{Gourgoulhon 2007}). On the other hand, however, notions of time and duration cannot themselves be divorced from those of dynamics and change. Indeed, the two are closely bound together.

To this end, in ED we follow Wheeler's maxim \cite{MTW 1973}: ``\emph{time is defined so that motion looks simple}." Since for short steps ED is dominated by fluctuations, eq.(\ref{Fluctuations}), the specification of the time interval is achieved through an appropriate choice of the multipliers $\alpha _{x}$. More specifically, we proceed by setting 
\begin{eqnarray}
\alpha _{x}=\frac{1}{ \delta \xi _{x}^{\bot }}\quad \text{so that}\quad
\left\langle \Delta w_{x}\Delta w_{x^{\prime }}\right\rangle =\frac{
\,\delta \xi _{x}^{\bot }}{g_{x}^{1/2}}\delta _{xx^{\prime }}~.
\label{6 Duration}
\end{eqnarray}%
With this the transition probability eq.(\ref{Trans Prob}) resembles a Wiener process, albeit in a rather unfamiliar context involving the propagation of fields on curved space.

\paragraph*{The local-time diffusion equations--- }

The dynamics expressed in integral form by (\ref{6 CK equation}) and (%
\ref{6 Duration}) can be rewritten in differential form. The result is \cite{Ipek et al 2017}\cite{Ipek et al 2019a},
\begin{eqnarray}
\delta \rho _{\sigma }\left[ \chi \right] =\int dx\frac{\delta \rho _{\sigma
}\left[ \chi \right] }{\delta \xi _{x}^{\bot }}\delta \xi _{x}^{\bot }=-\int
dx\frac{1}{g_{x}^{1/2}}\frac{\delta }{\delta \chi _{x}}%
\left( \rho _{\sigma }\left[ \chi \right] \frac{\delta \Phi _{\sigma }\left[
\chi \right] }{\delta \chi _{x}}\right) \delta \xi _{x}^{\bot }~, \label{FP b}
\end{eqnarray}%
where we have introduced
\begin{eqnarray}
\Phi_{\sigma }\left[ \chi \right] = \,\phi _{\sigma }\left[ \chi \right]
- \log \rho _{\sigma }^{1/2}\left[ \chi \right]~,\notag
\end{eqnarray}
which we refer to as the \emph{phase} functional.\footnote{Eventually, the phase functional $\Phi_{\sigma}[\chi] $ will be identified with the momentum conjugate to $\rho_{\sigma}$ and with a Hamilton-Jacobi type functional, or better yet, the phase of the wave functional in the quantum theory from which it inherits its name.}

For arbitrary choices of the infinitesimal lapse $\delta\xi_{x}^{\perp}$ we obtain an infinite set of local equations, one for each spatial point
\begin{eqnarray}
\frac{\delta \rho _{\sigma }}{\delta \xi _{x}^{\bot }}=-\frac{1}{g_{x}^{1/2}}\frac{%
\delta }{\delta \chi _{x}}\left( \rho _{\sigma }\,\frac{\delta \Phi _{\sigma
}}{\delta \chi _{x}}\right)~. \label{FP equation}
\end{eqnarray}%

To interpret these local equations, consider again the variation given in eq.(\ref{FP b}). In the special case where both surfaces $\sigma $ and $\sigma ^{\prime }$ happen to be flat then $g_{x}^{1/2}=1$ and $\delta \xi _{x}^{\bot }=dt$ are both spatial
\emph{constants} and eq.(\ref{FP b}) becomes equivalent to 
\begin{eqnarray}
\frac{\partial \rho _{t}\left[ \chi \right] }{\partial t}=-\int dx\frac{%
\delta }{\delta \chi _{x}}\left( \rho _{t}\left[ \chi \right] \frac{\delta
\Phi _{t}\left[ \chi \right] }{\delta \chi _{x}}\right) ~.  \label{FP c}
\end{eqnarray}%
We recognize this \cite{Caticha 2013}\cite{Caticha Ipek 2014} as a \emph{diffusion} or Fokker-Planck equation written as a
continuity equation for the flow of probability in configuration space $%
\mathcal{C}$. This suggests identifying the
\begin{eqnarray}
V_{x} = \frac{1}{g_{x}^{1/2}}\frac{\delta\Phi_{\sigma}[\chi]}{\delta\chi_{x}}~\notag
\end{eqnarray}
that appears in eq.(\ref{FP equation}) as the current velocity that regulates the flow of probability, which is valid for curved and flat spaces alike. Accordingly we will refer to (\ref{FP equation}) as the
\textquotedblleft local-time Fokker-Planck\textquotedblright\ equations
(LTFP). 


\section{The structure of surface deformations}\label{Embeddability}
Starting with Dirac \cite{Dirac 1951}\cite{Dirac 1958} and developed more fully by Hojman, Kucha\v{r}, and Teitelboim, a chief contribution of the DHKT program was the recognition that covariant dynamical theories had a rich structure that could be traced to the kinematics of surface deformations. Such structure can, however, itself be studied independently of any particular dynamics being considered. We give a brief review of the subject following the presentations of Kucha\v{r} \cite{Kuchar 1972} and Teitelboim \cite{Teitelboim 1972}.



For simplicity, we consider a generic functional $T\left[ X(x)\right] $ that assigns a real number to every surface defined by the four embedding variables $X^{\mu }(x)$. The variation in the functional $\delta T $ resulting from an arbitrary deformation $\delta \xi _{x}^{A}$ has the form 
\begin{eqnarray}
\delta T=\int dx\,\delta \xi _{x}^{\mu }\frac{\delta T}{\delta \xi _{x}^{\mu
}}=\int dx\,\left( \delta \xi _{x}^{\bot }G_{\bot x}+\delta \xi
_{x}^{i}G_{ix}\right) T~\,,  \label{delta T}
\end{eqnarray}%
where 
\begin{eqnarray}
G_{\bot x}=\frac{\delta }{\delta \xi _{x}^{\bot }}=n_{x}^{\mu }\frac{\delta 
}{\delta X_{x}^{\mu }}\quad \text{and}\quad G_{ix}=\frac{\delta }{\delta \xi
_{x}^{i}}=X_{ix}^{\mu }\frac{\delta }{\delta X_{x}^{\mu }}
\end{eqnarray}%
are the generators of normal and tangential deformations respectively. The generators of deformations $\delta /\delta \xi _{x}^{A}$ form a non-holonomic basis. Thus, unlike the vectors $\delta /\delta X_{x}^{\mu }$, which form a coordinate basis and therefore commute, the generators of deformations have a non-vanishing commutator ``algebra" given by
\begin{eqnarray}
\frac{\delta }{\delta \xi _{x}^{A}}\frac{\delta }{\delta \xi _{x^{\prime}}^{B}}-\frac{\delta }{\delta \xi _{x^{\prime }}^{B}}\frac{\delta }{\delta\xi _{x}^{A}}=\int dx^{\prime \prime }\,\kappa ^{C}{}_{BA}(x^{\prime \prime};x^{\prime },x)\frac{\delta }{\delta \xi _{x^{\prime \prime }}^{C}}
\label{commutator}
\end{eqnarray}%
where $\kappa ^{C}{}_{BA}$ are the \textquotedblleft structure
constants\textquotedblright\ of the \textquotedblleft
group\textquotedblright\ of deformations.

The previous quotes in \textquotedblleft group\textquotedblright\ and \textquotedblleft algebra\textquotedblright\ are a reminder that strictly, the set of deformations do not form a group. The composition of two successive deformations is itself a deformation, of course, but it also depends on the surface to which the first deformation is applied. As we will see below, the ``structure constants" $\kappa ^{C}{}_{BA}$ are not constant, they depend on the metric $g_{ij}$ of the initial surface.\footnote{In a dynamical approach to gravity the metric is itself a functional of the canonical variables. Thus its appearance in the ``algebra" of deformations is in part responsible for the rich structure of geometrodynamics.}

The calculation of $\kappa ^{C}{}_{BA}$ is given in \cite{Kuchar 1972}\cite{Teitelboim 1972}. The key idea is that of \emph{embeddability}, which proceeds as follows. Consider performing two successive infinitesimal deformations $\delta\xi^{A}$ followed by $\delta\eta^{A}$ on an initial surface $\sigma$: $\sigma \overset{\delta \xi }{\rightarrow }\sigma _{1}\overset{\delta \eta }{\rightarrow }\sigma ^{\prime }$. Performing now the very same deformations in the opposite order $\sigma \overset{\delta \eta }{\rightarrow }\sigma _{2}\overset{\delta \xi }{\rightarrow }\sigma ^{\prime \prime }$ yields a final surface $\sigma^{\prime\prime}$ that, in general, differs from $\sigma^{\prime}$. The key point is that since both $\sigma^{\prime}$ and $\sigma^{\prime\prime}$ are embedded in the very same space-time, then there exist be a third deformation $\delta\zeta^{A}$ that relates the two: $\sigma ^{\prime }\overset{\delta \zeta }{\rightarrow }\sigma ^{\prime \prime }$.

As shown by Teitelboim \cite{Teitelboim 1972}, however, the compensating deformation is not at all arbitrary, but can be determined entirely by geometrical arguments:
\begin{eqnarray}
\delta\zeta^{C}_{x^{\prime\prime}} = \int dx\int dx^{\prime}\kappa_{AB}^{C}(x^{\prime\prime};x,x^{\prime})\delta \xi^{A}_{x}\delta\eta^{B}_{x^{\prime}}~,\label{Compensating deformation}
\end{eqnarray}
where the only non-vanishing $\kappa$'s have the form
\begin{subequations}
\begin{eqnarray}
\kappa_{i\perp}^{\perp}(x^{\prime\prime};x,x^{\prime}) &=& -\kappa_{\perp i}^{\perp}(x^{\prime\prime};x^{\prime},x) = -\delta(x^{\prime\prime},x)\partial_{ix^{\prime\prime}}\delta(x^{\prime\prime},x^{\prime})\label{Kappa 1}\\
\kappa_{ij}^{k}(x^{\prime\prime};x,x^{\prime}) &=& -\kappa_{j i}^{k}(x^{\prime\prime};x^{\prime},x)\notag\\
 &=& \delta(x^{\prime\prime},x)\partial_{ix^{\prime\prime}}\delta(x^{\prime\prime},x^{\prime})\delta^{k}_{j}-\delta(x^{\prime\prime},x)\partial_{jx^{\prime\prime}}\delta(x^{\prime\prime},x^{\prime})\delta^{k}_{i}\label{Kappa 2}\\
 \kappa_{\perp\perp}^{i}(x^{\prime\prime};x,x^{\prime}) &=& -\kappa_{\perp \perp}^{i}(x^{\prime\prime};x^{\prime},x) \notag\\
 &=& - g^{ij}(x^{\prime\prime})\left (\delta(x^{\prime\prime},x^{\prime})\partial_{jx^{\prime\prime}}\delta(x^{\prime\prime},x)-\delta(x^{\prime\prime},x)\partial_{jx^{\prime\prime}}\delta(x^{\prime\prime},x^{\prime})\right )\label{Kappa 3}
\end{eqnarray}
\end{subequations}

Identification of the $\kappa$'s implies that the commutator in eq.(\ref{commutator}) satisfies the \textquotedblleft algebra\textquotedblright  \cite{Kuchar 1972}\cite{Teitelboim 1972},
\begin{eqnarray}
\left[G_{Ax},G_{Bx^{\prime}}\right] = \int dx^{\prime\prime}\kappa_{AB}^{C}(x^{\prime\prime};x,x^{\prime})G_{Cx^{\prime\prime}}~,\label{LB Generator Algebra }
\end{eqnarray}
which can now be written more explicitly as
\begin{subequations}
\begin{eqnarray}
\lbrack G_{\bot x},G_{\bot x^{\prime }}] &=&-(g_{x}^{ij}G_{jx}+g_{x^{\prime
}}^{ij}G_{jx^{\prime }})\partial _{ix}\delta (x,x^{\prime })~,  \label{LB1}
\\
\lbrack G_{ix},G_{\bot x^{\prime }}] &=&-G_{\bot x}\partial _{ix}\delta
(x,x^{\prime })~,  \label{LB2} \\
\lbrack G_{ix},G_{jx^{\prime }}] &=&-G_{ix^{\prime }}\,\partial _{jx}\delta
(x,x^{\prime })-G_{jx}\,\partial _{ix}\delta (x,x^{\prime })~,  \label{LB3}
\end{eqnarray}%
\end{subequations}
with all other brackets vanishing.

\section{Entropic geometrodynamics}\label{ND_ED}
In an \textit{entropic} dynamics, evolution is driven by information codified into constraints. An entropic geometrodynamics, it follows, consists of dynamics driven by a specific choice of constraints, which we discuss here. In \cite{Ipek et al 2017}\cite{Ipek et al 2019a}, quantum field theory in a curved space-time (QFTCS) was derived under the assumption that the geometry remains fixed. But such assumptions, we know, should break down when one considers states describing a non-negligible concentration of energy and momentum. Thus we must revise our constraints appropriately. A natural way to proceed is thus to allow the geometry itself to take part in the dynamical process: the geometry affects $\rho_{\sigma}[\chi]$ and $\phi_{\sigma}[\chi]$, they then act back on the geometry, and so forth. Our goal here is to make this interplay concrete.

\paragraph*{The canonical updating scheme--- }
A natural question that arises from the above discussion is how to implement the update of the geometry and drift potential $\phi_{\sigma}[\chi]$. Such a task involves two steps. The first is the proper identification of variables for describing the evolving geometry, while the other is the specific manner in which this joint system of variables, including the drift potential $\phi_{\sigma}[\chi]$, is updated. Fortunately, the two challenges can be dealt with quite independently of each other.

In devising a covariant scheme for updating we draw primarily from the work of IAC \cite{Ipek et al 2017}\cite{Ipek et al 2019a}. To review briefly, a primary assumption in the IAC approach was the adoption of a canonical framework for governing the coupled dynamics of $\rho_{\sigma}$ and $\phi_{\sigma}$, expressed more conveniently through the transformed variable $\Phi_{\sigma}$. Although this is certainly a strong assumption, it is one that has some justification. On one front, it can be argued that canonical structures seem to have a rather natural place in ED; arising from conservation laws in \cite{Bartolomeo et al 2014}\cite{Caticha Ipek 2014} and alternatively from symmetry considerations \cite{Caticha 2019b}. However, from another perspective entirely, the use of a canonical formalism in ED can also be traced to more pragmatic concerns, as it allows one to borrow from a roster of covariant canonical techniques designed in the context of classical physics (see e.g., \cite{Dirac Lectures}), but deployed for the purposes of ED. These together suggest that we view the canonical setting, and the symplectic symmetries that undergird it, as a central criterion for updating in ED. 

\paragraph*{The canonical variables--- }
Crucial to our updating scheme is an appropriate choice of variables. We pursue a conservative approach in which $\rho_{\sigma}$ and $\Phi_{\sigma}$ are packaged together as canonically conjugate variables following the prescription detailed in IAC \cite{Ipek et al 2019a}. The nontrivial task of choosing the geometric variables has long been the subject of a lively debate.\footnote{Just to name a few approaches, there are, of course, the original attempts at geometrodynamics from Dirac \cite{Dirac 1958} as well as Arnowitt, Deser, and Misner (ADM) \cite{ADM 1960} that start from the Einstein-Hilbert action and take the metric $g_{ij}$ as the fundamental building block. Somewhat more recently, due in part to the modern success of gauge theories, there has been some interest in taking, not the metric $g_{ij}$, but the Levi-Civita connection $\Gamma^{i}_{jk}$, as the fundamental gravitational object (see e.g., \cite{Ferraris Kijowski 1981}). In a similar spirit is the well-known discovery of Ashtekar \cite{Ashtekar 1986}, which uses triads and spin connections to rewrite GR.} Among these various approaches, however, the pioneering efforts of Hojman, Kucha\v{r}, and Teitelboim (HKT) \cite{Hojman Kuchar Teitelboim 1976} prove to be of special interest, as their work centers a purely canonical approach to the dynamics of geometry. Following HKT, it is possible to take the six components of the metric $g_{ij}$ to be the starting point for a geometrodynamics. The argument for this is grounded in simplicity. If one assumes, as they do, that evolution in local time should mirror the structure of surface deformations, then the metric appearing on the right hand side of (\ref{Kappa 3}) for the ``structure constant" $\kappa^{i}_{\perp\perp}$ must necessarily be a functional of the canonical variables. Clearly this is most easily satisfied if the metric is itself a canonical variable.

To complete the canonical framework, however, we must also introduce six variables $\pi^{ij}$ that are canonically conjugate to the $g_{ij}$. These variables, the conjugate momenta, are themselves \emph{defined} through the canonical Poisson bracket relations,\footnote{In particular, we do not assume the existence of a Legendre transformation from $\pi^{ij}$ to the ``velocity" of the metric $g_{ij}$.}
\begin{subequations}
\begin{eqnarray}
\big\{ g_{ijx},g_{klx^{\prime}}\big\} &=& \left\{\pi^{ij}_{x},\pi^{kl}_{x^{\prime}}\right \} = 0\\
\left\{g_{ijx},\pi^{kl}_{x^{\prime}}\right \} &=& \delta_{ij}^{kl}\delta(x,x^{\prime}) = \frac{1}{2}\left(\delta^{k}_{i}\delta^{l}_{j}+\delta^{k}_{j}\delta^{l}_{i}\right)\delta(x,x^{\prime})~.\label{6 canonical relations Grav}
\end{eqnarray}
\end{subequations}
That such relations are satisfied is made manifest by writing the Poisson brackets in local coordinates
\begin{eqnarray}
\left\{F,G\right\} &=& \int dx\left(\frac{\delta F}{\delta g_{ijx}}\frac{\delta G}{\delta \pi^{ij}_{x}}-\frac{\delta G}{\delta g_{ijx}}\frac{\delta F}{\delta \pi^{ij}_{x}}\right)\notag\\
&+&\int D\chi\left(\frac{\tilde{\delta} F}{\tilde{\delta}\rho[\chi] }\frac{\tilde{\delta} G}{\tilde{\delta} \Phi[\chi]}-\frac{\tilde{\delta} G}{\tilde{\delta}\rho[\chi] }\frac{\tilde{\delta} F}{\tilde{\delta} \Phi[\chi]}\right)~,\label{6 Poisson brackets}
\end{eqnarray}
where $F$ and $G$ are some arbitrary functionals of the phase space variables. (Note that $\pi^{ij}_{x}$ must be a tensor \emph{density} for the Poisson bracket to transform appropriately under a change of variables of the surface.)

\section{The canonical structure of space-time}\label{Path Independence}
A fully covariant dynamics requires that the updating in local time of all dynamical variables be consistent with the kinematics of surface deformations. Thus, the requirement that the deformed surfaces remain embedded in space-time, which amounts to imposing foliation invariance, translates into a consistency requirement of path independence: if the evolution from an initial instant into a final instant can occur along different paths, then all these paths must lead to the same final values for all dynamical quantities. The approach we adopt for quantum fields coupled to dynamical classical gravity builds on previous work by HKT \cite{Hojman Kuchar Teitelboim 1976} for classical geometrodynamics, and by IAC \cite{Ipek et al 2017}\cite{Ipek et al 2019a} for quantum field theory in a non-dynamical space-time.



Within this scheme the evolution of an arbitrary functional $F$ of the canonical variables is generated by application of these local Hamiltonians according to
\begin{eqnarray}
\delta F = \int dx \left \{F, H_{Ax} \right \}\delta\xi^{A}_{x} = \int dx \left (  \left \{F, H_{\perp x} \right \}\delta\xi^{\perp}_{x}+ \left \{F, H_{ix} \right \}\delta\xi^{i}_{x} \right )~,\label{Dyanmical law}
\end{eqnarray}
where parameters $\delta\xi^{A}_{x}$ with $A = (\perp , i = 1,2,3)$ describe an infinitesimal deformation, as per eqns.(\ref{6 deformation vector}) and (\ref{6 deformation components}), and $H_{Ax}$ are the corresponding generators. (Defined in this way, the $H_{Ax}$ turn out to be tensor densities.)

\paragraph*{Path independence--- }
The implementation of path independence \cite{Kuchar 1972}\cite{Teitelboim 1972} then rests on the idea that the Poisson brackets of the generators $H_{Ax}$ form an ``algebra" that closes in the same way, that is, with the same ``structure constants", as the ``algebra" of deformations in eqns.(\ref{LB1})-(\ref{LB3}),
\begin{subequations}
\begin{eqnarray}
\left\{ H_{\bot x},H_{\bot x^{\prime }}\right\} &=&(g_{x}^{ij}H_{jx}+g_{x^{\prime
}}^{ij}H_{jx^{\prime }})\partial _{ix}\delta (x,x^{\prime })~,  \label{6 PB 1}
\\
\left\{ H_{ix},H_{\bot x^{\prime }}\right \} &=&H_{\bot x}\partial _{ix}\delta
(x,x^{\prime })~,  \label{6 PB 2} \\
\left\{ H_{ix},H_{jx^{\prime }}\right \} &=&H_{ix^{\prime }}\,\partial _{jx}\delta
(x,x^{\prime })+H_{jx}\,\partial _{ix}\delta (x,x^{\prime })~.  \label{6 PB 3}
\end{eqnarray}%
\end{subequations}

We conclude this section with two remarks. First, we note that these equations have not been derived. Indeed, imposing (\ref{LB1})-(\ref{LB3}) as strong constraints constitutes the definition of what we mean by imposing consistency between the updating of dynamical variables and the kinematics of surface deformations. But this is not enough. As shown in \cite{Kuchar 1972}\cite{Teitelboim 1972}, path independence also demands that the initial values of the canonical variables must be restricted to obey the \emph{weak} constraints
\begin{eqnarray}
H_{\perp x} \approx 0\quad\text{and}\quad H_{ix} \approx 0 ~.\label{6 Hamiltonian constraints}
\end{eqnarray}
Furthermore, once satisfied on an initial surface $\sigma$ the dynamics will be such as to preserve these constraints for all subsequent surfaces of the foliation.


We also note that it is only by virtue of relating the conditions of ``integrability" to those of ``embeddability" that we can interpret the role of the local Hamiltonians $H_{\perp x}$ and $H_{ix}$ in relation to space and time. This is a crucial and highly non-trivial step. It is only once this is established that we can interpret $H_{\perp x}$ as a scalar density that is responsible for genuine dynamical evolution and $H_{ix}$ as a vector density that generates spatial diffeomorphisms; the former is called the super-Hamiltonian, while the latter is the so-called super-momentum.


\section{The canonical representation}\label{Canonical Rep}
We now turn our attention to the local Hamiltonian generators $H_{Ax}$, and more specifically, we look to provide explicit expressions for these generators in terms of the canonical variables. This problem was solved in the context of a purely classical geometrodynamics (with or without sources) by HKT in \cite{Hojman Kuchar Teitelboim 1976}. Here we aim to apply their techniques and methodology to a different problem: a geometrodynamics driven by ``quantum" sources. Fortunately, a considerable portion of the HKT formalism can be directly adopted for our purposes.
\subsection*{The super-momentum}
To determine the generators $H_{Ax}[\rho,\Phi;g_{ij},\pi^{ij}]$ it is easiest to begin with the so-called \textit{super-momentum} $H_{ix}[\rho,\Phi;g_{ij},\pi^{ij}]$. This is largely because the function of this generator is well understood: it pushes the canonical variables along the surface they reside on. Since there is no motion ``normal" to the surface, the action of this generator is purely kinematical.

Following HKT \cite{Hojman Kuchar Teitelboim 1976}, consider an infinitesimal tangential deformation such that a point originally labeled by $x^{i}$ is carried to the point previously labeled by $x^{i} + \delta\xi^{i}$. This will induce a corresponding change in any dynamical variables $F$ defined on that surface, $F\to F+\delta F$. This change $\delta F$ can then be computed in two distinct ways, which, of course, must agree. One is by calculating the Lie derivative along $\delta\xi$
\begin{eqnarray}
\delta F = \pounds_{\delta\xi} F~.\notag
\end{eqnarray}
and the other is using the super-momentum $H_{ix}$, so that
\begin{eqnarray}
\delta F = \int dx \left\{ F, H_{ix}\right \}\delta\xi^{i}_{x}~.\notag
\end{eqnarray}
\paragraph*{Gravitational super-momentum}
A straightforward example of this is shown for the metric $g_{ij}(x)$, which is a rank $(0,2)$ tensor. Its Lie derivative is given by \cite{Schutz}
\begin{eqnarray}
\pounds_{\delta\xi} g_{ij} = \partial_{k}g_{ij}\delta \xi^{k}_{x} + g_{ik}\partial_{j}\delta\xi^{k}_{x} + g_{kj}\partial_{i}\delta\xi^{k}_{x}~.\label{6 Lie derivative metric}
\end{eqnarray}
Alternatively, by using the Poisson brackets we obtain
\begin{eqnarray}
\delta g_{ijx} = \int dx^{\prime}\left\{g_{ijx}, H_{kx^{\prime}}\right \}\delta\xi_{x^{\prime}}^{k} = \int dx^{\prime}\frac{\delta H_{kx^{\prime}}}{\delta\pi^{ij}_{x}}\delta\xi_{x^{\prime}}^{k}~.\notag
\end{eqnarray}
Comparing the two and using the fact that $\delta\xi_{x}^{i}$ is arbitrary yields
\begin{eqnarray}
\frac{\delta H_{kx^{\prime}}}{\delta\pi^{ij}_{x}} = \partial_{kx}g_{ijx}\delta( x,x^{\prime})+ g_{ikx}\partial_{jx}\delta( x,x^{\prime}) + g_{kjx}\partial_{ix}\delta( x,x^{\prime}).\label{6 Tangential deformation metric}
\end{eqnarray}
The delta functions imply that $H_{ix}$ is local in the momentum $\pi^{ij}$.

To fix the dependence on $\pi^{ij}$, in fact, we can use a similar argument as above. Recalling that $\pi^{ij}$ is a rank $(2,0)$ tensor \emph{density} of weight one, we find that
\begin{eqnarray}
\pounds_{\delta\xi} \pi^{ij} = \partial_{kx}\left(\pi^{ij}\delta\xi^{k}_{x}\right) - \pi^{ik}\partial_{kx}\delta\xi^{j}_{x} - \pi^{kj}\partial_{kx}\delta\xi^{i}_{x}.\label{6 Tangential deformation momentum}
\end{eqnarray}
The same equation can be obtained through the use of a Hamiltonian,
\begin{eqnarray}
\delta\pi^{ij}_{x} = \int dx^{\prime}\left\{\pi^{ij}_{x},H_{kx^{\prime}}\right\}\delta\xi^{k}_{x^{\prime}} = -\int dx^{\prime}\frac{\delta H_{kx^{\prime}}}{\delta g_{ijx}}\delta\xi^{k}_{x^{\prime}} ~,\notag
\end{eqnarray}
so long as $H_{ix}$ satisfies
\begin{eqnarray}
 -\frac{\delta H_{kx^{\prime}}}{\delta g_{ijx}} = \partial_{kx}\left(\pi^{ij}_{x}\delta(x,x^{\prime})\right) - \pi^{il}_{x}\partial_{lx}\delta(x,x^{\prime}) \delta^{j}_{k}- \pi^{lj}_{x}\partial_{lx}\delta(x,x^{\prime})\delta^{i}_{k}~.\label{6 Tangential deformation momentum}
\end{eqnarray}
Integrating these equations for $H_{ix}$ yields
\begin{eqnarray}
H_{ix} =H_{ix}^{G} + \tilde{H}_{ix} ~,\label{6 Super-momentum grav + H}
\end{eqnarray}
where
\begin{eqnarray}
H_{ix}^{G} = -2 \partial_{jx}\left(\pi^{jk}\, g_{ik}\right) + \pi^{jk}\partial_{ix}g_{jk}~, \label{6 Super momentum grav}
\end{eqnarray}
is called the \emph{gravitational} super-momentum, and the functional $\tilde{H}_{ix} = \tilde{H}_{ix}[\rho,\Phi]$ is, at the moment, just an integration ``constant".

Some of these expressions can be simplified by introducing the \emph{covariant} derivative $\nabla_{i}$. For example, using $\nabla_{k} g_{ij} = 0$ we have
\begin{eqnarray}
\pounds_{\delta\xi} g_{ijx} = g_{jk}\nabla_{i}\delta\xi_{x}^{k} + g_{ik}\nabla_{j}\delta\xi_{x}^{k}~\label{6 Tangential deformation metric cov}
\end{eqnarray}
and
\begin{eqnarray}
\pounds_{\delta\xi} \pi^{ij}_{x} = \nabla_{k}\left(\pi^{ij}\delta\xi^{k}_{x}\right) - \pi^{ik}\nabla_{k}\delta\xi^{j}_{x} - \pi^{kj}\nabla_{k}\delta\xi^{i}_{x}~.\label{6 Tangential deformation momentum cov}
\end{eqnarray}
The gravitational super-momentum $H_{ix}^{G}$ also takes the particularly simple form
\begin{eqnarray}
H_{ix}^{G} = -2g_{ik} \nabla_{j}\pi^{jk}= -2\nabla_{j}\left(\pi^{jk}g_{ik}\right) = -2 \nabla_{j}\pi^{j}_{i}~.\label{6 Super momentum grav cov}
\end{eqnarray}

\paragraph*{The ``matter" super-momentum--- }
Next, we turn to the response of the variables $\rho_{\sigma}[\chi]$ and $\Phi_{\sigma}[\chi]$ under a relabeling of the surface coordinates $x^{i} \to x^{i}+\delta\xi^{i}_{x}$. As the coordinates are shifted, so too are the fields defined upon that surface so that
\begin{eqnarray}
\chi_{x} \to \chi_{x} + \pounds_{\delta\xi}\chi_{x}\quad\text{where}\quad \pounds_{\delta\xi}\chi_{x} = \partial_{ix}\chi_{x}\delta\xi^{i}_{x}\notag
\end{eqnarray}
is the Lie derivative for a scalar $\chi_{x}$. This induces a change in the probability, given by
\begin{eqnarray}
\delta\rho_{\sigma}[\chi] \equiv \rho_{\sigma}[\chi + \pounds_{\delta\xi}\chi_{x}] - \rho_{\sigma}[\chi] = \int dx \frac{\delta\rho_{\sigma}[\chi]}{\delta\chi_{x}}\partial_{ix}\chi_{x}\delta\xi^{i}_{x}~.
\end{eqnarray}

Alternatively, this same variation can be computed using the canonical framework
\begin{eqnarray}
\delta\rho_{\sigma}[\chi] = \int dx \left\{\rho_{\sigma}[\chi],H_{ix}\right\}\delta\xi^{i}_{x}~. \label{Tangential deformation rho a}
\end{eqnarray}
If we insert the super-momentum $H_{ix}$ given in eq.(\ref{6 Super-momentum grav + H}), we notice that only the $\tilde{H}_{ix}$ piece will contribute. And so we have
\begin{eqnarray}
\delta\rho_{\sigma}[\chi] = \int dx \left\{\rho_{\sigma}[\chi],\tilde{H}_{ix}\right\}\delta\xi^{i}_{x} = \int dx \frac{\tilde{\delta}\tilde{H}_{ix}}{\tilde{\delta}\Phi_{\sigma}[\chi]}\delta\xi_{x}^{i}~,\label{6 Tangential deformation rho b}
\end{eqnarray}
which for arbitrary $\delta\xi^{i}_{x}$ requires that
\begin{eqnarray}
\frac{\tilde{\delta}\tilde{H}_{ix}}{\tilde{\delta}\Phi_{\sigma}[\chi]} = \frac{\delta\rho_{\sigma}[\chi]}{\delta\chi_{x}}\partial_{ix}\chi_{x}~.\label{6 Tangential deformation rho c}
\end{eqnarray}
A similar argument for $\Phi_{\sigma}[\chi]$ shows that we must also have
\begin{eqnarray}
-\frac{\tilde{\delta}\tilde{H}_{ix}}{\tilde{\delta}\rho_{\sigma}[\chi]} = \frac{\delta\Phi_{\sigma}[\chi]}{\delta\chi_{x}}\partial_{ix}\chi_{x}~,\label{6 Tangential deformation Phi a}
\end{eqnarray}
so that
\begin{eqnarray}
\tilde{H}_{ix} = - \int D\chi \rho_{\sigma}\frac{\delta\Phi_{\sigma}}{\delta\chi_{x}}\partial_{ix}\chi_{x}~.\label{6 Super momentum ensemble}
\end{eqnarray}
Thus the total super-momentum, eq.(\ref{6 Super-momentum grav + H}),
\begin{eqnarray}
 H_{ix} = -2 \nabla_{j}\pi^{j}_{i} - \int D\chi \rho_{\sigma}\frac{\delta\Phi_{\sigma}}{\delta\chi_{x}}\partial_{ix}\chi_{x}\label{6 Super momentum}
\end{eqnarray}
contains two pieces, which we refer to as the gravitational and ``matter" contributions, respectively.\footnote{The division of generators into gravitational and ``matter" pieces established by DHKT is, strictly speaking, an abuse of language. The variables $\rho_{\sigma}$ and $\Phi_{\sigma}$ that constitute ``matter" are more properly understood as describing the statistical state of the material field $\chi_{x}$. Nonetheless, we stick with the convention as a useful shorthand.} Note that there is no gravitational dependence in the ``matter" side, or ``matter" dependence on the gravitational side, that is,
\begin{eqnarray}
H_{ix} = H_{ix}^{G}[g_{ij},\pi^{ij}] + \tilde{H}_{ix}[\rho,\Phi]~.\label{6 Super momentum split}
\end{eqnarray}

By equation (\ref{6 Hamiltonian constraints}) it is, of course, also understood that $H_{ix}$ is subject to the constraint
\begin{eqnarray}
H_{ix}\approx 0~.\label{6 Super momentum constraint}
\end{eqnarray}
Finally, although $H_{ix}$ was obtained, in essence, independently of the Poisson bracket (\ref{6 PB 3}), relating two tangential deformations, it nonetheless satisfies it automatically.\footnote{More technically, $H_{ix}$ is defined by our procedure up to an overall ``constant" vector density $f_{i}(x)$, which is independent of any canonical variables. This, it turns out, is required to vanish by eq.(\ref{6 PB 3}). See e.g., \cite{Hojman Kuchar Teitelboim 1976}.} It is therefore appropriate to view $H_{ix}$ as being completely determined; this is important as it allows us to treat equation (\ref{6 PB 1}) as a set of equations for $H_{\perp x}$ in terms of the known $H_{ix}$.

\subsection*{The super-Hamiltonian}
We now turn out attention to the generator $H_{\perp x}$ of local time evolution. Following Teitelboim \cite{Teitelboim thesis}, it is useful to decompose $H_{\perp x}$ into two distinct pieces
\begin{eqnarray}
H_{\perp x} = H_{\perp x}^{G}[g_{ij},\pi^{ij}] + \tilde{H}_{\perp x}[\rho,\Phi;g_{ij}, \pi^{ij}]~,\label{6 Super Hamiltonian split a}
\end{eqnarray}
consisting of a gravitational piece $H_{\perp x}^{G}$ depending only on the gravitational variables, and a ``matter" contribution that we suggestively denote by $\tilde{H}_{\perp x}$. It will sometimes be convenient to refer to these as the gravitational and ``matter" super-Hamiltonians, respectively.

As noted by Teitelboim, we make no assumptions in writing $H_{\perp x}$ in this way.\footnote{By construction, we can always identify a piece of the super-Hamiltonian that updates the geometry alone. The ``matter" super-Hamiltonian is then just defined as the difference between the total and gravitational pieces.} Given this splitting, however, we do make the following simplifying assumption: we require the ``matter" super-Hamiltonian $\tilde{H}_{\perp x}$ to be independent of the gravitational momentum $\pi^{ij}$, i.e.,
\begin{eqnarray}
\tilde{H}_{\perp x} = \tilde{H}_{\perp x}[\rho,\Phi;g_{ij}]~\notag
\end{eqnarray}
so that the super-Hamiltonian takes the form
\begin{eqnarray}
H_{\perp x} = H_{\perp x}^{G}[g_{ij},\pi^{ij}] + \tilde{H}_{\perp x}[\rho,\Phi;g_{ij}]~.\label{6 Super Hamiltonian split}
\end{eqnarray}

With this assumption in hand, it is possible to \emph{prove} \cite{Teitelboim thesis} that the metric appears in $\tilde{H}_{\perp x}$ only as a \emph{local} function of $g_{ij}$. That is, no derivatives of the metric are allowed, nor any other complicated functional dependencies; due to this fact, this was referred to as the  \emph{non-derivative} coupling assumption by Teitelboim.

\paragraph*{Modified Poisson brackets--- }
A particularly appealing aspect of this separation into gravitational and ``matter" pieces is that the Poisson bracket relations (\ref{6 PB 1})-(\ref{6 PB 3}) also split along similar lines. In fact, insert the decomposed generators $H_{Ax}$ from eqns.(\ref{6 Super momentum split}) and (\ref{6 Super Hamiltonian split}) into the Poisson bracket relations (\ref{6 PB 1})-(\ref{6 PB 3}). We find that the gravitational generators $H_{Ax}^{G} = (H_{\perp x}^{G},H_{ix}^{G})$ must satisfy a set of Poisson brackets
\begin{subequations}
\begin{eqnarray}
\left\{ H_{\bot x}^{G},H_{\bot x^{\prime }}^{G}\right\} &=&(g_{x}^{ij}H_{jx}^{G}+g_{x^{\prime
}}^{ij}H_{jx^{\prime }}^{G})\partial _{ix}\delta (x,x^{\prime })~,  \label{PB 1 G}
\\
\left\{ H_{ix}^{G},H_{\bot x^{\prime }}^{G}\right \} &=&H_{\bot x}^{G}\partial _{ix}\delta
(x,x^{\prime })~,  \label{PB 2 G} \\
\left\{ H_{ix}^{G},H_{jx^{\prime }}^{G}\right \} &=&H_{ix^{\prime }}^{G}\,\partial _{jx}\delta
(x,x^{\prime })+H_{jx}^{G}\,\partial _{ix}\delta (x,x^{\prime })~,  \label{PB 3 G}
\end{eqnarray}
\end{subequations}
which have closing relations \emph{identical} to those of (\ref{6 PB 1})-(\ref{6 PB 3}).

The ``matter" generators, which we collectively denote by $\tilde{H}_{Ax}$, satisfy a somewhat modified set of brackets
\begin{subequations}
\begin{eqnarray}
\left\{ \tilde{H}_{\bot x},\tilde{H}_{\bot x^{\prime }}\right\} &=&(g_{x}^{ij}\tilde{H}_{jx}+g_{x^{\prime
}}^{ij}\tilde{H}_{jx^{\prime }})\partial _{ix}\delta (x,x^{\prime })~,  \label{6 PB 1 matter}
\\
\left\{ H_{ix},\tilde{H}_{\bot x^{\prime }}\right \} &=&\tilde{H}_{\bot x}\partial _{ix}\delta
(x,x^{\prime })~,  \label{6 PB 2 matter} \\
\left\{ \tilde{H}_{ix},\tilde{H}_{jx^{\prime }}\right \} &=&\tilde{H}_{ix^{\prime }}\,\partial _{jx}\delta
(x,x^{\prime })+\tilde{H}_{jx}\,\partial _{ix}\delta (x,x^{\prime })~.  \label{6 PB 3 matter}
\end{eqnarray}
\end{subequations}
Here we call attention to the fact that eq.(\ref{6 PB 2 matter}) contains the total tangential generator $H_{ix}$, not just $\tilde{H}_{ix}$. This alteration occurs because the ``matter" super-Hamiltonian depends on the metric. That is, if we want to shift $\tilde{H}_{\perp x}$ along a given surface, we must shift the variables $(\rho,\Phi)$, as well as the metric $g_{ij}$; hence $H_{ix}$, not just $\tilde{H}_{ix}$, must appear.

Modulo the small distinction arising in eq.(\ref{6 PB 2 matter}), we see that the gravitational and ``matter" sectors decouple such that each piece forms an \emph{independent} representation of the ``algebra" of surface deformations. From a strategic point of view, the separation means that we can solve the Poisson bracket relations for geometry and ``matter" \emph{independently} of one another.

\paragraph*{The ``matter" super-Hamiltonian--- }
Our goal is to identify a family of ensemble super-Hamiltonians $\tilde{H}_{\perp x}[\rho,\Phi;g_{ij}]$ that are consistent with the Poisson brackets (\ref{6 PB 1 matter})-(\ref{6 PB 3 matter}). Let us briefly outline our approach. Thus far, we have completely determined the correct form of the ``matter" super-momentum $\tilde{H}_{ix}$ which is consistent with (\ref{6 PB 3 matter}). Moreover, the relation (\ref{6 PB 2 matter}) merely implies that $\tilde{H}_{\perp x}$ transforms as a scalar density under a spatial diffeomorphism. Thus we are left only to satisfy the first Poisson bracket (\ref{6 PB 1 matter}) for the unknown $\tilde{H}_{\perp x}$.

In addition to these considerations, however, the ED approach itself imposes additional constraints of a fundamental nature on the allowed $H_{\perp x}$, or more specifically $\tilde{H}_{\perp x}$. This is because, in ED, the introduction of a symplectic structure and its corresponding Hamiltonian formalism is not meant to replace the entropic updating methods that yield the LTFP equations, but to \emph{augment} them, appropriately. As a consequence, we demand, as a matter of principle, that the ``matter" super-Hamiltonian $\tilde{H}_{\perp x}$ be \emph{defined} so as to reproduce the LTFP equations of (\ref{FP equation}).

More explicitly, we require $\tilde{H}_{\perp x}$ to be such that its action on $\rho_{\sigma}$ generates the LTFP equations
\begin{eqnarray}
 \big\{ \rho_{\sigma}[\chi],H_{\perp x} \big\} =  \frac{\delta\rho_{\sigma}[\chi]}{\delta\xi^{\perp}_{x}}~, \label{FP equation H a}
\end{eqnarray}
which translates to
\begin{eqnarray}
\frac{\tilde{\delta}\tilde{H}_{\perp x}}{\tilde{\delta}\Phi_{\sigma}[\chi]}= -\frac{1}{g_{x}^{1/2}}\frac{\delta}{\delta\chi_{x}}\left (\rho_{\sigma}[\chi]\frac{\delta\Phi_{\sigma}[\chi]}{\delta\chi_{x}}    \right )~.\label{FP equation H b}
\end{eqnarray}
It is simple to check that the $\tilde{H}_{\perp x}$ that satisfies this condition is given by
\begin{eqnarray}
\tilde{H}_{\perp } = \int D\chi \rho_{\sigma} \frac{1}{2g_{x}^{1/2}}\left (  \frac{\delta\Phi_{\sigma}}{\delta\chi_{x}}\right )^{2} +F_{x}[\rho;g_{ij}]~.\label{e Hamiltonian a}
\end{eqnarray}
The first term in (\ref{e Hamiltonian a}) is \emph{fixed} by virtue of consistency with the LTFP equations, whereas $F_{x}[\rho;g_{ij}]$ is a yet undetermined ``constant" of integration, which may depend on $\rho$ as well as the metric. However, $F_{x}$ is not entirely arbitrary, its functional form is restricted by the Poisson bracket, eq.(\ref{6 PB 1 matter}).

Before proceeding, note that, up until this point, our discussion of path independence has been developed on a formal level, largely independent of ED itself. Such a formalism on its own, however, is necessarily devoid of many crucial physical ingredients. For instance, in ED we \emph{define} local time as a measure of the field fluctuations through (\ref{Fluctuations}), which leads to the LTFP equations. In principle, this has nothing to do with abstract parameters $\delta\xi^{A}_{x}$ introduced as part of local updating. It is only once we say that the entropic updating of ED must agree with the local time evolution generated by $\tilde{H}_{\perp x}$, made explicit in (\ref{FP b}), that the two notions coincide. In ED, the clock that measures the local proper time $\delta\xi^{\perp}_{x}$ is nothing but the field fluctuations themselves.



Continuing with our task at hand, we seek a family of models that are consistent with eq.(\ref{6 PB 1 matter}). This is accomplished for suitable choices of $F_{x}$. We pursue this in manner similar to \cite{Ipek et al 2019a}. Begin by rewriting $\tilde{H}_{\perp x}$ as
\begin{eqnarray}
\tilde{H}_{\perp x} = \tilde{H}_{\perp x}^{0}+ F_{x}~, \label{6 e-H}
\end{eqnarray}
where we have introduced
\begin{eqnarray}
\tilde{H}_{\perp x}^{0} = \int D\chi \,\rho_{\sigma}\left ( \frac{1}{%
2g_{x}^{1/2}}\left( \frac{\delta \Phi_{\sigma} }{\delta \chi _{x}}\right) ^{2}+\frac{g^{1/2}_{x}}{2}g^{ij}\partial_{ix}\chi_{x}\partial_{jx}\chi_{x}\right ) ~.  \label{6 e-H b}
\end{eqnarray}%
This amounts simply to a redefinition of the arbitrary $F_{x}$ in (\ref{e Hamiltonian a}). An advantage of this definition, however, is that the newly defined $\tilde{H}_{\perp x}^{0}$ automatically satisfies
\begin{eqnarray}
\left\{ \tilde{H}^{0}_{\bot x},\tilde{H}^{0}_{\bot x^{\prime }}\right\} =(g_{x}^{ij}\tilde{H}_{jx}+g_{x^{\prime
}}^{ij}\tilde{H}_{jx^{\prime }})\partial _{ix}\delta (x,x^{\prime })~.  \label{6 PB 1 matter Ho}
\end{eqnarray}
Finding a suitable $F_{x}$ is therefore accomplished by satisfying\footnote{Since $F_{x}[\rho; g_{ij}]$ is independent of $\Phi$, we have that $\left\{ F_{x},F_{x^{\prime}}\right\} = 0 $, identically, from which eq.(\ref{PB 1 matter necessary}) follows.}
\begin{eqnarray}
\left\{ \tilde{H}^{0}_{\bot x},F_{x^{\prime}}\right\} = \left\{ \tilde{H}^{0}_{\bot x^{\prime}},F_{x}\right\}~.\label{PB 1 matter necessary}
\end{eqnarray}
Clearly a necessary condition for an acceptable $F_{x}$ is that the Poisson bracket $\left\{ \tilde{H}^{0}_{\bot x},F_{x^{\prime}}\right\} $ must be symmetric upon exchange of $x$ and $x^{\prime}$. Since $\tilde{H}^{0}_{\perp x}$ must itself reproduce the LTFP equations, the condition (\ref{PB 1 matter necessary}) translates to \cite{Ipek et al 2017}\cite{Ipek et al 2019a}
\begin{eqnarray}
\frac{1}{g_{x}^{1/2}}\frac{\delta}{\delta\chi_{x}}\left (\rho_{\sigma} \frac{\delta}{\delta\chi_{x}} \frac{\tilde{\delta}F_{x^{\prime}}}{\tilde{\delta}\rho_{\sigma}}   \right ) = \frac{1}{g_{x^{\prime}}^{1/2}}\frac{\delta}{\delta\chi_{x^{\prime}}}\left (\rho_{\sigma} \frac{\delta}{\delta\chi_{x^{\prime}}} \frac{\tilde{\delta}F_{x}}{\tilde{\delta}\rho_{\sigma}}   \right )~,\label{PB 1 matter necessary b}
\end{eqnarray}
which is an equation linear in $F_{x}$.

A complete description of solutions to eq.(\ref{PB 1 matter necessary b}) lies outside the scope of the current work. However, a restricted family of solutions, which are nonetheless of physical interest, can be found for $F_{x}$'s of the form
\begin{eqnarray}
F_{x}[\rho] = \int D\chi f_{x}\left(\rho,\frac{\delta\rho}{\delta\chi_{x}}; g_{ijx}\right)~,\label{6 Local potentials}
\end{eqnarray}
where $f_{x}$ is a \emph{function}, not functional, of its arguments. For such a special type of $F_{x}$ one can check by substitution into (\ref{PB 1 matter necessary b}) that
\begin{eqnarray}
f_{x} \sim g_{x}^{1/2}\rho \chi_{x}^{n}\quad \text{(integer n)}\quad \text{and}\quad f_{x}\sim \frac{\rho}{g_{x}^{1/2}}\left (\frac{\delta\log\rho}{\delta\chi_{x}}\right )^{2}\notag
\end{eqnarray}
are acceptable solutions. Since eq.(\ref{PB 1 matter necessary b}) is \emph{linear} in $F_{x}$, solutions can be superposed so that a suitable family of $\tilde{H}_{\perp x}$'s is given by
\begin{equation}
\tilde{H}_{\perp x} = \int D\chi \, \rho_{\sigma} \left ( \frac{1}{%
2g_{x}^{1/2}}\left( \frac{\delta \Phi_{\sigma} }{\delta \chi _{x}}\right) ^{2}+\frac{g^{1/2}_{x}}{2}g^{ij}\partial_{ix}\chi_{x}\partial_{jx}\chi_{x} + g_{x}^{1/2}V_{x}(\chi_{x})+ \frac{\lambda}{g_{x}^{1/2}}\left (\frac{\delta\log\rho_{\sigma}}{\delta\chi_{x}}\right )^{2}\right )~,\label{e Hamiltonian final}
\end{equation}
where $V_{x} = \sum_{n}\lambda_{n}\chi_{x}^{n}$ is a function that is polynomial in $\chi_{x}$.

As discussed in \cite{Ipek et al 2019a}, the last term can be interpreted as the ``local quantum potential". To see this we recall that in flat space-time the quantum potential is given by \cite{Caticha Ipek 2014}
\begin{eqnarray}
Q =\int d^{3}x \int D\chi \rho \, \lambda \left (\frac{\delta\log\rho}{\delta\chi_{x}}\right )^{2}~.\label{Quantum Potential Flat}
\end{eqnarray}
The transition to a curved space-time is made by making the substitutions
\begin{eqnarray}
d^{3}x \to g_{x}^{1/2}d^{3}x\quad\text{and}\quad \frac{\delta}{\delta\chi_{x}}\to\frac{1}{g_{x}^{1/2}}\frac{\delta}{\delta\chi_{x}}~,\notag
\end{eqnarray}
yielding the result
\begin{eqnarray}
Q_{\sigma} = \int d^{3}x \int D\chi \rho  \frac{\lambda}{g_{x}^{1/2}} \left (\frac{\delta\log\rho}{\delta\chi_{x}}\right )^{2}~.\notag
\end{eqnarray}
The term inside the spatial integral is exactly what appears in (\ref{e Hamiltonian final}), which justifies the name. The contribution of the local quantum potential to the energy is such that those states that are more smoothly spread out in configuration space tend to have lower energy. The corresponding coupling constant $\lambda > 0$ controls the relative importance of the quantum potential; the case $\lambda < 0$ is excluded because it leads to instabilities.

\paragraph*{The gravitational super-Hamiltonian--- }
We now proceed to determining the last remaining element of our scheme, the gravitational super-Hamiltonian $H_{\perp x}^{G}$. Fortunately, under the assumption of non-derivative coupling mentioned above, it is possible to completely separate the Poisson bracket relations (\ref{6 PB 1})-(\ref{6 PB 3}) into pieces that are purely gravitational and those that are pure ``matter". Consequently, to determine $H_{\perp x}^{G}$ it suffices merely to solve the Poisson bracket relations (\ref{PB 1 G})-(\ref{PB 3 G}), which involve only the gravitational variables $g_{ij}$ and $\pi^{ij}$.

Such a task, however, is mathematically equivalent to determining the generators of \emph{pure} geometrodynamics, in which the coupling to ``matter" is absent. But it is precisely this latter challenge which was, in fact, addressed by the efforts of HKT in \cite{Hojman Kuchar Teitelboim 1976} --- they proposed a solution to exactly those brackets that appear in (\ref{PB 1 G})-(\ref{PB 3 G}). Their main result was an important one: the only time-reversible solution to equations (\ref{PB 1 G})-(\ref{PB 3 G}) is nothing less than Einstein's GR in vacuum. We briefly review their argument here, and adapt their solution to our current work in ED.

To begin, since $g_{ij}$ is the intrinsic metric of the surface, its behavior under a purely normal deformation is known \cite{Gourgoulhon 2007} to be
\begin{eqnarray}
\delta g_{ij}(x) = \pounds_{\delta \xi^{\perp}}g_{ijx} = -2 K_{ijx}~,\label{Normal deformation metric a}
\end{eqnarray}
where $K_{ijx}$ is a symmetric $(0,2)$ tensor that is called the \emph{extrinsic} curvature. Note that identifying $K_{ijx}$ with the response of $g_{ijx}$ under a normal deformation is a geometric requirement, not a dynamical one; without this, we could not interpret $g_{ij}$ as residing on a space-like cut of space-time. That being said, $K_{ijx}$ is at this juncture an undetermined functional of the canonical variables, and more to the point, we do not assume at the moment any simple relationship with the momenta $\pi^{ij}_{x}$ --- this must be derived.

Alternatively, the deformation in (\ref{Normal deformation metric a}) must also be attainable using the normal generator $H_{\perp x}^{G}$,
\begin{eqnarray}
\delta g_{ijx} = \int dx^{\prime}\left\{ g_{ijx},H_{\perp x^{\prime}}^{G} \right \}\delta\xi^{\perp}_{x^{\prime}}~.\notag
\end{eqnarray}
More explicitly $H_{\perp x}^{G}$ should satisfy
\begin{eqnarray}
\frac{\delta H_{\perp x^{\prime}}^{G}}{\delta\pi^{ij}_{x}} =-2 K_{ijx}\delta(x,x^{\prime})~. \label{6 Normal deformation metric}
\end{eqnarray}
The appearance of the Dirac delta in eq.(\ref{6 Normal deformation metric}) implies that $H_{\perp x}^{G}$ is \emph{local} in $\pi^{ij}_{x}$. Therefore $H_{\perp x}^{G}$ is a \emph{function}, not a functional, of $\pi^{ij}_{x}$.

Such a simplification turns out to be important: once $H_{\perp x}^{G}$ is a function of $\pi^{ij}$, it is possible to produce an \emph{ansatz} for $H_{\perp x}^{G}$ in terms of powers of $\pi^{ij}$. HKT then supplement this with an additional simplifying assumption: that geometrodynamics be \emph{time-reversible}. This has the advantage of removing all terms in $H_{\perp x}^{G}$ that have odd powers of $\pi^{ij}_{x}$.


Incorporating these ingredients, it is then possible to consider $H_{\perp x}^{G}$'s of the form
\begin{eqnarray}
H_{\perp x}^{G} = \sum_{n = 0}^{\infty}G^{(2n)}_{i_{1}j_{1}i_{2}j_{2}\cdots i_{2n}j_{2n}x}\pi^{i_{1}j_{1}}_{x}\pi^{i_{2}j_{2}}_{x}\cdots\pi^{i_{2n}j_{2n}}_{x}~,\label{6 Super hamiltonian grav ansatz}
\end{eqnarray}
where the coefficients $G^{(2n)}_{ij\cdots}$ are functionals of the metric $g_{ij}$, but depend on the point $x$. Furthermore, since $H_{\perp x}^{G} $ is scalar density and $\pi^{ij}$ a tensor density, $G_{ij\cdots}^{(2n)}$ must transform as a tensor density of weight $1-2n$. Since $\pi^{ij}$ is a symmetric tensor we expect $G^{(2n)}_{ij\cdots}$ to be symmetric under exchange of indices $i_{a}j_{a}\leftrightarrow j_{a}i_{a}$ for any pair $a$; also, $G^{(2n)}_{ij\cdots}$ should be symmetric on interchange of any pair $i_{a}j_{a}\leftrightarrow i_{b}j_{b}$ as this just corresponds to exchanging the $\pi$'s.

Naturally, one narrows the allowable $H_{\perp x}^{G}$ by inserting the ansatz (\ref{6 Super hamiltonian grav ansatz}) into the Poisson bracket (\ref{PB 1 G}). Without delving too far into the details (which can be found in \cite{Hojman Kuchar Teitelboim 1976}) we quote their solution. Only the $n = 0, 1$ terms survive, giving
\begin{eqnarray}
H_{\perp x}^{G} = \kappa \, G_{ijkl}\pi^{ij}\pi^{kl} - \frac{g^{1/2}}{2\kappa}\left(R-\Lambda\right)~,
 \label{6 Super hamiltonian grav cosmo}
\end{eqnarray}
where we have introduce the super metric
\begin{eqnarray}
G_{ijkl} =  \frac{1}{g_{x}^{1/2}}\left(g_{ik}g_{jl}+g_{il}g_{jk}-g_{ij}g_{kl}\right)~,
\end{eqnarray}
and where $R$ is the Ricci scalar for the metric $g_{ij}$.

The constant $\kappa$ is a coefficient that, eventually, determines the coupling to ``matter". We follow standard convention in identifying it as $\kappa = 8\pi G$, where $G$ is Newton's constant. The other parameter $\Lambda$, of course, is the cosmological constant. For simplicity, going forward we set $\Lambda = 0$.

Putting everything together, we have that $H_{\perp x}^{G}$ takes the form
\begin{eqnarray}
H_{\perp x}^{G} = \frac{\kappa}{g_{x}^{1/2}}\left (2\pi^{ij}\pi_{ij}-\pi^{2}\right ) - \frac{g^{1/2}}{2\kappa}R~,\label{6 Super hamiltonian grav}
\end{eqnarray}
where $\pi = \pi^{ij}g_{ij} = \text{Tr}(\pi^{ij}) $. This is exactly the standard gravitational super-Hamiltonian obtained by Dirac \cite{Dirac 1958} and ADM \cite{ADM 1960} by starting from the Einstein-Hilbert Lagrangian.

\paragraph*{Total super-Hamiltonian--- }
Putting together the ingredients of this section, the total super-Hamiltonian is thus
\begin{eqnarray}
H_{\perp x} = H^{G}_{\perp x}+\tilde{H}_{\perp x} ~,\label{6 Super-Hamiltonian Total}
\end{eqnarray}
where $H_{\perp x}^{G}$ is given above by eq.(\ref{6 Super hamiltonian grav}) and where a suitable family of $\tilde{H}_{\perp x}$'s have been identified in eq.(\ref{6 e-H}). With this, the super-Hamiltonian constraint is then just
\begin{eqnarray}
H_{\perp x} = H^{G}_{\perp x}+\tilde{H}_{\perp x} \approx 0~.\label{6 Super-Hamiltonian constraint}
\end{eqnarray}

\section{The dynamical equations}\label{ED_DyEqns}

As argued by HKT, although the role of space-time was crucial to the developing the equations (\ref{6 PB 1})-(\ref{6 Hamiltonian constraints}), one notices that all signs of the enveloping space-time have dropped out in the closing relations (\ref{6 PB 1})-(\ref{6 PB 3}).\footnote{For instance, they depend on the surface metric $g_{ij}$, but not on its extrinsic curvature $K_{ij}$.} Thus in geometrodynamics we can dispense with the notion of an \emph{a priori} given space-time and instead consider the three-dimensional Riemannian manifold $\sigma_{t}$ as primary.

The idea is a simple one. The canonical variables are evolved with the generators $H_{Ax}$, satisfying eqns.(\ref{6 PB 1})-(\ref{6 Hamiltonian constraints}). Such an evolution will cause both the ``matter" and geometry, to change. We might then give this manifold with updated intrinsic geometry a new name, $\sigma_{t'}$. Repeating this procedure results in an evolution of the dynamical variables, and consequently, what one might view as a succession of manifolds $\{\sigma_{t}\}$, parameterized by the label $t$. Thus this iterative process constructs a space-time, step by step. We now investigate the dynamical equations that result from this procedure.

\subsection*{Some formalism}
Consider the evolution of an arbitrary functional $T_{t}$ of the dynamical variables defined on $\sigma_{t}$, 
\begin{eqnarray}
\delta T_{t} = T_{t+dt} - T_{t} = \int dx \left \{T_{t}, H_{Ax} \right \}\delta\xi^{A}_{x} =  \int dx \, N^{A}_{xt} \left \{T_{t}, H_{Ax} \right \}dt~,\label{6 Evolution lapse shift}
\end{eqnarray}
where we have introduced four arbitrary functions
\begin{eqnarray}
N_{xt} = \frac{\delta\xi^{\perp}_{x}}{dt}\quad\text{and}\quad N^{i}_{xt} = \frac{\delta\xi^{i}_{x}}{dt}~,\label{6 Def Lapse Shift}
\end{eqnarray}
which are the \emph{lapse} and vector \emph{shift}, respectively. Just as the evolution parameters $\delta\xi^{A}_{x}$ were completely arbitrary, these functions can be freely specified, and amount \emph{eventually} to picking a particular foliation of space-time.\footnote{At the moment, the lapse and shift have no definite geometrical meaning. But as is well known, we can eventually identify the lapse $N$ and shift $N^{i}$ as being related to components of the space-time metric $^{4}g_{\mu\nu} \equiv \gamma_{\mu\nu}$. (We changed the symbol temporarily to avoid confusion.) More specifically, we would have \cite{Hojman Kuchar Teitelboim 1976}
\begin{eqnarray}
N = \left(- \gamma^{00}\right)^{-1/2}\quad\text{and}\quad N_{i} = \gamma_{0i}~.\notag
\end{eqnarray}
}

As in the Dirac approach to geometrodynamics, the $N^{A}_{xt} = (N_{xt}, N^{i}_{xt})$ are \emph{not} functionals of the canonical variables, therefore we can rewrite eq.(\ref{6 Evolution lapse shift}) as
\begin{eqnarray}
\delta T_{t}  = \left \{T_{t}, H[N,N^{i}] \right \}dt~,\label{6 Evolution smeared}
\end{eqnarray}
where we have introduced the notion of a ``smeared" Hamiltonian
\begin{eqnarray}
H[N,N^{i}]=\int dx\,\left( N_{xt}H_{\bot x}+N_{xt}^{i}H_{ix}\right) \label{6 Smeared Hamiltonian}
\end{eqnarray}%
that, for given $N^{A}_{xt}$, generates an evolution parameterized by $t$.\footnote{Had the $N^{A}$ been functionals of the canonical variables then the Poisson bracket in eq.(\ref{6 Evolution smeared}) would have generated extra terms from their action on $N^{A}$ and thus eq.(\ref{6 Evolution smeared}) would not have been equivalent to eq.(\ref{6 Evolution lapse shift}).} This \emph{global} $H[N,N^{i}]$ (note the spatial integral) conforms more naturally to our typical notions of a Hamiltonian. Thus, the derivative with respect to the parameter $t$ is
\begin{eqnarray}
\partial_{t} T_{t} \equiv \frac{\delta T_{t}}{dt} = \left \{ T_{t}, H[N,N^{i}]    \right \}~,\label{6 Def time derivative}
\end{eqnarray}
or more explicitly by
\begin{eqnarray}
\partial_{t} T_{t} = \int dx\,\left( N_{xt}\left \{T_{t},H_{\perp x}\right \}+N_{xt}^{i}\left \{T_{t},H_{ix}\right \}\right)~,\label{6 Def time derivative b}
\end{eqnarray}
and more succinctly
\begin{eqnarray}
\partial_{t} T_{t} = \pounds_{m} T_{t} + \pounds_{N^{i}} T_{t}~, \label{6 Def time derivative c}
\end{eqnarray}
where the vector $m^{\mu}_{x} = N_{xt} n^{\mu}_{x}$ is the so-called evolution vector.

\subsection*{The evolution of the ``matter" sector}
The goal is to determine the evolution of the probability distribution $\rho_{t}[\chi]$ and phase functional $\Phi_{t}[\chi]$, given an initial state $(\rho_{t}, \Phi_{t}; g_{ij\, t}, \pi^{ij}_{t})$ that satisfies the initial value constraints (\ref{6 Hamiltonian constraints}).
\paragraph*{Dynamical equations for the probability and phase--- }
Given the basic dynamical law, eq.(\ref{6 Def time derivative}), the time evolution of the variables $\rho_{t}$ and $\Phi_{t}$ are given by
\begin{eqnarray}
\partial_{t}\rho_{t} = \left \{\rho_{t}, H   \right \}\quad\text{and}\quad \partial_{t}\Phi_{t} = \left \{\rho_{t}, H   \right \}~,
\end{eqnarray}
where $H$ is the smeared Hamiltonian given above. Thus, in accordance with eq.(\ref{6 Def time derivative b}), we have the result
\begin{subequations}
\begin{eqnarray}
\partial_{t}\rho_{t}  &=& \int dx \, \left ( N_{xt}\left \{\rho_{t}, H_{\perp x}\right \}+ N^{i}_{xt}\left \{\rho_{t}[\chi], H_{i x}\right \} \right ) \\ 
\partial_{t}\Phi_{t}  &=& \int dx \, \left ( N_{xt}\left \{\Phi_{t}, H_{\perp x}\right \}+ N^{i}_{xt}\left \{\Phi_{t}, H_{i x}\right \} \right )~. \label{6 Evolution Phi}
\end{eqnarray}
\end{subequations}

Using the family of $H_{\perp x}$'s that we identified in eq.(\ref{6 Super-Hamiltonian Total}) and the super-momentum $H_{ix}$ in eq.(\ref{6 Super momentum}) we can compute all of the necessary Poisson brackets. From the super momentum in eq.(\ref{6 Super momentum}) we can compute the tangential pieces
\begin{subequations}
\begin{eqnarray}
\left \{\rho_{t}[\chi], H_{i x}\right \} &=& \frac{\tilde{\delta}\tilde{H}_{ix}}{\tilde{\delta}\tilde{\Phi_{t}}} = \frac{\delta\rho_{t}}{\delta\chi_{x}}\partial_{ix}\chi_{x}~,\\
\left \{\Phi_{t}[\chi], H_{i x}\right \} &=& -\frac{\tilde{\delta}\tilde{H}_{ix}}{\tilde{\delta}\tilde{\rho_{t}}} = \frac{\delta\Phi_{t}}{\delta\chi_{x}}\partial_{ix}\chi_{x}~.\label{6 Tangential PB}
\end{eqnarray}
\end{subequations}
Of course, the entire family of $\tilde{H}_{\perp x}$'s were designed to reproduce the LTFP equations, thus by construction we have
\begin{eqnarray}
\left \{\rho_{t}[\chi], H_{\perp x}\right \} = \frac{\tilde{\delta}\tilde{H}_{\perp x}}{\tilde{\delta}\tilde{\Phi_{t}}} = -\frac{1}{g_{x}^{1/2}}\frac{\delta}{\delta\chi_{x}}\left ( \rho_{t}\frac{\delta\Phi_{t}}{\delta\chi_{x}}  \right )~,\label{FP equation final}
\end{eqnarray}
in agreement with eq.(\ref{FP equation}). The remaining Poisson bracket determines the local time evolution of the phase functional $\Phi_{t}$ and is given by
\begin{align}
\left \{\Phi_{t}[\chi], H_{\perp x}\right \} = - \frac{\tilde{\delta}\tilde{H}_{\perp x}}{\tilde{\delta}\tilde{\rho_{t}}}  = \frac{1}{2 g_{x}^{1/2}}\left(\frac{\delta\Phi_{t}[\chi]}{\delta\chi_{x}}\right)^{2}+\frac{g_{x}^{1/2}}{2}g^{ij}\partial_{i}\chi_{x}\partial_{j}\chi_{x}+ \frac{\tilde{\delta}F_{x}[\rho]}{\tilde{\delta}\rho_{t}[\chi]}~,\label{6 LTHJ equations}
\end{align}
where $F_{x}$ is of the form specified by eq.(\ref{6 e-H}).
\paragraph*{The local time Hamilton-Jacobi equations--- }
To interpret the local equations (\ref{6 LTHJ equations}), we write the full time evolution for the phase functional by inserting eqns.(\ref{6 LTHJ equations}) and (\ref{6 Tangential PB}) into eq.(\ref{6 Evolution Phi}), yielding
\begin{align}
-\partial_{t}\Phi_{t} =& \int dx \, \left [N_{xt} \left (\frac{1}{2 g_{x}^{1/2}}\left(\frac{\delta\Phi_{t}}{\delta\chi_{x}}\right)^{2} +\frac{g_{x}^{1/2}}{2}g^{ij}\partial_{i}\chi_{x}\partial_{j}\chi_{x} + \frac{\tilde{\delta}F_{x}}{\tilde{\delta}\rho_{t}}\right )+ N^{i}_{xt}\frac{\delta\Phi_{t}}{\delta\chi_{x}}\partial_{ix}\chi_{x}\right ]~.
\end{align}
To bring this equation into a more familiar form we consider the special case of flat space-time by setting the metric to be a Kroenecker delta $g_{ij} = \delta_{ij}$, so that  $g_{x}^{1/2} = 1$, and we let  $N = 1$, $N^{i} = 0$. Moreover, for simplicity we also make the assignment $F_{x} = 0$. This results in a time evolution for $\Phi_{t}$ that has the form
\begin{eqnarray}
-\partial_{t}\Phi_{t} = \int dx \left (\frac{1}{2}\left(\frac{\delta\Phi_{t}}{\delta\chi_{x}}\right)^{2} +\frac{1}{2}\delta^{ij}\partial_{i}\chi_{x}\partial_{j}\chi_{x} \right )~,\notag
\end{eqnarray}
which is exactly the classical Hamilton-Jacobi equation for a massless Klein-Gordon field in flat space-time. Thus, in analogy with the LTFP equations, we refer to eq.(\ref{6 LTHJ equations}) as the \emph{local time Hamilton-Jacobi} (LTHJ) equations, as there is one equation of the Hamilton-Jacobi type for every spatial point.

The LTFP and LTHJ equations, with the tangential equations (\ref{6 Tangential PB}), and the evolution equations (\ref{6 Evolution Phi}), give us the ability to evolve an appropriately chosen initial state $(\rho_{t},\Phi_{t})$. In general, this is a coupled non-linear evolution driven by a dependence on the metric $g_{ij}$. To a large extent, this completes our discussion of how the epistemic variables evolve. In a subsequent section, however, we discuss in some detail the dynamics of a specific class of models, those that involve the local quantum potential.

\subsection*{The evolution of the geometrical variables}
We now review the content of  the well-known Einstein's equations written within the canonical language. (A good review of these equations is given, for example, by ADM \cite{ADM 2008}.)
\paragraph*{Evolution of metric--- }
The goal is to determine the evolution of the geometrical variables $(g_{ij},\pi^{ij})$. Beginning with the metric $g_{ij}$ defined on some initial three-space $\sigma_{t}$, we wish to determine how it evolves in response to the generators $H_{Ax}$. Applying  eq.(\ref{6 Def time derivative b}) for the time derivative, we have that
\begin{eqnarray}
\partial_{t} g_{ijx} = \int dx^{\prime} \left\{g_{ijx} , H_{Ax'} \right \}N^{A}_{x^{\prime}t} =  \int dx^{\prime} \frac{\delta H_{Ax'}}{\delta\pi^{ij}_{x}}N^{A}_{x^{\prime}t}  ~.\label{6 Metric time deriv a}
\end{eqnarray}
To compute this, recall that the tangential piece is known from equations (\ref{6 Tangential deformation metric}) and (\ref{6 Tangential deformation metric cov}). This gives us
\begin{eqnarray}
\pounds_{N^{i}}\, g_{ijx} =  \int dx^{\prime} \, \frac{\delta H_{ix'}}{\delta\pi^{ij}_{x}} \,  N^{i}_{x^{\prime}t} = \nabla_{i}N_{jxt}+\nabla_{j}N_{ixt}~, \label{6 Lie derivative shift  metric}
\end{eqnarray}
where $\pounds_{N^{i}} g_{ijx}$ is just the Lie derivative along the vector field defined by the shift $N^{i}$, and where $N_{ixt} = g_{ijx}N^{j}_{xt}$.

To obtain the remaining piece, first differentiate the $H_{\perp x}$ given in eq.(\ref{6 Super hamiltonian grav})
\begin{eqnarray}
 \frac{\delta H_{\perp x^{\prime}}}{\delta\pi^{ij}_{x}} = \frac{4\kappa}{g_{x}^{1/2}}\left(\pi_{ijx} - \frac{1}{2}\pi_{x} g_{ijx}\right)\delta(x,x^{\prime})~,\notag
\end{eqnarray}
where $\pi_{ij}$ is the conjugate momentum with its indices lowered, and $\pi = g_{ij}\pi^{ij} = \text{Tr}(\pi^{ij})$ is the trace of the gravitational momentum. Putting these terms together, eq.(\ref{6 Metric time deriv a}) reads as
\begin{subequations}
\begin{eqnarray}
\partial_{t}g_{ijx} &=& \pounds_{m}g_{ijx} + \pounds_{N^{i}}g_{ijx} \label{6 Metric time deriv b}\\
 \pounds_{m}g_{ijx} &=&  \frac{2\kappa}{g_{x}^{1/2}}\left(2\pi_{ijx} - \pi_{x} g_{ijx}\right)N_{xt} \\
  \pounds_{N^{i}}g_{ijx} &=& \nabla_{i}N_{jxt}+\nabla_{j}N_{ixt}~.
\end{eqnarray}
\end{subequations}
This gives us the evolution of the metric with foliation parameter $t$.

We can also now identify the extrinsic curvature tensor
\begin{eqnarray}
K_{ij} = \frac{\kappa}{g^{1/2}_{x}}\left(\pi_{x} g_{ijx} - 2\pi_{ijx}\right) ~.
\end{eqnarray}
Inverting this relationship we get $\pi^{ij}$ in terms of $K^{ij}$, which yields
\begin{eqnarray}
\pi^{ij}_{x} = \frac{g^{1/2}_{x}}{2\kappa}\left (Kg^{ij}_{x} - K^{ij}_{x} \right )~. \label{6 Extrinsic curvature pi b}
\end{eqnarray}
where $K = K_{ij}g^{ij} = \text{Tr}(K_{ij})$ is the trace of $K_{ij}$. This is of some interest if one wishes to compare the canonical formulation to the standard so-called ``Lagrangian" approach (see e.g., \cite{Carroll 2004}).

\paragraph*{Evolution of conjugate momentum--- }
To this point, the dynamics of the geometry has not differed from a purely classical geometrodynamics, such as that of HKT. This is because the ``matter" super-Hamiltonian and super-momentum were, by definition, completely independent of $\pi^{ij}$ (due, of course, to the non-derivative coupling assumption) therefore the evolution of $g_{ij}$ did not receive contributions from the ``matter" sector. This changes when we consider the dynamics of the conjugate momentum $\pi^{ij}$.

The evolution of $\pi^{ij}$ is determined \emph{via} the equation
\begin{eqnarray}
\partial_{t} \pi_{x}^{ij} = \int dx^{\prime} \left\{\pi^{ij}_{x} , H_{Ax'} \right \}N^{A}_{x^{\prime}t} =  - \int dx^{\prime} \frac{\delta H_{Ax'}}{\delta g_{ijx}}N^{A}_{x^{\prime}t}  ~.\label{6 Pi time deriv a}
\end{eqnarray}
Recalling now that both the gravitational and ``matter" super-Hamiltonians $H_{\perp x} = H_{\perp x}^{G}[g_{ij},\pi^{ij}] + \tilde{H}_{\perp x}[\rho,\Phi;g_{ij}]$ depend explicitly on the metric, but that $\tilde{H}_{ix}$ does not, this expression slightly simplifies to
\begin{eqnarray}
\partial_{t}\pi^{ij}_{x}= -\int dx^{\prime}\left(N_{x^{\prime}t}\frac{\delta H^{G}_{\perp x^{\prime}}}{\delta g_{ijx}}+ N_{x^{\prime}t}\frac{\delta \tilde{H}_{\perp x^{\prime}}}{\delta g_{ijx}} +N^{i}_{x^{\prime}t}\frac{\delta H_{ix^{\prime}}^{G}}{\delta g_{ijx}}\right)~,
\end{eqnarray}
where we have have separated the contributions from the gravitational and ``matter" sectors. In the notation of eq.(\ref{6 Def time derivative c}) we write this as
\begin{eqnarray}
\partial_{t}\pi^{ij}_{x} =  \pounds_{m}\pi^{ij}_{x} +\pounds_{N^{i}} \pi^{ij}_{x}~,
\end{eqnarray}
where
\begin{eqnarray}
\pounds_{m}\pi^{ij}_{x} = \pounds_{m}^{G}\pi^{ij}_{x}+ \pounds_{m}^{M}\pi^{ij}_{x}~,
\end{eqnarray}
and
\begin{subequations}
\begin{eqnarray}
\pounds_{m}^{G}\pi^{ij}_{x} &=& -\int dx^{\prime}\left(N_{x^{\prime}t}\frac{\delta H^{G}_{\perp x^{\prime}}}{\delta g_{ijx}}\right)~,  \label{6 Pi Lie normal G} \\ 
\pounds_{m}^{M}\pi^{ij}_{x} &=& -\int dx^{\prime}\left(N_{x^{\prime}t}\frac{\delta \tilde{H}_{\perp x^{\prime}}}{\delta g_{ijx}}\right)~, \label{6 Pi Lie normal M} \\
\pounds_{N^{i}}\pi^{ij}_{x} &=& -\int dx^{\prime}\left(N^{i}_{x^{\prime}t}\frac{\delta H^{G}_{i x^{\prime}}}{\delta g_{ijx}}\right)~.\label{6 Pi Lie tangential G}
\end{eqnarray}
\end{subequations}

From (\ref{6 Tangential deformation momentum cov}) the last term is fairly easy to determine,
\begin{eqnarray}
\pounds_{N^{i}}\pi^{ij}_{x} =  \nabla_{kx}\left(\pi^{ij}_{x}N^{k}_{x}\right) - \pi^{ik}_{x}\nabla_{kx}N^{j}_{x} - \pi^{kj}_{x}\nabla_{kx}N^{i}_{x}
~.\label{6 Pi Lie tangential G b}
\end{eqnarray}

The calculation of the other two terms is much more involved. Fortunately the expression for $\pounds_{m}^{G}\pi^{ij}$ is already well known \cite{Gourgoulhon 2007}\cite{ADM 2008} and we merely quote the result,
\begin{align}
\pounds_{m}^{G}\pi^{ij}_{x} &=-\frac{g^{1/2}}{2\kappa}\left(R^{ij}_{x}-\frac{1}{2}g^{ij}_{x}R_{x}\right )N_{xt}+ \frac{\kappa}{g^{1/2}_{x}} \, g^{ij}_{x}\left(\pi^{kl}_{x}\pi_{klx}-\frac{1}{2}\pi^{2}_{x}\right)N_{xt}\notag\\
 &-\frac{4\kappa}{g^{1/2}_{x}}\left(\pi^{ik}_{x}\pi^{j}_{kx}-\frac{1}{2}\pi_{x}\pi^{ij}_{x}\right)N_{xt}+\frac{g^{1/2}_{x}}{2\kappa}\left(\nabla^{i}_{x}\nabla^{j}_{x}N_{xt} - g^{ij}_{x}\nabla^{k}_{x}\nabla_{kx}N_{xt}\right)~.\label{Lie derivative normal pi G}
\end{align}

To calculate the remaining piece $\pounds_{m}^{M}\pi^{ij}_{x}$ we first recall that having assumed a \emph{non-derivative} coupling of gravity to matter, the metric $g_{ij}$ appears in $\tilde{H}_{\perp x}$ as an undifferentiated function (not a functional) and without any derivatives \cite{Teitelboim thesis}, which implies that
\begin{eqnarray}
\frac{\delta \tilde{H}_{\perp x^{\prime}}}{\delta g_{ijx}} =\frac{\partial \tilde{H}_{\perp x}}{\partial g_{ijx}} \,  \delta(x,x') ~,\notag
\end{eqnarray}
obtaining
\begin{eqnarray}
\pounds_{m}^{M}\pi^{ij}_{x} = - \frac{\partial \tilde{H}_{\perp x}}{\partial g_{ijx}} N_{xt}~.\notag
\end{eqnarray}
Using eq.(\ref{6 e-H}), for an arbitrary choice of $F_{x}$, but one that still satisfies eq.(\ref{6 PB 1 matter}), the ``matter" source has the form
\begin{eqnarray}
\pounds_{m}^{M}\pi^{ij}_{x}  = N_{xt} \left (  \frac{\partial \tilde{H}^{0}_{\perp x}}{\partial g_{ijx}} + \frac{\partial F_{x}}{\partial g_{ijx}}  \right )~,\label{6 Matter source}
\end{eqnarray}
with
\begin{align}
 \frac{\partial \tilde{H}^{0}_{\perp x}}{\partial g_{ijx}} =& -\frac{1}{2}\int D\chi \rho \left [\left (\frac{1}{g_{x}^{1/2}}\left (\frac{\delta\Phi}{\delta\chi_{x}}\right )^{2}  -g^{1/2}_{x}g^{kl}_{x}\partial_{kx}\chi_{x}\partial_{lx}\chi_{x}    \right )g^{ij}_{x}- g_{x}^{1/2}\partial_{x}^{i}\chi_{x}\partial^{j}_{x}\chi_{x}\right ]~,\label{6 Matter source b}
\end{align}
where $\tilde{H}_{\perp x}^{0}$ was given in eq.(\ref{6 e-H b}), and where $\partial^{i}_{x}\chi_{x} = g^{ij}_{x}\partial_{jx}\chi_{x}$.

In total, the equations of motion for the gravitational field follow Hamilton's equations for the gravitational variables $(g_{ijx},\pi^{ij}_{x})$, given by eq.(\ref{6 Metric time deriv b}) for the evolution of the metric, and for the conjugate momentum $\pi^{ij}_{x}$ we have
\begin{eqnarray}
\partial_{t}\pi^{ij}_{x} - \pounds_{n}^{G}\pi^{ij}_{x} - \pounds_{N^{i}}\pi^{ij}_{x} = -N_{xt}\frac{\partial \tilde{H}_{\perp x}}{\partial g_{ijx}} ~.\label{6 Pi time deriv b}
\end{eqnarray}
This is an equation in which geometrical variables on the left-hand side are sourced by the variables $(\rho,\Phi)$, which contain all the information available about the field $\chi_{x}$ on the right-hand side. Below we will discuss this equation in the presence of ``quantum matter."

\section{Quantum sources of gravitation}\label{ED_QT}
The transition to what may be termed a \emph{quantum} form of dynamics amounts to an appropriate choice of the functional $F_{x}[\rho;g_{ij}]$ (see e.g., \cite{Caticha Ipek 2014}\cite{Ipek et al 2017}\cite{Ipek et al 2019a}). In particular, for $F_{x}[\rho;g_{ij}]$ we choose exactly the local quantum potential introduced as part of (\ref{e Hamiltonian final}). A convenient choice for the coupling constant is $\lambda = 1/8$.\footnote{As argued, for instance, in \cite{Bartolomeo et al 2014}, there is no loss of generality in making this choice. For the case of nonrelativistic particles it can be proved \cite{Caticha 2019b} that an ED that preserves the appropriate symplectic and metric structures implies the presence of a quantum potential with the correct coefficient.} (This choice of $\lambda$ makes plain that we work with a system of units where $\hbar = c = 1$.)

The connection to conventional quantum theory is made explicit by a change of variables from the probability $\rho$ and phase $\Phi$ to the complex variables
\begin{eqnarray}
\Psi= \rho^{1/2}e^{i\Phi}\quad\text{and}\quad \Psi^{*} = \rho^{1/2}e^{-i\Phi}~.\label{6 Define Psi}
\end{eqnarray}
Such a change of variables is, in fact, a canonical transformation, and so, the new variables form a canonical pair given by $(\Psi, i\Psi^{*})$, which obey a natural generalization of the standard Poisson bracket relations
\begin{eqnarray}
\left \{ \Psi[\chi], i\Psi^{*}[\chi^{\prime}]  \right \} = \delta[\chi - \chi^{\prime}]~,
\end{eqnarray}
where $ \delta[\chi - \chi^{\prime}]$ is a Dirac delta \emph{functional}.

\subsection*{Quantum operators and geometrodynamics revisited}
Having chosen an $F_{x}[\rho;g_{ij}]$ of the type described above, the ensemble generators $\tilde{H}_{Ax}$ take a particularly special form:
\begin{eqnarray}
\tilde{H}_{Ax} = \int D\chi \Psi^{*}\hat{H}_{Ax}\Psi = \left \langle \hat{H}_{Ax}  \right \rangle~, \label{6 Quantum Super Hamiltonian ensemble}
\end{eqnarray}
which is the expected value of the local Hamiltonian operators $\hat{H}_{Ax} = (\hat{H}_{\perp x},\hat{H}_{ix})$, given by
\begin{subequations}
\begin{align}
\hat{H}_{\perp x} &= - \frac{1}{2g^{1/2}}\frac{\delta^{2}}{\delta\chi^{2}_{x}}+\frac{g^{1/2}}{2}g^{ij}\partial_{i}\chi_{x}\partial_{j}\chi_{x}+g^{1/2}V_{x}(\chi_{x};g_{ij}) \label{6 Quantum Super Hamiltonian operator} \\
\hat{H}_{i x} &= i\,\partial_{i}\chi_{x}\frac{\delta}{\delta\chi_{x}}~. \label{6 Quantum Super momentum operator}
\end{align}
\end{subequations}
To obtain the ``matter" contribution in eq.(\ref{6 Pi time deriv b}) for the conjugate momentum $\pi^{ij}_{x}$, note that the metric appears in $\tilde{H}_{\perp x}$ through the density $g_{x}^{1/2}$ and the inverse metric $g^{ij}_{x}$. Variations of these quantities with respect to the metric $g_{ijx}$ are given by \cite{Carroll 2004}
\begin{eqnarray}
\delta g_{x} = g  \,g^{ij}_{x} \, \delta g_{ijx}\quad\text{and}\quad \delta g^{ij}_{x} = g^{ik}g^{jl}\delta g_{kl}~.\notag
\end{eqnarray}
Then
\begin{eqnarray}
  \frac{\partial \tilde{H}_{\perp x}}{\partial g_{ijx}} = \int D\chi \, \Psi^{*}  \frac{\partial \hat{H}_{\perp x}}{\partial g_{ijx}} \Psi~,\notag
\end{eqnarray}  
where we have defined the operator
\begin{eqnarray}
  \frac{\partial \hat{H}_{\perp x}}{\partial g_{ijx}} = \partial^{i}_{x}\chi_{x}\partial^{j}_{x}\chi_{x}+g^{ij}\left (\frac{1}{2g_{x}^{1/2}}\frac{\delta^{2}}{\delta\chi_{x}^{2}}+\frac{g^{1/2}_{x}}{2}g^{kl}\partial_{k}\chi_{x}\partial_{l}\chi_{x}+V_{x}(\chi_{x})\right ).\label{6 Quantum stress operator}
\end{eqnarray}

\paragraph*{Geometrodynamics with quantum sources--- }
Our goal here is to rewrite the main equations of geometrodynamics with sources given by quantum matter. We begin first with the constraint equations. With the aid of the local operators introduced in (\ref{6 Quantum Super Hamiltonian operator}) and (\ref{6 Quantum Super momentum operator}), the total Hamiltonian generators take the explicit form
\begin{subequations}
\begin{align}
H_{\perp x} &= \frac{\kappa}{g_{x}^{1/2}}\left (2\pi^{ij}_{x}\pi_{ijx} -\pi^{2}_{x}\right) - \frac{g^{1/2}_{x}}{2\kappa}R_{x} + \int D\chi \, \Psi^{*} \hat{H}_{\perp x} \Psi , \label{6 Quantum Hamiltonian}\\
H_{i x} &= -2\nabla_{kx}\left (\pi^{kj}_{x}g_{ijx}\right )+ \int D\chi \, \Psi^{*} \hat{H}_{i x} \Psi ~,\label{6 Quantum momentum}
\end{align}
\end{subequations}
which are, of course, subject to the constraints
\begin{eqnarray}
H_{\perp x}\approx 0\quad \text{and}\quad H_{i x}\approx 0~. \label{Quantum constraints}
\end{eqnarray}
As relations (\ref{6 Quantum Hamiltonian}) and (\ref{6 Quantum momentum}) coupled together with (\ref{Quantum constraints}) make abundantly clear, the quantum state and the geometrical variables can no longer be treated as independent. This has important consequences for the time evolution of $\Psi$.

Moving on, note that the dynamical equation for the metric, given in eq.(\ref{6 Metric time deriv b}), does not depend directly on the choice of $F_{x}$ and therefore remains unchanged in the quantum context. The dynamical equation for the conjugate momentum $\pi^{ij}_{x}$, however, is modified. Using the operator introduced in (\ref{6 Quantum stress operator}), this becomes
\begin{eqnarray}
\partial_{t}\pi^{ij}_{x} - \pounds_{n}^{G}\pi^{ij}_{x} - \pounds_{N^{i}}\pi^{ij}_{x} = -N_{xt}\int D\chi \, \Psi^{*}\frac{\partial \hat{H}_{\perp x}}{\partial g_{ijx}}\Psi~,\label{6 Quantum Pi evolution}
\end{eqnarray}
where $\pounds_{m}^{G}\pi^{ij}_{x} = $ and $\pounds_{N^{i}}\pi^{ij}_{x}$ are given in eqns.(\ref{6 Pi Lie normal G}) and (\ref{6 Pi Lie tangential G b}), respectively. This is the crucial equation in which the dynamical geometry is itself affected by the epistemic state $\Psi$.

Putting it all together, the eqns.(\ref{6 Quantum Hamiltonian})-(\ref{6 Quantum Pi evolution}), and eq.(\ref{6 Metric time deriv b}) for the metric, constitute a system of equations that are \emph{formally} equivalent to the semi-classical Einstein equations (SCEE) (see e.g., \cite{Wald 1994}) put in the canonical form. However, that is where the similarities end. On several key issues of interpretation, in particular, the ED approach is vastly different from the SCEE as they are normally understood. Far from being trivial, these distinctions turn out to be quite important since many objections (see e.g., \cite{Kibble et al 1980}\cite{Unruh 1984}) to the usual SCEE are, in fact, based on such considerations.

\subsection*{Quantum dynamics}
One advantage of the complex variables $(\Psi,\Psi^{*})$ is that the dynamics takes a \emph{familiar} form. Indeed, since $\Psi$ and $\Psi^{*}$ are just functions of our canonical variables $(\rho,\Phi)$ we can just use eq.(\ref{6 Def time derivative c}) to determine their evolution along a time parameter $t$, which gives
\begin{eqnarray}
\partial_{t}\Psi_{t}[\chi] =  \int dx \left \{\Psi_{t}[\chi],H_{Ax}\right \}N^{A}_{xt}~.\label{6 Quantum Evolution Psi a}
\end{eqnarray}

The tangential component is obtained in a straightforward fashion by
\begin{eqnarray}
 \left \{\Psi_{t}[\chi],H_{ix}\right \} = \partial_{i}\chi_{x}\frac{\delta\Psi_{t}[\chi]}{\delta\chi_{x}}~,\label{6 Quantum Tangential Psi}
\end{eqnarray}
which is reasonable since this is just the Lie derivative of $\Psi[\chi]$ along the surface.\footnote{Compare, for example, to eqns.(\ref{6 Tangential deformation rho c}) and (\ref{6 Tangential deformation Phi a}) for $\rho[\chi]$ and $\Phi[\chi]$.} The local normal evolution of $\Psi$, on the other hand, is given by
\begin{eqnarray}
 \left \{\Psi_{t}[\chi],\tilde{H}_{\perp x}\right \} =-i \, \hat{H}_{\perp x}\Psi_{t}[\chi]~.
\end{eqnarray}
Inserting these results into eq.(\ref{6 Quantum Evolution Psi a}) for a general evolution of $\Psi_{t}[\chi]$, we then have
\begin{eqnarray}
i \, \partial_{t}\Psi_{t}[\chi] = \int dx\left (N_{xt}\hat{H}_{\perp x}+N^{i}_{xt}\hat{H}_{ix}\right )\Psi_{t}[\chi]~.\label{6 Quantum Schrodinger equation}
\end{eqnarray}
Finally, substituting eqns.(\ref{6 Quantum Super Hamiltonian operator}) and (\ref{6 Quantum Super momentum operator}) in for $\hat{H}_{\perp x}$ and $\hat{H}_{i x}$, respectively, yields the equation
\begin{align}
i \, \partial_{t}\Psi_{t} &= \int dx \left [ N_{xt}\left(-\frac{1}{2g^{1/2}}\frac{\delta^{2}}{\delta\chi^{2}_{x}}+\frac{g^{1/2}}{2}g^{ij}\partial_{i}\chi_{x}\partial_{j}\chi_{x}+g^{1/2}V_{x}\right ) - N^{i}_{xt}\partial_{i}\chi_{x}\frac{\delta}{\delta\chi_{x}}\right ]\Psi_{t}
\end{align}
which is, ostensibly, just a linear differential operator for the complex variable $\Psi_{t}$, which \emph{suggests} calling it a Schr\"{o}dinger functional equation. We discuss below.

\paragraph*{But is it quantum?--- }
The question of which criteria a theory must obey to deserve being called a quantum theory is a matter of taste and of convention. If the criteria involve the presence of $\hbar$, of a wave equation for a complex wave function, with an uncertainty principle and non-local correlations, the reconstruction of the standard formalism of quantum field theory in the limit of say, a flat space-time, then the ED model formulated above definitely qualifies as being quantum. On the other hand, if the defining criterion is the existence of a superposition principle, then our ED model is not a quantum theory.

To see that the coupling to classical gravity implies violations of the superposition principle we note that geometrodynamics is a constrained dynamical system, where the operative constraints are given by the Hamiltonian constraints in (\ref{Quantum constraints}). Solving these constraint equations often involves solving for components of the metric in terms of the quantum sources, which then gets fed back into evolution equations for $\Psi$.\footnote{This dynamic can most readily be seen within the linearized gravity regime, or the weak field limit. In fact, there are some arguments (see e.g., \cite{Bahrami et al 2014}) that the so-called Newton-Schr\"{o}dinger equation, which is a non-linear equation, results from the Newtonian limit of the semi-classical Einstein equations. Investigation of this result coming from ED is underway.} This feedback leads to a non-linear time evolution (see e.g., \cite{Kibble et al 1980}) in which the $\Psi$ itself appears as a potential in eq.(\ref{6 Quantum Schrodinger equation}).


\section{Concluding remarks}
\label{conclusion}

The ED developed here couples quantum ``matter" to a dynamical background on the basis of three key principles: (1) A properly entropic setting wherein the dynamics of probability is driven by information encoded into \emph{constraints}. (2) The preservation of a symplectic structure as a primary criterion for updating the evolving constraints. (3) Imposing the Poisson bracket ``algebra" of DHKT as a representation of the kinematics of surface deformations.

Our approach results in several interesting features.
Although written in the relatively less common language of geometrodynamics,
the eqns.(\ref{6 Metric time deriv b}) and
(\ref{Quantum constraints})-(\ref{6 Quantum Pi evolution}) are
formally equivalent to the so-called semi-classical\ Einstein equations
(SCEE)
\begin{eqnarray}
G_{\mu\nu}=8\pi G\langle \Psi | \hat{T}_{\mu\nu}| \Psi\rangle
~,\label{6 Semi classical equations}%
\end{eqnarray}
with classical Einstein tensor $G_{\mu\nu}$ (see e.g., \cite{Carroll 2004}),
but sourced by the expected value of the quantum stress-energy
tensor.\footnote{The components of the quantum stress-energy tensor are
related in a simple manner to the ensemble quantities in
eqns.(\ref{6 Quantum Super Hamiltonian ensemble}) and
(\ref{6 Quantum stress operator}). The derivation of the SCEE from an action
by Kibble and Randjbar-Daemi in \cite{Kibble et al 1980} makes this
relationship more precise.} Such a theory of gravity has long been seen as a
desirable step intermediate to a full theory of QG, in part because it
contains well-established physics --- QFTCS and classical GR
--- in the limiting cases where they are valid. But there has been much debate
(see e.g., \cite{Unruh 1984}\cite{Eppley Hannah 1977}\cite{Page Geilker 1981}\cite{Duff 1981}), on the other hand, as to the status of
semi-classical theories as true QG candidate; with many harboring a negative view. In the next chapter we conclude by tackling these criticisms head on, discussing what's good, what's bad, and what's ugly about the ED approach to semi-classical gravity.


\chapter{The entropic dynamics approach to semi-classical gravity --- is it viable? }\label{CH conclusion}

\begin{center}
\epigraph{\textit{``Today there is no interpretation of quantum mechanics that does not have serious flaws, and that we ought to take seriously the possibility of finding some more satisfactory other theory, to which quantum mechanics is merely a good approximation."}}{--- \textup{Steven Weinberg,} \cite{Weinberg 2015}}
\end{center}

Over the past century experimentalists and engineers have managed to extract a wide array of new predictions and technologies out of quantum theory. For all practical purposes, quantum theory, whatever that might mean, appears to be working just fine. Why is it then that a certain segment of theorists keep squawking about the problems of quantum theory, like canaries in a coal mine that never seems to collapse? Indeed, what is it about textbook quantum theory that is \emph{so} difficult for some theorists to accept? Are they just being pedants with a penchant for intellectual tidiness, or is there something valid in their criticism of quantum theory? Is it rigor, or is it just many words and hot air?

Here we have taken the view that there are serious problems with quantum theory that cannot be ignored. The prevailing view that interpretations of quantum theory have no consequences for physics is not just wrong, you might say it is not even wrong. The standard Copenhagen approach, complete with its division of the universe into microscopic/macroscopic, quantum/classical, system/apparatus, readily admits that it cannot describe nature as an undivided whole. Surely seeking to bridge this gap is in the purview of physics. Indeed, what, after all, is the projection postulate? Is it a physical law? If it is, then it is the job of physicists to \emph{explain} this.\footnote{If it is not, we have argued here, based on \cite{Caticha/Johnson}\cite{Dave Johnson thesis}, that it is an instance of entropic updating in the form of Bayes rule.} From another point of view, if one adopts a Many Worlds, or Everettian \cite{Everett 1957} interpretation, then what is to be made of the apparent adherence of experimental outcomes to the Born rule? Can this be derived? If not, does it make sense to look for violations? Alternatively, postulating additional microscopic variables to make sense of quantum theory, such as in the de Broglie-Bohm approach, introduces new questions, but also new opportunities, that are not present in the other dominant interpretations (see e.g., \cite{Valentini 2010}). The point is that understanding quantum theory, successful though it might be, is not an empty pursuit. Just the contrary, it is by truly fleshing out its conceptual framework that we gain the ability to ask those questions which may, in time, probe the blemishes of quantum theory and not just reify its virtues.

\begin{center}
\epigraph{\textit{``Ultimately, we need not just to calculate a theory's predictions for familiar types of experiment, but to understand the theory's properties. The reason is that this is crucial in deciding the extent to which a theory has been tested, and hence in designing new tests."}}{--- \textup{Adrian Kent,} \cite{Kent 1990}}
\end{center}

In this work we have advocated for using the Entropic Dynamics (ED) framework for the purpose of constructing the quantum formalism. The benefit of such an approach is that there is no need to interpret the theory, its interpretation is built into its construction. In ED, we presume the existence of an external world whose features are not always perfectly known, but are, in principle, knowable. Our incomplete information about nature is therefore codified in probabilities and in constraints. Thus in ED the quantum state $\Psi$ is fully epistemic in character, i.e. it is a tool for reasoning about nature. However, it is a tool for reasoning about nature on the basis of some very particular pieces of information. An advantage of the ED approach is that it makes explicit that information which does lead to a fully quantum dynamics, and hence also that which does not.





One of the chief developments in ED over the past few years, motivated in large part by the work in this thesis, has been the recognition that the constraints that result in a quantum theory are constraints based in symmetry principles. In the context of quantum scalar fields in curved space-time, which has been our primary focus here, there have been two constraints that have proved particularly crucial: (I) Conservation of a symplectic form, (II) Foliation invariance symmetry. A consequence of the ED approach is the insight that these constraints apply, not directly to the ontic microstates themselves, but rather, to the epistemic macrostates $\rho$ and $\Phi$. From this perspective, it is not quite appropriate to address whether the fields themselves propagate in accordance with relativity, instead it is the case that the \emph{statistical} predictions made on the basis of these constraints will not violate relativity.


As mentioned previously, however, the constraints (I) and (II) do not quite fix the unitary time evolution characteristic of quantum dynamics. For this, some additional inputs are required; whether that be in the form of a quantum potential, as we have employed here, or through additional constraints besides that (see e.g.,  \cite{Caticha 2019b}). In any case, the ability in ED to separate models into quantum/non-quantum should be considered a feature of the approach, as this allows an assessment of the conditions under which that linear structure may emerge, and those situations in which it does not. To this end, among the central results of this thesis is that quantum field theory in a curved space-time emerges rather naturally on the basis of the constraints (I) and (II), plus a quantum potential, so long as the background is fixed. The situation of a dynamical background, on the other hand, is qualitatively different. The  ingredients that lead to a full quantum theory in the context of a fixed background, no longer do so when the background is allowed to evolve. It is, in fact, a fundamental aspect of the coupling of quantum scalar fields to a dynamical background that the theory will instead be nonlinear. While surely one expects a regime in which these nonlinearities can be neglected and we recover the standard quantum field theory in curved space-time, this must be regarded only as an approximation to the full dynamics. In this approach there is thus no unitary time evolution, nor any quantum superposition principle. Period.

What then should we make of this? Do we trust this model, or are these features so unsavory that the whole thing should be dispensed with? That is to say, is the ED approach to semi-classical gravity viable?


\section*{Problems and non-problems in semi-classical gravity}

\begin{center}
\epigraph{\textit{``On principle, it is quite wrong to try founding a theory on observable magnitudes alone. In reality the very opposite happens. It is the theory which decides what we can observe."}}{--- \textup{Albert Einstein}, \cite{Kumar 2008}}
\end{center}

Semi-classical gravity often refers to the coupling of relativistic quantum fields to a dynamical background that evolves classically and has typically been evoked phenomenologically, as a means to model processes where quantum effects are present, but gravity cannot be ignored. On a formal level there appear to be clear similarities between the usual semi-classical gravity considered in the literature (hereby simply semi-classical gravity) and the ED model developed here, right down to the semi-classical Einstein equations, which are the operative dynamical equations that are shared by both. There are therefore a whole host of mathematical features that are common to ED and semi-classical gravity alike. However, the conceptual underpinnings of semi-classical gravity, to the extent that they are ever seriously considered, deviate wildly from those of ED. Here we detail some of the standard charges leveled against semi-classical gravity and discuss whether similar criticisms apply to ED, pointing to future work in situations where such concerns are valid.

\paragraph*{On divergences--- }
One area, in particular, where our development of ED has been rather silent pertains to the infrared and ultraviolet divergences that are prevalent in all local quantum field theories, and thus to semi-classical gravity, as well. To this end, ED can take advantage of the extensive work done in semi-classical gravity the past few decades by adapting to ED methods meant to treat technical issues encountered first in semi-classical gravity, such as those relating to regularization and subsequent renormalization. To the extent that these methods are unsatisfactory in semi-classical gravity, perhaps because they require particular symmetries or that the differing techniques give similarly differing answers, they will also likely fail to be optimal in ED, as well.

For ED there is also the added difficulty that it is formulated most naturally in the Schr\"{o}dinger representation, which has not proven to be the most convenient setting for demonstrating renormalizability in many quantum field theory applications. While one may conceive of first transforming to the Heisenberg or interaction pictures for computational purposes, such a transformation would technically require a well-defined $ \hat{T}_{\mu\nu}$ to do so; nonetheless, an attempt in this direction for simple semi-classical models is given in \cite{Randjbar-Daemi et al 1980}, and for semi-classical gravity itself in \cite{Randjbar-Daemi thesis}. Other schemes based firmly in the Schr\"{o}dinger representation have also been explored (see e.g., \cite{Symanzik 1981}\cite{Floreanini et al 1987}\cite{Leuscher 1985}), but it would be rather misleading to suggest that the literature on this subject is nearly as rich as in the usual approaches.

Nevertheless, it is possible that what is needed to address the issue of divergences that plague relativistic quantum field theories is not necessarily a technical breakthrough, but some additional physical insights. One option in this direction is the introduction of a minimum length. Widely viewed as a plausible consequence of most quantum gravity candidates, such a notion would certainly remedy many of the issues with ultraviolet divergences. That being said, implementation of this notion often itself comes accompanied with several issues; either by requiring a fully quantized quantum gravity theory, such as in loop quantum gravity or string theory, which have their own obvious technical hurdles, or an explicitly discretized rendering of space-time, which consequently also breaks invariance under the full Lorentz group. Drawing on the work of R.L. Ingraham \cite{Ingraham 1962}\cite{Ingraham 1964}, and later A. Caticha \cite{Caticha 2016}\cite{Caticha 2019c}, another possibility is that the statistical fluctuations of space itself may induce a minimum length that would still be fully compatible with the usual relativistic symmetries. Such an approach would clearly be within the spirit of an ED  driven by information codified into constraints, and thus suggests a promising avenue of future research.



\paragraph*{On the coupling to gravity--- } Criticism of semi-classical models of gravity often, quite rightfully, point to the coupling of gravity to matter in (\ref{6 Semi classical equations}) as being rather \emph{ad hoc}. Indeed, the usual argument for the expectation value $\langle \hat{T}_{\mu\nu} \rangle$ on the right hand side of eq.(\ref{6 Semi classical equations}) is at best a guess. In ED, on the other hand, the coupling of geometry to the expected value of matter is \emph{derived} on the basis of well-defined assumptions and constraints. In other words, here we derive semi-classical gravity from first principles. Moreover, such principles have already been tested elsewhere: not only do they provide a reconstruction of general relativity through the work of Dirac, Hojman, Kucha\v{r}, and Teitelboim, but they also provide a reconstruction of non-relativistic quantum mechanics and relativistic quantum field theory.

Related to this is the issue of identifying an appropriate quantum stress energy tensor $\hat{T}_{\mu\nu}$ in the first place. Even in classical general relativity there are a couple of definitions one may choose for the stress-energy tensor $T_{\mu\nu}$; either by simple variation of the action, or possibly through the construction of symmetrized Noether currents. But this problem is made considerably worse in quantum field theory, especially in a curved space-time. One chief reason for this is that, in the standard approaches, the stress-energy operator $\hat{T}_{\mu\nu}$ consists of products of operators, or more precisely, operator-valued distributions. As is familiar from just ordinary quantum mechanics, there is an ambiguity in how to order these operators, as compared to their classical analogues, owing to their lack of commutativity. Besides this, however, products of distributions are also know to be poorly-defined mathematically, an issue that is compounded in curved space-time since the usual techniques for regularization available in Minkowski space are no longer available, and there is no suitable replacement valid for generically curved space-times.

Some of these challenges are avoided in ED. First, because the approach, inspired by Dirac, Hojman, Kucha\v{r}, and Teitelboim, relies from the outset on the canonical formalism there are no problems with identifying the sources of gravitation due to matter. Indeed, such sources are constructed explicitly through ``algebraic" methods. Since ED also dispenses with the notion of fields as operators, and the Hilbert spaces on which they act, there are no issues with operator ordering, either; a marked improvement on the usual approaches. As remarked above, however, ED shares with standard approaches the complication of constructing a consistent, finite value of $\langle \hat{T}_{\mu\nu} \rangle$, readily definable in all space-times of interest. This is a matter of great importance and must absolutely be addressed in some way, shape, or form. Again, the work of A. Caticha in devising fully relativistic models with a minimum length appears to be a promising direction.


\paragraph*{On some features of semi-classical Einstein equations--- }
With the semi-classical equations (\ref{6 Semi classical equations}) in hand, it is reasonable to inspect the character of their solutions. We may be interested with, for example, whether the resulting space-times are similar to that of classical general relativity, that is, whether the solutions have some qualitatively new features. But a major difficulty is that the semi-classical Einstein equations are thoroughly daunting. One reason for this, as we discussed above, is that eqns.(\ref{6 Semi classical equations}) are not quite correct on their own, they need to be fixed to take care of the divergences implicit in $\hat{T}_{\mu\nu}$. Addressing this issue in the standard way induces additional terms in eqns.(\ref{6 Semi classical equations}) that are, for instance, quadratic in the Ricci curvature tensor $R_{\mu\nu}$. Thus while the classical Einstein equations, which contain derivatives of up second order, have a well-posed initial value formulation, the corresponding semi-classical equations may not owing to the counterterms introduced by renormalization. Such systems of equations are therefore often unstable \cite{Ford 2005}; and indeed, there are hints that the Minkowski vacuum state is not a stable solution to these equations (see e.g., \cite{Horowitz 1980}\cite{Randjbar-Daemi 1981}).

Related to this, in classical physics it is well understood that Einstein's equations do not have much content without imposing some reasonable conditions on the distribution of energy-momentum (see e.g., \cite{Carroll 2004}). These are typically called energy conditions and they come in a variety of flavors; a common one is the so-called weak energy condition $T_{\mu\nu}u^{\mu}u^{\nu}\geq 0$, for time-like $u^{\mu}$, but there are other, such as the strong energy condition, null energy condition, dominant energy condition, and more besides these (see e.g., \cite{Curiel 2014}). However many of these conditions appear to be violated by quantum field theory quite generically \cite{Epstein et al 1965}, and certainly therefore in semi-classical gravity as well (see e.g., \cite{Barcelo Visser 2002}). On its face this appears problematic. First since many trusted results in gravitational physics, as well as black hole physics and thermodynamics, seem to rely on these conditions. But also because there are arguments that violations of these conditions lead to issues with causality (see e.g., \cite{Barcelo Visser 2002}\cite{Morris et al 1988}\cite{Ford Roman 1996}). Nevertheless, it is possible that these problems can be tamed somewhat by resorting to alternative energy conditions (see e.g., \cite{Ford 1978}\cite{Fewster 2012}\cite{Kontou Sanders 2020}).

It is not our goal here to address these issues in any significant way other than simply to acknowledge that they are there and need to be investigated. ED, of course, is not a cure all for problems such as these and we do not expect that the specific model laid out here will be particularly advantageous in settling all of these difficulties. What ED does provide, however, is the kind of conceptual clarity that may allow future progress. How, for instance, does a minimum length model change the issues of stability? Alternatively, below we discuss how ED avoids the common claim that semi-classical gravity results in superluminal signaling. Can the issues of causality resulting from energy condition violations be similarly avoided in ED? Indeed, in what sense are these conditions violated? Are these violations on the expected values of stress-energy? Or are they violations on the level of the eigenstate spectrum, and thus connected to measurement? Indeed, the ED perspective suggests we view the latter with a healthy dose of suspicion.

\paragraph*{On ontology/epistemology---}
In ED issues of interpretation are carefully addressed, while a similar level of conceptual rigor is often severely lacking in most standard accounts of semi-classical gravity. Indeed, it is entirely plausible that the exact same mathematical framework may be wielded to produce entirely different answers to the very same question. Thus a direct comparison of the ED approach with the semi-classical gravity encountered in the literature can be somewhat misguided without additional considerations.

One area where the ED approach proves particularly valuable is in maintaining a clear distinction between the ontic/epsitemic aspects of semi-classical gravity. In ED the field variables are ontic, or ``beable", whereas their probabilities are fully Bayesian, and thus epistemic. Somewhat less clear, however, is the status of the additional variables, which include the geometry and the drift potential. But here they too are taken to be of epistemic origin. From this viewpoint, while these variables do characterize information relevant for the prediction of the ontic microstates, they themselves are not subject to direct measurement in an experiment. Instead, it is rather more appropriate to claim that such variables are \emph{inferred} from an ensemble of measurements, much like parameters entering into a statistical model, suggesting the name ``inferrables". 

On its face this is, of course, a rather shocking (and perhaps criminal) stance to take, given that it seems to grant the gravitational field a secondary physical role. But note crucially that the gravitational field being epistemic is not to say it lacks physical content. Indeed, similar concerns could be raised about the role of temperature in statistical physics. Is temperature physical? Clearly the notion of temperature enters in our description of physics, but we are under no illusions that it is temperature itself that we are observing in a column of mercury. Therefore, just as we may regard the notion of temperature as being ``physical" but not ontic, a similar interpretation may be given to the gravitational field in the ED approach to semi-classical gravity.

This interpretation complicates the typical analysis of semi-classical gravity. In fact, even the name is severely misleading. It implies that there is something ``classical" in the first place. What though is ``classical" about Entropic Dynamics? One could, perhaps, point to the fact that gravity in our model shares with classical mechanics a common mathematical formalism. But this is hardly useful.  Indeed, in ED the quantum state $\Psi$ is similarly described within the canonical framework --- is $\Psi$ classical, too? Clearly not. What truly distinguishes classical gravity from the gravitational field in our approach is that we do not adopt the, often tacit, interpretive postulate of classical mechanics, namely, that the points of phase space are \emph{ontic} degrees of freedom. Thus the classification of variables into classical/quantum on the basis of their mathematical description appears unsatisfactory in ED. Likely better is the division of variables into ontic/epistemic, which is mirrored by the division into beable/inferrable.

\paragraph*{On superluminal signaling--- }
Owing to the absence of experiments that could test semi-classical gravity (or any other theory of quantum gravity) on an empirical level, considerations of interpretation have come to have great weight in assessing its viability. In the past few decades there have, in fact, been several influential arguments, first by Eppley and Hannah \cite{Eppley Hannah 1977}, later by Page and Geilker \cite{Page Geilker 1981}, as well as by Unruh \cite{Unruh 1984}, that sought to rule out semi-classical gravity on purely theoretical grounds.

The basic idea is that, while we tend to think of classical or quantum mechanics on their own as being, in some sense, self-consistent (a claim that is itself highly questionable), the status of hybrid models where gravity is classical and matter is quantum, are in comparison less understood. Thus it appears reasonable to investigate aspects of such theories, particularly with regards to measurement, to understand whether they conflict with well-established physical principles, such as energy-momentum conservation, the uncertainty principle, or even relativity. Since many objections to semi-classical gravity utilize a similar reasoning, it is useful to address their criticisms.

Take the argument of Eppley and Hannah (EH), for instance, whose work is frequently relied upon. We follow the treatment due to Huggett and Callender \cite{Callender Huggett 2001}. EH consider a gedanken experiment in which there is an electron in a box that is later partitioned by a barrier so that the electron is equally likely to be on either side of the divide. The state of the electron is therefore given by
\begin{equation}
\psi = \frac{1}{\sqrt{2}}(\psi_{L}+\psi_{R})~,\notag
\end{equation}
where $\psi_{L}$ and $\psi_{R}$ are wavefunctions of identical shape but with domains of support on opposite sides of the box. Each half of the box is then carried far apart by two friends, Armin and Bahar, who take the left and right box, respectively.

In a standard version of such an experiment, Bahar looks in her box and, for instance, finds it empty. According to the usual Copenhagen interpretation, Bahar has performed an experiment that ``collapses" the wavefunction $\psi \to \psi_{L}$, which thereby instantaneously ``produces" an electron in Armin's box. If quantum mechanics is a complete description of reality then this, in the words of Einstein, corresponds to some ``spooky action at a distance". Nevertheless, as is well known, this non-locality cannot be used to send superluminal signals from Bahar to Armin. This is because when Armin performs a single trial he is not aware of whether Bahar has already performed her measurement or not, so finding (or not finding) an electron is not informative. Even in the long run, after Armin and Bahar have repeated many trials, each will see the electron roughly about half the time regardless of whether the other has looked in their box or not (see e.g., \cite{Albers et al 2008}).

In the analysis of EH things are changed by allowing gravity to interact with a quantum system. They consider two possibilities: either (a) the interaction will collapse the wavefunction, or (b) the interaction will not collapse the wavefunction. Before continuing, it is worthwhile to note that EH must proceed like this, by classifying modes of collapse, because they do not have a proper model of collapse themselves. This turns out to be problematic, first because ED is just such a theory that it does not fall into either of the categories laid out by EH, and thus evades their conclusions. But second, Albers et al \cite{Albers et al 2008} make the point that if the interaction does, in fact, collapse the wavefunction, corresponding to option (a), then the details of that interaction will be important in examining the behavior of the system. But there have been many attempts in recent years to model the collapse of the quantum state due to gravity, with varying degrees of success. (For recent reviews of the subject, see e.g., \cite{Bassi et al 2013}\cite{Bassi et al 2017}.) So long as such models are actively being improved, and not experimentally ruled out, the arguments by EH cannot be conclusive on this front.

We are thus left to consider option (b), that the gravitational interaction does not cause the state to collapse. Here more definite complications appear to arise. If gravity can interact with a quantum system without causing collapse then one could consider a situation where an incident gravitational wave ``scatters" off of a quantum system. Presumably, in such a scenario, how the wave behaves depends on the state of the quantum system. But it is exactly at this point that one confronts problems with relativity. If Armin can always monitor his box, and therefore probe the quantum state, he can determine if and when Bahar has performed a measurement, and even perhaps the outcome of her measurement. However, if the collapse occurs instantaneously, as it is typically assumed, then Armin will likewise have \emph{instant} access to the actions of Bahar, who we have already assumed is far away. Thus Huggett and Callender conclude that semi-classical gravity leads to superluminal signaling, subject to (i) the standard interpretation of quantum mechanics, (ii) the division of ``the world", or \emph{ontology}, into gravitational and quantum parts, and (iii) that gravity does not force a collapse.

ED avoids this conclusion in essence because it avoids all three of these criteria. First, as mentioned above, in ED there is a clear delineation between variables that are ontic and those that are epistemic. In the analysis of EH, they always assume that ``semi-classical" means not just that the gravitational field has a certain mathematical form, but that it is a separate physical resource that can be controlled, manipulated, and detected in a measuring device. By contrast, in our model, while the gravitational field always has a precise value, this value cannot elicited in the process of measurement. It must instead be inferred from reading a material detector that is itself quantum mechanical. Thus the detector and the electron must be treated together by a joint quantum state. Furthermore, ED does have a clear interpretation, and it is certainly distinct from the usual Copenhagen interpretation. This allows ED to satisfactorily resolve the problem of measurement \cite{Caticha/Johnson}\cite{Caticha Vanslette 2017}. Indeed, in ED a measuring device is not described as a black box subject to rules that violate the Schr\"{o}dinger equation, but as a physical process subject to the very same laws that describe the rest of the world. As a result, in a fully covariant ED, such as the model developed in this thesis, the process of measurement is described by the same causal flow of probability that characterizes any other physical process. Claims, like those of EH and others, that semi-classical gravity leads to superluminal signaling are thus, more properly, viewed as an artifact of a rather primitive interpretation of quantum theory. Given a satisfactory relativistic model of the measuring device itself, no such issues are expected occur in the ED approach.

\paragraph*{On macroscopic superpositions--- }
Another source of confusion with semi-classical gravity is the issue of what happens when the location of a macroscopic source is uncertain. (See \emph{e.g.}, \cite{Unruh 1984}\cite{Bahrami et al 2014}.) Consider, for example, a
macroscopic mass $m$ that is equally likely to be at $x_{1}$ or at $x_{2}$.
Will the gravitational field itself be equally likely to point either towards
$x_{1}$ or towards $x_{2}$? Or, will the gravitational field be as if
generated by a mass $m$ located at the expected position $(x_{1}+x_{2})/2$? The resolution of this paradox in ED demonstrates the advantages of having a derivation from first principles. The $\langle\hat{T}_{\mu\nu}\rangle$ was derived, or better, it
was inferred from a very specific type of information that leads, by an
abuse of language, to what one might call a pure quantum state. The issue then is what
is the gravitational field generated by a pure state that happens to be a
macroscopic "Schr\"{o}dinger cat"? To the extent that this is a
\textquotedblleft pure\textquotedblright\ state then ED gives a sharp
prediction: the gravitational field is generated by an $\langle\hat{T}_{\mu
\nu}\rangle$ centered at the average position.


But as highlighted many years ago by Unruh \cite{Unruh 1984}, such a prediction seems to contradict our everyday expectation that the gravitational field is produced from a source at the location $x_{1}$ or $x_{2}$, not their average $(x_{1}+x_{2})/2$. It is worth revisiting Unruh's argument, albeit in a slightly modified form.\footnote{To save space, we adapt Unruh's argument to the situation described above. The key difference is that in Unruh's experiment $\langle\hat{T}_{\mu\nu}\rangle$ was time dependent, which is not a crucial fact in the argument.} He considers the situation where the macroscopic superposition above was prepared in a box with a Cavendish balance just outside of it which would swing depending on which side of the box the mass actually resided. Naturally, one would not expect to observe the balance and see it at the center, rather it should presumably swing to one side or the other. Since semi-classical gravity seems to predict the former, Unruh goes on to argue that it must be wrong, or in some sense incomplete. But this conclusion does not apply to the ED approach, so long as one is careful to include all available information into the analysis. The key insight is that the balance is itself a material detector which, if it is useful at all, will be correlated (through interactions) with the matter inside. Observing the position of the balance is therefore no passive activity since it gives you information about the state of matter inside the box.\footnote{In fact, one does not even need ED to resolve this issue. The notion that the measuring device should also be treated quantum mechanically is one that goes back to von Neumann and is a firmly established practice even in the Copenhagen interpretation.} 

Thus a full accounting of the available information, including the state of the balance and its correlations with the matter inside the box, leads to an objective assignment for $\Psi$ where $ \langle \hat{T}_{\mu\nu} \rangle$ is centered on either $x_{1}$ or $x_{2}$. On the other hand, if you include the system and detector into the analysis such that the state of neither is known with certainty, then semi-classical gravity might predict a gravitational field sourced at the center position (or something else entirely, depending on the detector itself). Therefore we conclude that there is no paradox. What we have are two predictions drawn on the basis of two distinct informational states, each of which is entirely compatible with the ED approach to semi-classical gravity.

\paragraph*{On quantum coherence--- }

Following the above discussion, a more problematic issue for semi-classical gravity, as we have described it thus far, is the extent to which it is even possible to incorporate all of the relevant information. After all, real experimental setups are rather complicated and messy. Just take the scenario with the macroscopic ``Schr\"{o}dinger cat" states described above and place it in a terrestrial laboratory, for example. The matter in our ``cat" state is correlated with the balance, and both are correlated with the matter that makes up the lab and planet it resides on. What is the gravitational field, now?

Echoing our earlier position, in ED we recognize the role of information. In practice we cannot keep track of all the variables affecting our system, and this is particularly true when the system in question reaches macroscopic proportions. Thus such a Schr\"{o}dinger cat state could not be physically realized anyway: it would immediately suffer decoherence. To analyze such a situation the ED framework would need to be extended to incorporate information (\emph{i.e.}, additional constraints) that describes additional sources of uncertainty. In this extended ED it is conceivable that the gravitational field itself would be uncertain. This is an avenue of research that is currently under investigation.

\appendix 

\chapter{Formalism and notation\label{ch:formalism}}
We deal here with a set of dynamical variables $\rho(x)$ and $\Phi(x)$ that are fields over a configuration space $x\in\mathbf{X}_{N}$, such that
\begin{equation}
\rho(x):\mathbf{X}_{N}\to\mathbb{R}\quad\text{and}\quad\Phi(x):\mathbf{X}_{N}\to\mathbb{R}.
\end{equation}
The notation that we introduce here deals with the particle theory introduced in chapter 3 to maintain brevity. However, the shift to fields does not introduce any major changes. Indeed, the major distinction is that integrals for fields will (often) take the form,
\begin{equation}
\int_{\mathbf{X}_{N}} d^{3N}x \left(\cdot\right)\quad\longleftrightarrow\quad \int_{\mathcal{C}} D\chi \, \int_{\mathbb{R}^{3}}d^{3}x \left(\cdot\right).
\end{equation}
This is true unless otherwise noted. 

Let the space of probabilities $\rho(x)$ be $\mathcal{E}_{\rho}$ and the space of phase functions $\Phi(x)$ be $\mathcal{E}_{\Phi}$ so that the joint space can be written as\footnote{We are not concerned here with the detailed structure of $\mathcal{E}_{\rho}$ and $\mathcal{E}_{\Phi}$, however, it is interesting to note that the space $\mathcal{E}_{\rho}$ \textit{does} have some interesting properties. For one, it has a Riemannian metric, the Fisher-Rao metric,
\begin{equation}
 dl^{2} = C \int dx\frac{\delta\rho_{x}^{2}}{\rho_{x}} = 4C\int dx \left(\delta\rho^{1/2}_{x}\right)^{2}~.
 \end{equation}
This structure turns out to be valuable for reconstructing the full formalism of quantum theory (see e.g., \cite{Caticha 2019b}).}
\begin{equation}
\Gamma = \mathcal{E}_{\rho}\otimes\mathcal{E}_{\Phi}.
\label{Particles Gamma defined}
\end{equation}
The space $\Gamma$ will be called the \textit{ensemble} phase space, or \textit{e}-phase space for short. We will see shortly that this is an appropriate name since the pair $\rho(x)$ and $\Phi(x)$ will form a canonical pair with $\rho(x)$ being the canonical coordinate and $\Phi(x)$ the conjugate momentum.

Continuing on, a point $X\in\Gamma$ will be given by coordinates $X^{\alpha x} = \left(\rho(x),\Phi(x)\right)$, where $\alpha$ takes on either of two values, $\rho$ or $\Phi$. (We use notation throughout that is consistent with \cite{Caticha 2019b}.) Moreover, it is clear from the notation that $\Gamma$ is a doubly \textit{infinite}-dimensional manifold: two-dimensions --- one for $\rho(x)$ and another for $\Phi(x)$ --- for each point $x\in\mathbf{X}_{N}$. As such, the dynamical system in question is also \textit{infinite}-dimensional. This fact brings with it certain technical niceties (see e.g., \cite{Chernoff Marsden 2006}), but by-in-large, we will assume that things are well-defined for the objects we are interested; when such subtleties are, in fact, relevant, we will be sure to mention them.
\paragraph*{Ensemble functionals}
Perhaps the simplest object that can be defined on $\Gamma$ is an \textit{ensemble} functional, or \textit{e}-functional, which is formally a rank-zero tensor. Such tensors $\tilde{F}[\rho,\Phi]=\tilde{F}[X]$ are mappings satisfying
\begin{equation}
\tilde{F}[\rho,\Phi]:\Gamma \to \mathbb{R},
\label{Particles e-functional mapping}
\end{equation}
which produce a real number at every point $X\in\Gamma$. Related to this is the notion of a functional derivative. Consider an infinitesimal variation of an e-functional
\begin{equation}
\delta\tilde{F}[\rho,\Phi] = \int dx \left(\frac{\delta\tilde{F}}{\delta\rho(x)}\delta\rho(x)+\frac{\delta\tilde{F}}{\delta\Phi(x)}\delta\Phi(x)\right),
\label{Particles e-functional variation}
\end{equation}
which follows from the rules of calculus. The coefficients of the variations of $\delta\rho(x)$ and $\delta\Phi(x)$ are then identified as the functional derivatives with respect to $\rho(x)$ and $\Phi(x)$, respectively; one can think of these as the generalization of a partial derivative to spaces of infinite dimensions (see e.g., ch.2 of \cite{Greiner Field Quant} for an introduction).
\paragraph*{Vectors}
Continuing on, vectors $\bar{\mathbf{V}}$ are rank $(0,1)$ tensors that are elements of $T_{X}$, the tangent space of $\Gamma$ at the point $X$. (We use the notation that all tensors, other than e-functionals, are identified with bold-face font. The components of tensors, however, do not since the index placement makes the tensor stucture clear. Also, vectors and one-forms being special, receive a ``bar" and ``tilde" overtop, respectively.) These tensors have components given by $V^{\alpha x}$, using a notation analogous to that used for $X^{\alpha x}$.

The coordinates $X^{\alpha x}$ also establish a set of \textit{coordinate} basis vectors $\bar{\mathbf{e}}_{\alpha x}$ on $\Gamma$ (see e.g., \cite{Schutz}), given by\footnote{Although it is unusual to think of basis vectors $\bar{e}_{\alpha x}$ as linear differential operators $\delta/\delta X^{\alpha x}$, the index notation makes it clear that they do, in fact, transform in exactly the same way. And, this is nonetheless a common convention, so we employ it here as well.}
\begin{equation}
\bar{\mathbf{e}}_{\alpha x} = \frac{\delta}{\delta X^{\alpha x}} = \left(\frac{\delta}{\delta \rho(x)},\frac{\delta}{\delta \Phi(x)}\right).
\label{Particles coordinate basis}
\end{equation}
In this coordinate basis $\bar{\mathbf{e}}_{\alpha x}$, given by the variables $\rho(x)$ and $\Phi(x)$, a generic vector $\bar{\mathbf{V}}$ reads
\begin{equation}
\bar{\mathbf{V}} = V^{\alpha x}\bar{\mathbf{e}}_{\alpha x} = \int dx \left(V^{\rho x}\frac{\delta}{\delta\rho(x)} + V^{\Phi x} \frac{\delta}{\delta\Phi(x)}\right),
\label{Particles Vectors Gamma}
\end{equation}
where we have introduced the convention that repeated indices $\alpha$ and $x$ are summed and integrated over, respectively. We will use this notation for other operations as well.

A vector field is then a rule for assigning a vector $\bar{\mathbf{V}}$ to every point $X\in\Gamma$. That is, to each point $X\in\Gamma$ there is a tangent space $T_{X}$, and a vector field (or any tensor field, for that matter) is a rule for choosing one element of $T_{X}$ for each point $X$. Conversely, at each point along a curve in $\Gamma$ there is a tangent to that curve. For a given vector field $\bar{\mathbf{V}}$, the set of curves for which their tangents correspond to $\bar{\mathbf{V}}$ at each point $X$ are called the integral curves of that vector field. This is important because for a certain distinguished vector field, the Hamiltonian vector field for dynamics, the integral curves correspond the trajectories of $\rho(x)$ and $\Phi(x)$ through $\Gamma$.
\paragraph*{One-forms}
A one-form is then defined in terms of vectors and e-functionals as that linear mapping that takes a vector as an input and produces an e-functional. That is, a one form $\tilde{\mathbf{B}}$ is a $(1,0)$ rank tensor,\footnote{Note that we have abused the ``tilde" notation since it used both for one-forms $\tilde{\mathbf{B}}$ and e-functionals $\tilde{F}$. However while e-functionals receive plain font, the one-forms are bold-faced. The ``tilde" notation is standard for one-forms in differential geometry, but it is also standard in ED to use them for e-functionals. We think the current notation is able to satisfy both conventions.} with components $B_{\alpha x}$, such that for any vector $\bar{\mathbf{V}}\in T_{X}$ we have
\begin{equation}
\tilde{\mathbf{B}}\left(\bar{\mathbf{V}}\right):T_{X}\to\mathbb{R}\quad\text{where}\quad \tilde{\mathbf{B}}\left(\bar{\mathbf{V}}\right)  =V^{\alpha x}B_{\alpha x}.
\label{Particles one-form vector map}
\end{equation}
Conversely, as the notation in (\ref{Particles one-form vector map}) suggests, we may also view a vector $\bar{\mathbf{V}}$ itself as a linear mapping of one-forms to real numbers: $\tilde{\mathbf{B}}\left(\bar{\mathbf{V}}\right) = \bar{\mathbf{V}}\left(\tilde{\mathbf{B}}\right) = V^{\alpha x}B_{\alpha x}$. Vectors are then said to be \textit{dual} to one-forms, and \textit{vice versa}. One forms are thus elements of $T^{*}_{X}$, the \textit{cotangent} space at the point $X$, i.e. the space dual to the tangent space $T_{X}$.

A well-known example of a one-form is the gradient of a function, here the gradient of an e-functional. For an arbitrary e-functional $\tilde{F}[X]$ this is obtained by application of the \textit{exterior derivative},
\begin{equation}
\tilde{\mathbf{d}}\tilde{F}[X] = \int dx \frac{\delta\tilde{F}[X]}{\delta X^{\alpha x}}\tilde{\mathbf{d}}X^{\alpha x} = \int dx \left(\frac{\delta\tilde{F}[X]}{\delta \rho(x)}\tilde{\mathbf{d}}\rho(x)+\frac{\delta\tilde{F}[X]}{\delta \Phi(x)}\tilde{\mathbf{d}}\Phi(x)\right).
\label{Particles gradient one-form}
\end{equation}
Let us explain this a bit. First of all, the notation is, perhaps, misleading. The exterior derivative $\tilde{\mathbf{d}}$ is not itself a tensor, it is a tensor operation valid on most ``nice" manifolds with the property that $\tilde{\mathbf{d}}^{2} = 0$. The true tensor here is the object $\left(\tilde{\mathbf{d}}\tilde{F}\right)$ since it combines a valid tensor $\tilde{F}$ with a valid tensor operation $\tilde{\mathbf{d}}$; in fact, $\left(\tilde{\mathbf{d}}\tilde{F}\right)$ is an \textit{exact} one-form.

From (\ref{Particles gradient one-form}) it is possible to identify the set of coordinate basis one-forms as being
\begin{equation}
\tilde{\mathbf{b}}^{\alpha x} = \tilde{\mathbf{d}}X^{\alpha x} = \left(\tilde{\mathbf{d}}\rho(x),\tilde{\mathbf{d}}\Phi(x)\right),
\end{equation}
which satisfy
\begin{equation}
\tilde{\mathbf{b}}^{\alpha x}\left(\bar{\mathbf{e}}_{\beta x'}\right) = \delta^{\alpha}_{\beta}\delta(x,x'),
\end{equation}
for the basis vectors $\bar{\mathbf{e}}_{\beta x'}$ given in (\ref{Particles coordinate basis}). The components of the gradient one-form are then given by
\begin{equation}
\left(\tilde{\mathbf{d}}\tilde{F}\right)_{\alpha x} =\frac{\delta\tilde{F}[X]}{\delta X^{\alpha x}} = \left(\frac{\delta\tilde{F}[X]}{\delta \rho(x)},\frac{\delta\tilde{F}[X]}{\delta \Phi(x)}\right),
\end{equation}
so that
\begin{equation}
\tilde{\mathbf{d}}\tilde{F} =\left(\tilde{\mathbf{d}}\tilde{F}\right)_{\alpha x}\tilde{\mathbf{b}}^{\alpha x} = \left(\tilde{\mathbf{d}}\tilde{F}\right)_{\alpha x}\tilde{\mathbf{d}}X^{\alpha x}.
\end{equation}
Similarly, a generic one-form $\tilde{\mathbf{B}}$ is written as
\begin{equation}
\tilde{\mathbf{B}} = B_{\alpha x} \tilde{\mathbf{b}}^{\alpha x}.
\end{equation}
\paragraph*{Higher rank tensors}
Additional tensors can also be introduced on $\Gamma$. For example, a $(0,2)$ rank tensor $\mathbf{G}:\bar{\mathbf{V}},\bar{\mathbf{U}}\to\mathbb{R}$, takes two vectors $\bar{\mathbf{V}}$ and $\bar{\mathbf{U}}$ and produces a real number. Such a tensor can be written in a coordinate free notation as $\mathbf{G}\left(\cdot,\cdot\right)$, or in a coordinate dependent notation with components $G_{\alpha x,\beta x'}$ such that
\begin{equation}
\mathbf{G}\left(\bar{\mathbf{U}},\bar{\mathbf{V}}\right) = G_{\alpha x,\beta x'}V^{\alpha x}U^{\beta x'}.
\label{Particles Tensors Gamma a}
\end{equation}

We will also sometimes need a $(2,0)$ rank tensor $\mathbf{C}(\cdot,\cdot)$, with components $C^{\alpha x,\beta x'}$. This object takes two one-forms $\tilde{\mathbf{A}}$ and $\tilde{\mathbf{B}}$ as inputs such that
\begin{equation}
\mathbf{C}\left(\tilde{\mathbf{A}},\tilde{\mathbf{B}}\right) = C^{\alpha x,\beta x'}A_{\alpha x}B_{\beta x'}.
\label{Particles Tensors Gamma b}
\end{equation}
Higher rank tensors may be introduced in analogous ways, however this is not necessary for our purposes.

\chapter{The local-time Fokker-Planck equations}\label{CH LTFP}

\label{appendix FP}To rewrite the dynamical equation (\ref{5Evolution
equation}) in differential form consider the probability $P[\chi ,\sigma
|\chi _{0},\sigma _{0}]$ of a \emph{finite} transition from a field
configuration $\chi _{0}$ at some early surface $\sigma _{0}$ to a
configuration $\chi $ at a later $\sigma $. The result of a further
evolution from $\sigma $ to a neighboring $\sigma ^{\prime }$ obtained from $%
\sigma $ by an infinitesimal normal deformation $\delta \xi _{x}^{\bot }$ is
given by (\ref{5Evolution equation}), 
\begin{equation}
P\left[ \chi ^{\prime },\sigma ^{\prime }|\chi _{0},\sigma _{0}\right] =\int
D\chi \,P\left[ \chi ^{\prime },\sigma ^{\prime }|\chi ,\sigma \right] P%
\left[ \chi ,\sigma |\chi _{0},\sigma _{0}\right] ~.
\end{equation}%
To obtain a differential equation one cannot just Taylor expand as $\delta
\xi _{x}^{\bot }\rightarrow 0$ because $P\left[ \chi ^{\prime },\sigma
^{\prime }|\chi ,\sigma \right] $ becomes a very singular object --- a delta
functional. Instead, we multiply by an arbitrary smooth test functional $T%
\left[ \chi ^{\prime }\right] $ and integrate 
\begin{eqnarray}
\int D\chi ^{\prime }\,P\left[ \chi ^{\prime },\sigma ^{\prime }|\chi
_{0},\sigma _{0}\right] T\left[ \chi ^{\prime }\right]  
= \int D\chi \,P\left[ \chi ,\sigma |\chi _{0},\sigma _{0}\right] \int
D\chi ^{\prime }\,T\left[ \chi ^{\prime }\right] \,P\left[ \chi ^{\prime
},\sigma ^{\prime }|\chi ,\sigma \right] \,.  \label{Test Function a}
\end{eqnarray}%
Next expand the test function $T\left[ \chi ^{\prime }\right] =T[\chi
+\Delta \chi ]$ in powers of $\Delta \chi =\chi ^{\prime }-\chi $. Since $%
\chi $ is Brownian to obtain $T\left[ \chi ^{\prime }\right] $ to first
order in $\delta \xi _{x}^{\bot }$ we need to keep second order in $\Delta
\chi _{x}$, 
\begin{eqnarray}
T\left[ \chi ^{\prime }\right] =T\left[ \chi \right] +\int dx\frac{\delta T%
\left[ \chi \right] }{\delta \chi _{x}}\Delta \chi _{x} 
+\frac{1}{2}\int dx\,dx^{\prime }\frac{\delta ^{2}T\left[ \chi \right] }{%
\delta \chi _{x}\delta \chi _{x^{\prime }}}\Delta \chi _{x}\Delta \chi
_{x^{\prime }}+\cdots .  \label{Test Function b}
\end{eqnarray}%
Use this expansion together with (\ref{5Exp Step 1}) and (\ref{5Fluctuations})
to obtain 
\begin{eqnarray*}
\int D\chi ^{\prime }\,T\left[ \chi ^{\prime }\right] \,P\left[ \chi
^{\prime },\sigma ^{\prime }|\chi ,\sigma \right] 
= T\left[ \chi \right] +\int dx\,\frac{\eta \delta \xi _{x}^{\bot }}{%
g_{x}^{1/2}}\left\{ \frac{\delta T\left[ \chi \right] }{\delta \chi _{x}}\,%
\frac{\delta \phi \left[ \chi \right] }{\delta \chi _{x}}+\frac{1}{2}\frac{%
\delta ^{2}T\left[ \chi \right] }{\delta \chi _{x}^{2}}\right\} \,.
\end{eqnarray*}%
Substituting back into eq.(\ref{Test Function a}), leads to 
\begin{eqnarray}
&&\int D\chi \,\left\{ P\left[ \chi ,\sigma ^{\prime }|\chi _{0},\sigma _{0}%
\right] -P\left[ \chi ,\sigma |\chi _{0},\sigma _{0}\right] \right\} T\left[
\chi \right]  \notag \\
&=&\int dx\,\frac{\eta \delta \xi _{x}^{\bot }}{g_{x}^{1/2}}\int D\chi P%
\left[ \chi ,\sigma |\chi _{0},\sigma _{0}\right] \left\{ \frac{\delta T%
\left[ \chi \right] }{\delta \chi _{x}}\,\frac{\delta \phi \left[ \chi %
\right] }{\delta \chi _{x}}+\frac{1}{2}\frac{\delta ^{2}T\left[ \chi \right] 
}{\delta \chi _{x}^{2}}\right\}  \label{Test Function c}
\end{eqnarray}%
Since $T[\chi ]$ is arbitrary, after some integrations by parts we get 
\begin{eqnarray}
&&P\left[ \chi ,\sigma ^{\prime }|\chi _{0},\sigma _{0}\right] -P\left[ \chi
,\sigma |\chi _{0},\sigma _{0}\right] =\int dx\frac{\delta P\left[ \chi
,\sigma |\chi _{0},t_{0}\right] }{\delta \xi _{x}^{\bot }}\delta \xi
_{x}^{\bot }  \notag \\
&=&\int dx\,\frac{\eta \delta \xi _{x}^{\bot }}{g_{x}^{1/2}}\left\{ -\frac{%
\delta }{\delta \chi _{x}}\left( \,P\left[ \chi ,\sigma |\chi _{0},\sigma
_{0}\right] \frac{\delta \phi \left[ \chi \right] }{\delta \chi _{x}}\right)
+\frac{1}{2}\frac{\delta ^{2}}{\delta \chi _{x}^{2}}P\left[ \chi ,\sigma
|\chi _{0},\sigma _{0}\right] \right\} ~.  \notag \\
&&  \label{Evol eq b}
\end{eqnarray}%
Finally, for a finite evolution from $\sigma _{0}$ to $\sigma $, (\ref%
{5Evolution equation}) reads, 
\begin{equation}
\rho _{\sigma }\left[ \chi \right] =\int D\chi _{0}\,P\left[ \chi ,\sigma
|\chi _{0},\sigma _{0}\right] \rho _{\sigma _{0}}\left[ \chi _{0}\right] ~.
\end{equation}%
A further infinitesimal normal deformation $\sigma \rightarrow \sigma
^{\prime }$ by $\delta \xi _{x}^{\bot }$ gives 
\begin{eqnarray}
\rho _{\sigma ^{\prime }}\left[ \chi \right] -\rho _{\sigma }\left[ \chi %
\right] = \int dx\frac{\delta \rho _{\sigma }\left[ \chi \right] }{\delta
\xi _{x}^{\bot }}\delta \xi _{x}^{\bot }  
= \int D\chi _{0}\,\left( \int dx\frac{\delta P\left[ \chi ,\sigma |\chi
_{0},t_{0}\right] }{\delta \xi _{x}^{\bot }}\delta \xi _{x}^{\bot }\right)
\rho _{\sigma _{0}}\left[ \chi _{0}\right]
\end{eqnarray}%
which, using (\ref{Evol eq b}) and the fact that $\sigma ^{\prime }$ (or $%
\delta \xi _{x}^{\bot }$) can be freely chosen leads to the local
Fokker-Planck equations, 
\begin{equation}
\frac{\delta \rho _{\sigma }\left[ \chi \right] }{\delta \xi _{x}^{\bot }}=%
\frac{\eta }{g_{x}^{1/2}}\left\{ -\frac{\delta }{\delta \chi _{x}}\left(
\,\rho _{\sigma }\left[ \chi \right] \frac{\delta \phi \left[ \chi \right] }{%
\delta \chi _{x}}\right) +\frac{1}{2}\frac{\delta ^{2}}{\delta \chi _{x}^{2}}%
\rho _{\sigma }\left[ \chi \right] \right\} ~,  \label{FP 3}
\end{equation}%
which is the LTFP equations (\ref{5FP equation}) we set out to derive.

\chapter{Solutions to the normal-normal Poisson Bracket}\label{CH NN-PB}

\label{CH NN-PB}

\subsubsection*{Setup}
We consider solutions to the Poisson bracket in (\ref{55 PB of HF}). By virtue of the assumption we made in section \ref{55S Canonical representation}, this means we have already restricted ourselves to solutions where the surface variables $X^{\mu}_{x}$ appear through functions of the metric $g_{ij}$.

The allowed solutions for the bracket eq.(\ref{55 PB of HF}) are most conveniently examined by considering instead the Jacobi identity
\begin{equation}
\left\{\left\{\rho ,\tilde{H}_{\perp x}\right\},\tilde{H}_{\perp y}\right\}-\left\{\left\{\rho ,\tilde{H}_{\perp y}\right\},\tilde{H}_{\perp x}\right\}=\left\{\rho,\left\{\tilde{H}_{\perp y} ,\tilde{H}_{\perp x}\right\}\right\}~.
\end{equation}
Using the Poisson bracket relation (\ref{5PB 1}) we have
\begin{equation}
\left\{\left\{\rho ,\tilde{H}_{\perp x}\right\},\tilde{H}_{\perp y}\right\}-\left\{\left\{\rho ,\tilde{H}_{\perp y}\right\},\tilde{H}_{\perp x}\right\}=\left\{\rho,\left(\tilde{H}_{x}^{i}+\tilde{H}_{y}^{i}\right)\partial_{ix}\delta\left(x,y\right)\right\},
\label{5Necessary condition}
\end{equation}
where
\begin{equation}
\tilde{H}_{x}^{i}=g_{x}^{ij}\tilde{H}_{jx}\hspace{.75 cm}\text{and}\hspace{.75 cm}\tilde{H}_{y}^{i}=g_{y}^{ij}\tilde{H}_{jy}
\end{equation}
are the contravariant tangential generators that are given by eq.(\ref{5e-Momentum}). If we split $\tilde{H}_{x}$ as in (\ref{5e-Hp b}) then eq.(\ref{5Necessary condition}) implies that $F_{x}^{0}[X,\rho;\chi]$ must satisfy
\begin{equation}
\frac{1}{g_{x'}^{1/2}}\frac{\delta}{\delta\chi_{x'}}\left(\rho\,\frac{\delta}{\delta\chi_{x'}} \, \frac{\tilde{\delta}F_{x}^{0}}{\tilde{\delta}\rho}\right)=\frac{1}{g_{x}^{1/2}}\frac{\delta}{\delta\chi_{x}}\left(\rho\,\frac{\delta}{\delta\chi_{x}} \, \frac{\tilde{\delta}F_{x^{\prime}}^{0}}{\tilde{\delta}\rho}\right),
\label{5Necessary condition b}
\end{equation}
which, again, is an equation linear in $F_{x}^{0}$. This implies that a general solution is simply the sum of individual solutions. For brevity and conciseness we denote the linear functional differential operator in eq.(\ref{5Necessary condition b}) to be
\begin{equation}
\hat{O}_{x} = \frac{1}{g_{x}^{1/2}}\frac{\delta}{\delta\chi_{x}}\left(\rho\,\frac{\delta}{\delta\chi_{x}} \right) = \frac{1}{g_{x}^{1/2}}\left(\frac{\delta\rho}{\delta\chi_{x}}\frac{\delta}{\delta\chi_{x}} +\rho \frac{\delta^{2}}{\delta\chi_{x}^{2}} \right).
\label{5Necessary condition Op}
\end{equation}
\paragraph*{On Notation:}
For conciseness, we will sometimes write the domain for these e-functionals, consisting  of the variables $\rho$, $X^{\mu}_{x}$, and indirectly, $\chi_{x}$, as just $D$ so that $D = (\rho, X^{\mu}_{x}, \chi)$. Thus we may rewrite these e-functionals as just $F_{x}^{0}[D]$ or $f_{x}^{(n)}(D)$, respectively, when there is no risk of confusion.

\paragraph*{Ansatz--- }
We consider here a family of possible e-functionals for the unknown $F_{x}^{0}$. The subsequent solutions will not therefore be the most \emph{general} solution possible, but hopefully the family comprises a physically reasonable subset of solutions. In particular, consider $F_{x}^{0}$ of the form
\begin{equation}
F_{x}^{0}[\rho, X;\chi] = \prod_{n = 1}^{N} \mathcal{F}^{(n)}_{x}[\rho, X;\chi],
\label{5Fx ansatz}
\end{equation}
which is an $N$-fold product ($N$ is as large as desired) of e-functionals that are integrals over \textit{functions} $f^{(n)}_{x}$ of the arguments $\rho$, $\chi$, $X^{\mu}$, and their derivatives:
\begin{equation}
\mathcal{F}_{x}^{(n)}[\rho, X;\chi] = \int D\chi_{n} f_{x}^{(n)}\left(\rho,\frac{\delta^{(m)}\rho}{\delta\chi_{nx}^{m}}; X^{\mu}_{x},\chi_{nx}\right).
\label{5Fx Local Definition}
\end{equation}
The dependence of these functions $f_{x}^{(n)}$ can be as complicated as desired, so long as $F_{x}^{0}$ is a scalar density; in particular, dependence on $\rho$ could include derivatives up to any order, the field $\chi_{x}$ can appear as a polynomial to arbitrary order, and the geometry can be any complicated scalar function of the metric components.\footnote{What does not appear, for instance, are the analogue of kernel operators inside the integrals or non polynomial functions of $\chi_{x}$.}

\paragraph*{Identities}
A couple of identities will prove useful in our analysis here. The first takes advantages of the peculiar pathological nature of the Dirac delta
\begin{equation}
a(x)\delta(x,x')b(x') = a(x')\delta(x,x')b(x') = a(x)\delta(x,x')b(x).
\label{5Dirac identity}
\end{equation}
Essentially, once a Dirac delta is present we can freely switch the spatial dependences of the fields, functionals, etc.

Another useful fact is that the field functional derivative of an e-functional vanishes
\begin{equation}
\frac{\delta}{\delta\chi_{x}} M_{x}[\rho, X; \chi] = 0.
\label{5Functional deriv identity}
\end{equation}
We can see this result most clearly by noting that, strictly speaking, writing these functionals as $M_{x}[\rho, X; \chi]$ is an abuse of notation since we have actually integrated over all $\chi$. When we compute such a derivative, we let $\chi\to\chi+\epsilon$ so that $M_{x}'= M_{x}[\rho,\chi+\epsilon]$. But we could always change variables in the integral from $\chi\to\chi+\epsilon$ to obtain the original expression; thus the difference $M_{x}'-M_{x} = 0$, always.
\subsubsection*{Solution: $N\geq 2$}
Now we look to insert the ansatz (\ref{5Fx ansatz}) into the condition eq.(\ref{5Necessary condition b}) and deduce the restrictions on $F_{x}^{0}$. Before doing this, let us compute
\begin{align}
\frac{\tilde{\delta}F_{x}^{0}}{\tilde{\delta}\rho} &= \frac{\tilde{\delta}\mathcal{F}_{x}^{(1)}}{\tilde{\delta}\rho}\prod_{n \neq 1}\mathcal{F}^{(n)}_{x}+\frac{\tilde{\delta}\mathcal{F}_{x}^{(2)}}{\tilde{\delta}\rho}\prod_{n \neq 2}\mathcal{F}^{(n)}_{x}+\cdots \notag\\
&= \sum_{m}\frac{\tilde{\delta}\mathcal{F}_{x}^{(m)}}{\tilde{\delta}\rho}\prod_{n \neq m}\mathcal{F}^{(n)}_{x}.
\label{5Fx fderiv rho}
\end{align}
Note that the e-functional derivatives acting on the ``local" e-functionals $\mathcal{F}_{x}^{(m)}$, as defined in eq.(\ref{5Fx Local Definition}), produces a function
\begin{equation}
\frac{\tilde{\delta}\mathcal{F}_{x}^{(m)}[\rho, X;\chi]}{\tilde{\delta}\rho[\chi]} = a_{x}^{(m)}\left (\rho, \frac{\delta^{(n)}\rho}{\delta\chi_{x}^{n}}, X_{x}^{\mu}, \chi_{x}\right )\equiv a_{x}^{(m)}(D_{x})\label{5ax definition} 
\end{equation}
of the arguments, not another e-functional; thus it depends explicitly on $\chi_{x}$.

Feeding (\ref{5Fx fderiv rho}) into eq.(\ref{5Necessary condition b}), and noting that the operator $\hat{O}_{x}$ passes through the $\mathcal{F}_{x}^{(n)}[D]$ due to the identity eq.(\ref{5Functional deriv identity}), we have that
\begin{equation}
\hat{O}_{x^{\prime}}\frac{\tilde{\delta}F_{x}^{0}[D]}{\tilde{\delta}\rho[\chi]} = \hat{O}_{x}\frac{\tilde{\delta}F_{x^{\prime}}^{0}[D]}{\tilde{\delta}\rho[\chi]} 
\end{equation}
implies
\begin{equation}
\sum_{m}\prod_{n\neq m}\mathcal{F}_{x}^{(n)}[D]\,\hat{O}_{x^{\prime}}\,a_{x}^{(m)}(D_{x}) = \sum_{m}\prod_{n\neq m}\mathcal{F}_{x^{\prime}}^{(n)}[D]\,\hat{O}_{x}\,a_{x^{\prime}}^{(m)}(D_{x^{\prime}}).
\end{equation}
Introduce now the compact notation
\begin{equation}
\prod_{n\neq m}\mathcal{F}_{x}^{(n)}[D] \equiv A_{x}^{(m)}[D]
\label{5Ax definition}
\end{equation}
so that we have the condition above rewritten as
\begin{equation}
\sum_{m}A_{x}^{(m)}[D] \, \hat{O}_{x^{\prime}}\, a_{x}^{(m)}(D_{x}) = \sum_{m}A_{x^{\prime}}^{(m)}[D] \, \hat{O}_{x}\, a_{x}^{(m)}(D_{x^{\prime}}).\label{5Necessary condition c} 
\end{equation}
As was apparent in eq.(\ref{5Necessary condition b}), for a particular $F_{x}^{0}[D]$ to work it is necessary that each side of eq.(\ref{5Necessary condition c}) be symmetric in $x$ and $x^{\prime}$.

This is not, however, easily accomplished. One route to satisfying this condition is for $\hat{O}_{x} a_{x^{\prime}}^{(m)}(D_{x^{\prime}})$ to introduce a Dirac delta; this could then be used via the identity eq.(\ref{5Dirac identity}) to manipulate the spatial dependence and make both sides equal. This suggests that any acceptable function $a_{x}^{(m)}(D_{x})$ must have an explicit $\chi_{x}$ dependence since the only way for a Dirac delta to appear is from the identity: $\delta\chi_{x}/\delta\chi_{x^{\prime}} = \delta(x,x^{\prime})$. (Note that for \textit{arbitrary} choices of $\rho[\chi]$, successive applications of the field functional derivatives on $\rho[\chi]$ need not produce a Dirac delta, and so, such situations are not considered.) But, without this mechanism, it is impossible for the left hand and right hand sides to agree since the e-functionals $A_{x}^{(m)}$ and $A_{x^{\prime}}^{(m)}$ have a different spatial dependence. This implies that we \textit{cannot} have a dependence $a_{x}^{(m)} = a_{x}^{(m)}(\rho,X^{\mu})$ that does not expressly include the fields. (There is an exception to this conclusion that we will get to shortly.)\footnote{To clarify things, consider, for example, the case $N = 2$. Using the definition (\ref{5Ax definition}), we then have
\begin{align}
&\mathcal{F}_{x}^{(2)}[D]\hat{O}_{x^{\prime}} a_{x}^{(1)}(D_{x}) + \mathcal{F}_{x}^{(1)}[D]\hat{O}_{x^{\prime}} a_{x}^{(2)}(D_{x})\notag\\
 = & \mathcal{F}_{x^{\prime}}^{(2)}[D]\hat{O}_{x} a_{x^{\prime}}^{(1)}(D_{x^{\prime}}) + \mathcal{F}_{x^{\prime}}^{(1)}[D]\hat{O}_{x} a_{x^{\prime}}^{(2)}(D_{x^{\prime}}).
\end{align}
Clearly, no choice of $a$'s or $F$'s (or equivalently, $A$'s) can make this work. Even if, for example, we choose the $a$'s so that the $\hat{O}_{x}a_{x^{\prime}}$'s are symmetric in $x$ and $x^{\prime}$, we have no way to make the $A$'s match. }

More exactly, a guaranteed solution of $F_{x}^{0}$ includes terms like
\begin{equation}
F_{x}^{0} = \prod_{n}\int D\chi_{n}\rho[\chi_{n}] f_{x}(\chi_{nx}; X_{x}^{\mu})~,\label{5Fx gen solution n>= 2}
\end{equation}
which takes the form of products of expected values of the field $\chi_{x}$, and linear combinations there of. This can easily be checked substituting into eq.(\ref{5Necessary condition b}).
\begin{claim}
So long as the number of products $N\geq 2$, then eq.(\ref{5Fx gen solution n>= 2}) is the \textbf{only} possible form for $\mathcal{F}_{x}^{(n)}$.
\end{claim}
\paragraph*{Proof.}
The goal is to determine the allowed $a$'s (although, recall that the $a$'s are related to the integrands of the $F$'s in a complicated way). A general $a_{x}^{(m)}(D_{x})$ can always be written in powers of the field $\chi_{x}$ as
\begin{equation}
a_{x}^{(m)}(D_{x}) = \sum_{n} \chi_{x}^{n} \, \alpha_{x}^{(mn)}\left(\rho,\frac{\delta^{(k)}\rho}{\delta\chi_{x}^{k}}, X_{x}^{\mu}\right ),
\label{5ax ansatz}
\end{equation}
where the $\alpha_{x}^{(mn)}$ is a generic function of its inputs for every $m$ and $n$ that can be as complicated and general as desired. If we act the operator $\hat{O}_{x^{\prime}}$ on this, then some of the derivatives in $\hat{O}_{x^{\prime}}$ will hit the field $\chi_{x}$ and produce a $\delta(x,x')$. However, due to the product rule, the derivatives will also hit the $\alpha_{x}^{(mn)}$, and no Dirac delta will be produced. (Again, this analysis should hold for \textit{arbitrary} choices of $\rho[\chi]$.) We write this symbolically as 
\begin{equation}
\hat{O}_{x^{\prime}}a_{x}^{(m)}(D_{x}) = \delta(x,x')\beta_{x}^{(m)}(D_{x}) + B_{xx^{\prime}}^{(m)}
\end{equation}
where \[B_{xx'} = \sum_{n}\chi_{x}^{n}\,\hat{O}_{x^{\prime}}\alpha_{x}^{(mn)}(D_{x})\]
(and the form of $\beta_{x}$ doesn't matter) so that the condition eq.(\ref{5Necessary condition c}) has the structure
\begin{equation}
\sum_{m}A_{x}^{(m)}[D] \, B_{xx^{\prime}}^{(m)} = \sum_{m}A_{x^{\prime}}^{(m)}[D] \, B_{x^{\prime}x}^{(m)}.\label{5Necessary condition d} 
\end{equation}
Again, no suitable choice of $B_{xx^{\prime}}^{(m)}$ --- symmetric in $x$, $x^{\prime}$, or not --- will make the two sides equal, owing to the e-functionals $A_{x}^{(m)}$ and $A_{x^{\prime}}^{(m)}$. In fact, eq.(\ref{5Necessary condition d}), together with the ansatz eq.(\ref{5ax ansatz}) imply that the condition
\begin{equation}
\sum_{m}A_{x}^{(m)}[D]\sum_{n}\chi_{x}^{n}\,\hat{O}_{x^{\prime}}\alpha_{x}^{(mn)}(D_{x})=\sum_{m}A_{x^{\prime}}^{(m)}[D]\sum_{n}\chi_{x^{\prime}}^{n}\,\hat{O}_{x}\alpha_{x^{\prime}}^{(mn)}(D_{x^{\prime}})
\end{equation}
should be met, after accounting for the delta functions. Since no further delta's will be produced by $\hat{O}\alpha$, each side is only equal if they are each equal to a constant
\begin{equation}
\sum_{n}\chi_{x}^{n}\left[\sum_{m}A_{x}^{(m)}[D]\,\hat{O}_{x^{\prime}}\alpha_{x}^{(mn)}(D_{x})\right] = c,
\end{equation}
where $c$ is independent of any spatial dependence, and hence doesn't depend at all on the fields or geometry (but might depend, perhaps, on $\rho[\chi]$, for instance). The term in brackets, however, has no explicit $\chi_{x}$ dependence,\footnote{It depends on the fields only indirectly through functional derivatives of $\rho[\chi]$, but we only care about the explicit form of the potential $F_{x}^{0}$ for arbitrary choices of $\rho[\chi]$.} so the only way to maintain this constraint, without completely discarding all dependence on $\chi_{x}$ in $F_{x}^{0}$ --- which is absolutely required for a non-trivial $F_{x}^{0}$ --- is for the term in brackets to vanish. This, in turn, implies that\footnote{Consider a special case. If all the $A_{x}^{(m)}$ are equal, then $A_{x}^{(m)} = \prod_{n\neq m}\mathcal{F}_{x}^{(n)} = \prod_{n\neq m'}\mathcal{F}_{x}^{(n)} \equiv \left(\mathcal{F}_{x}\right)^{N-1} $, for all $m$, $m^{\prime}$. This then requires that the $\alpha$'s are related by $\alpha_{x}^{(mn)} = \alpha_{x}^{(m^{\prime}n)}$. We then have $\sum_{m}\hat{O}_{x^{\prime}}\alpha_{x}^{(mn)} = 0 \longrightarrow \hat{O}_{x^{\prime}}\alpha_{x}^{(mn)} = 0$. In the general case, if $A_{x}^{(m)}\neq A_{x}^{(m')}$ then $\sum_{m}A_{x}^{(m)}\hat{O}_{x^{\prime}}\alpha_{x}^{(mn)} = 0$. We can always rearrange this condition so that $\hat{O}_{x^{\prime}}\alpha_{x}^{(in)} = -(A_{x}^{(i)})^{-1}\sum_{m\neq i} A_{x}^{(m)}\hat{O}_{x^{\prime}}\alpha_{x}^{(mn)}$. For example, with $N=2$ we have
\begin{equation}
\hat{O}_{x^{\prime}}\alpha_{x}^{(1n)} = - \frac{A_{x}^{(2)}}{A_{x}^{(1)}}\hat{O}_{x^{\prime}}\alpha_{x}^{(2n)}.
\end{equation}
While the LHS has a function-like (i.e. ``local") dependence on $\rho[\chi]$, etc., the RHS has an e-functional and function dependence. Now, unless the $A$'s are constant e-functionals (which would defeat our purpose) then an expression like this only makes sense if the $\hat{O}\alpha = 0$. A similar argument then holds for arbitrary $N$, and thus eq.(\ref{5O alpha = 0}).}
\begin{equation}
\hat{O}_{x^{\prime}}\alpha_{x}^{(mn)} = \frac{1}{g_{x^{\prime}}}\frac{\delta}{\delta\chi_{x^{\prime}}}\left(\rho\frac{\delta\alpha_{x}^{(mn)}}{\delta\chi_{x^{\prime}}}\right) = 0~.\label{5O alpha = 0}
\end{equation}
For this to hold, however, it must mean that the term in parentheses is independent of $\chi_{x^{\prime}}$. But $\delta\alpha_{x}^{(mn)}/\delta\chi_{x^{\prime}}$ can only be independent of $\chi_{x^{\prime}}$ if $\alpha_{x}^{(mn)}$ doesn't depend at all on $\rho[\chi]$ since, if it did, it would be guaranteed to introduce such a dependence through
\begin{equation}
\frac{\delta\rho[\chi]}{\delta\chi_{x^{\prime}}} = g[\chi;\chi_{x^{\prime}})
\end{equation}
and related relations for higher derivatives. Thus,
\begin{equation}
\alpha_{x}^{(mn)} = c_{x}^{(mn)}(X_{x}^{\mu})
\end{equation}
where $c_{x}^{(mn)}$ depends only on the geometry, so that
\begin{equation}
a_{x}^{(m)} = \sum_{n}c_{x}^{(mn)}(X_{x}^{\mu})\,\chi_{x}^{n}.
\end{equation}
Furthermore, using the definition of $a_{x}^{(m)}$ from eq.(\ref{5ax definition}) we have that
\begin{equation}
a_{x}^{(m)} = \frac{\tilde{\delta}\mathcal{F}_{x}^{(m)}}{\tilde{\delta}\rho} = \sum_{n}c_{x}^{(mn)}(X_{x}^{\mu})\,\chi_{x}^{n}
\end{equation}
which can be easily integrated to yield
\begin{equation}
\mathcal{F}_{x}^{(m)} =\int D\chi_{m}\,\rho[\chi_{m}]\sum_{n}c_{x}^{(mn)}(X_{x}^{\mu})\,(\chi_{mx})^{n} .\label{5Fx gen solution}
\end{equation}
This is exactly the same form as eq.(\ref{5Fx gen solution n>= 2}) and concludes our proof.
\paragraph*{Some comments}
Clearly the $\mathcal{F}_{x}^{(m)}$ can be interpreted as expected values of the field $\chi$, with some additional contributions from the geometrical variables. From this point on, however, we can offer no further restriction on the appearance of the $\chi$'s and geometrical quantities that appear. In principle, almost anything is possible; again, so long as $\mathcal{F}_{x}^{(m)}$ is a scalar density.

For completeness, we write down the expression for $F_{x}^{0}$. Since it must be a scalar density, it should be $\propto g_{x}^{1/2}$. Other than that, we define \[V_{x}^{(m)}(\chi_{mx},X_{x}^{\mu}) = \sum_{n}c_{x}^{(mn)}(X_{x}^{\mu})\,(\chi_{mx})^{n}\] so that $F_{x}^{0}$ takes the form
\begin{equation}
F_{x}^{0} = g_{x}^{1/2} \prod_{m}\int D\chi \, \rho[\chi ]\, V_{x}^{(m)}(\chi_{x},X_{x}^{\mu}),
\end{equation}
where it is assumed that the $V_{x}^{(m)}$'s should be scalar valued.
\subsubsection*{Solutions: $N = 1$}
Something that perhaps avoided attention in our treatment above was a certain special case that requires additional care. In the ansatz for $F_{x}^{0}$ we considered this ``potential" term to be a product of e-functionals $\mathcal{F}_{x}^{(m)}$. The analysis above appears correct so long as the number of products is $N\geq 2$. For just a single e-functional, we require a separate argument to determine its general form.

In the argument above, when $F_{x}^{0}$ is a product of e-functionals, then the $\tilde{\delta}F_{x}^{0}/\tilde{\delta}\rho$ introduces a sum of terms; each of these terms must be symmetric in $x$, $x^{\prime}$, but each of which also comes preceded by a product of e-functionals that depend on $x$ (or $x^{\prime}$) exclusively, and one requires a $\delta(x,x^{\prime})$ to appear --- \textit{in every term} --- and thus we are led to the $\mathcal{F}_{x}^{(m)}$ in eq.(\ref{5Fx gen solution}). When there is only one e-functional, on the other hand, this situation does not arise and so more analysis is needed.

Let $F_{x}^{0}[\rho]$ be just a single e-functional
\begin{equation}
F_{x}^{0}[\rho] = \int D\chi f_{x}(D_{x}).
\end{equation}
Inserting this term into eq.(\ref{5Necessary condition b}) yields
\begin{equation}
\hat{O}_{x^{\prime}}a_{x}(D_{x}) = \hat{O}_{x}a_{x^{\prime}}(D_{x})\quad\text{where}\quad a_{x}(D_{x}) = \frac{\tilde{\delta}F_{x}^{0}}{\tilde{\delta}\rho}.\label{5Necessary condition local a}
\end{equation}
Now, $a_{x}(D_{x})$ is a function of $\rho$ (and its derivatives), $\chi$, and the geometry. We provide an ansatz for $a_{x}$
\begin{equation}
a_{x}(D_{x}) = \sum_{n}\chi_{x}^{n}\alpha_{x}^{(n)}(\rho,X_{x}^{\mu}),
\end{equation}
which is exactly as in eq.(\ref{5ax ansatz}); here $\alpha_{x}^{(n)}$ is a distinct function of $\rho$ (and its functional derivatives with respect to $\chi_{x}$, to any order) and the geometry $X_{x}^{\mu}$ for each $n$. And, again, as the operator $\hat{O}_{x^{\prime}}$ acts on $a_{x}$ we have that a Dirac delta is produced for some terms and not for others so that there is the general form
\begin{equation}
\hat{O}_{x^{\prime}}a_{x} = b_{x} \, \delta(x,x^{\prime}) + \sum_{n}\chi_{x}^{n}\hat{O}_{x^{\prime}}\alpha_{x}^{(n)}.
\end{equation}
The terms with the delta function vanish from (\ref{5Necessary condition local a}). What remains is
\begin{equation}
\sum_{n}\chi_{x}^{n}\hat{O}_{x^{\prime}}\alpha_{x}^{(n)} = \sum_{n}\chi_{x^{\prime}}^{n}\hat{O}_{x}\alpha_{x^{\prime}}^{(n)}.
\end{equation}

Since $\hat{O}_{x^{\prime}}$ acts on $\alpha_{x}(\rho,X_{x}^{\mu})$ but produces no delta functions, the $\chi_{x}$ (or $\chi_{x^{\prime}}$) on each side present an obstacle to symmetrizing this equation in $x$, $x^{\prime}$. This implies two options: (1) functions with no $\chi_{x}$ dependence at all, (2) each side identically vanishes since the functions $\alpha_{x}^{(n)}=\alpha_{x}^{(n)}(X_{x}^{\mu})$ depend only on the geometry, that is, they are constants with respect to $\chi$ (i.e. $\hat{O}\alpha = 0$) so that \[a_{x} = a_{x}(\chi_{x},X_{x}^{\mu})\equiv g_{x}^{1/2}\, V_{x}(\chi_{x},X_{x}^{\mu}).\] Using the definition in (\ref{5Necessary condition local a}), this latter option implies \[F_{x}^{0}[\rho] = g_{x}^{1/2}\int D\chi \,\rho \, V_{x}(\chi_{x},X_{x}^{\mu}).\] This is just the expected value of a ``local potential" $V_{x}(\chi_{x}, X_{x}^{\mu})$ and no further restrictions can be imposed on functions of this type (other than they be scalar valued). Thus this $V_{x}$ can be identified with the one introduced in section \ref{55S Canonical representation}, in eq.(\ref{5local potential V}).

On the other hand, the first option implies functions \[a_{x} = \alpha_{x}\left(\rho,\frac{\delta^{(m)}\rho}{\delta\chi_{x}^{m}}, X_{m}^{\mu}\right)\]
which satisfy $\hat{O}_{x^{\prime}}\alpha_{x} = \hat{O}_{x}\alpha_{x^{\prime}}$. Or more explicitly,
\begin{equation}
\frac{1}{g_{x^{\prime}}^{1/2}}\left(\frac{\delta\rho}{\delta\chi_{x^{\prime}}}\frac{\delta\alpha_{x}}{\delta\chi_{x^{\prime}}} +\rho \frac{\delta^{2}\alpha_{x}}{\delta\chi_{x^{\prime}}^{2}} \right) = \frac{1}{g_{x}^{1/2}}\left(\frac{\delta\rho}{\delta\chi_{x}}\frac{\delta\alpha_{x^{\prime}}}{\delta\chi_{x}} +\rho \frac{\delta^{2}\alpha_{x^{\prime}}}{\delta\chi_{x}^{2}} \right).
\label{5Necessary condition local b}
\end{equation}
How can this condition be satisfied? First, recall that our objective is to determine a functional form for $\alpha_{x}$ such that both sides match. In other words, an $\alpha_{x}$ is acceptable when both sides of this equation ``look the same". Next, notice that $\alpha_{x}$ depends on $\rho$, its functional derivatives $\delta^{(m)}\rho/\delta\chi_{x}^{m}$, and the geometry.

Inspecting eq.(\ref{5Necessary condition local b}), we see that $\hat{O}_{x^{\prime}}$ on the LHS will introduce derivatives of $\rho[\chi]$ with respect to $\chi_{x^{\prime}}$ of up to second order, while the RHS will do the same but for $\chi_{x}$. This \textit{requires} us to consider only those $\alpha_{x}$ that are at most second order in the functional derivatives, i.e.\[\alpha_{x} = \alpha_{x}\left(\rho,\frac{\delta\rho}{\delta\chi_{x}},\frac{\delta^{2}\rho}{\delta\chi_{x}^{2}}; X_{x}^{\mu}\right).\] And, moreover, each term in the function must have only two functional derivatives. The most general such function can be written as\footnote{We do not need to begin with this truncated ansatz, in fact. We can begin with arbitrarily high orders of derivatives of $\rho$, but it should be apparent that they will never produce an $\alpha_{x}$ such that the functional form of each side in (\ref{5Necessary condition local b}) are the same.}
\begin{equation}
\alpha_{x} = \lambda_{x}^{(1)} \, m_{1}(\rho)\left(\frac{\delta\rho}{\delta\chi_{x}}\right)^{2} + \lambda_{x}^{(2)} \, m_{2}(\rho)\left(\frac{\delta^{2}\rho}{\delta\chi_{x}^{2}}\right)
\end{equation}
where $\lambda_{x}^{(1)}$, $\lambda_{x}^{(2)}$ are ``constants" that depend on the geometry, but not $\chi$ or $\rho$. The functions $m_{i}(\rho)$ depend only on $\rho[\chi]$ (i.e. the probability at a single point in configuration space) and are not yet determined; indeed, it is our goal to find their functional form.

Another restriction comes from requiring that $F_{x}^{0}$ be a scalar density (since it is a piece of the local e-Hamiltonian $\tilde{H}_{x}$), which implies that $\alpha_{x}$ similarly be a scalar density. However, because each functional derivative itself is a scalar density, and there are two such derivatives in each term of $\alpha_{x}$, then $\lambda_{x}^{(i)}$ must be an inverse scalar density $\propto g_{x}^{-1/2}$ for $\alpha_{x}$ to have the correct tensor weight. To account for this, we redefine the $\lambda_{x}^{(i)}$'s so that \[\lambda_{x}^{(i)} = \frac{\lambda^{(i)}}{g_{x}^{1/2}},\]
where the $\lambda^{(i)}$'s are now just coupling constants, independent of the geometry or anything else.\footnote{The outcome of our calculation will show that that this step is justified.} In fact, we can just absorb them into the $m_{i}$'s, which gives us
\begin{equation}
\alpha_{x} =\frac{1}{g_{x}^{1/2}}\left(m_{1}(\rho)\left(\frac{\delta\rho}{\delta\chi_{x}}\right)^{2} + m_{2}(\rho)\left(\frac{\delta^{2}\rho}{\delta\chi_{x}^{2}}\right)\right)\label{5alpha x ansatz}
\end{equation}

Before continuing, it is important to note that functional derivatives commute, i.e. \[\frac{\delta^{2}}{\delta\chi_{x}\delta\chi_{x^{\prime}}} = \frac{\delta^{2}}{\delta\chi_{x^{\prime}}\delta\chi_{x}} \]
and that the functional derivatives acting on the $m_{i}(\rho)$ behave as \[\frac{\delta m_{i}(\rho)}{\delta\chi_{x}} = \frac{dm_{i}(\rho)}{d\rho[\chi]}\frac{\delta \rho[\chi]}{\delta\chi_{x}}~.\] We now plug the $\alpha_{x}$ in (\ref{5alpha x ansatz}) into the condition eq.(\ref{5Necessary condition local b}). Many terms will automatically be symmetric in $x$, $x^{\prime}$, and thus drop out. The remaining terms can only be made symmetric if the following set of equations between the undetermined $m_{i}(\rho)$ are satisfied:\footnote{In fact, we have that the LHS of eq.(\ref{5Necessary condition local b}) becomes:
\begin{subequations}
\begin{align}
\hat{O}_{x^{\prime}}\alpha_{x}\longrightarrow &\rho\left(\frac{dm_{1}}{d\rho}\frac{\delta^{2}\rho}{\delta\chi_{x^{\prime}}^{2}}\left(\frac{\delta\rho}{\delta\chi_{x}}\right)^{2}+2m_{1}\frac{\delta\rho}{\delta\chi_{x}}\frac{\delta^{3}\rho}{\delta\chi_{x}\delta\chi_{x^{\prime}}^{2}}\right)\label{5Necessary condition local LHS 1}\\
&+\left(\frac{dm_{2}}{d\rho}+ \rho\frac{d^{2}m_{2}}{d\rho^{2}}\right)\left(\frac{\delta\rho}{\delta\chi_{x^{\prime}}}\right)^{2}\frac{\delta^{2}\rho}{\delta\chi_{x}^{2}} \label{5Necessary condition local LHS 2}\\
&+\left(m_{2}+2\rho\frac{dm_{2}}{d\rho}\right)\frac{\delta\rho}{\delta\chi_{x^{\prime}}}\frac{\delta^{3}\rho}{\delta\chi_{x}\delta\chi_{x^{\prime}}^{2}} ,\label{5Necessary condition local LHS 3}
\end{align}
\end{subequations}
where the ``$\longrightarrow$" symbolizes what remains after the symmetric pieces from the LH and RH sides cancel. This expression can itself be made symmetric in $x$, $x^{\prime}$. For instance, the first term in (\ref{5Necessary condition local LHS 1}) and the expression in (\ref{5Necessary condition local LHS 2}) could be symmetric if eq.(\ref{5Necessary condition final 2}) was satisfied --- just equate the ``coefficients" with $m_{1}$, $m_{2}$, etc. On the other hand, the second term in (\ref{5Necessary condition local LHS 1}) and (\ref{5Necessary condition local LHS 3}) could be made symmetric if eq.(\ref{5Necessary condition final 2}) is satisfied by, again, setting the undetermined coefficients equal.}
\begin{equation}
m_{1} = \frac{1}{2}\left(\frac{1}{\rho}\frac{dm_{2}}{d\rho} - \frac{m_{2}}{\rho^{2}} + 2 \frac{d^{2}m_{2}}{d\rho^{2}}\right)\label{5Necessary condition final 1}
\end{equation}
and
\begin{equation}
\frac{dm_{1}}{d\rho} = \frac{1}{\rho}\frac{dm_{2}}{d\rho}+\frac{d^{2}m_{2}}{d\rho^{2}}.\label{5Necessary condition final 2}
\end{equation}
Since the functions $m_{1}$ and $m_{2}$ are coupled, it means neither term in eq.(\ref{5alpha x ansatz}) is on its own a solution to eq.(\ref{5Necessary condition local b}) --- they must appear together.

The coupled eqns.(\ref{5Necessary condition final 1}) and (\ref{5Necessary condition final 2}) can be combined into the single equation for $m_{2}$ that reads
\begin{equation}
\frac{dm_{2}(\rho)}{d\rho} = - \frac{m_{2}(\rho)}{\rho},
\end{equation}
which has the solution
\begin{equation}
m_{2}(\rho) = \frac{1}{\rho}.\label{5m2}
\end{equation}
This can then be used to determine $m_{1}$, which is most easily achieved through eq.(\ref{5Necessary condition final 1}), which yields
\begin{equation}
m_{1}(\rho) =- \frac{1}{2}\,\frac{1}{\rho^{2}}.\label{5m1}
\end{equation}
Thus $\alpha_{x}$ has the form
\begin{equation}
\alpha_{x} = \frac{\lambda}{g_{x}^{1/2}}\left(\frac{1}{\rho}\frac{\delta^{2}\rho}{\delta\chi_{x}^{2}}- \frac{1}{2\rho^{2}}\left(\frac{\delta\rho}{\delta\chi_{x}}\right)^{2}\right) = \frac{\tilde{\delta}F_{x}^{0}}{\tilde{\delta}\rho}
\end{equation}
where $\lambda$ is a yet arbitrary coupling constant. It is now possible to integrate this expression, giving us
\begin{equation}
F_{x}^{0} = -\frac{\lambda}{2g_{x}^{1/2}}\int D\chi \frac{1}{\rho}\left(\frac{\delta\rho}{\delta\chi_{x}}\right)^{2}.
\end{equation}
Upon redefinition of the constant $\lambda$ we find that this potential agrees with the ``local" quantum potential introduced in (\ref{5PB HF e}) and (\ref{5Q}), or more explicitly,
\begin{equation}
Q_{x} = \frac{\lambda}{g_{x}^{1/2}}\int D\chi \frac{1}{\rho}\left(\frac{\delta\rho}{\delta\chi_{x}}\right)^{2}.
\end{equation}

Thus the full class of acceptable relativistic potentials can thus be written as
\begin{equation}
F_{x}^{0} = g_{x}^{1/2}\int D\chi \, \rho V_{x}(\chi_{x},X_{x}^{\mu})+Q_{x} + g_{x}^{1/2} \prod_{m}\int D\chi_{m}\rho[\chi_{m}]V_{x}^{(m)}.
\label{5PB Full solution}
\end{equation}


\bibliographystyle{plain}

\cleardoublepage
\ifdefined\phantomsection
  \phantomsection  
\else
\fi
\addcontentsline{toc}{chapter}{Bibliography}

\bibliography{thesis}

\end{document}